\documentclass[nofootinbib,aps,prd,preprintnumbers,showpacs,showkeys]{revtex4-1}

\usepackage{amssymb}
\usepackage{amsmath}
\usepackage{graphicx}
\usepackage{dcolumn}
\usepackage{bm}
\usepackage{epsfig}
\usepackage{hyperref}

\def\figdir{}

\newcommand\figwidth{.48\textwidth}
\newcommand\Eq[1]{Eq.~\ref{eq:#1}}
\newcommand\Fig[1]{Fig.~\ref{fig:#1}}
\newcommand\Sec[1]{Sec.~\ref{sec:#1}}
\newcommand\Appendix[1]{Appendix~\ref{appendix:#1}}
\newcommand\Tab[1]{Table~\ref{tab:#1}}

\newcommand\calC{\mathcal C}
\newcommand\calG{\mathcal G}
\newcommand\calO{\mathcal O}

\newcommand\calH{\mathcal H}
\newcommand\calP{\mathcal P}
\newcommand\calQ{\mathcal Q}
\newcommand\calR{\mathcal R}
\newcommand\calN{\mathcal N}
\newcommand\Tr{\textrm{Tr}}

\begin{document}

\preprint{MIT-CTP/4545}

\title{Signal/noise enhancement strategies for \\ stochastically estimated correlation functions}
\author{William Detmold}
\email{wdetmold@mit.edu}
\author{Michael G. Endres}
\email{endres@mit.edu}
\affiliation{Center for Theoretical Physics, Massachusetts Institute of Technology, Cambridge, Massachusetts 02139, USA}


\date{\today}

\begin{abstract}
We develop strategies for enhancing the signal/noise ratio for stochastically sampled correlation functions.
The techniques are general and offer a wide range of applicability.
We demonstrate the potential of the approach with a generic two-state system, and then explore the practical applicability of the method for single hadron correlators in lattice quantum chromodynamics.
In the latter case, we determine the ground state energies of the pion, proton, and delta baryon, as well as the ground and first excited state energy of the rho meson using matrices of correlators computed on an exemplary ensemble of anisotropic gauge configurations.
In the majority of cases, we find a modest reduction in the statistical uncertainties on extracted energies compared to conventional variational techniques.
However, in the case of the delta baryon, we achieve a factor of three reduction in statistical uncertainties.
The variety of outcomes achieved for single hadron correlators illustrates an inherent dependence of the method on the properties of the system under consideration and the operator basis from which the correlators are constructed. 
\end{abstract}

\maketitle

\section{Introduction}
\label{sec:introduction}

Monte Carlo calculations provide one of the few reliable tools for investigating the properties of quantum field theories nonpertubatively.
The majority of zero-temperature calculations rely on an accurate numerical estimation of correlation functions at large Euclidean-time separations in order to garner information about the spectrum of the theory, and about matrix elements of operators in the eigenbasis of the Hamiltonian.
For systems with a discrete spectrum (such as those confined to a finite box), correlation functions will decay exponentially at late times.
The rate of the decay is determined by the lowest energy states carrying the same quantum numbers as those of the interpolating operators used to construct the correlation function.
The square root of the variance of such correlators, however, will often have an exponential decay which, by contrast, falls off at a slower rate in time.
For correlators constructed from underlying bosonic degrees of freedom, this rate is determined by the vacuum energy, whereas for those involving fermionic degrees of freedom, it is set by the energy of the lightest state with appropriate valence quantum numbers \cite{Lepage:1989hd}.

The relevant figure of merit for the numerical estimation of any observable in a Monte Carlo simulation is the signal to noise ratio (signal/noise), where the noise is taken to be the square root of the variance divided by the total number of samples.
Because of the exponential degradation of signal/noise with time separation, numerical studies of the zero temperature properties of a theory can be quite formidable.
In the case of lattice quantum chromodynamics (QCD), for example, a correlator for a single proton with mass $m_p$ will have a signal/noise which falls off exponentially in time separation with a decay rate of $m_p - \frac{3}{2} m_\pi$, where $m_\pi$ is the mass of the pion (taking into account the finite temporal extent, the degradation can be even worse \cite{Beane:2009gs}).
Since the physical proton mass is nearly seven times that of the pion, a signal/noise problem is expected.
In this instance, the problem is found to be challenging, but manageable.
However, for systems involving $A$ nucleons, the decay rate is approximately $A$ times that of the proton.
Thus, numerical study of multinucleon systems beyond that of a few are forbiddingly costly in terms of computational resources.
Furthermore, the difficulty only worsens when considering correlation functions designed to extract the energies of hadronic excited states.

Since the exponential degradation of signal to noise becomes most severe at late times, a natural strategy for combating the problem is to consider operators possessing a greater overlap with the states of interest, thereby suppressing excited state contamination at earlier time separations, where correlators exhibit a larger ratio of signal to noise.
With that mindset, along with other motivations, considerable attention has been devoted to understanding how to extract ground and excited state energies variationally from Hermitian matrices of correlation functions \cite{Michael:1982gb,Michael:1985ne,Luscher:1990ck,Blossier:2009kd} and from non-Hermitian matrices of correlation functions \cite{Fleming:2004hs,Beane:2009kya,Beane:2009gs}.
From a practical standpoint, these methods (hereafter referred to as ``source optimization'' methods) are all rooted in finding solutions to a generalized eigenvalue problem, but do not account for the statistical uncertainties of the correlators involved.
This could be particularly problematic if the correlator possess some anomalously noisy matrix elements which may spoil the analysis.
Alternatively, analysis is possible by performing multiexponential (matrix) fits to the correlation functions in order to properly account for the presence of excited state contamination.
For large correlator matrices and for the small ensembles often encountered in lattice QCD, however, correlated multiexponential matrix fits are often encumbered by the fact that they would require inversion of a singular (or nearly singular) correlation matrix.

In this paper, we propose a complementary tack for mitigating signal/noise degradation in correlation functions to the methods mentioned above.
As we discuss below, it is not possible to completely eliminate the exponential degradation of signal/noise at late times in most correlation functions.
Accepting this, we explore whether or not an enhancement of signal/noise is possible by exploiting the interplay between the time-independent overlap factors that enter into the signal/noise ratio.
Theoretically, large enhancements are possible, as we expose in a toy model.
In real numerical data from lattice QCD calculations, we find that in most cases, a significant improvement in the signal/noise of correlators is possible at intermediate and late times.
For the examples we consider, however, this improvement typically translates into only a moderate reduction in uncertainties in the energies determined from them.
Despite this, our study exposes a fascinating geometry underlying signal/noise, which may serve as a useful tool for further exploring its enhancement.
Although the strategies introduced here are primarily applied to two-point correlators, we expect that similar approaches can yield a reduction in uncertainties for three-point correlators as well, and thereby a reduction in the uncertainties in extracted matrix elements.

The organization of this paper is as follows.
In \Sec{signal_noise_conventions}, we establish the notational conventions used throughout, as well as frame the problem we will address.
We argue that in addition to source optimization via variational techniques, a different avenue for optimization is available for exploration, which we refer to as ``signal/noise optimization.''
In \Sec{toy_model}, we demonstrate how signal/noise can be enhanced for the case of a simple two state system.
Despite its simplicity, this example cleanly illustrates some of the general features expected of the strategy, as well as some of its deficiencies.
In \Sec{signal_noise_optimization}, we provide explicit formulas for optimizing signal/noise which are applicable to any system and valid for all time separations.
In \Sec{correlator_optimization_strategies}, we propose strategies for correlator optimization which utilize either one, or a combination, of the source- and signal/noise-optimization methods.
In \Sec{applications}, we evaluate the various strategies, applied to examples of single hadron correlation functions in lattice QCD.
A figure summarizing the efficacy of some of the methods presented in this work, for an exemplary lattice QCD data set, can be found in \Sec{comparison_of_strategies}.
In \Sec{conclusion}, we summarize our findings and discuss them in a broader context.

Source optimization methods, such as the variational method \cite{Michael:1982gb,Michael:1985ne,Luscher:1990ck,Blossier:2009kd}, will play an important role in our analysis of the signal/noise landscape for correlation functions.
To facilitate the discussion, we review the approach, which is generally applicable to Hermitian matrices of correlation functions, in \Appendix{source_optimization}.
Our presentation of the method is framed as an optimization problem, in the same spirit as signal/noise optimization, with extrema determined by solutions to a well-known generalized eigenvalue problem.
Finally, in \Appendix{app1}, we provide additional details on the derivation of a result in \Sec{sn_basis}.

\section{Signal/noise for a generic correlation function}
\label{sec:signal_noise_conventions}

Consider an $N^\prime\times N$ matrix of correlation functions, $C$, in Euclidean spacetime with matrix elements given by
\begin{eqnarray}
C_{ij}(\tau) = \langle \Omega | \hat\calO^\prime_i e^{- \hat H \tau} \hat\calO^\dagger_j| \Omega \rangle
             = \sum_n Z^\prime_{in} Z^*_{jn} e^{-E_n \tau}\ ,
\label{eq:correlator}
\end{eqnarray}
where $Z^\prime_{in} = \langle \Omega | \hat\calO^\prime_i | n \rangle$ and $Z_{jn} = \langle \Omega | \hat\calO_j | n \rangle$ are overlap factors (``Z factors'') associated with the respective sink and source interpolating operators $\hat\calO_i^\prime$ and $\hat\calO_j$, $|n\rangle$ are the energy eigenstates of the Hamiltonian $\hat H$ with associated energies $E_n$ (ordered such that $E_n\le E_m$ for $n<m$), and $|\Omega \rangle$ is the vacuum state.
Throughout, we adopt a notation where primed quantities are associated with the sink operators, and unprimed quantities are associated with the source operators.
The indices $i$ and $j$ run over the values $i = 0,\ldots,N^\prime-1$, and $j=0,\cdots, N-1$.
We furthermore assume that the operators $\hat\calO_i^\prime$ and $\hat\calO_j$ have the same quantum numbers for all values of $i$ and $j$.
In the special case where $N^\prime=N$ and $\hat\calO^\prime_i = \hat\calO_i$ for all $i=0,\cdots,N-1$, then the matrix of correlation functions is Hermitian.
A generic correlation function can be constructed by taking various linear combinations of source and sink interpolating operators, with relative weights specified by a complex $N$-dimensional source vector $\psi$ and a complex $N^\prime$-dimensional sink vector ${\psi^\prime}$.
Without loss of generality, one can assume that the source and sink vectors are normalized such that $\psi^\dagger \psi = 1$ and ${\psi^\prime}^\dagger {\psi^\prime} = 1$.

In the path-integral language, any fermionic degrees of freedom may be ``integrated out'' leaving a path-integral over purely bosonic degrees of freedom (and if necessary, appropriate auxiliary fields) weighted by an appropriate effective action.
The correlation functions, in turn, may be expressed as an expectation value, $C = \langle \calC \rangle$, over individual correlators $\calC$ which depend only on the remaining bosonic degrees of freedom; the expectation value $\langle\cdots \rangle$ is understood as an average over the bosonic degrees of freedom, weighted by the effective action.
For any given choice of source and sink vectors, the ``signal'' in a Monte Carlo simulation is estimated by an ensemble average of ${\psi^\prime}^\dagger C \psi$, and the associated uncertainty on the estimate (or ``noise'') is given by $\sigma_c({\psi^\prime},\psi) /\sqrt{\calN}$, where $\calN$ is the size of the ensemble,
\begin{eqnarray}
\sigma_c^2({\psi^\prime},\psi) = \sigma^2({\psi^\prime},\psi) - |{\psi^\prime}^\dagger C \psi|^2
\end{eqnarray}
is the second central moment of the correlator distribution, and
\begin{eqnarray}
\sigma^2({\psi^\prime},\psi) = \left({\psi^\prime} \otimes{\psi^\prime}^* \right)^\dagger \Sigma^2 \left( \psi \otimes\psi^* \right)\ ,\qquad  \Sigma^2 = \langle \calC \otimes \calC^* \rangle \ .
\label{eq:Sigma}
\end{eqnarray}
Note that ${\psi^\prime}\otimes {\psi^\prime}^*$ is an ${N^\prime}^2$ dimensional vector, $\psi\otimes\psi^*$ is an $N^2$ dimensional vector, and  $\Sigma^2$ is an ${N^\prime}^2\times N^2$ matrix.
In the proceeding discussion, it is useful to also define the positive definite matrices
\begin{eqnarray}
\sigma_\psi^2 =  \langle \calC \psi \psi^\dagger \calC^\dagger \rangle \ ,\qquad \sigma_{\psi^\prime}^2 = \langle \calC^\dagger \psi^\prime {\psi^\prime}^\dagger \calC \rangle\ .
\end{eqnarray}
These matrices are $N^\prime\times N^\prime$ and $N\times N$ dimensional, respectively, and satisfy the relations
\begin{eqnarray}
\sigma^2({\psi^\prime},\psi) = {\psi^\prime}^\dagger \sigma_\psi^2 {\psi^\prime} = \psi^\dagger \sigma_{\psi^\prime}^2 \psi \ .
\end{eqnarray}

The matrix $\Sigma^2$ is itself a correlation function (hereafter referred to as a ``noise correlator''), which may be expanded as
\begin{eqnarray}
\Sigma_{ik;jl}^2(\tau) = \sum_n \tilde Z^\prime_{ik,n} \tilde Z^*_{jl,n} e^{-\tilde E_n\tau}\ ,
\end{eqnarray}
where $\tilde E_n$ and $\tilde Z_n$ ($\tilde Z_n^\prime$) are the energies and overlap factors associated with the ``noise states,'' $|n\rangle$.
In this expression, the sums on $i$ and $k$ run from $1,\cdots,N^\prime$, $j$ and $l$ run from $1,\cdots,N$, and $n$ labels the noise states carrying the appropriate valence quantum numbers\footnote{In the case of fermions, the states may possess nontrivial valence quantum numbers, attributed to the fact that the variance of the correlator is taken {\it after} integrating out the fermions.}.
As with the states and energies governing the signal correlator, the states and energies governing the noise correlator are determined purely by the (discretized, finite volume) Hamiltonian of the system.
The signal/noise ratio at a given time slice is expressed up to an overall root-$\calN$ scaling factor by
\begin{eqnarray}
\theta_c({\psi^\prime},\psi) = \frac{\left| {\psi^\prime}^\dagger C \psi\right| }{ \sigma_c({\psi^\prime},\psi) } \ ,
\end{eqnarray}
or equivalently by,
\begin{eqnarray}
\theta_c({\psi^\prime},\psi) = \left[ \frac{1}{ \theta^2({\psi^\prime},\psi)} -1  \right]^{-1/2}\ ,
\end{eqnarray}
where
\begin{eqnarray}
\theta({\psi^\prime},\psi) = \frac{\left| {\psi^\prime}^\dagger C \psi\right| }{ \sigma({\psi^\prime},\psi) }\ .
\label{eq:signal_noise}
\end{eqnarray}
Note that $0\leq \theta({\psi^\prime},\psi) \leq 1$ for all $\psi^\prime$ and $\psi$ and that $\theta_c({\psi^\prime},\psi)$ is a monotonically increasing function of $\theta({\psi^\prime},\psi)$.
To leading order in $\theta$, $\theta_c({\psi^\prime},\psi) = \theta({\psi^\prime},\psi)$ for $\theta({\psi^\prime},\psi) \ll 1$, and $\theta_c({\psi^\prime},\psi)$ diverges as $\theta({\psi^\prime},\psi) \to 1$.
Since we are primarily interested in the former case, we will often refer to both $\theta_c({\psi^\prime},\psi)$ and  $\theta({\psi^\prime},\psi)$ as the signal/noise.

Let us now consider a correlator constructed using a fixed source and sink vector.
The late-time behavior of such a correlation function is given by
\begin{eqnarray}
{\psi^\prime}^\dagger C \psi \sim {\psi^\prime}^\dagger Z^\prime_0  Z^\dagger_0 \psi  e^{-E_0 \tau}\ ,
\end{eqnarray}
up to relative corrections of order $\Delta = e^{-(E_1-E_0)\tau}$, where $E_0$ ($E_1$) is the energy of the lightest (first excited) state carrying the quantum numbers of $\hat \calO_i$ and $\hat\calO_j^\prime$, and $Z_0$ ($Z^\prime_0$) is a vector of overlap factors of the states created by the various interpolating operators onto the ground state.
By comparison, the late-time behavior of the variance is given by 
\begin{eqnarray}
\sigma^2({\psi^\prime},\psi) \sim \left( {\psi^\prime}^\dagger \tilde Z^\prime_0 {\psi^\prime} \right) \left(  \psi^\dagger \tilde Z_0 \psi \right) e^{-\tilde E_0 \tau} \ ,
\end{eqnarray}
up to relative corrections of order $\tilde\Delta = e^{-(\tilde E_1-\tilde E_0)\tau}$, where $\tilde E_0$ ($\tilde E_1$) is the energy of the lightest (first excited) noise state created with the appropriate valence quantum numbers, and $\tilde Z_0$ ($\tilde Z^\prime_0$) is the associated overlap with the lightest noise state.
Note that, since $\sigma^2({\psi^\prime},\psi)$ is positive-definite for all $\psi$ (${\psi^\prime}$), it follows that $\tilde Z_0$ ($\tilde Z^\prime_0$) is positive-definite when viewed as a two-index matrix.
Combining these observations, we see that the leading late-time scaling of the signal/noise falls off exponentially as
\begin{eqnarray}
\theta({\psi^\prime},\psi) \sim \frac{| {\psi^\prime}^\dagger Z^\prime_0|}{ \sqrt{ {\psi^\prime}^\dagger \tilde Z^\prime_0 {\psi^\prime}} }  \frac{ | Z^\dagger_0 \psi | }{ \sqrt{ \psi^\dagger \tilde Z_0 \psi  } } \, e^{-\left(E_0 - \frac{1}{2} \tilde E_0 \right) \tau}\ .
\label{eq:late_time_sn}
\end{eqnarray}
In particular, note that the signal/noise at late times is factorizable.
Returning to the example of the proton, discussed in the introduction, one has $E_0 = m_p$, whereas $\tilde E_0 = 3 m_\pi$ up to interaction effects, so the exponential degradation of the proton signal/noise is as argued in \Sec{introduction}.

Although the time-dependent exponential fall-off of the signal/noise is an inherent property of the correlators (i.e., it is a property of the system and is independent of the choice of source and sink), we nonetheless retain some control over the ratio through the overlap factors.
In particular, a short calculation (see \Sec{fixed_source_unconstrained_sink_sn} for details) indicates that the signal/noise is maximized at late times for source and sink vectors given by
\begin{eqnarray}
\psi^\prime_0 \propto (\tilde Z_0^\prime)^{-1} Z_0^\prime \ , \qquad \psi_0 \propto (\tilde Z_0)^{-1} Z_0\ ,
\label{eq:sn_soln}
\end{eqnarray}
and takes the maximum value 
\begin{eqnarray}
\theta(\psi^\prime_0,\psi_0) \sim \sqrt{ {Z_0^\prime}^\dagger (\tilde Z_0^\prime)^{-1} Z_0^\prime   } \sqrt{ Z_0^\dagger (\tilde Z_0)^{-1} Z_0 }  \, e^{-\left(E_0 - \frac{1}{2} \tilde E_0 \right) \tau}\ .
\label{eq:max_sn}
\end{eqnarray}
Note that in the weak-coupling limit, $\Sigma^2\to C\otimes C^*$, and therefore one finds $\tilde Z_0 \to Z_0 Z_0^\dagger$, $\tilde Z^\prime_0 \to Z^\prime_0 {Z^\prime_0}^\dagger$, and $\tilde E_0 \to 2 E_0$ up to additive perturbative corrections in the coupling.
Additionally, one may confirm that $(\tilde Z_0)^{-1} \to Z_0 Z_0^\dagger/|Z_0|^4$ and $(\tilde Z^\prime_0)^{-1} \to Z^\prime_0 {Z^\prime_0}^\dagger/|Z^\prime_0|^4$, up to terms which are divergent in the coupling, but orthogonal to $Z_0$ and $Z_0^\prime$, respectively.
As a consequence, $\theta(\psi^\prime_0,\psi_0)$ tends to unity in the limit, indicating that the fluctuations in the correlator vanish as one might expect from turning off the interactions.

Away from weak coupling, and particularly when $\theta \ll 1$, one does not know {\it a priori} whether source and sink vectors chosen to maximize the overlap with a desired state (i.e., the practice often adopted by practitioners) exhibit a signal/noise ratio that is comparable to that of the maximal value, $\theta(\psi^\prime_0,\psi_0)$, or whether it is significantly suppressed by comparison.
The degree to which the signal/noise becomes attenuated as a function of the distance (appropriately defined) away from the optimal choices, $\psi^\prime_0$ and $\psi_0$, is an open question, and to our knowledge has never been explored.
In the following sections, we address this question in detail and propose several strategies for exploiting the interplay between the overlap factors in this ratio which are valid not only at late times where excited state contamination is absent, but also at earlier times where is it present.
We start by considering the signal/noise for correlators in a general two-state system, and then apply what we have learned to hadronic correlation functions in QCD.
We demonstrate that, counter to intuition, it is in indeed advantageous in some cases to forgo optimizing the overlap of interpolating operators onto eigenstates in favor of signal/noise optimization via tuning of the ratios of overlap factors that appear in \Eq{late_time_sn}.
Although our study focuses specifically on QCD, the methods are general and applicable to stochastically sampled correlators for any relativistic or nonrelativistic quantum theory.

\section{Toy model: two state system}
\label{sec:toy_model}

Before considering the general problem of signal/noise optimization, let us first examine the signal/noise properties for a two-state system.
Without loss of generality, we may consider a $2 \times 2$ matrix of correlation functions
\begin{eqnarray}
C \propto Z_0 Z_0^\dagger + \Delta Z_1 Z_1^\dagger \ ,
\label{eq:two_state_corr}
\end{eqnarray}
expressed in a basis where
\begin{eqnarray}
Z_0 =
\left( \begin{array}{c}
1 \\
0 
\end{array} \right) \ ,\qquad
Z_1 =
\left( \begin{array}{c}
0 \\
1  
\end{array} \right) \ .
\end{eqnarray}
Note that if $Z_n$ and $Z_n^\prime$ differ in \Eq{correlator}, one could always perform a change of basis such that they are equal, provided the source and sink operator basis is complete.
We assume such is the case for the two-state toy model we consider here.
The associated noise correlator is given by
\begin{eqnarray}
\Sigma^2 \propto \tilde Z_0 \tilde Z_0^\dagger \left[ 1+ \calO(\tilde \Delta)\right]
\end{eqnarray}
at late times $\tau > (\tilde E_1 - \tilde E_0)^{-1}$, with a positive-definite ground state noise overlap factor given by
\begin{eqnarray}
\tilde Z_0 =
\left( \begin{array}{cc}
a & b \\
b^* & c 
\end{array} \right) \ ,
\end{eqnarray}
for some unknown parameters of $a$, $b$, and $c$, which are system-dependent.
Note that positivity of $\tilde Z_0$ requires that $ac > |b|^2$, $a>0$ and $c>0$.
Also note that the signal correlator and the noise correlator may have different quantum numbers, and therefore $\Delta$ need not equal $\tilde \Delta$.
Furthermore, the number of noise states need not be the same as the number of signal states.

We may begin by studying the signal/noise associated with this system for arbitrary but equal source and sink vectors parametrized by $\psi^\prime(\omega,\delta) = \psi(\omega,\delta)$, and
\begin{eqnarray}
\psi(\omega,\delta) =
\left( \begin{array}{c}
\cos \omega \\
\sin \omega \,  e^{i\delta}
\end{array} \right) \ ,
\label{eq:two_vector}
\end{eqnarray}
up to an overall irrelevant phase factor chosen so as to make the upper component real.
In this parametrization, $\omega\in[0,\pi)$ and $\delta\in[-\pi/2,\pi/2)$; all values of $\delta$ outside of the specified domain can be mapped back into the domain with an accompanied shift in $\omega$ and an overall phase rotation.
We explore the behavior of the signal/noise not only as a function of these parametrization angles, but also as a function of temporal extent $\tau$, which is implicit in the parameter $\Delta$ appearing in \Eq{two_state_corr}.

The signal/noise ratio for this simple model, $\theta_\star(\omega,\delta) = \theta({\psi^\prime}(\omega,\delta),\psi(\omega,\delta))$, has the general functional form
\begin{eqnarray}
\theta_\star(\omega,\delta) = \theta_\star(0,0) \frac{1 + \Delta \tan^2\omega}{ 1 + \frac{c}{a} \tan^2\omega + 2 \frac{|b|}{a} \cos\left(  \arg(b)+\delta \right) \tan\omega}\ ,
\end{eqnarray}
up to relative corrections in the noise of order $\tilde \Delta$.
A short calculation shows that the signal/noise has a global maximum $\theta_\star(\omega_\star,\delta_\star) = R_\star \,\theta_\star(0,0)$ located at a critical point $(\omega_\star,\delta_\star)$ which satisfies
$\arg(b)+\delta_\star = \pi$ and
\begin{eqnarray}
\frac{c}{a} = \frac{R_\star-1+\Delta \tan^2\omega_\star}{R_\star \tan^2 \omega_\star} \ ,\qquad \frac{|b|}{a} = \frac{R_\star-1}{R_\star \tan\omega_\star} \ .
\label{eq:toy_critical_points}
\end{eqnarray}
Constraints on the parameters $R_\star$, and $\omega_\star$ follow from the positivity requirements imposed on $\tilde Z_0$; specifically, one finds that $R_\star\ge1$ and $\tan\omega_\star\ge0$ (i.e., $0\le \omega_\star \le \pi/2$).
The signal/noise for an arbitrary source and sink vector can be expressed in terms of the parameters $R_\star$, $\omega_\star$, $\delta_\star$ and $\Delta$.
Introducing the notation $\hat\theta_\star(\omega,\delta) = \theta_\star(\omega,\delta) / \theta_\star(\omega_\star,\delta_\star)$ for the signal/noise normalized by its maximum value, one finds
\begin{eqnarray}
\hat\theta_\star(\omega,\delta) = \frac{1 + \rho_\star x^2_\star(\omega)}{ R_\star + (R_\star-1) x_\star(\omega) \left[ x_\star(\omega) -2 \cos (\delta - \delta_\star) \right] + \rho_\star x^2_\star(\omega) }\ ,
\label{eq:sn_parametrization}
\end{eqnarray}
where $\rho_\star = \Delta \tan^2\omega_\star$ and $x_\star(\omega) = \tan\omega/\tan\omega_\star$.
With this parametrization, the normalized signal/noise ratios for the ground and excited states are given by
\begin{eqnarray}
\hat\theta_\star(0,0) = \frac{1}{R_\star}\ ,\qquad \hat\theta_\star(\pi/2,0) = \frac{\rho_\star}{R_\star-1+\rho_\star}\ ,
\label{eq:sn_parametrization_eigstate}
\end{eqnarray}
respectively.

The parameter $\rho_\star$ quantifies the amount of excited state contamination that is present in a correlator with source and sink vectors evaluated at the maximum of the signal/noise, and depends on the temporal extent, $\tau$, through $\Delta$.
In the limit $\rho_\star \ll 1$, the ground state provides the dominant contribution to the correlation function, either due to a sufficiently small $\omega_\star$ or due to an exponential suppression of the excited state at late times.
Interestingly, the functional form of $\hat\theta_\star(\omega,\delta_\star)$ in this regime is that of a Breit-Wigner distribution with a peak located at $x_\star(\omega)=1$, and a half-width at half maximum given by $(R_\star-1)^{-1/2}$.
For a fixed $R_\star$, the signal/noise associated with the excited state is given by
\begin{eqnarray}
\theta_\star(\pi/2,0) = \theta_\star(0,0) \frac{R_\star}{R_\star-1+\rho_\star} \rho_\star \ ,
\end{eqnarray}
and is suppressed by $\calO(\Delta)$ compared to the ground state.
Since the parameter $R_\star$ is only bounded from below by unity, one finds from \Eq{sn_parametrization_eigstate} that there always exists an optimal source/sink vector which possesses better signal/noise than one having perfect overlap with the ground state.
Furthermore, the maximum achievable enhancement (as determined by $R_\star$ for the ground state) could theoretically be quite large, depending on the intrinsic properties of the system as characterized by $a$, $b$, $c$, and $\Delta$.

Notice from \Eq{toy_critical_points} that in the limit $\rho_\star\to 0$, the matrix elements of $\tilde Z_0$ satisfy
\begin{eqnarray}
|b|^2 = ac \left(1-\frac{1}{R_\star} \right)\ ,
\label{eq:noise_overlap_condition}
\end{eqnarray}
and therefore the limit $R_\star\to\infty$ implies the limit $\det \tilde Z_0 \to 0$.
The requirement that $\tilde Z_0$ be near-singular in order to achieve a large signal/noise enhancement is in fact consistent with the form of \Eq{max_sn}.
To better understand the origins of a large enhancement in the toy model, let us study the structure of $\tilde Z_0$ in greater detail.
The eigenvalues of $\tilde Z_0$, to leading order in $1/R_\star$, are given by
\begin{eqnarray}
\lambda_0 &=& a \left[ \frac{\cos^2\omega_\star}{R_\star} + \cdots \right] \ , 
\end{eqnarray}
and
\begin{eqnarray}
\lambda_1 = \frac{a}{\sin^2\omega_\star} \left[ 1 -  \frac{\cos^2\omega_\star(1 +\sin^2\omega_\star)}{R_\star} +\cdots\right] \ ,
\end{eqnarray}
respectively, where $\theta_\star(0,0) = a^{-1} e^{-(E_0 - \frac{1}{2} \tilde E_0)\tau}$, or equivalently, $a^{-1} = \lim_{\tau\to 0} \theta_\star(0,0)$. 
The corresponding eigenvectors are given by
\begin{eqnarray}
v_0 = \left( 1+ \frac{\cos^2\omega_\star \sin^2\omega_\star}{R_\star} \right) \psi(\omega_\star,\delta_\star) - \frac{\cos\omega_\star \sin^2\omega_\star}{R_\star} Z_0 + \cdots \ ,
\end{eqnarray}
and
\begin{eqnarray}
v_1 = \left( 1- \frac{\sin^4\omega_\star}{R_\star}  \right) \psi\left(\omega_\star+\frac{\pi}{2},\delta_\star\right) - \frac{\sin^3\omega_\star}{R_\star} Z_0 + \cdots\ .
\end{eqnarray}
Note that the eigenvectors are orthonormal up to corrections of order $\calO(1/R_\star^2)$.
Expressing the noise overlap factor and its inverse in terms of the eigenvectors and eigenvalues, one finds to leading order in $1/R_\star$ the expressions:
\begin{eqnarray}
\tilde Z_0 = \frac{a}{\sin^2\omega_\star} \psi\left(\omega_\star+\frac{\pi}{2},\delta_\star\right) \psi^\dagger\left(\omega_\star+\frac{\pi}{2},\delta_\star\right) + \cdots
\end{eqnarray}
and 
\begin{eqnarray}
(\tilde Z_0)^{-1} = \frac{R_\star}{ a \cos^2\omega_\star  } \psi(\omega_\star,\delta_\star) \psi^\dagger(\omega_\star,\delta_\star) + \cdots \ .
\label{eq:noise_overlap_decomposition}
\end{eqnarray}
For the two-state system, we find that a large signal/noise enhancement is possible at late times provided that $(\tilde Z_0)^{-1}$ is expressible as an outer product of vectors, up to relative corrections in $1/R_\star$.
It is no coincidence that the outer product that is formed in \Eq{noise_overlap_decomposition} involves the optimal vectors, $\psi(\omega_\star,\delta_\star)$; the result is generally true for systems at late time, whenever the enhancement factor is large, and follows directly from \Eq{sn_soln}.
Referring back to \Eq{max_sn}, one finds that in addition to the requirement that $\tilde Z_0$ possess a near-zero eigenvalue, the corresponding eigenvector must also have an overlap onto $Z_0$ which is parametrically larger than that eigenvalue by comparison, in order to achieve large signal/noise enhancements.
This qualitative requirement not only holds for the two-state model, but also extends trivially to many state systems.

It is instructive to also consider the effects on the signal/noise enhancement at late times due to the subleading correction to the noise correlation function, which is of order $\tilde \Delta$.
Using the eigenvalue and eigenvector results above, we may explicitly compute the finite $\tilde \Delta$ corrections to the signal/noise, evaluated at the critical angles $(\omega_\star,\delta_\star)$.
In particular, in the regime where $\rho_\star \ll 1$, it is given by
\begin{eqnarray}
\theta_\star(\omega_\star,\delta_\star) = \frac{R_\star \theta_\star(0,0)}{\left[ 1 +  \frac{ R^2_\star}{a^2 \cos^4\omega_\star}  |\psi^\dagger(\omega_\star,\delta_\star) \tilde Z_1 \psi(\omega_\star,\delta_\star)|^2 \tilde \Delta + \ldots \right]^{1/2}}\ .
\end{eqnarray}
From this expression, we see that the signal/noise enhancement is set by $R_\star$ only in the regime where
\begin{eqnarray}
R_\star \psi^\dagger(\omega_\star,\delta_\star) \tilde Z_1 \psi(\omega_\star,\delta_\star) \sqrt{\tilde \Delta} \ll a \cos^2 \omega_\star \ .
\end{eqnarray}
This inequality is always satisfied at sufficiently late times, provided $\omega_\star \neq \pi/2$.
However, for earlier times, and for sufficiently large $R_\star$, this inequality may be violated, and the signal/noise enhancement will effectively be cut off by the subleading contribution to the noise correlator.
In that regime, the signal/noise will instead behave as 
\begin{eqnarray}
\theta_\star(\omega_\star,\delta_\star) \sim \frac{ \cos^2\omega_\star }{ \psi^\dagger(\omega_\star,\delta_\star) \tilde Z_1 \psi(\omega_\star,\delta_\star) } e^{-(E_0-\frac{1}{2} \tilde E_1)\tau} \ .
\end{eqnarray}
In other words, in the limit of sufficiently large $R_\star$, it is possible to choose the source/sink vectors so that the leading contribution to the noise correlator is projected away.
Consequently, it is the subleading contribution (the first excited noise state) that determines exponential degradation of the signal/noise.

\begin{figure} 
\includegraphics[width=\figwidth]{\figdir 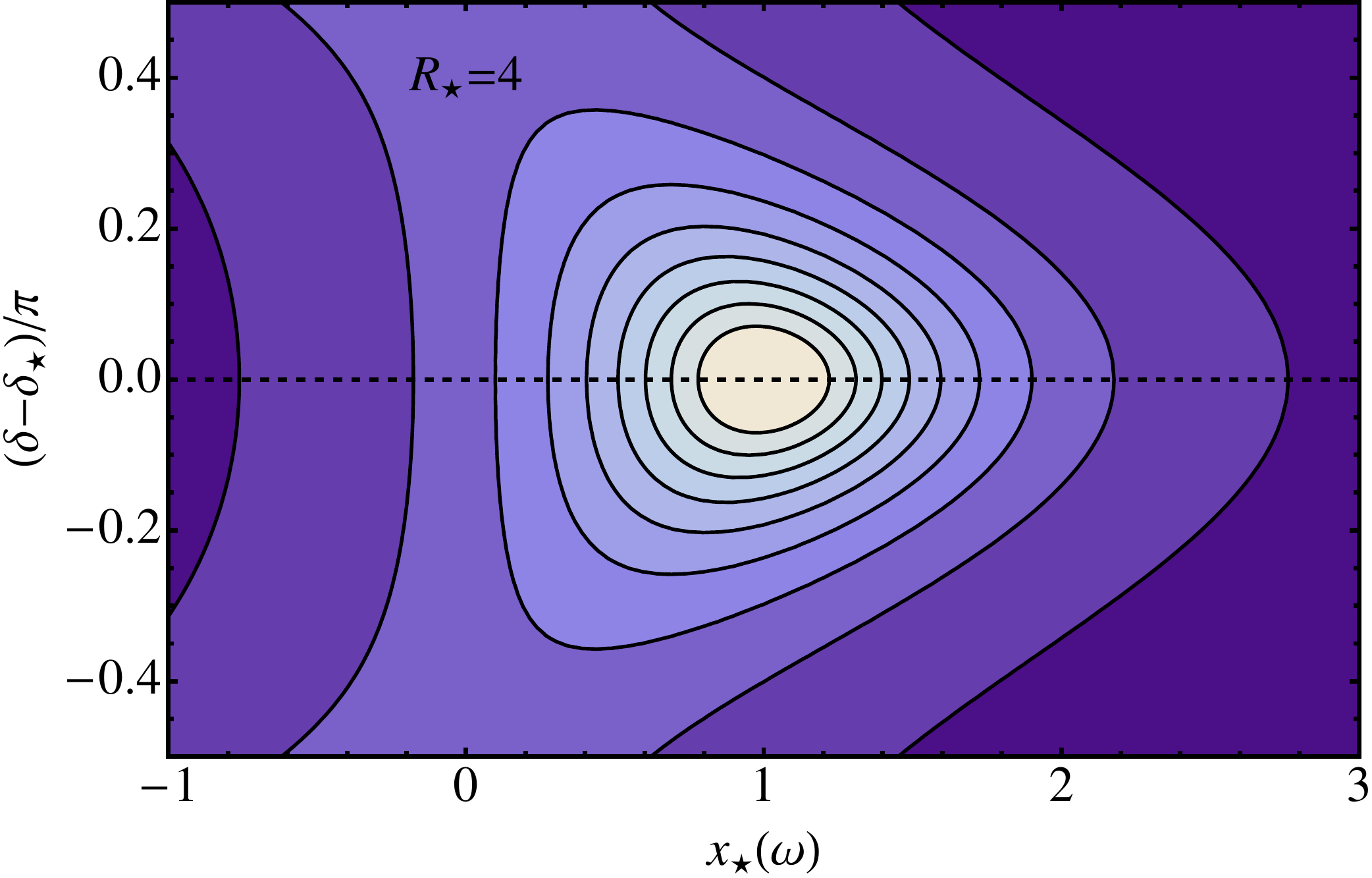}
\includegraphics[width=\figwidth]{\figdir 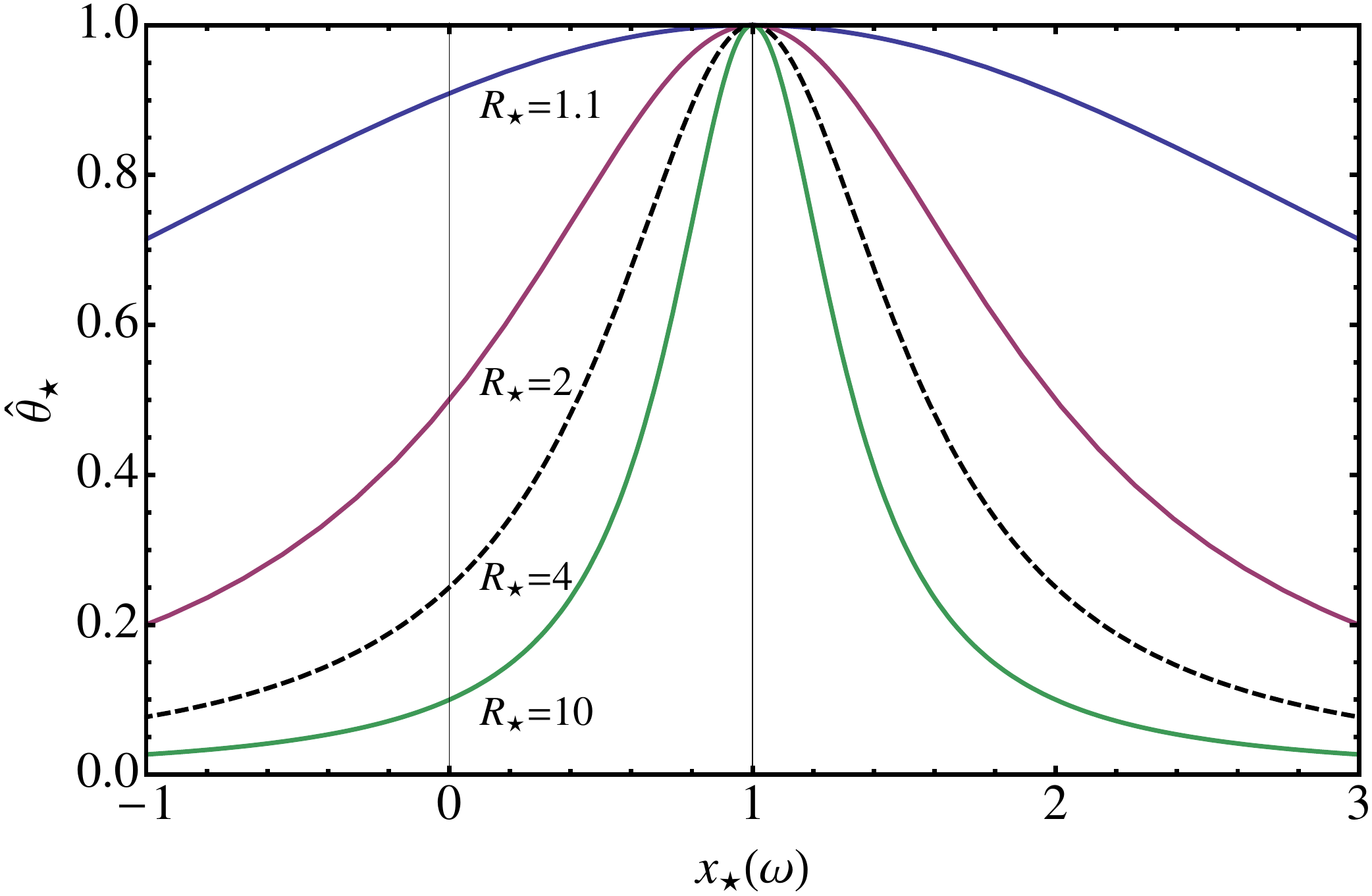}
\caption{\label{fig:toy_plots1}%
Left: Density plot of the signal/noise, $\hat\theta_\star$, as a function of the source and sink vector parametrization angles $\omega$ and $\delta$ for $R_\star=4$ and $\rho_\star=0$ (lighter regions correspond to larger values of $\hat\theta_\star$).
The signal/noise depends implicitly on $\omega$ through the parameter $x_\star(\omega)$, defined in \Eq{sn_parametrization}.
Right: Signal/noise as a function of $x_\star(\omega)$, for $R_\star = 1.1, 2, 4$ and $10$, $\rho_\star=0$ and $\delta=\delta_\star$.
A source with perfect overlap with the ground state corresponds to $x_\star(\omega) = 0$; a correlator with maximal signal/noise corresponds to $x_\star(\omega) = 1$.
Dashed curve in each plot correspond to the same values of $R_\star$ and $\delta_\star$.
}
\end{figure}

Turning now to the regime $\rho_\star \gg 1$, the optimal source and sink is by happenstance such that the excited state is the dominant contribution to the correlator whereas the ground state is suppressed.
Since $\Delta$ decays exponentially in time, this regime can only be sustained for a short period before an eventual transition into the former regime.
Since $\omega_\star$ is a parameter intrinsic to the system, there is no guarantee that a regime satisfying $\rho_\star\gg1$ exists, as it would require an exponential fine-tuning of $\omega_\star$ in the vicinity of $\pi/2$ at late times (the former regime by contrast is always guaranteed for sufficiently late times).
One might conclude, as a general rule, that signal/noise enhancement for a given excited state should therefore only involve mixing with states that are higher in energy than the target state, rather than lower in energy.
In cases where the energy splitting is sufficiently small and times are moderate, however, the degree of fine-tuning required is relaxed making the realization of such a regime more probable.
Should such a regime be realized for the system, then a significant enhancement in signal/noise for the exited state is only possible provided $R_\star \gtrsim \rho_\star$.

\begin{figure} 
\includegraphics[width=\figwidth]{\figdir 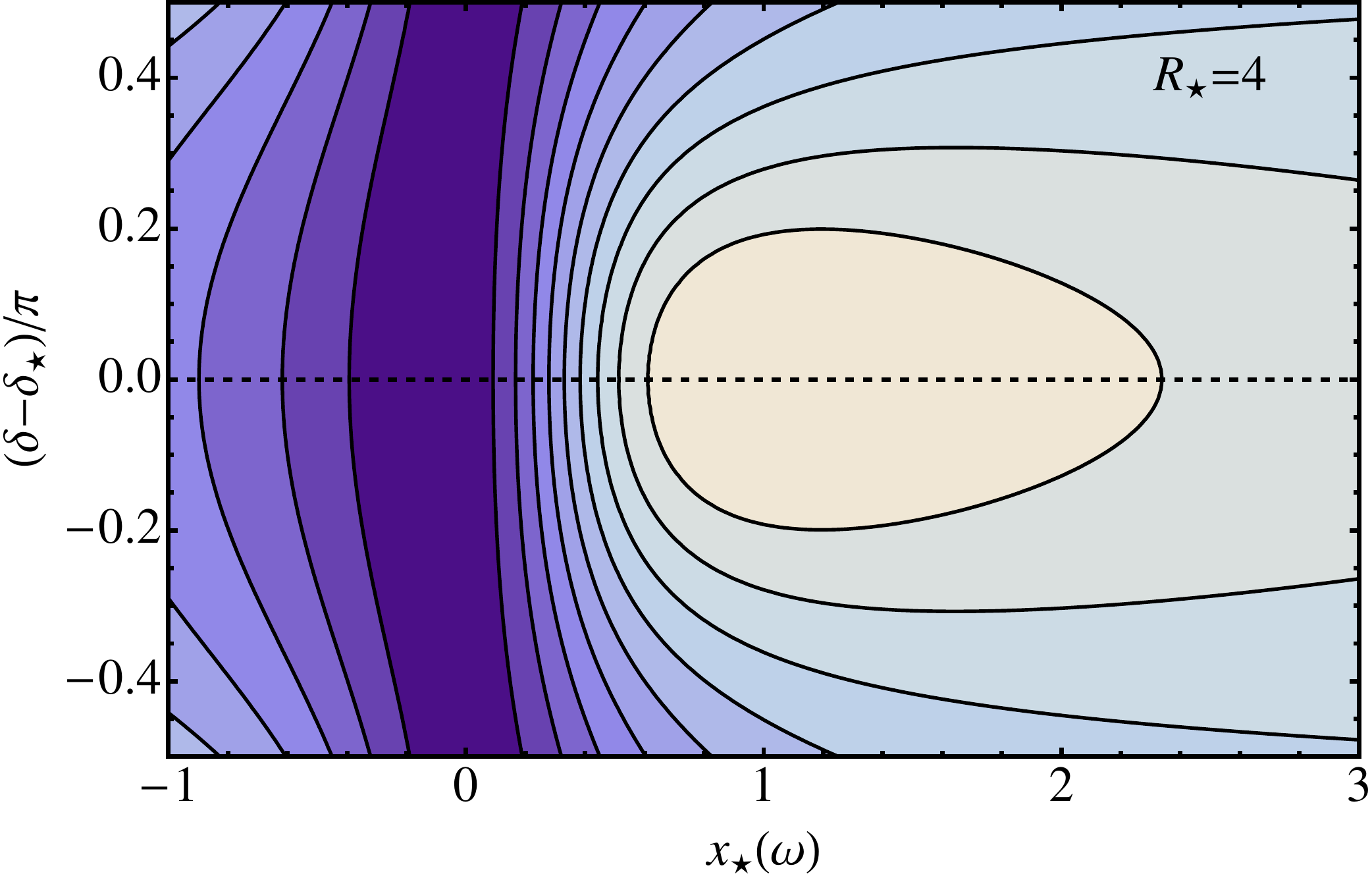}
\includegraphics[width=\figwidth]{\figdir 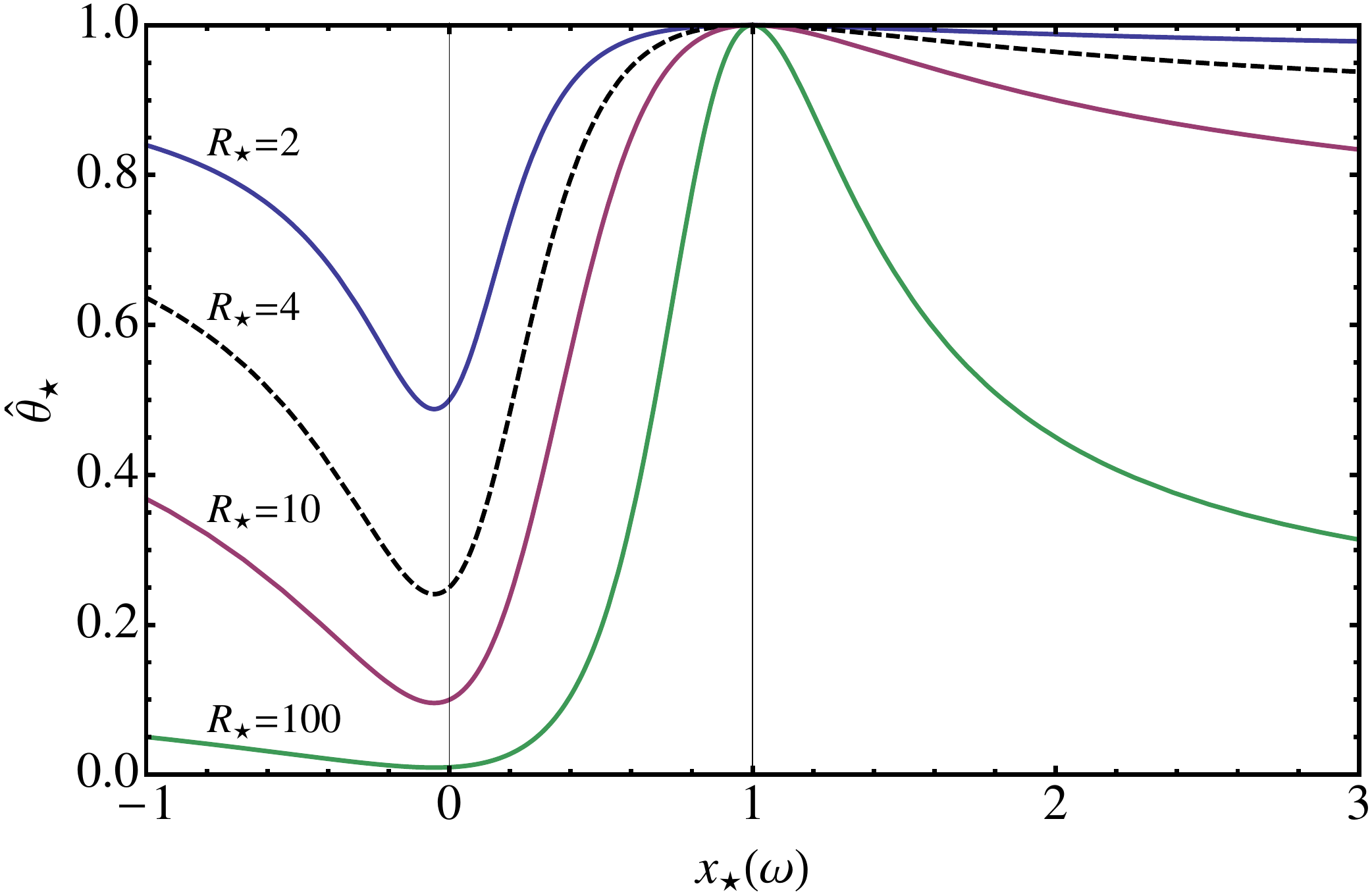}
\caption{\label{fig:toy_plots2}%
Left: Same as \Fig{toy_plots1} (left) for $R_\star=4$ and $\rho_\star=20$.
Right: Same as \Fig{toy_plots1} (right) for $R_\star = 2, 4, 10$ and $100$, $\rho_\star=20$ and $\delta=\delta_\star$.
}
\end{figure}

Let us now consider the signal/noise landscape in these two regimes as a function of $\omega$ and $\delta$, given the parameters $R_\star$, $\omega_\star$, $\delta_\star$ and $\rho_\star$.
A density plot of $\hat\theta_\star(\omega,\delta)$ is provided in \Fig{toy_plots1} (left) for $R_\star=4$ and $\rho_\star=0$, and is representative of the regime $\rho_\star\ll1$.
The $\omega$ dependence of the signal/noise along a curve of constant $\delta = \delta_\star$ is also plotted for various values of $R_\star$ in \Fig{toy_plots1} (right).
The ground state in each figure is located at $x_\star(\omega) = 0$, whereas the excited state corresponds to the limits $x_\star(\omega) \to \pm\infty$.
These states have a normalized signal/noise given by \Eq{sn_parametrization_eigstate}.
Similar plots are shown in \Fig{toy_plots2} (left and right) for the case where $\rho_\star = 20$, and are representative of the regime $\rho_\star \gg 1$.
Here, ones finds only a modest signal/noise enhancement for the excited state when $R_\star=4$.
The normalized signal/noise increases with $R_\star$, however, and results in significant enhancement once $R_\star \gtrsim \rho_\star$.
The normalized signal/noise for the ground state is suppressed compared to the excited state in this regime as a result of the optimal state having poor overlap with the ground state.

Frequently, one is interested in the regime $\rho_\star\approx 1$, in which correlators receive comparable contributions from both the ground and excited states.
In this regime, there is an ambiguity as to whether the signal/noise enhancement can be attributed to the ground state or the excited state.
One can crisply retain such a distinction, however, by fixing the source vector so as to produce an eigenstate of the Hamiltonian and then optimizing the sink vector ${\psi^\prime}(\omega,\delta)$ with respect to its arguments.
Particularly, let us consider a source vector of the form $\psi_n = (\delta_{n,0},\delta_{n,1})$.
Since the correlator receives no contributions from states orthogonal to the source in such cases, the temporal behavior will remain that of a pure exponential rather than a sum of exponentials.
Contrary to intuition, by allowing the sink vector to have a nonzero overlap with both states, an enhancement in the signal/noise $\theta_n(\omega,\delta) = \theta({\psi^\prime}(\omega,\delta),\psi_n)$ is nonetheless possible.

\begin{figure} 
\includegraphics[width=\figwidth]{\figdir 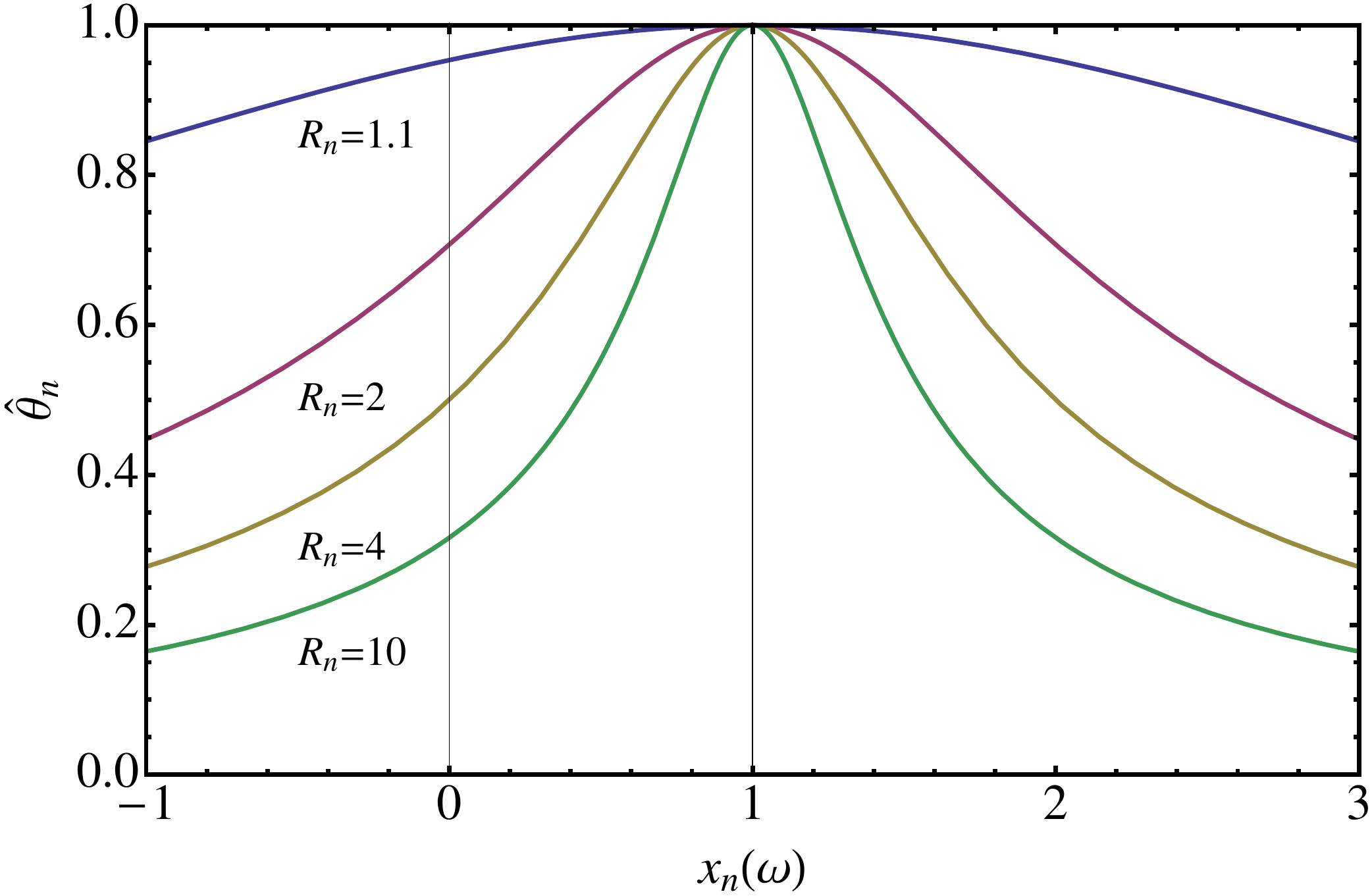}
\includegraphics[width=\figwidth]{\figdir 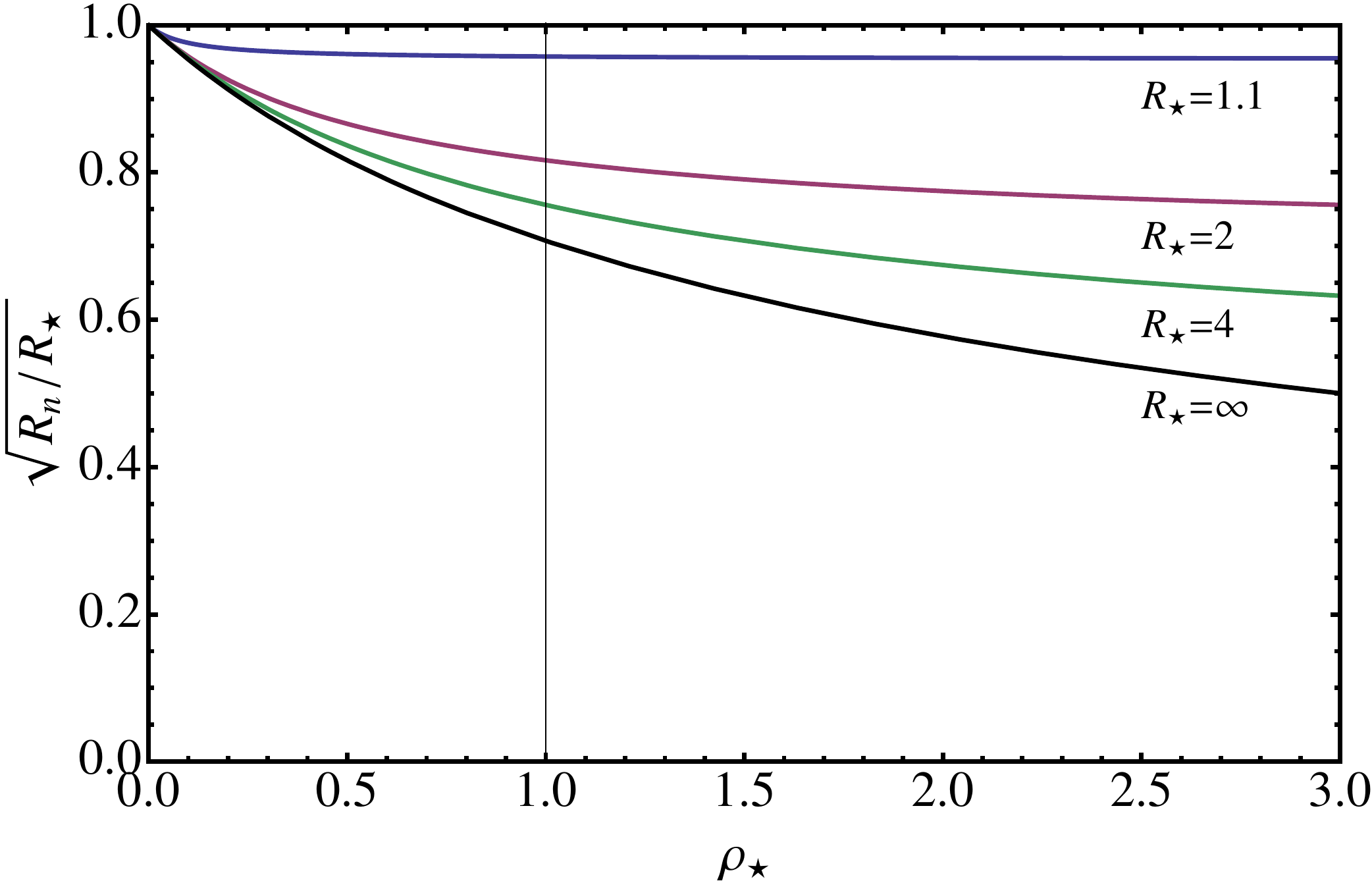}
\caption{\label{fig:toy_ratios}%
Left: Normalized signal/noise $\hat \theta_n$ as a function of $x_n(\omega)$ for various values of $R_n$.
Right: $\sqrt{R_n/R_\star}$ as a function of $\rho_\star$ for various values of $R_\star$.
}
\end{figure}

As was the case for equal source and sink vectors, we may define a normalized signal/noise by $\hat \theta_n(\omega,\delta) = \theta_n(\omega,\delta)/\theta_n(\omega_n,\delta_n)$, where $(\omega_n,\delta_n)$ maximizes $\theta_n(\omega,\delta)$.
The normalized signal/noise may then be expressed as:
\begin{eqnarray}
\hat \theta_n(\omega,\delta) = \frac{1}{\sqrt{R_n + (R_n-1) x_n(\omega) \left[ x_n(\omega) -2 \cos (\delta - \delta_n) \right] }}\ ,
\label{eq:sn_parametrization_partial}
\end{eqnarray}
where $\theta_n(\omega_n,\delta_n) = \sqrt{R_n} \theta(\psi_n,\psi_n)$, $R_n\ge1$, and
\begin{eqnarray}
x_0(\omega) = \frac{\tan\omega}{\tan\omega_0} \ ,\qquad x_1(\omega) = \frac{\cot\omega}{\cot\omega_1}  \ .
\end{eqnarray}
A plot of this quantity as a function of $x_n(\omega)$ is provided in \Fig{toy_ratios} (left) for various values of $R_n$ and $\delta = \delta_n$.
The parameters $R_n$, $\omega_n$, and $\delta_n$ are related to $R_\star$, $\omega_\star$, $\delta_\star$ and $\rho_\star$ in a well-defined way.
Specifically, one obtains $\delta_n = \delta_\star$,
\begin{eqnarray}
R_n = R_\star \frac{R_\star -1 + \rho_\star}{R_\star (1+\rho_\star)-1},
\label{eq:R_ratio}
\end{eqnarray}
and
\begin{eqnarray}
\tan\omega_0 = \tan\omega_\star \frac{R_\star -1}{ R_\star -1 + \rho_\star} \ , \qquad \tan\omega_1 = \tan\omega_\star \frac{R_\star}{ R_\star -1}\ .
\label{eq:omega_relations}
\end{eqnarray}
To understand how the enhancement compares between strategies, it is instructive to consider the ratio
\begin{eqnarray}
\frac{1}{\sqrt{R_\star}} \frac{\theta_n(\omega_n,\delta_n)}{\theta(\psi_n,\psi_n)} = \sqrt{ \frac{R_n}{R_\star}} \ ,
\label{eq:sn_ratio}
\end{eqnarray}
which may be expressed completely in terms of $R_\star$ and $\rho_\star$ using \Eq{R_ratio}.
This ratio, plotted in \Fig{toy_ratios} (right) as a function of $R_\star$ for various $\rho_\star$, provides a measure of the amount of signal/noise enhancement achieved using a fixed source (projected to an eigenstate), but expressed in terms of the optimization parameters for an equal source and sink.
Note that $\sqrt{R_n/R_\star}$ is bounded from above by unity, and from below by $\max(1/\sqrt{R_\star},1/\sqrt{1+\rho_\star})$.
These bounds lead to the inequalities
\begin{eqnarray}
\sqrt{R_\star} \theta(\psi_n,\psi_n) \ge  \theta_n(\omega_n,\delta_n) \ge \max\left(1,\sqrt{\frac{R_\star}{1+\rho_\star}} \right) \theta(\psi_n,\psi_n) \ ,
\end{eqnarray}
which imply that the signal/noise enhancement for each state ($n=0,1$) is always greater than unity, but at best $1/\sqrt{R_\star}$ that which is possible by considering equal source and sink vectors.
Furthermore, the signal/noise enhancement inequalities are saturated in the late-time limit when $\rho_\star \ll 1$. 
It is important to note that in exchange for the sacrifice in signal/noise enhancement compared to the equal source and sink case, the correlators under consideration here remain completely free of contamination from states other than the target state $|n\rangle$ at early and intermediate times.
Consequently, they may actually lead to better extractions of the eigenstates energies.

Finally, let us address an interesting puzzle which arises from the results presented thus far.
Imagine that we were to make independent measurements of correlators using source and sink vectors which have perfect overlap with the ground and excited states.
The state with optimal signal/noise may then be constructed as a linear combination of those independently measured correlators.
Naive error propagation tells us that the uncertainties should be added in quadrature and therefore an enhancement in signal/noise should not be possible.
Yet the signal/noise analysis presented above provides just the opposite conclusion.
This apparent paradox highlights the critical importance of correlations between correlator matrix elements as a source for ``noise non-conservation'' and suggests an important lesson.
The success of the signal/noise enhancement strategies discussed here require that correlators be measured on the same ensemble, such that the correlations in their statistical fluctuations are retained and therefore may be exploited.
The importance of correlations between correlator matrix elements is explicitly demonstrated in \Sec{hadron_landscape} for QCD correlators involving the rho meson.

\section{Signal/noise optimization}
\label{sec:signal_noise_optimization}

Having discussed a simple system as motivation, we now return to the general case of signal/noise optimization.
There are a number of different approaches that one can take, as we set out in the following.

\subsection{Fixed source and unconstrained sink}
\label{sec:fixed_source_unconstrained_sink_sn}

Assuming a fixed source vector, $\psi$, one may ask what choice of sink vector, ${\psi^\prime}$, will maximize the signal/noise for any given time slice.
This is a straight-forward optimization problem; to solve it, we introduce a Lagrange multiplier $\xi^\prime$ in order to enforce the constraint ${\psi^\prime}^\dagger {\psi^\prime} = 1$.
Since the signal/noise is positive, the critical points of $\theta({\psi^\prime},\psi)$ subject to the constraint coincide with those of the function
\begin{eqnarray}
\Xi({\psi^\prime},\psi,\xi^\prime) = \log\theta^2({\psi^\prime},\psi) + \xi^\prime \left( {\psi^\prime}^\dagger {\psi^\prime} - 1\right)\ .
\label{eq:xi_func}
\end{eqnarray}
Differentiating $\Xi$ with respect to ${\psi^\prime}^\dagger$ and $\xi^\prime$ one obtains
\begin{eqnarray}
\Xi_{{\psi^\prime}^\dagger}({\psi^\prime},\psi,\xi^\prime)  &=& \frac{ C \psi }{  {\psi^\prime}^\dagger C \psi } - \frac{\sigma^2_\psi {\psi^\prime} }{ {\psi^\prime}^\dagger \sigma^2_\psi {\psi^\prime} } + \xi^\prime {\psi^\prime} \ ,\cr
\Xi_{\xi^\prime}({\psi^\prime},\psi,\xi^\prime)  &=& {\psi^\prime}^\dagger {\psi^\prime} -1 \ ,
\end{eqnarray}
where we have introduced the short-hand notation
\begin{eqnarray}
\Xi_{{\psi^\prime}^\dagger}({\psi^\prime},\psi,\xi^\prime) &=& \frac{\partial }{\partial {\psi^\prime}^\dagger}  \Xi({\psi^\prime},\psi,\xi^\prime)\ , \cr
\Xi_{\xi^\prime}({\psi^\prime},\psi,\xi^\prime)  &=& \frac{\partial }{\partial \xi^\prime}  \Xi({\psi^\prime},\psi,\xi^\prime)\ .
\end{eqnarray}
The critical points $\psi^\prime_0$ and $\xi^\prime_0$ are determined by solutions to the equations
\begin{eqnarray}
\Xi_{{\psi^\prime}^\dagger}(\psi^\prime_0,\psi,\xi^\prime_0) = 0\ ,\qquad \Xi_{\xi^\prime}(\psi^\prime_0,\psi,\xi^\prime_0) = 0.
\end{eqnarray}
Noting that
\begin{eqnarray}
{\psi^\prime}^\dagger \Xi_{{\psi^\prime}^\dagger}({\psi^\prime},\psi,\xi^\prime) = \xi^\prime
\end{eqnarray}
for all $\psi^\prime$, one finds the solutions
\begin{eqnarray}
\xi^\prime_0 = 0 \ ,\qquad \psi^\prime_0 = A^\prime_0(\psi) \sigma^{-2}_\psi C \psi\ ,
\end{eqnarray}
where
\begin{eqnarray}
A^\prime_0(\psi)^{-2} = \psi^\dagger C^\dagger \sigma_\psi^{-4} C \psi 
\end{eqnarray}
is determined from the normalization condition on $\psi^\prime_0$.

In order to determine if this solution corresponds to a maximum of $\theta({\psi^\prime},\psi)$, we must compute the second derivatives of $\Xi({\psi^\prime},\psi,\xi^\prime)$ taken with respect to ${\psi^\prime}$ and $\xi^\prime$ (and their conjugates) and evaluated at the critical point.
Introducing the notation
\begin{eqnarray}
\Xi_{{\psi^\prime} {\psi^\prime}^\dagger}({\psi^\prime},\psi,\xi^\prime)
&=&
\frac{\partial^2}{\partial {\psi^\prime} \partial {\psi^\prime}^\dagger} \Xi({\psi^\prime},\psi,\xi^\prime)\ , \cr
\Xi_{{\psi^\prime}^\dagger {\psi^\prime}^\dagger}({\psi^\prime},\psi,\xi^\prime)
&=&
\frac{\partial^2}{\partial {\psi^\prime}^\dagger \partial {\psi^\prime}^\dagger} \Xi({\psi^\prime},\psi,\xi^\prime)\ , \cr
\Xi_{\xi^\prime {\psi^\prime}^\dagger}({\psi^\prime},\psi,\xi^\prime)
&=&
\frac{\partial^2}{\partial \xi^\prime \partial {\psi^\prime}^\dagger} \Xi({\psi^\prime},\psi,\xi^\prime)\ , \cr
\Xi_{\xi^\prime \xi^\prime}({\psi^\prime},\psi,\xi^\prime)
&=&
\frac{\partial^2}{\partial \xi^\prime \partial \xi^\prime} \Xi({\psi^\prime},\psi,\xi^\prime) \ ,
\end{eqnarray}
the second derivatives evaluated at the critical point are given by
\begin{eqnarray}
\Xi_{{\psi^\prime} {\psi^\prime}^\dagger}(\psi^\prime_0,\psi,\xi^\prime_0)
= -\frac{1}{{\psi^\prime_0}^\dagger \sigma^2_\psi \psi^\prime_0}  \sigma_\psi  \left[ 1 - \frac{ \sigma_\psi \psi^\prime_0 {\psi^\prime_0}^\dagger \sigma_\psi }{ {\psi^\prime_0}^\dagger \sigma^2_\psi \psi^\prime_0 } \right] \sigma_\psi\ , 
\end{eqnarray}
and
\begin{eqnarray}
\Xi_{{\psi^\prime}^\dagger {\psi^\prime}^\dagger}(\psi^\prime_0,\psi,\xi^\prime_0) = 0 \ ,\quad
\Xi_{\xi^\prime {\psi^\prime}^\dagger}(\psi^\prime_0,\psi,\xi^\prime_0) = \psi^\prime_0\ ,\quad
\Xi_{\xi^\prime \xi^\prime}(\psi^\prime_0,\psi,\xi^\prime_0) = 0\ .
\end{eqnarray}
Note that the optimization problem is constrained, so the Hessian evaluated at the critical point is bordered, taking the form
\begin{eqnarray}
\calH = \left( \begin{array}{cc}
h & \psi^\prime_0 \\
{\psi^\prime_0}^\dagger & 0
\end{array} \right)\ , \qquad h \equiv \Xi_{{\psi^\prime} {\psi^\prime}^\dagger}(\psi^\prime_0,\psi,\xi^\prime_0)\ .
\end{eqnarray}
Since the upper left block of $\calH$ satisfies the relation $h \psi^\prime_0 = 0$, the eigenvectors of $\calH$ are given by
\begin{eqnarray}
W_{\pm} = \left( \begin{array}{c}
\psi^\prime_0  \\
\pm 1
\end{array} \right)\ ,\ \qquad
W^{\perp}_{k} = \left( \begin{array}{c}
w^{\perp}_{k}  \\
0
\end{array} \right)\ ,
\end{eqnarray}
where ${\psi^\prime_0}^\dagger w^{\perp}_{k} = 0$ ($k=1,\cdots, N^\prime-1$).
The eigenvalues associates with $W_{\pm}$ are $\pm1$; one can argue that the eigenvalues associated with $W^{\perp}_{k}$ are all negative-definite due to the fact that the subspace of $h$ orthogonal to $\psi^\prime_0$ is negative-definite (this can be seen, for example, by expressing $h$ in a basis where one of the basis vectors is proportional to $\sigma_\psi \psi^\prime_0$).

Given that the optimization problem involves a single constraint, and the observation that there is only one positive eigenvalue of $\calH$, it follows that the critical point corresponds to a local maximum.
Since the domain of $\theta({\psi^\prime},\psi)$ is compact (i.e., it is the space of complex unit vectors ${\psi^\prime}$ for fixed $\psi$) and there are no other candidates for a critical point, the local maximum is in fact a global maximum.
Given that the domain is compact, one might wonder why, for a fixed $\psi$, there is not a second critical point corresponding to a minimum of the signal/noise.
This is easily explained, however, by observing that the signal/noise vanishes for an $N^\prime-1$ dimensional complex subspace defined by the vectors ${\psi^\prime}$ which are orthogonal to $C\psi$.
Since the signal/noise is not differentiable at the boundary of this domain, there can be no critical points associated with this region.
Since there is only a single critical point, corresponding to a maximum, it follows that the signal/noise is a decreasing function of the sink vector; the maximum signal/noise attainable, given a specified source $\psi$, is therefore 
\begin{eqnarray}
\theta(\psi^\prime_0,\psi) = \sqrt{ \psi^\dagger C^\dagger \sigma^{-2}_\psi  C \psi }\ .
\end{eqnarray}
Note that in the late time limit, $\psi^\prime_0$ is expected to become independent of $\psi$ due to the factorization found in \Eq{late_time_sn}.
This independence is spoiled at earlier times, however, due to the presence of excited state contamination and uncertainties in the estimates of $C$ and $\Sigma^2$.

\subsection{Unconstrained source and sink}
\label{sec:unconstrained_source_sink_sn}

Extremization of $\theta(\psi^\prime,\psi)$ with respect to the sink, $\psi^\prime$, and source, $\psi$, proceeds in a similar fashion.
One must now introduce an additional Lagrange multiplier to \Eq{xi_func} to enforce the normalization constraint $\psi^\dagger \psi = 1$ on the source vector.
An analysis similar to the previous section yields the critical points $\psi^\prime_0$ and $\psi_0$ satisfying the dual relations
\begin{eqnarray}
\frac{ C \psi_0 }{  {\psi^\prime_0}^\dagger C \psi_0 } = \frac{\sigma^2_{\psi_0} \psi^\prime_0 }{ {\psi^\prime_0}^\dagger \sigma^2_{\psi_0} \psi^\prime_0 }
\ ,\qquad
\frac{ C^\dagger \psi^\prime_0 }{  \psi_0^\dagger C^\dagger \psi^\prime_0 } = \frac{\sigma^2_{\psi^\prime_0} \psi_0 }{ \psi_0^\dagger \sigma^2_{\psi^\prime_0} \psi_0 }\ ,
\label{eq:dual_relations}
\end{eqnarray}
subject to the constraints ${\psi^\prime_0}^\dagger \psi^\prime_0 = 1$ and $\psi_0^\dagger \psi_0 = 1$.
Each vector may be expressed as a nonlinear function of the other
\begin{eqnarray}
\psi^\prime_0 = A^\prime_0(\psi_0) \sigma^{-2}_{\psi_0} C \psi_0 \ ,\qquad
\psi_0 = A_0(\psi^\prime_0) \sigma^{-2}_{\psi^\prime_0} C^\dagger \psi^\prime_0\ ,
\label{eq:dual_relation_soln}
\end{eqnarray}
with normalization factors given by
\begin{eqnarray}
A^\prime_0(\psi_0)^{-2} = \psi_0^\dagger C^\dagger \sigma^{-4}_{\psi_0} C \psi_0 \ ,\qquad
A_0(\psi^\prime_0)^{-2} = {\psi^\prime_0}^\dagger C \sigma^{-4}_{\psi^\prime_0} C^\dagger \psi^\prime_0 \ ,
\end{eqnarray}
respectively.
An examination of the second derivatives of $\theta(\psi^\prime_0,\psi_0)$ follows in an analogous manner to the previous analysis.
One can confirm that the critical points (there may be more than one in the presence of statistical uncertainties and/or for early time extents) satisfying the above relations indeed corresponds to a local maximum of $\theta({\psi^\prime},\psi)$.

Determining $\psi_0$ and $\psi^\prime_0$ analytically from \Eq{dual_relation_soln} is rather nontrivial because of the nonlinearity of the expressions, however, the optimal vectors may be found numerically using an iterative technique.
Starting from an arbitrary but appropriately normalized initial source/sink vector pair $({\psi^\prime}^{[0]}, \psi^{[0]})$ one generates new vectors using the iterative procedure
\begin{eqnarray}
{\psi^\prime}^{[n+1]} &=& A^\prime_0(\psi^{[n]}) \sigma^{-2}_{\psi^{[n]}} C \psi^{[n]} \ , \cr
\psi^{[n+1]} &=& A_0({\psi^\prime}^{[n]}) \sigma^{-2}_{{\psi^\prime}^{[n]}} C^\dagger {\psi^\prime}^{[n]}\ .
\label{eq:iterative_dual_relation_soln}
\end{eqnarray}
If the procedure converges (a condition that must be confirmed explicitly; we have found that this procedure works well for the data sets we have explored numerically), then the solution given by
\begin{eqnarray}
\psi^\prime_0 &=& \lim_{n\to\infty} {\psi^\prime}^{[n]} \ , \cr
\psi_0 &=& \lim_{n\to\infty} \psi^{[n]} \ ,
\label{eq:local_max}
\end{eqnarray}
is a local maximum of $\theta({\psi^\prime},\psi)$.
At late times, the convergence is expected to be rapid (in principle after only a single iteration), since $\psi^\prime_0$ and $\psi_0$ become independent of each other up to excited state contamination and fluctuations in the estimates of $C$ and $\sigma^2$.
Alternatively, one may find a global maximum of $\theta({\psi^\prime},\psi)$ numerically using a Monte Carlo technique such as simulated annealing, or a combination of the two methods.

Finally, let us consider the special case where $N=M$, and $C = C^\dagger$.
An estimate of the correlator matrix will only satisfy such a relation up to statistical uncertainties unless the individual correlators contributing to the estimate also satisfy $\calC = \calC^\dagger$.
Normally this situation does not occur, but imposing such a condition on $\calC$ (i. e., by taking only the Hermitian part of $\calC$) is permitted, as it has no effect on the ensemble average in the $\calN\to\infty$ limit.
Assuming such a condition is imposed on $\calC$, one may confirm that $\psi_0 = \psi^\prime_0$ is a solution to \Eq{dual_relations}.
At sufficiently late times, this solution must correspond to a global maximum of $\theta$, as a result of \Eq{late_time_sn}, although for intermediate times it need not be.
We will return to this important solution in later sections.

\subsection{Fixed source and constrained sink}
\label{sec:fixed_source_constrained_sink_sn}

Next, let us briefly consider a constrained optimization problem for the sink vector $\psi^\prime$, given a fixed source vector $\psi$.
The constraints we would like to impose are of the form ${\phi^\prime}^\dagger \psi^\prime = 0$ for some arbitrary $N^\prime\times P^\prime$ matrix $\phi^\prime$ of rank $P^\prime<N^\prime$.
The constraints on $\psi^\prime$ may be imposed by introducing a $P^\prime$-dimensional vector of Lagrange multipliers $\zeta^\prime$.
Extremizing the signal/noise subject to these constraints yields the critical point $\psi^\prime_0$ which satisfies the relations
\begin{eqnarray}
\frac{ C \psi }{  {\psi^\prime_0}^\dagger C \psi } - \frac{\sigma^2_\psi \psi^\prime_0 }{ {\psi^\prime_0}^\dagger \sigma^2_\psi \psi^\prime_0 } + \xi^\prime \psi^\prime_0 + \phi^\prime \zeta^\prime = 0 \ ,
\label{eq:constrained_eqns}
\end{eqnarray}
where ${\psi^\prime_0}^\dagger \psi^\prime_0 = 1$ and  ${\phi^\prime}^\dagger \psi^\prime_0 = 0$.
After left multiplying \Eq{constrained_eqns} by ${\phi^\prime}^\dagger \sigma_\psi^{-2}$, one may solve for $\zeta^\prime$ in terms of $\phi^\prime$ and $\psi$, up to an irrelevant constant of proportionality.
Inserting the expression for $\zeta^\prime$ back into \Eq{constrained_eqns}, yields the solution 
\begin{eqnarray}
\psi^\prime_0 = A^\prime_0(\psi)  \sigma_\psi^{-1} \calP^\prime_\psi \sigma_\psi^{-1}  C \psi \ ,
\label{eq:constrained_sink_soln}
\end{eqnarray}
where
\begin{eqnarray}
\calP^\prime_\psi = 1 - \sigma_\psi^{-1} \phi^\prime  \left({\phi^\prime}^\dagger \sigma_\psi^{-2} \phi^\prime \right)^{-1} {\phi^\prime}^\dagger \sigma_\psi^{-1} \ ,
\end{eqnarray}
is a Hermitian projection operator satisfying ${\calP^\prime_\psi}^2 = \calP^\prime_\psi$, $\calP^\prime_\psi  = {\calP^\prime_\psi}^\dagger$,  and $\calP^\prime_\psi \sigma_\psi^{-1} {\phi^\prime} = 0$.
The normalization factor for the solution is given by
\begin{eqnarray}
A^\prime_0(\psi)^{-2} = \psi^\dagger C^\dagger \sigma_\psi^{-1} \calP^\prime_\psi \sigma_\psi^{-2} \calP^\prime_\psi \sigma_\psi^{-1}  C \psi\ .
\end{eqnarray}
These solutions may be of use when optimizing the signal/noise of an excited state at asymptotically late times, where maintaining orthogonality of the optimized sink with lower energy eigenstates is desirable, as was discussed in \Sec{toy_model} for the toy model.
For example, by choosing $\phi^\prime$ equal to the source-optimized vector(s) for the ground state estimated at one or more time slices, one may then optimize the signal/noise of the first excited state while introducing only minimal ground state contamination.

\subsection{Constrained source and sink}
\label{sec:constrained_source_constrained_sink_sn}

The results of the previous subsection may be trivially extended to the case of a constrained sink vector, ${\psi^\prime}$, and a constrained source vector, $\psi$.
The constraints are imposed by the conditions ${\phi^\prime}^\dagger \psi^\prime = 0$ for some arbitrary $N^\prime\times P^\prime$ matrix $\phi^\prime$ of rank $P^\prime<N^\prime$ and $\phi^\dagger \psi = 0$ for some arbitrary $N\times P$ matrix $\phi$ of rank $P<N$.
Extremizing the signal/noise with respect to $\psi^\prime$ and $\psi$, subject to the constraints yields the dual relations
\begin{eqnarray}
\psi^\prime_0 = A^\prime_0(\psi_0)  \sigma_{\psi_0}^{-1} \calP^\prime_{\psi_0} \sigma_{\psi_0}^{-1}  C \psi_0 \ ,\qquad
\psi_0 = A_0(\psi^\prime_0)  \sigma_{\psi^\prime_0}^{-1} \calP_{\psi^\prime_0} \sigma_{\psi^\prime_0}^{-1}  C^\dagger \psi^\prime_0\ ,
\label{eq:constrained_source_sink_soln}
\end{eqnarray}
where
\begin{eqnarray}
\calP^\prime_{\psi_0} = 1 - \sigma_{\psi_0}^{-1} \phi^\prime  \left({\phi^\prime}^\dagger \sigma_{\psi_0}^{-2} \phi^\prime \right)^{-1} {\phi^\prime}^\dagger \sigma_{\psi_0}^{-1} \ ,\qquad
\calP_{\psi^\prime_0} = 1 - \sigma_{\psi^\prime_0}^{-1} \phi  \left(\phi^\dagger \sigma_{\psi^\prime_0}^{-2} \phi \right)^{-1} \phi^\dagger \sigma_{\psi^\prime_0}^{-1} \ ,
\end{eqnarray}
and
\begin{eqnarray}
A^\prime_0(\psi_0)^{-2} &=& \psi_0^\dagger C^\dagger \sigma_{\psi_0}^{-1} \calP^\prime_{\psi_0} \sigma_{\psi_0}^{-2} \calP^\prime_{\psi_0} \sigma_{\psi_0}^{-1}  C {\psi_0} \ , \cr
A_0(\psi^\prime_0)^{-2} &=& {\psi^\prime_0}^\dagger C \sigma_{\psi^\prime_0}^{-1} \calP_{\psi^\prime_0} \sigma_{\psi^\prime_0}^{-2} \calP_{\psi^\prime_0} \sigma_{\psi^\prime_0}^{-1}  C^\dagger \psi_0^\prime \ .
\end{eqnarray}
These relations may be solved using the iterative approach discussed in \Sec{unconstrained_source_sink_sn}.

\subsection{Steepest ascent}
\label{sec:steepest_ascent}

The methods of \Sec{fixed_source_unconstrained_sink_sn} - \Sec{constrained_source_constrained_sink_sn} provide a linear algebra prescription for finding source and sink vectors which extremize the signal/noise, subject to possible constraints $(\phi^\prime,\phi)$.
Alternatively, we may explore signal/noise optimization as a continuous process by considering a trajectory of steepest ascent along the signal/noise landscape, starting from some initial sink and source vectors ${\psi^\prime}^{[0]}$ and $\psi^{[0]}$.
Infinitesimal steps of size $\epsilon$ taken along the path of steepest ascent are enumerated as follows:
\begin{eqnarray}
{\psi^\prime}^{[n+1]} = \frac{ {\psi^\prime}^{[n]} + \epsilon \eta^\prime({\psi^\prime}^{[n]},\psi^{[n]}) }{\left|{\psi^\prime}^{[n]} + \epsilon \eta^\prime({\psi^\prime}^{[n]},\psi^{[n]} ) \right|}
\ ,\qquad 
\psi^{[n+1]} = \frac{ \psi^{[n]} + \epsilon \eta({\psi^\prime}^{[n]},\psi^{[n]}) }{ \left|\psi^{[n]} + \epsilon \eta({\psi^\prime}^{[n]},\psi^{[n]}) \right|  } \ ,
\label{eq:steepest_ascent}
\end{eqnarray}
where
\begin{eqnarray}
\eta^\prime(\psi^\prime,\psi) = \calR^\prime \left(  \frac{ C \psi }{  {\psi^\prime}^\dagger C \psi } - \frac{\sigma^2_{\psi} \psi^\prime }{ {\psi^\prime}^\dagger \sigma^2_{\psi} \psi^\prime } \right)
\ ,\qquad
\eta(\psi^\prime,\psi) = \calR \left( \frac{ C^\dagger \psi^\prime }{  \psi^\dagger C^\dagger \psi^\prime } - \frac{\sigma^2_{\psi^\prime} \psi }{ \psi^\dagger \sigma^2_{\psi^\prime} \psi } \right) \ ,
\label{eq:steepest_ascent_forces}
\end{eqnarray}
and
\begin{eqnarray}
\calR^\prime = 1 - \phi^\prime\left( {\phi^\prime}^\dagger \phi^\prime \right)^{-1} {\phi^\prime}^\dagger
\ ,\qquad
\calR = 1 - \phi\left( \phi^\dagger \phi \right)^{-1} \phi^\dagger\ , 
\label{eq:projectors}
\end{eqnarray}
provided ${\phi^\prime}^\dagger {\psi^\prime}^0 = 0$ and $\phi^\dagger\psi^0 = 0$.
As in \Sec{unconstrained_source_sink_sn}, this procedure yields a local maximum given by \Eq{local_max}.
Note that the trajectory of steepest ascent provides a natural means for interpolating between the initial vectors ${\psi^\prime}^{[0]}$ and $\psi^{[0]}$ and the signal/noise optimized vectors $\psi^\prime_0$ and $\psi_0$.
This may be particularly useful in the case where the initial vectors correspond to source-optimized vectors, and allows for a more detailed investigation of the interplay between signal/noise enhancement and excited state contamination, as will be explained below.
Note that this strategy also extends trivially to the case of a fixed source vector; there one need only update ${\psi^\prime}^{[n]}$ according to \Eq{steepest_ascent}, while updating the source vector according to $\psi^{[n+1]}=\psi^{[n]}$.

\subsection{Construction of an optimal basis}
\label{sec:sn_basis}

The optimal signal/noise ratio, given by \Eq{max_sn}, is invariant under $C \to U C V^\dagger$, where $U$ and $V$ are unitary transformations corresponding to rotations of the source and sink interpolating operators.
One may exploit this property in order to construct a unique basis for classifying operators according to their signal/noise.
The orthonormal basis vectors, $\psi^\prime_\alpha$ ($\alpha = 0,\cdots, N^\prime-1$) and $\psi_\alpha$ ($\alpha = 0,\cdots,N-1$), are defined such that the following properties are satisfied:
\begin{eqnarray}
\theta(\psi^\prime_\alpha, \psi_0) &=& \left. \max_{\psi^\prime}{\theta(\calQ^\prime_{\alpha-1}{\psi^\prime},\psi_0)} \right|_{\psi^\prime=\psi^\prime_\alpha} \ ,\qquad \alpha = 0,\cdots, N^\prime-1\ , \cr
\theta(\psi^\prime_0, \psi_\alpha) &=& \left. \max_\psi{\theta(\psi^\prime_0,\calQ_{\alpha-1}\psi)} \right|_{\psi=\psi_\alpha} \ ,\qquad \alpha = 0,\cdots, N-1 \ ,
\label{eq:sn_basis_defn}
\end{eqnarray}
where
\begin{eqnarray}
\calQ^\prime_\alpha &=& 1-\sum_{\gamma=0}^\alpha  \psi^\prime_\gamma {\psi^\prime_\gamma}^\dagger \ ,\qquad \calQ^\prime_{-1} = 1\ ,\cr
\calQ_\alpha &=& 1-\sum_{\gamma=0}^\alpha  \psi_\gamma \psi_\gamma^\dagger \ ,\qquad \calQ_{-1} = 1\ ,
\end{eqnarray}
are Hermitian projection operators satisfying ${\calQ^\prime_\alpha}^2 = \calQ^\prime_\alpha$ and $\calQ_\alpha^2 = \calQ_\alpha$, and where
\begin{eqnarray}
{\psi^\prime}_\alpha^\dagger {\psi^\prime}_\beta = \delta_{\alpha\beta} \ ,\qquad
\psi_\alpha^\dagger \psi_\beta = \delta_{\alpha\beta} \ .
\end{eqnarray}
The right-hand-side of \Eq{sn_basis_defn}, for each component $\alpha$, should be interpreted as finding the maximum value of $\theta({\psi^\prime},\psi)$ by varying ${\psi^\prime}$ (at fixed $\psi_0$) or $\psi$ (at fixed $\psi^\prime_0$) within a subspace orthogonal to basis vectors labeled by $\beta<\alpha$.
Following this convention, the basis vectors will be ordered such that 
\begin{eqnarray}
\theta(\psi^\prime_0,\psi_\alpha) &\geq& \theta(\psi^\prime_0,\psi_{\alpha+1})\ , \cr
\theta(\psi^\prime_\alpha,\psi_0) &\geq& \theta(\psi^\prime_{\alpha+1},\psi_0)\ ,
\end{eqnarray}
for each $\alpha$.
The basis vectors, $\psi^\prime_\alpha$, and $\psi_\alpha$, may be found by solving a generalized set of equations for the critical points:
\begin{eqnarray}
\frac{ C \psi_0 }{  {\psi^\prime_\alpha}^\dagger C \psi_0 } &=& \frac{\sigma^2_{\psi_0} \psi^\prime_\alpha }{ {\psi^\prime_\alpha}^\dagger \sigma^2_{\psi_0} \psi^\prime_\alpha } + \sum_{\beta=0}^\alpha \xi^\prime_{\alpha\beta} \psi^\prime_\beta\ , \cr
\frac{ C^\dagger \psi^\prime_0 }{  \psi_\alpha^\dagger C^\dagger \psi^\prime_0 } &=& \frac{\sigma^2_{\psi^\prime_0} \psi_\alpha }{ \psi_\alpha^\dagger \sigma^2_{\psi^\prime_0} \psi_\alpha } + \sum_{\beta=0}^\alpha \xi_{\alpha\beta} \psi_\beta\ ,
\label{eq:sn_basis_eqns}
\end{eqnarray}
where $\xi^\prime_{\alpha\beta}$ and $\xi_{\alpha\beta}$ are Lagrange multipliers introduced to enforce the orthonormality constraints on $\psi^\prime_\alpha$ and $\psi_\alpha$, respectively.
Despite the formidable appearance of these equations, an explicit solution exists (the derivation is nontrivial; see \Appendix{app1} for details), and is given recursively by
\begin{eqnarray}
\psi^\prime_\alpha &=& A_\alpha(\psi_0) \calQ^\prime_{\alpha-1} \sigma^{-2}_{\psi_0} \psi^\prime_{\alpha-1} \ ,\qquad \alpha>0 \cr
\psi_\alpha &=& B_\alpha(\psi^\prime_0) \calQ_{\alpha-1} \sigma^{-2}_{\psi^\prime_0} \psi_{\alpha-1} \ ,\qquad \alpha>0 \ ,
\label{eq:sn_basis_solns}
\end{eqnarray}
where $A_\alpha(\psi_0)$ and $B_\alpha(\psi^\prime_0)$ are normalization factors given by
\begin{eqnarray}
A^{-2}_\alpha(\psi_0) &=& {\psi^\prime_{\alpha-1}}^\dagger \sigma^{-2}_{\psi_0} \calQ^\prime_{\alpha-1} \sigma^{-2}_{\psi_0} \psi^\prime_{\alpha-1} \ ,\cr 
B^{-2}_\alpha(\psi^\prime_0) &=& \psi_{\alpha-1}^\dagger \sigma^{-2}_{\psi^\prime_0} \calQ_{\alpha-1} \sigma^{-2}_{\psi^\prime_0} \psi_{\alpha-1} \ ,
\end{eqnarray}
for $\alpha>0$.
Provided $\psi^\prime_0$ and $\psi_0$ are initially determined, an orthonormal basis can then be constructed iteratively using the relations provided above.

Interestingly, despite the fact that $\theta(\psi^\prime,\psi)$ is itself a nonlinear function of $\psi^\prime$ and $\psi$, under the transformation $C \to U C V^\dagger$, the basis transforms linearly as $\psi^\prime_\alpha \to U \psi^\prime_\alpha$ and $\psi_\alpha \to V \psi_\alpha$.
Furthermore, regardless of the basis in which $C$ is presented, working in the signal/noise basis, the matrix elements
\begin{eqnarray}
\theta_{\alpha\beta} = \theta(\psi^\prime_\alpha,\psi_\beta)
\end{eqnarray}
are uniquely determined.
Given the above ordering of basis vectors, it is reasonable to expect the hierarchy in signal/noise
\begin{eqnarray}
\theta_{\alpha\beta} &\gtrsim& \theta_{\alpha+1,\beta} \ ,\cr
\theta_{\alpha\beta} &\gtrsim& \theta_{\alpha,\beta+1} \ ,\cr
\theta_{\alpha\beta} &\gtrsim& \theta_{\alpha+1,\beta+1} \ ,
\end{eqnarray}
although this would need to be confirmed explicitly in each case.
The advantage of working in the optimal signal/noise basis is that it allows one to easily classify and identify which elements in a matrix of correlators have the least significant signal/noise.
One might then excise poorly determined components of the correlation function before further analysis.
Such a strategy might be particularly useful for analysis using variational methods for extracting excited states, which, in practice, requires well-determined matrix elements to work effectively.
Finally, we note that although a maximal signal/noise basis was constructed for both the source and the sink, one could develop a similar construction for just the sink (given a fixed source) or just the source (given a fixed sink).

\subsection{Maximal signal/noise}
\label{sec:max_sn}

Throughout the discussion thus far, we have considered signal/noise optimization of correlation functions of the particular form ${\psi^\prime}^\dagger C \psi$.
The most general linear combination of the correlation functions under consideration, however, takes an even more generic form, namely $\Tr \Phi^\dagger C $, where $\Phi$ is an arbitrary complex $N^\prime\times N$ matrix.
Optimization of the signal/noise for this correlator can be achieved using the results of the previous sections by combining the source and sink indices together into a single collective index.
In particular, one may regard the $N^\prime\times N$ dimensional matrix $C$ instead as an $N^\prime N\times1$ dimensional matrix, and $\Phi$ as an $N^\prime N$ dimensional ``sink'' vector.
The ``source'' vector in this case is just a one-dimensional vector normalized to unity.
The full (unconstrained) optimization of the signal/noise of this correlator will yield the maximum achievable signal/noise of all the methods presented, and we use this to define $\theta_\textrm{max}$.
However, the combination of correlation matrix elements so determined is not a bi-local correlation function.
Interestingly, one may easily prove that at asymptotically late times, the critical value for $\Phi_0$ which maximizes the signal/noise is given by
\begin{eqnarray}
\Phi_0 \propto (\tilde Z_0^\prime)^{-1} Z_0^\prime Z_0^\dagger (\tilde Z_0)^{-1} \ ,
\end{eqnarray}
and does yield a bi-local correlation function.
Furthermore, the signal/noise $\theta_\textrm{max}$ is identical to that of \Eq{max_sn}.
This need not be the case, however, at earlier time separations.

We may regard $\theta_\textrm{max}$, evaluated at every time slice, as a measure by which to normalize the signal/noise achieved by all other optimization strategies.
For convenience, we introduce a normalized signal/noise ratio,
\begin{eqnarray}
\bar \theta(\psi^\prime,\psi) = \frac{\theta(\psi^\prime,\psi)}{\theta_\textrm{max}}\ ,
\end{eqnarray}
for bilocal correlators constructed with source and sink vectors $\psi^\prime$ and $\psi$.
This quantity is bounded from above by unity and from below by zero, and at sufficiently late times is independent of time separation (i.e., the time-dependent exponentials cancel exactly between numerator and denominator) provided there is a gap in the noise spectrum.
In our analysis of hadron correlators in lattice QCD presented in \Sec{comparison_of_strategies}, this measure of signal/noise will be used extensively. 

\subsection{Multiple time slices}
\label{sec:multiple_time_slices}

The signal/noise optimization methods presented thus far may be extended over multiple time slices, labeled by $\tau_k$ for $k=0,\ldots,Q-1$ (which need not be consecutive) in order to account for temporal correlations that are often found to be important in correlator data.
The individual correlators can be expressed as a $Q\times Q$ block-diagonal matrix
\begin{eqnarray}
{\bm \calC} =
\left( \begin{array}{cccc}
\calC(\tau_0) & & & \\
& \calC(\tau_1) & & \\
&  &  \ddots & \\
&  &  & \calC(\tau_{Q-1})
\end{array} \right) \ ,
\end{eqnarray}
with each block component of size $N^\prime \times N$.
The expectation value of this correlator is given by ${\bm C} = \langle {\bm \calC} \rangle$ and the associated $Q^2 \times Q^2$ block noise matrix is given by
\begin{eqnarray}
{\bm \Sigma}^2 = \langle {\bm \calC} \otimes {\bm \calC}^* \rangle - {\bm C} \otimes {\bm C}^* \ ,
\end{eqnarray}
where each block component is of size ${N^\prime}^2\times N^2$.
Note that, although $\bm C$ is block-diagonal, ${\bm \Sigma}^2$ has nontrivial off-diagonal block matrix elements because of temporal correlations.
Since correlators formed from $\bm C$ will involve multiple time slices, we must impose constraints on the sink and source vectors,
\begin{eqnarray}
{\bm \psi}^\prime =
\left( \begin{array}{c}
\psi^\prime(\tau_0) \\
\psi^\prime(\tau_1) \\
\vdots \\
\psi^\prime(\tau_{Q-1})
\end{array} \right) \ ,\qquad
{\bm \psi} =
\left( \begin{array}{c}
\psi(\tau_0) \\
\psi(\tau_1) \\
\vdots \\
\psi(\tau_{Q-1})
\end{array} \right) \ ,
\end{eqnarray}
in order to preserve their overall normalization as well as the time-independence of the orientation of the subvectors for each block.
Specifically, they must satisfy the normalization conditions
\begin{eqnarray}
{{\bm\psi}^\prime}^\dagger {\bm \psi^\prime} = \frac{Q}{ { {\bm w}^\prime }^2}  \ ,\qquad {\bm \psi}^\dagger {\bm \psi} = \frac{Q}{ {\bm w}^2}  \ ,
\end{eqnarray}
for some ${{\bm w}^\prime}^2$ and ${\bm w}^2$ (defined later) and
\begin{eqnarray}
{\psi^\prime}^\dagger(\tau_0) \psi^\prime(\tau_0) = 1 \ ,\qquad \psi^\dagger(\tau_0) \psi(\tau_0) = 1\ .
\end{eqnarray}
The constraints on the time-independent orientation of the source and sink vectors are imposed by the conditions
\begin{eqnarray}
{{\bm \phi}^\prime}^\dagger {\bm \psi}^\prime = 0 \ ,\qquad {\bm \phi}^\dagger \bm \psi = 0\ ,
\label{eq:temporal_constraints}
\end{eqnarray}
where ${\bm\phi}^\prime$ and $\bm \phi$ are $Q\times (Q-1)$ bidiagonal matrices of the form
\begin{eqnarray}
{{\bm \phi}^\prime}^\dagger  =
\left( \begin{array}{rrrrrr}
1 & -w^\prime(\tau_1) & & &&\\
  & 1 & -w^\prime(\tau_2) & & &\\
&  &  \ddots & \ddots & &\\
&  & & 1 & -w^\prime(\tau_{Q-1})
\end{array} \right) \ ,
\end{eqnarray}
and
\begin{eqnarray}
{\bm \phi}^\dagger  =
\left( \begin{array}{rrrrrr}
1 & -w(\tau_1) & & &&\\
  & 1 & -w(\tau_2) & & &\\
&  &  \ddots & \ddots & &\\
&  & & 1 & -w(\tau_{Q-1})
\end{array} \right) \ , 
\end{eqnarray}
and where each entry is proportional to a unit block matrix of appropriate dimensionality (either $N^\prime\times N^\prime$ for ${\bm \phi}^\prime$ or $N\times N$ for $\bm \phi$).
Note that the second set of conditions enforces agreement of source and sink vectors between neighboring time slices up to the arbitrary, real, and time-dependent proportionality constants $w(\tau_k)$.
In particular, the constraints require $\psi(\tau_{k-1}) = w(\tau_k) \psi(\tau_k)$ for the source vectors (for each value of $k$) and similar relations for the sink vectors.
Combining these with the normalization conditions, one finds 
\begin{eqnarray}
\frac{1}{ {{\bm w}^\prime}^2} = \frac{1}{Q} \sum_{k=0}^{Q-1} \prod_{j=0}^k \frac{1}{ {w^\prime}^2(\tau_j) }\ , \qquad
\frac{1}{ {\bm w}^2} = \frac{1}{Q} \sum_{k=0}^{Q-1} \prod_{j=0}^k \frac{1}{ w^2(\tau_j) }\ ,
\end{eqnarray}
with $w^\prime(\tau_0) = w(\tau_0) \equiv 1$.
Generally the weights may be specified arbitrarily, however, a particularly simple but useful choice is one which satisfies
\begin{eqnarray}
w^\prime(\tau_k) w(\tau_k) = \frac{{\psi^\prime}^\dagger(\tau_k) C(\tau_k) \psi^\prime(\tau_k) }{{\psi^\prime}^\dagger(\tau_k) C(\tau_{k-1}) \psi^\prime(\tau_k) }
\label{eq:weights}
\end{eqnarray}
so that the correlator evaluated at each time slice contributes equal weight in the optimization.

The fixed source and unconstrained sink optimization of \Sec{fixed_source_unconstrained_sink_sn} and the unconstrained source and sink optimization of \Sec{unconstrained_source_sink_sn} may be extended to multiple time slices using the results of \Sec{fixed_source_constrained_sink_sn} and \Sec{constrained_source_constrained_sink_sn}.
In particular, one may use the solutions \Eq{constrained_sink_soln} and \Eq{constrained_source_sink_soln}, respectively, to impose the temporal constraints given by \Eq{temporal_constraints}.
In the former case, one should take $w(\tau_k)=1$ for all $k$ (and $w^\prime(\tau_k)$ chosen arbitrarily), whereas in the latter one may take $w^\prime(\tau_k)=w(\tau_k)$.
In either case, to a first approximation, one may take $w^\prime(\tau_k) w(\tau_k) \approx e^{E_0}$ for the optimization.
This approximation may then be refined using the optimized solutions and \Eq{weights} in successive iterations of the optimization procedure.
Further constraints on $\psi^\prime(\tau_k)$ and/or $\psi(\tau_k)$, such as those of \Sec{fixed_source_constrained_sink_sn} and \Sec{constrained_source_constrained_sink_sn}, may be imposed by considering direct product constraint conditions of the form $({\phi^\prime}^\dagger\otimes {{\bm\phi}^\prime}^\dagger) {\bm\psi}^\prime = 0$ and $(\phi^\dagger\otimes {\bm\phi}^\dagger) {\bm\psi} = 0$, where ${\bm \phi}^\prime$ ($\bm \phi$) are the $Q\times (Q-1)$ block matrices defined above, and $\phi$ ($\phi^\prime$) are $N\times N$ ($N^\prime \times N^\prime$) time-independent constraint matrices associated with each time slice.

\subsection{General properties of optimization}
\label{sec:sn_summary}

Finally, let us summarize some intuitive and easily proved properties of the results of this section, as well as their implications.
Within each of the optimization strategies discussed above, the maximum achievable signal/noise for any $N^\prime\times N$ matrix of correlators $C$ is greater than or equal to the maximum signal/noise achieved by considering any of its sub-matrices.
This result follows from the fact that the critical vectors found for any sub-matrix lie within a subspace of the possible vectors of the full matrix.
It is therefore always advantageous to consider the optimization strategies discussed here using the largest possible basis of observables allowed within the confines of the available computational resources.
In the context of lattice QCD, fermion propagators generated from a given quark source are computationally costly, so obtaining a large basis of source observables may not be practical.
However, highly efficient algorithms have recently been developed for performing quark-contractions at the sink (with an eye towards performing first principles nuclear physics calculations), making the generation of a large basis of sink observables not only feasible, but computationally negligible compared to the other stages of the lattice simulation \cite{PhysRevD.81.111504,PhysRevD.82.014511,PhysRevD.86.054507,Doi:2012xd,PhysRevD.87.094513,PhysRevD.87.114512}.

As with any extremization problem, imposing constraints on the search domain can only decrease the maximum achievable signal/noise.
It follows that the maximum signal/noise is achieved via the maximal signal/noise optimization described in \Sec{max_sn}, followed by that of \Sec{unconstrained_source_sink_sn} (as previously discussed, these two approaches converge in the late time limit).
Imposing further constraints by either fixing the source, as in \Sec{fixed_source_unconstrained_sink_sn}, or by imposing orthogonality conditions on the sink, as in \Sec{fixed_source_constrained_sink_sn}, can only further diminish the signal/noise enhancement of the correlator.
It is important to note that imposing such constraints may ultimately lead to advantages, however, when it comes to fitting correlators and extracting energies, as they allow for greater control over the excited state contamination introduced by signal/noise optimization.
In \Sec{correlator_optimization_strategies}, we will discuss further how various strategies may be combined to yield the greatest signal/noise of the final result, namely, the energies extracted from correlators.

\section{Correlator optimization strategies}
\label{sec:correlator_optimization_strategies}

Let us now consider the application of the optimization methods introduced in \Sec{signal_noise_optimization} toward enhancement of signal/noise in correlation functions, and ultimately to the reduction of uncertainties in the energies extracted from them.
In the above discussions, we have shown how one may optimize the signal/noise based on a particular time slice (or multiple time slices), however in practice, carrying out such an optimization may, in turn, introduce excited state contamination at early times, potentially nullifying any gains in signal/noise of the extracted energies.
It is therefore crucial that we understand the interplay between signal/noise enhancement and excited state contamination in the various strategies we consider.
Here, we discuss possible approaches, which combine the ideas of signal/noise-optimization in \Sec{signal_noise_optimization} and source-optimization reviewed in \Appendix{source_optimization}, with the mindset of balancing such considerations.
In \Sec{applications}, we will explore the performance of some of these methods using numerical data from lattice QCD.

\subsection{Signal/noise optimized source and sink}

In this first scheme, one begins by choosing reference times, $\tau_\textrm{s}$ and $\tau_\textrm{n}$, at which to evaluate the signal and noise correlators ($C$ and $\Sigma^2$), appearing in \Eq{signal_noise}.
Although the signal/noise of the correlator is, strictly speaking, only properly defined for $\tau_\textrm{s}=\tau_\textrm{n}$, these times may be chosen independently for the purpose of signal/noise optimization.
When contributions from higher energy states in $C$ and $\Sigma^2$ (evaluated at $\tau_\textrm{s}$ and $\tau_\textrm{n}$, respectively) are negligible, one may verify that the choice $\tau_\textrm{s}\neq\tau_\textrm{n}$ leads at most to an overall time-dependent re-scaling of $\theta$, which has little impact on the solutions $\psi^\prime_0$ and $\psi_0$ in the limit of infinite statistics.

The advantage of using separate time slices to define the signal and noise becomes apparent when one considers the statistical uncertainties in $C$ and $\Sigma^2$, separately. 
As was argued in \Sec{introduction}, the signal/noise associated with the stochastic estimate of the signal decays exponentially with time at a rate governed by $ E_0 - \frac{1}{2} \tilde E_0$.
On the other hand, the signal/noise associated with the stochastic estimate of the noise itself (i.e., the second moment of the correlator distribution divided by the square root of the fourth moment) is often approximately independent of time at late times.
In the case of QCD, for example, according to an argument attributed to Savage (see \cite{Nicholson:2012xt} and references therein), both the second and fourth moments of the individual correlator distribution have a fall-off governed by pions, the latter of which falls off at twice the rate of the former, up to corrections attributed to interactions.
Thus the signal/noise decay rate associated with a stochastic estimate of the variance of a baryonic correlator is approximately constant.

From a practical standpoint, one may find it advantageous to choose $\tau_\textrm{s} < \tau_\textrm{n}$ in order to ensure that the estimates for both $C$ and $\Sigma^2$ individually possess relative statistical uncertainties which are much smaller than unity.
If $\tau_\textrm{s}$ were chosen so large that the uncertainties in $C$ are comparable to $C$ itself, then the errors on the signal/noise optimized source and sink vectors will possess $\calO(1)$ errors or larger, rendering them unreliable.
Since the signal/noise associated with the variance estimate is often independent of time, however, $\tau_\textrm{n}$ may in principle be chosen arbitrarily large, and in particular large enough so as to remove the excited state corrections, $\tilde\Delta$, defined in \Sec{signal_noise_conventions}.

Accounting for such considerations, one may use \Eq{iterative_dual_relation_soln} to obtain source and sink vectors, $\psi^\prime_0$, and $\psi_0$, which may be used to construct a single signal/noise-optimized correlator.
Using the methods of \Sec{sn_basis}, one may also develop an entire source and sink basis in which the signal/noise of the correlation matrix is ordered.
As previously noted, although this strategy generally achieves an enhanced signal/noise over all times (despite the analysis being carried out at a particular $\tau_\textrm{s}$ and $\tau_\textrm{n}$), it provides little control over the degree of excited state contamination introduced by the optimization at earlier times.
Such excited state contamination may be removed by imposing constraints on the optimization, or accounted for by a subsequent multiexponential fit to the various sub-matrices (or individual matrix elements), with emphasis placed on those with greatest signal/noise.

\subsection{Variational source and signal/noise optimized sink }

Let us assume that for a given correlator, $C$, we have obtained a source vector, $\psi_n$, which has a maximal overlap with the ground state, or a particular excited state labeled by $n$.
In the case where $C=C^\dagger$, this vector may have been obtained by solving the generalized eigenvalue problem represented by \Eq{generalized_ev} for some reference time $\tau_0$, and then selecting a source vector solution $\psi_n(\tau_1)$ at a second time slice $\tau_1>\tau_0$ as a representative source-optimized vector.
The time slice $\tau_1$ should be chosen such that the source vector provides the best possible approximation of the $n$th eigenstate of the Hamiltonian, up to corrections associated with excited state contamination and errors which may be attributed to statistical fluctuations in the estimate.
For non-Hermitian matrices $C$, the Matrix-Prony method or some other technique \cite{Fleming:2004hs,Beane:2009kya,Beane:2009gs} may provide a viable alternative for finding such a source with optimal overlap onto a particular state.

If the source vector $\psi_n(\tau_1)$ had a perfect overlap with the ground state, then the correlator ${\psi^\prime}^\dagger C(\tau) \psi_n(\tau_1)$ for any $\psi^\prime$ would be monotonically decreasing function of $\tau$, despite the fact that the sink vector differs from the source vector.
In practice this is not the case, however, since a loss in positivity can enter through the exponentially small excited state contamination and through the statistical noise entering into the estimate of $\psi_n(\tau_1)$.
For a sufficiently large basis, and for a well-chosen source vector, such a loss in positivity is expected to be negligible until late time separations, at which point a small residual ground state overlap could dominate%
\footnote{For an excited state, the correlator will decay exponentially for intermediate times with a decay rate equal to the excited state energy. However, one might expect an eventual (and potentially sudden) transition to an exponential with decay rate equal to the ground state energy arising from a small but finite overlap attributed to statistical fluctuations.
If one already has a precise estimate of the ground state, this contribution can be incorporated into the fit model.
In the case of the ground state, the finite temporal extent of the lattice geometry may give rise to thermal contamination in which hadronic states propagate both forward and backward in time.
Depending on interactions, such states may appear to have energies below the ground state but with exponentially suppressed overlap factors \cite{Beane:2009gs}}.
Assuming such an optimized source vector is found, one may use the methods of \Sec{fixed_source_unconstrained_sink_sn} to optimize the sink vector at reference times $\tau_s$ and $\tau_n$ so as to maximize the signal/noise of ${\psi^\prime}^\dagger C(\tau) \psi_n(\tau_1)$.
Alternatively, one may optimize the sink vector using the methods of \Sec{fixed_source_constrained_sink_sn} in order to impose constraints on $\psi^\prime$, such as that it should be orthogonal to $C(\tau_0) \psi_m(\tau_1)$ for $m<n$ (up to fluctuations in the estimate).
As argued in \Sec{toy_model}, such a strategy may offer a compromise between a reduction of excited state contamination, and enhancement of signal/noise that is favorable for extracting energies with minimal uncertainty.

\subsection{Variational method with noisy correlator projection}
\label{sec:noisy_excisions}

Let us next consider individual correlators that are Hermitian, and given by
\begin{eqnarray}
\tilde \calC(\tau) = C(\tau_0)^{1/2} \calC(\tau) C(\tau_0)^{1/2} \ , 
\end{eqnarray}
for some reference time $\tau_0$, such that $\tilde C(\tau) = \langle \tilde \calC(\tau) \rangle$.
Note that the eigenvalues of $\tilde C(\tau)$, denoted by $\lambda_n(\tau)$, will coincide with the principle correlators obtained from a variational analysis of the correlator $C(\tau)$ (e.g., the solutions to the generalize eigenvalue problem, defined by \Eq{generalized_ev}).
Note that the normalized correlator is defined so that $\tilde C(\tau_0) = 1$ and $\lambda_n(\tau_0)=1$.
We may further consider the critical vectors $\psi_0 = \psi^\prime_0$ corresponding to a solution to \Eq{dual_relations} applied to $\tilde C$ at time slices $\tau_\textrm{s}$ and $\tau_\textrm{n}$ and the signal/noise basis developed from them.
Let us define the individual correlators, $\tilde \calG(\tau)$, in this basis and the expectation value, $\tilde G(\tau) = \langle \tilde \calG(\tau) \rangle$, which retains the property that $\tilde G(\tau_0)=1$.
Since the transformation to the signal/noise basis is unitary, $\tilde G(\tau)$ has eigenvalues $\tilde \lambda_n(\tau) = \lambda_n(\tau)$ which coincide with those of $\tilde C(\tau)$.

Note that the rows and columns of $\tilde \calG(\tau)$ are structured according to signal/noise.
One may therefore define a $(N-R)\times (N-R)$ dimensional ``reduced'' correlator matrix $\tilde \calG^{(r)}$ by eliminating the last $R$ rows and columns, which are associated with the matrix elements of $\tilde G(\tau)$ that have the smallest signal/noise.
Diagonalizing the reduced correlator matrix produces eigenvalues
\begin{eqnarray}
\tilde \lambda^{(r)}_n(\tau) = e^{-E_n(\tau-\tau_0)}\ ,
\end{eqnarray}
up to relative corrections to the $n$th eigenvalue, scaling as $e^{-(E_{N-R}-E_n)\tau}$ in the range $\tau_0<\tau<2 \tau_0$.
Although the systematic effects from excited state contamination in this approach are parametrically larger than those of the original matrix, a reduction in statistical uncertainties attributed to excision of the noisiest components of $\tilde G(\tau)$ may offer some advantages in situations where the systematic errors are smaller than statistical errors.
In such situations, the method offers an avenue for placing both sources of uncertainty on an equal footing, and is in some sense a generalization of the pruning techniques discussed in \cite{Lichtl:2006dt}.

\section{Applications: lattice QCD correlation functions}
\label{sec:applications}

In this section, an exploration of the optimization strategies outlined in \Sec{correlator_optimization_strategies} is performed for matrices of single hadron correlation functions obtained from lattice QCD studies.
Particularly, we consider a Hermitian $26\times26$ matrix of rho meson correlators obtained from the Hadron Spectrum Collaboration, and approximately Hermitian $5\times 5$ matrices of pion, proton and delta correlators. 
The former data was used in a study reported in \cite{PhysRevD.87.034505}, whereas the latter data are closely related to those used in \cite{Beane:2009kya}.
We refer the reader to those studies (also see related works \cite{PhysRevD.82.034508}) for complete details of the measurements, and here describe only the relevant aspects which are required for this exploration.

All correlator measurements were performed on anisotropic gauge-field configurations that were generated by the Hadron Spectrum Collaboration using an $N_f=2+1$ flavor tadpole-improved clover fermion action and Symanzik-improved gauge action \cite{PhysRevD.78.054501,PhysRevD.79.034502}.
The rho meson correlators were measured on lattices of size $24^3 \times 128$, whereas the other correlators were measured on lattices of size $20^3\times 128$.
All lattices were generated with an anisotropy factor $b_s/b_\tau \approx 3.5$, where $b_s=0.1227(8)\,\textrm{fm}$ and $b_\tau$ are the spatial and temporal lattice spacings, respectively, and with quark masses corresponding to a pion mass, $m_\pi \approx 390\textrm{MeV}$, and a kaon mass, $m_K \approx 546\textrm{MeV}$.
For the remainder of this section, we work in lattice units, explicitly setting $b_\tau$ to unity.

Rho correlators were measured on an ensemble of $\calN=566$ configuration using a basis of $26$ zero-momentum projected operators belonging to the $T_1^-$ irreducible representation of the octahedral group with parity, $O_h^D$.
The other correlators were generated on an ensemble of $\calN=305$ configurations using interpolating operators composed of $\calO(30)$ randomly placed Gaussian-smeared sources and zero-momentum projected Gaussian-smeared sinks, leading to a stochastically approximated wall source (these correlator matrices are therefore not strictly Hermitian, but are numerically close to Hermitian).
Throughout the study, we take $\calC = \calC^\dagger$ for each individual correlator in the ensemble, as is allowed by the symmetries.

\subsection{Analysis details}
\label{sec:analysis_details}

To illustrate the interplay between signal/noise enhancement and excited state contamination, it is helpful to consider plots of the effective mass, defined by
\begin{eqnarray}
m_{\textrm{eff}}(\tau) = -\frac{1}{\Delta\tau} \log \frac{{\psi^\prime}^\dagger C(\tau+\Delta\tau) \psi}{ {\psi^\prime}^\dagger C(\tau) \psi}\ .
\end{eqnarray}
In the case where ${\psi^\prime}^\dagger Z_n^\prime \neq 0$ and $Z_n^\dagger \psi \neq 0$ for all $n$, the effective mass has a late-time behavior given by
\begin{eqnarray}
m_{\textrm{eff}}(\tau) \sim E_0 + \left( \frac{{\psi^\prime}^\dagger Z_1^\prime Z_1^\dagger \psi}{{\psi^\prime}^\dagger Z_0^\prime Z_0^\dagger \psi } \right) \frac{1-e^{-(E_1-E_0)\Delta\tau}}{\Delta\tau} e^{-(E_1-E_0)\tau} + \ldots\ ,
\end{eqnarray}
up to exponentially small corrections attributed to excited state contamination.
For source-optimized correlators constructed such that ${\psi^\prime}^\dagger Z_m^\prime = 0$ or $Z_m^\dagger \psi = 0$ for states $m<n$, the late-time limit of $m_{\textrm{eff}}(\tau)$ will converge to the energy $E_n$, rather than $E_0$.
For this study, we always take $\Delta\tau =1$.

Fully correlated, multiexponential, $\chi^2$ minimizing fits are performed on correlation functions over a time range $\tau\in[\tau_\textrm{min},\tau_\textrm{max}]$ in order to extract the low-lying energies of the system.
Statistical uncertainties on all quantities (e.g., correlation functions, effective masses, extracted energies, etc.) are determined using bootstrap resampling.
Systematic uncertainties associated with excited state contamination in the correlator fits are accounted for by varying the minimum time value $\tau_\textrm{min}$ of the temporal range over which the fit was performed.
Similarly, systematic errors associated with finite temperature effects (attributed to ``around the world'' propagation of states in time) are studied by varying the maximum time value $\tau_\textrm{max}$ in the temporal range over which the fits were performed.
For the proton, delta, and rho, such finite temperature effects are found to be negligible; for the pion, however, such effects are significant and accounted for by including a backward-propagating pion in the multiexponential fit.

\subsection{Signal/noise landscape for the rho}
\label{sec:hadron_landscape}

Before proceeding with the full analysis of hadron data, it is instructive to explore the signal/noise landscape for an approximately diagonal  $2\times 2$ rho correlator matrix, the results of which may be directly compared with those of the toy model introduced in \Sec{toy_model}.
Starting from the original $26\times 26$ matrix of rho correlation functions, we use the variational approach of \Appendix{source_optimization} with $\tau_0=3$ to find approximate eigenstates of the Hamiltonian.
The associated source-optimized source and sink vectors are selected at time slice $\tau_1 = 14$.
We then consider the signal/noise associated with correlators evaluated at time slices $\tau_\textrm{s} = \tau_\textrm{n} = 14$, working in a basis where the estimated correlation function is diagonal.
For this analysis, we study the signal/noise as a function of mixing angles between only two approximate eigenstates; this is achieved by truncating the $26\times 26$ individual correlator matrices $\calC$ to a $2\times 2$ sub-matrix in the diagonal basis at time slice $\tau_\textrm{s} = \tau_\textrm{n}$.
Note that in this basis, the individual correlators $\calC$ themselves are not diagonal, but the expectation value $C=\langle \calC \rangle$ is diagonal up to statistical fluctuations and systematic errors associated with a finite basis for $\tau \gtrsim \tau_0$.

\begin{figure} 
\includegraphics[width=\figwidth]{\figdir 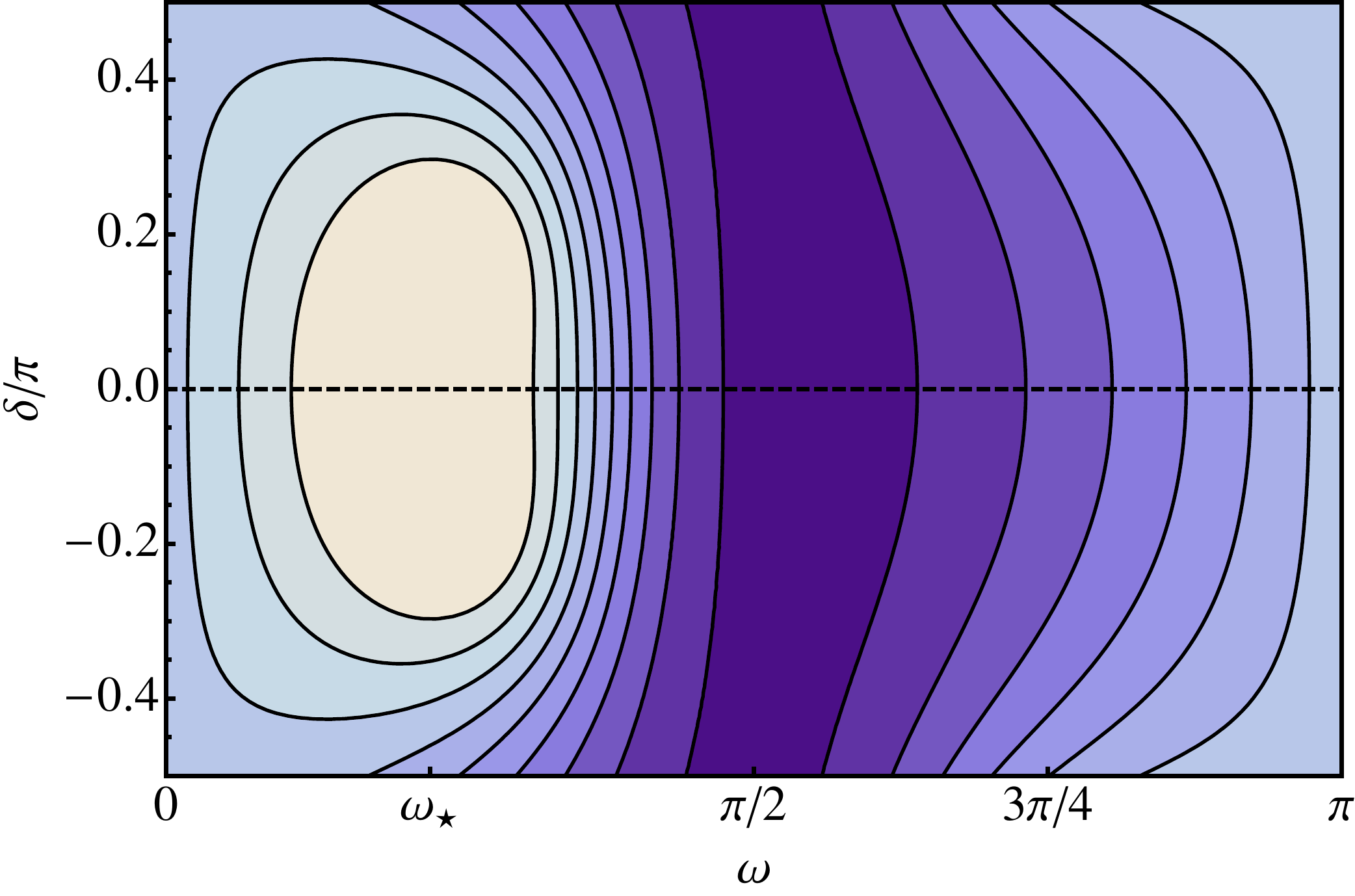}
\includegraphics[width=\figwidth]{\figdir 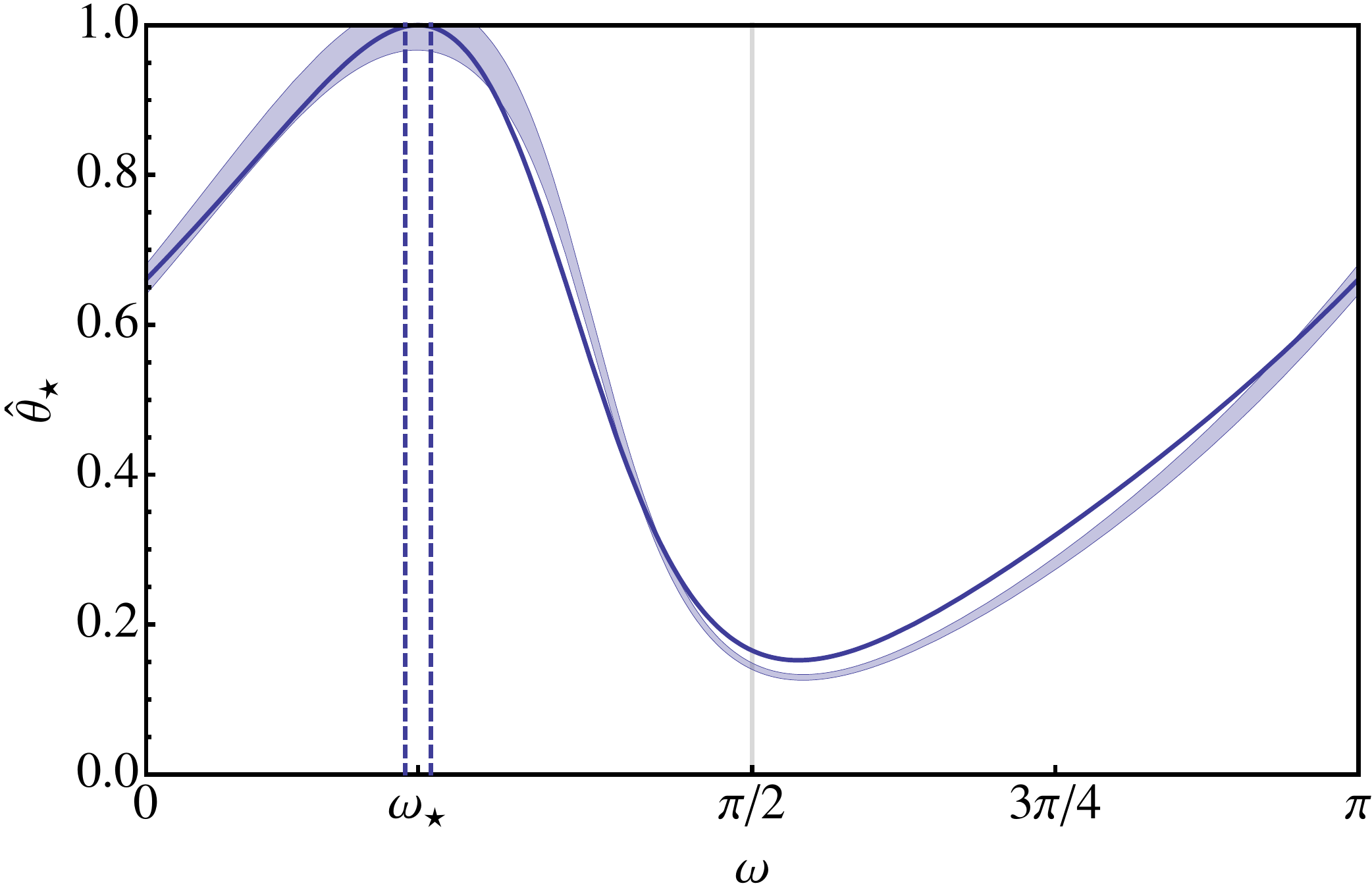}
\caption{\label{fig:rho_InterpNA_landscape_Star}%
Left: Landscape of the normalized signal/noise $\hat\theta_\star(\omega,\delta)$ for a $2\times 2$ rho correlator with mixing between the ground state and second excited state.
The stochastically estimated ground and excited states corresponds to $\omega=0$ and $\omega=\pi/2$, respectively.
Right: Normalized signal/noise (indicated by $1\sigma$ error bands) as a function of the mixing angle $\omega$ for fixed $\delta=\delta_\star$.
The maximum signal noise is achieved at the angle $\omega_\star$, indicated by $1\sigma$ error bands (dashed, vertical lines).
The solid curve corresponds to the two-state theoretical model for $\hat\theta_\star(\omega,0)$ using \Eq{sn_parametrization}.
}
\end{figure}

\begin{figure} 
\includegraphics[width=\figwidth]{\figdir 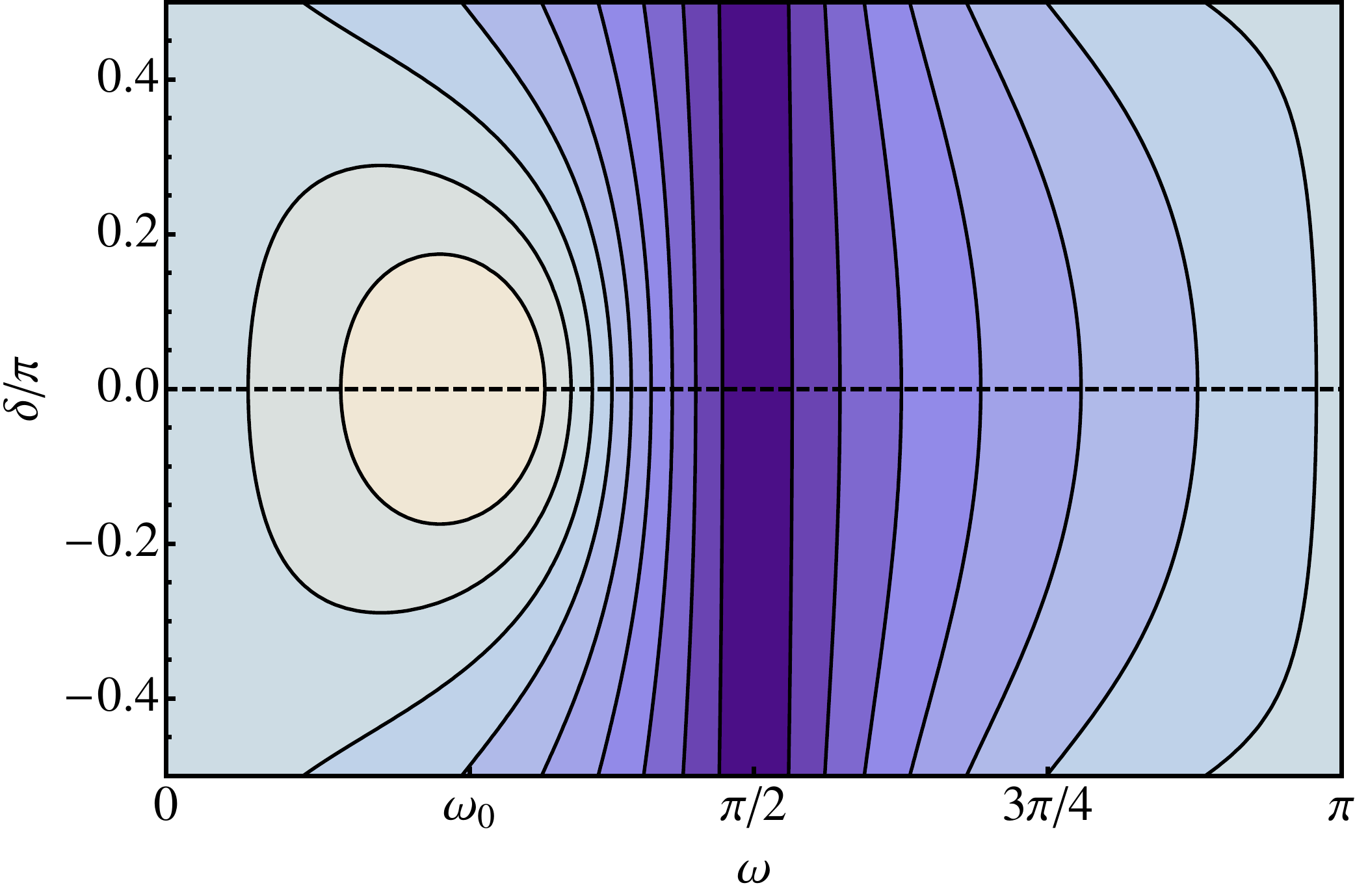}
\includegraphics[width=\figwidth]{\figdir 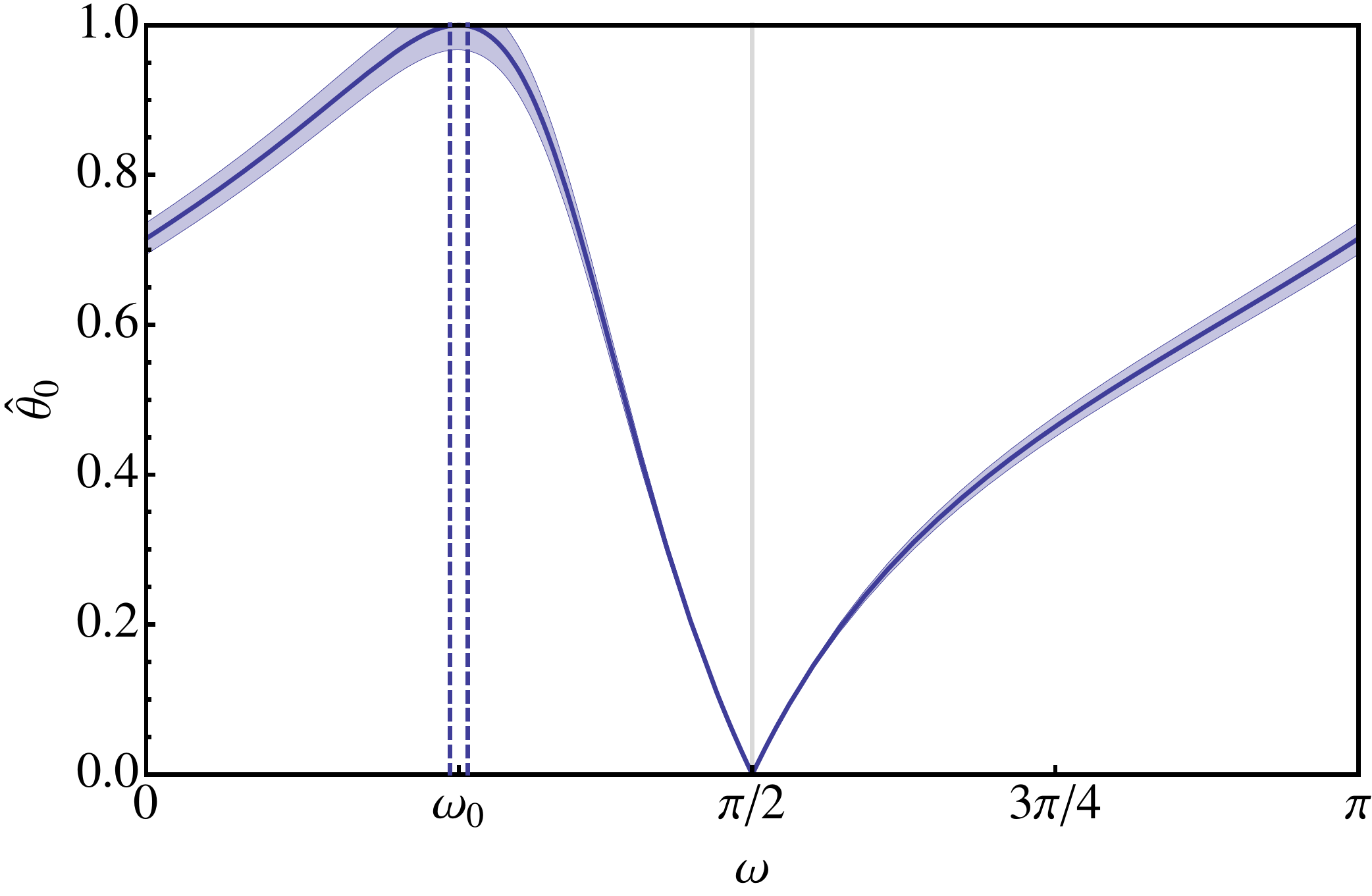}
\caption{\label{fig:rho_InterpNA_landscape_0}%
Same as \Fig{rho_InterpNA_landscape_Star} for $\hat\theta_0(\omega,\delta)$.
}
\end{figure}

\begin{figure} 
\includegraphics[width=\figwidth]{\figdir 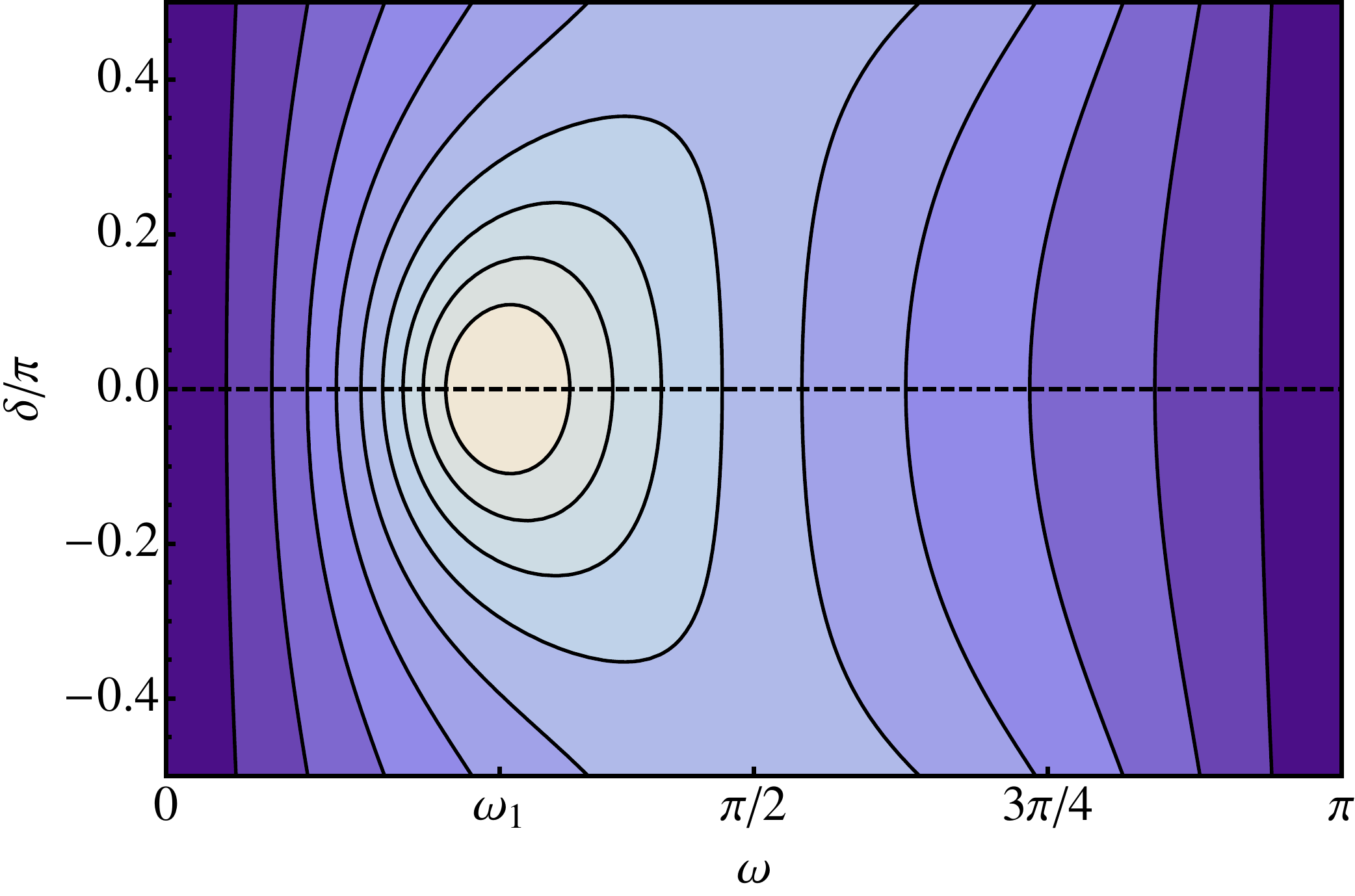}
\includegraphics[width=\figwidth]{\figdir 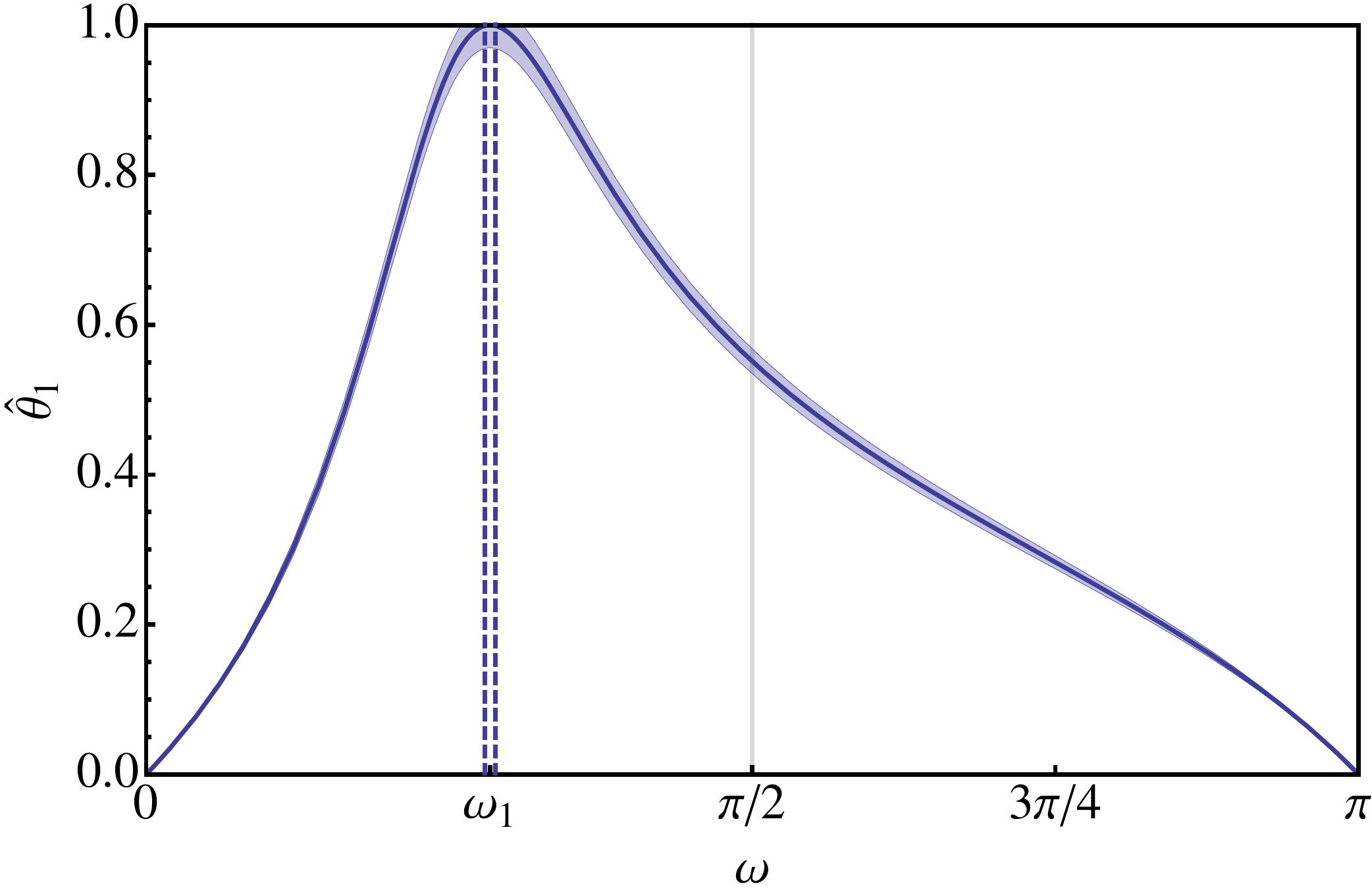}
\caption{\label{fig:rho_InterpNA_landscape_1}%
Same as \Fig{rho_InterpNA_landscape_Star} for $\hat\theta_1(\omega,\delta)$.
}
\end{figure}

For a given $2\times 2$ truncation of the rho correlator matrix, and a sink vector parametrized by \Eq{two_vector}, we explore how the signal/noise varies as a function of the mixing angles $(\omega,\delta)$.
As in \Sec{toy_model}, we consider cases where the source vector is equal to either the sink vector, ground state vector ($\psi_0$), or excited state vector ($\psi_1$).
For the correlator matrix considered, we find that the signal/noise is only marginally enhanced by allowing mixing between the ground and first excited state, whereas the enhancement is considerably more significant when mixing the ground and second excited state.
For the purpose of demonstrating the method, and visualization of the signal/noise landscape, we therefore concentrate for the moment on the latter case.
In more practical applications, such as those considered in the following subsection, mixing between all $26$ low-lying states is considered in order to achieve a maximal enhancement in signal/noise.

A density plot of the signal/noise landscape $\theta_\star$ for the rho is presented in \Fig{rho_InterpNA_landscape_Star} (left) for the case of equal source and sink vectors.
In this example, a maximum enhancement $R_\star = 1.51(6)$ is achieved at $\omega_\star = 0.70(3)$ when $\delta_\star=0$.
A plot of the signal/noise as a function of the mixing angle $\omega$ for fixed $\delta=0$ is plotted in \Fig{rho_InterpNA_landscape_Star} (right).
The error bands in this plot indicate $1\sigma$ uncertainties associated with the signal/noise.
Note that at $\omega=\pi/2$, the normalized signal/noise is nonzero due to the presence of excited state contamination.
By extremizing the signal/noise, one may determine both $\omega_\star$ and $R_\star$, and by comparing the relative magnitudes of the diagonal elements of the correlator at $\tau_\textrm{s}=\tau_\textrm{n}$, one may determine the amount of excited state contamination $\Delta$ in the signal.
Using these values as input parameters for the two-state model function given by \Eq{sn_parametrization}, we obtain a curve (solid line in \Fig{rho_InterpNA_landscape_Star} (right)) that is qualitatively consistent with the signal/noise error bands shown in the figure.
Deviations are presumably attributed to finite $\tilde \Delta$ effects (excited states) in the noise, which have been omitted from the toy-model analysis.
Similar plots are displayed in \Fig{rho_InterpNA_landscape_0} (\Fig{rho_InterpNA_landscape_1}) for the case where the source is chosen to have maximal overlap with the ground (second excited) state.
The normalized signal/noise $\hat\theta_0$ ($\hat\theta_1$) has a maximum enhancement $\sqrt{R_0}=1.40(4)$ ($\sqrt{R_1}=1.82(6)$) at $\omega_0=0.81(2)$ ($\omega_1=0.89(1)$); using the results obtained by extremizing the signal/noise as input parameters for the model function, \Eq{sn_parametrization_partial}, we obtain curves which are also consistent with the signal/noise error bands shown in the figures.

Note that the numerical values obtained for $(R_\star,\omega_\star)$ and $(R_n,\omega_n)$ fail to satisfy the relations given by \Eq{R_ratio} and \Eq{omega_relations}.
This may be traced to the fact that corrections to the noise, governed by $\tilde \Delta$, were omitted from the model.
If $\tilde \Delta$ is not negligible, the presence of such effects may lead to significant deviations in the relations.
The size of the effects may be estimated by diagonalizing the noise correlator and comparing the lowest two eigenvalues; carrying out such an analysis yields $\tilde\Delta \approx 1/5$ for the choice of $\tau_\textrm{n}$ under consideration.
Despite the presence of $\tilde \Delta$ corrections, the functional form of the toy-model results given by \Eq{sn_parametrization} and \Eq{sn_parametrization_partial} nonetheless appear to provide a good qualitative description for the enhancement properties of the system as a function of $\omega$.
Evidently most of the finite $\tilde \Delta$ effects can be absorbed into the parameters $(R_\star,\omega_\star)$ and $(R_n,\omega_n)$ rather than leading to corrections to the functional form of the toy model expression for the signal/noise.

\begin{figure} 
\includegraphics[width=\figwidth]{\figdir 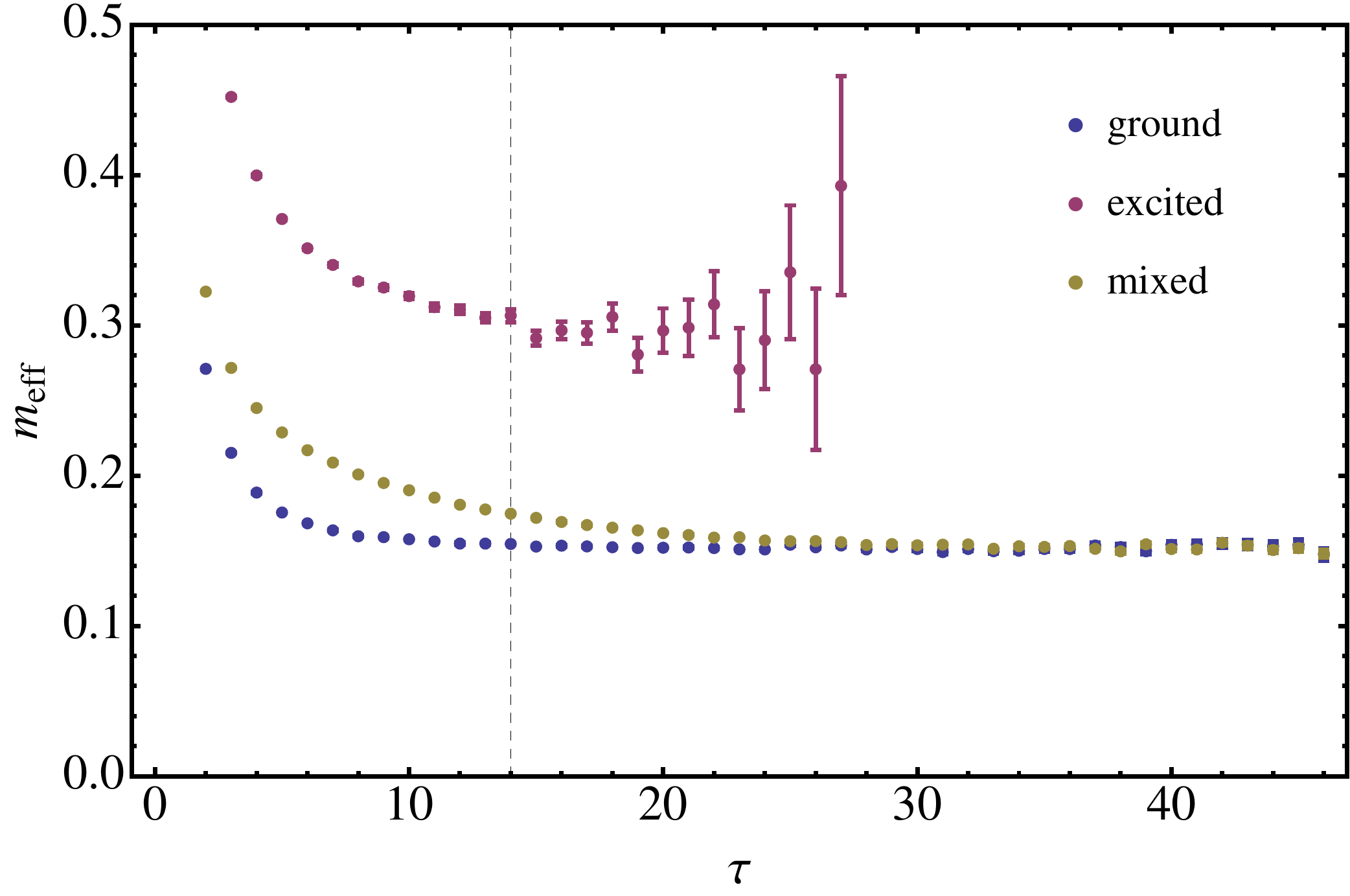}
\includegraphics[width=\figwidth]{\figdir 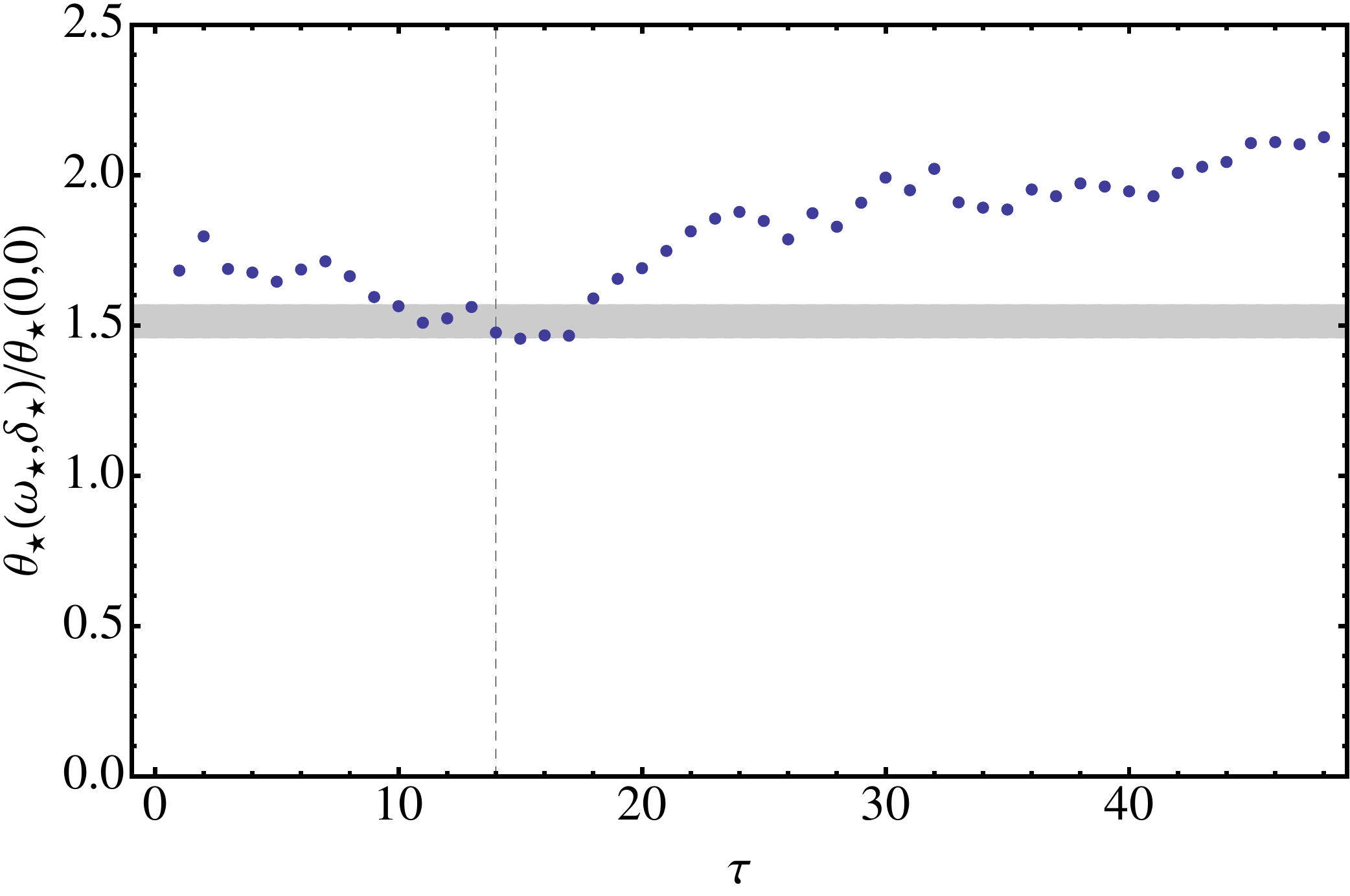}
\caption{\label{fig:rho_mEff_Star}%
Left: Effective mass plots associated with the correlators $\psi_0^\dagger C \psi_0$ (ground), $\psi^\dagger_1 C \psi_1$ (excited), and  $\psi^\dagger(\omega_\star,\delta_\star) C \psi(\omega_\star,\delta_\star)$ (mixed), where $\psi(\omega,\delta)$ is parametrized by \Eq{two_vector} and $C$ is an approximately diagonal $2\times 2$ rho correlator.
The optimal angles $(\omega_\star, \delta_\star)$ were obtained by optimizing $\theta_\star(\omega,\delta)$ at a fixed time slice $\tau_\textrm{s}=\tau_\textrm{n}$, indicated by the dashed line.
Right: Ratio of the optimized signal/noise $\theta_\star(\omega_\star,\delta_\star)$ and ground state signal/noise $\theta_\star(0,0)$ as a function of time, for time-independent $(\omega_\star,\delta_\star)$.
The solid band indicates the $1\sigma$ uncertainties on the enhancement factor $R_\star$ obtained from optimizing $\hat \theta_\star(\omega,\delta)$.
}
\end{figure}

\begin{figure} 
\includegraphics[width=\figwidth]{\figdir 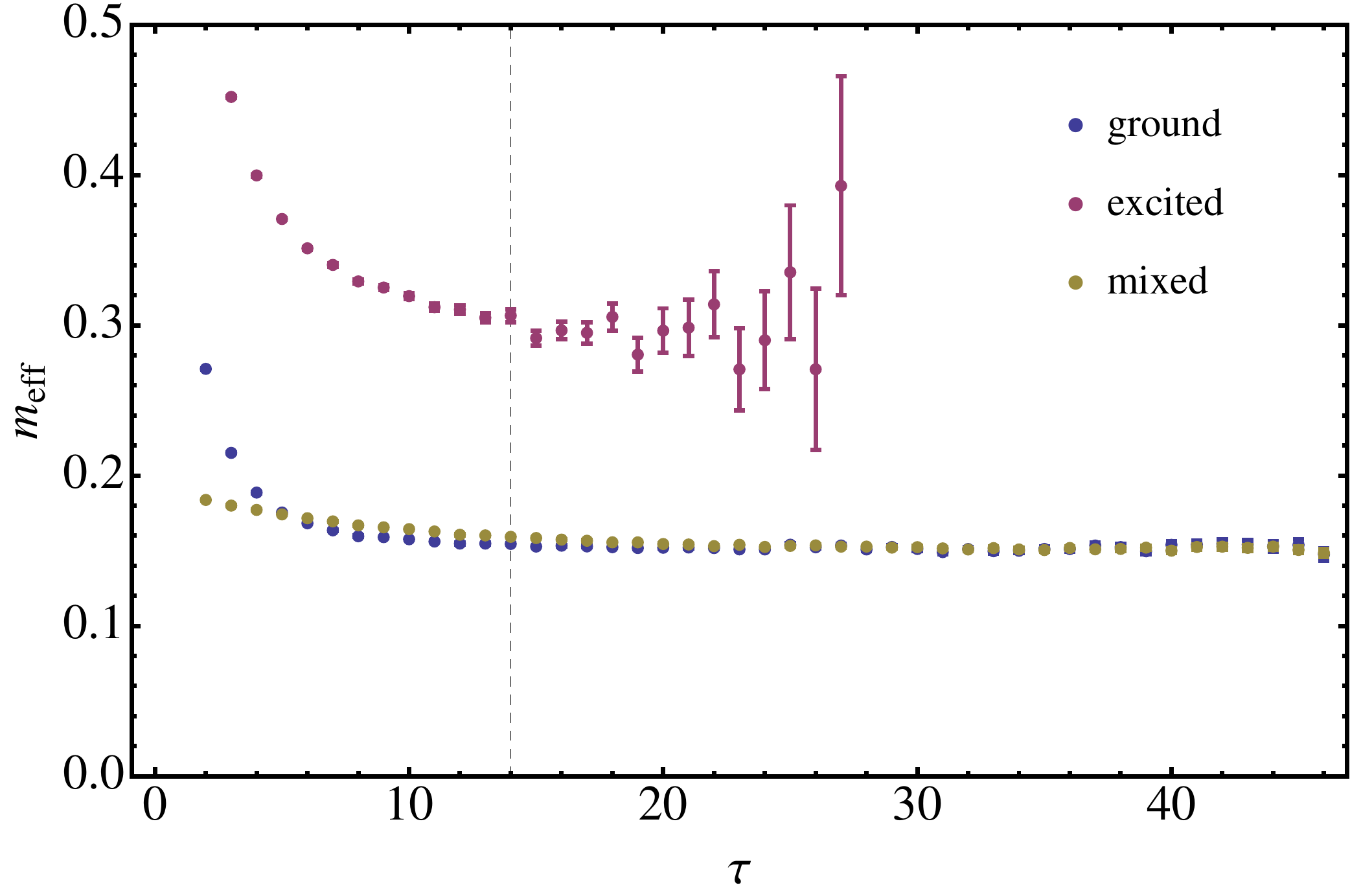}
\includegraphics[width=\figwidth]{\figdir 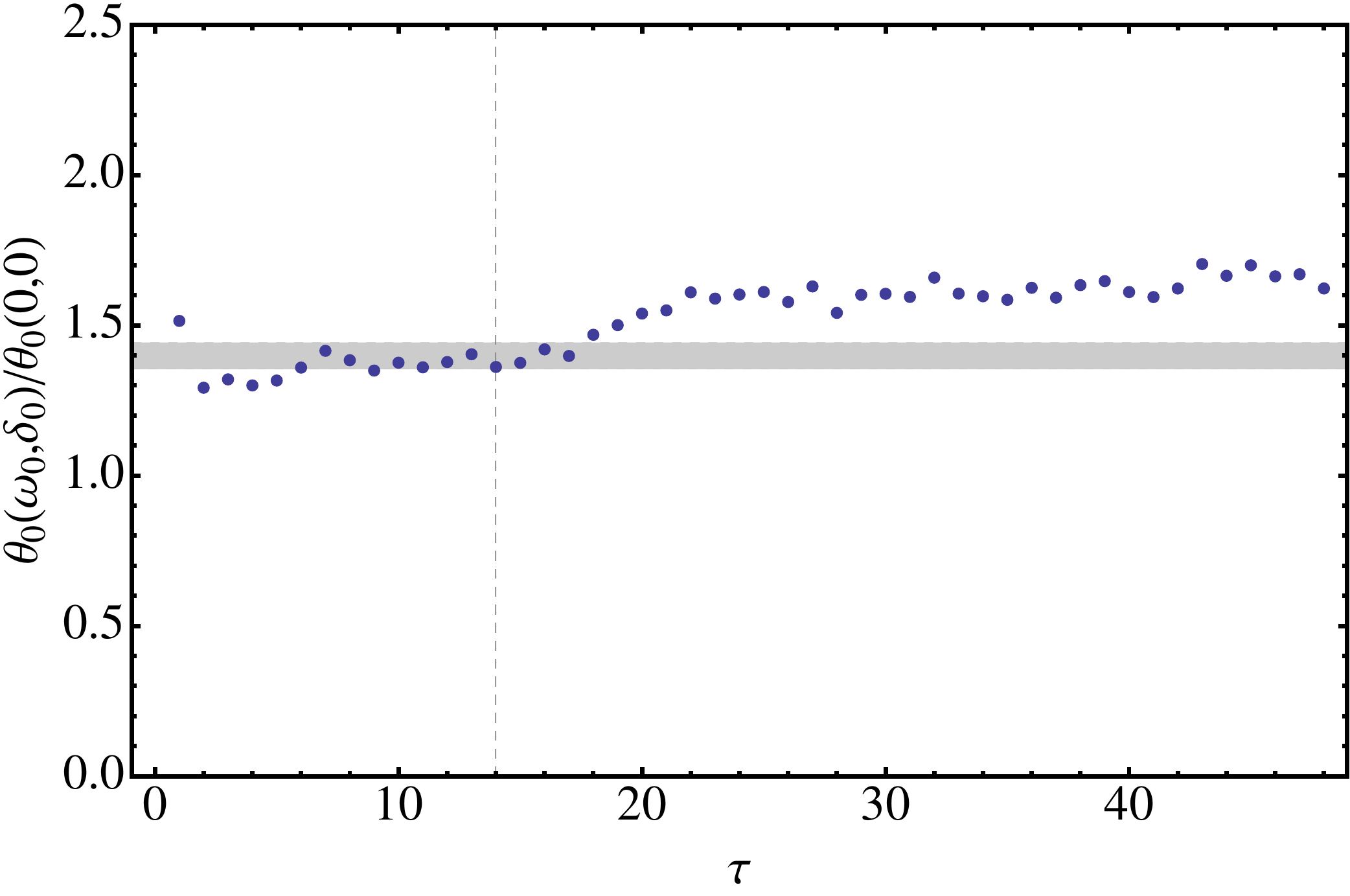}
\caption{\label{fig:rho_mEff_0}%
Same as \Fig{rho_mEff_Star} for the correlators $\psi^\dagger_0 C \psi_0$ (ground), $\psi^\dagger_1 C \psi_1$ (excited), and $\psi^\dagger(\omega_0,\delta_0) C \psi_0$ (mixed).
}
\end{figure}

\begin{figure} 
\includegraphics[width=\figwidth]{\figdir 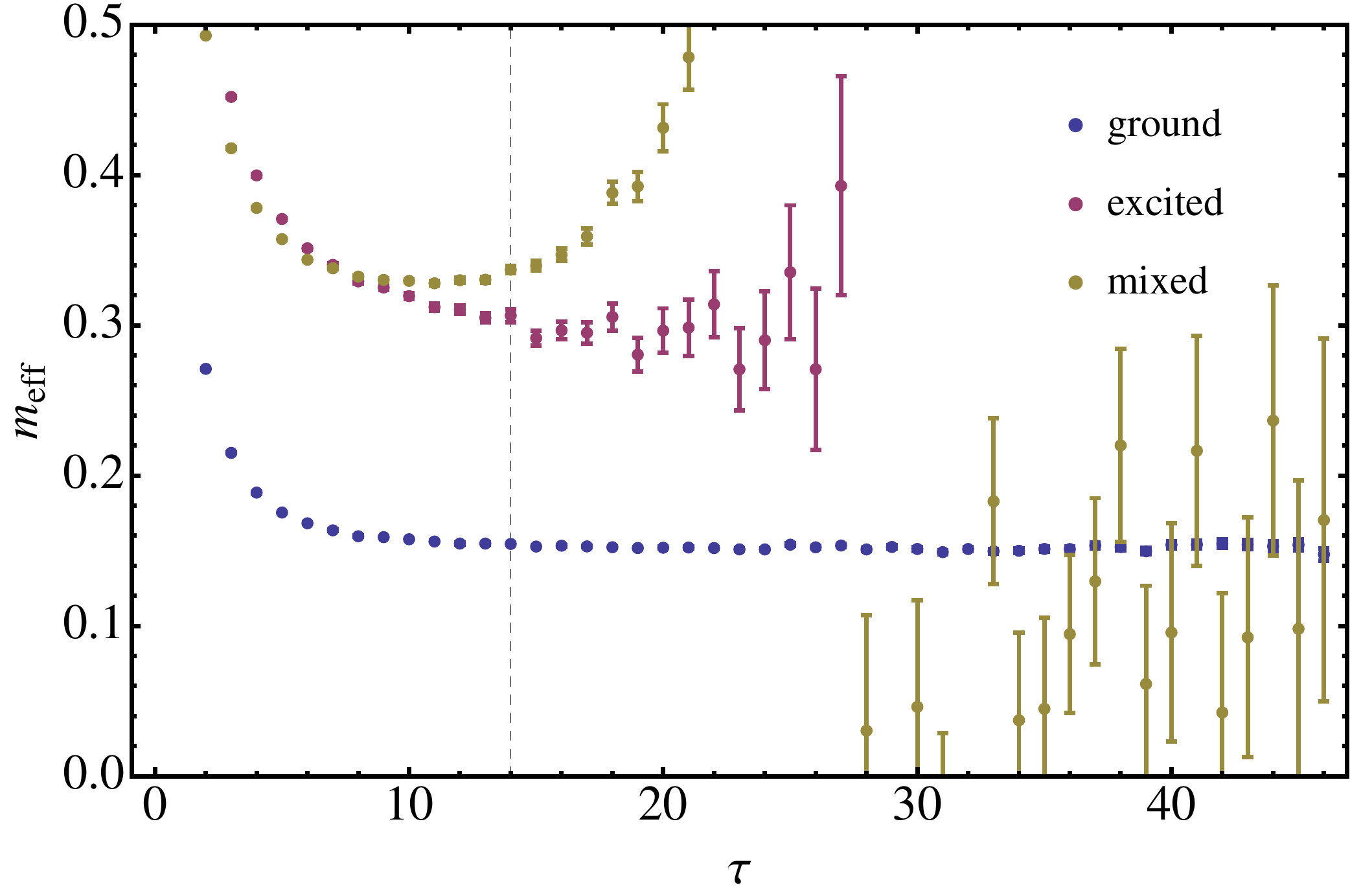}
\includegraphics[width=\figwidth]{\figdir 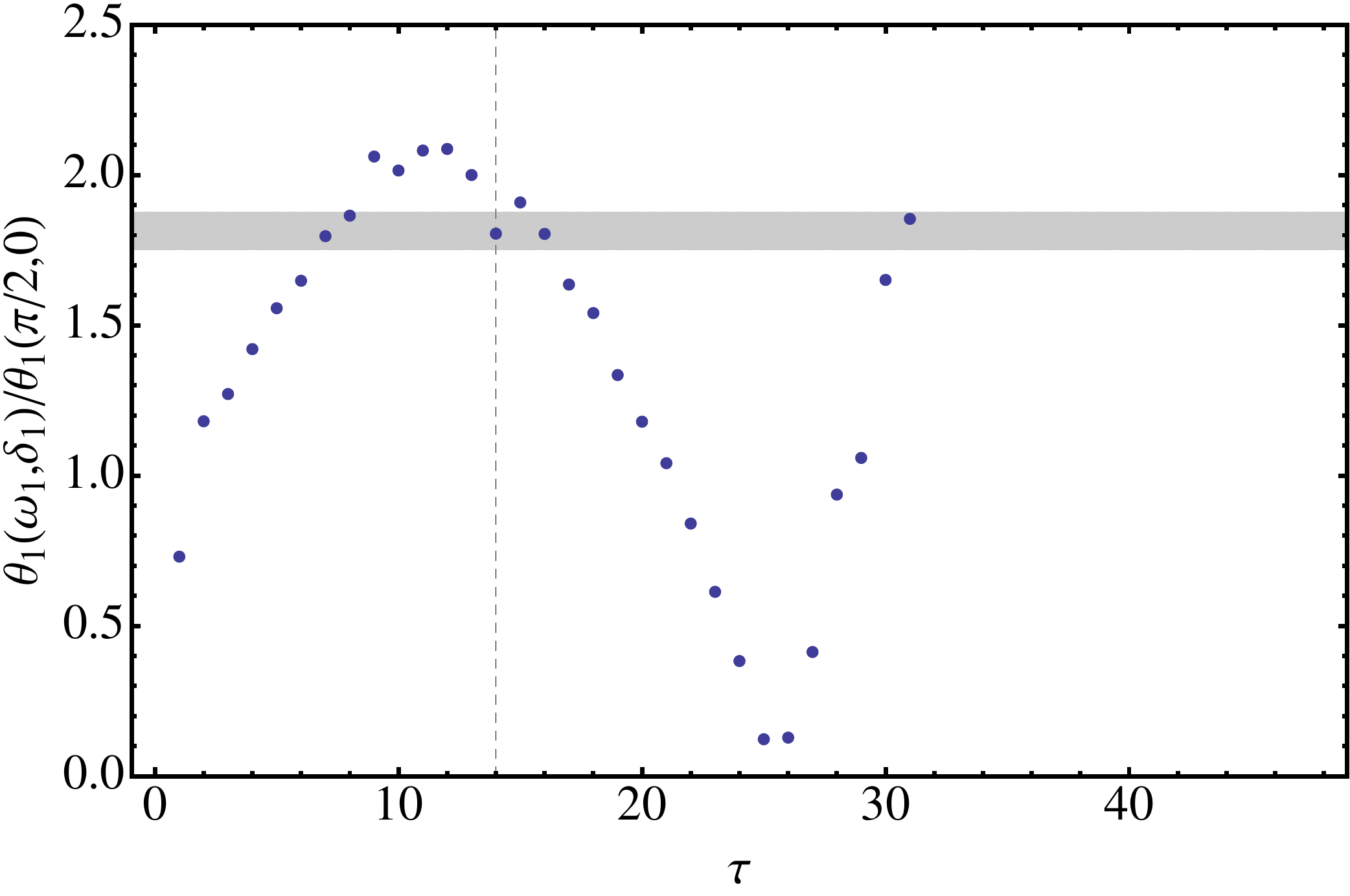}
\caption{\label{fig:rho_mEff_1}%
Same as \Fig{rho_mEff_Star} for the correlators $\psi^\dagger_0 C \psi_0$ (ground), $\psi^\dagger_1 C \psi_1$ (excited), and $\psi^\dagger(\omega_1,\delta_1) C \psi_1$ (mixed).
}
\end{figure}

In \Fig{rho_mEff_Star}, \Fig{rho_mEff_0} and \Fig{rho_mEff_1} (left), we demonstrate the impact of mixing ground and excited states on correlation functions over the full time extent.
In particular, we show the effective masses obtained for both the ground and second excited states.
These were constructed from source-optimized source and sink vectors obtained from a variational analysis at time slice $\tau_1$.
Overlaid on the same plots are effective masses for correlators constructed from the signal/noise-optimized vectors determined at $\tau_\textrm{s}=\tau_\textrm{n}$ (indicated by a vertical dashed line) using each of the three optimization strategies discussed.
A signal/noise-optimized correlator obtained for equal source and sink vectors is displayed in \Fig{rho_mEff_Star} (left).
In this case, one finds considerable excited state contamination in the signal/noise-optimized correlator at intermediate time extents compared to the source-optimized correlators, but also a significant enhancement in the signal/noise over the entire time range of the data.
Signal/noise-optimized correlators obtained with fixed, source-optimized sources are shown in \Fig{rho_mEff_0} and \Fig{rho_mEff_1} (left) for the ground and excited states, respectively.
In the former case, the excited state contamination is in large part removed.
In the latter case, however, small levels of ground state contamination in the estimate of the excited state source vector (presumably because of statistical fluctuations) becomes the dominant contribution to the correlator at late times.
The signal/noise optimized effective mass in \Fig{rho_mEff_1} (left) diverges at some intermediate time slice prior to converging to the value for the  ground state energy because of a relative sign in the ground and excited state overlap factors appearing in the correlator (for this construction, the source and sink differ, so positivity is not guaranteed).
For the signal/noise enhancement of the excited state correlator to be profitable in this case, one requires either an extremely precise estimate of the excited state source vector so as to minimize the ground state contamination, or a precise estimate of the ground state so that that its role may be included into the energy-extracting fit.
The conclusion drawn from this example is perhaps not so surprising: signal/noise enhancement of correlators associated with a given state is most profitable when mixing with states of higher energy.
To mix with lower energy states requires precision determinations of the source-optimized excited-state source vector.

In \Fig{rho_mEff_Star}, \Fig{rho_mEff_0}, and \Fig{rho_mEff_1} (right), we show the signal/noise enhancement achieved with signal/noise-optimized correlator, compared to either the ground or excited state correlators, as a function of time separation.
Indicated on these plots (horizontal bands indicating $1\sigma$ uncertainties) are the enhancement factors ($R_\star$, $\sqrt{R_n}$, respectively) found by the signal/noise optimization procedure at time slice $\tau_\textrm{s} = \tau_\textrm{n}$.
Although the signal/noise enhancement is performed at a single time-slice in these examples (indicated by a vertical dashed line), we find that it is nonetheless sustained over a wide range of time slices.
This behavior follows naturally from our arguments that the enhancement is associated with an optimization of time-independent overlap factors.

\begin{figure} 
\includegraphics[width=\figwidth]{\figdir 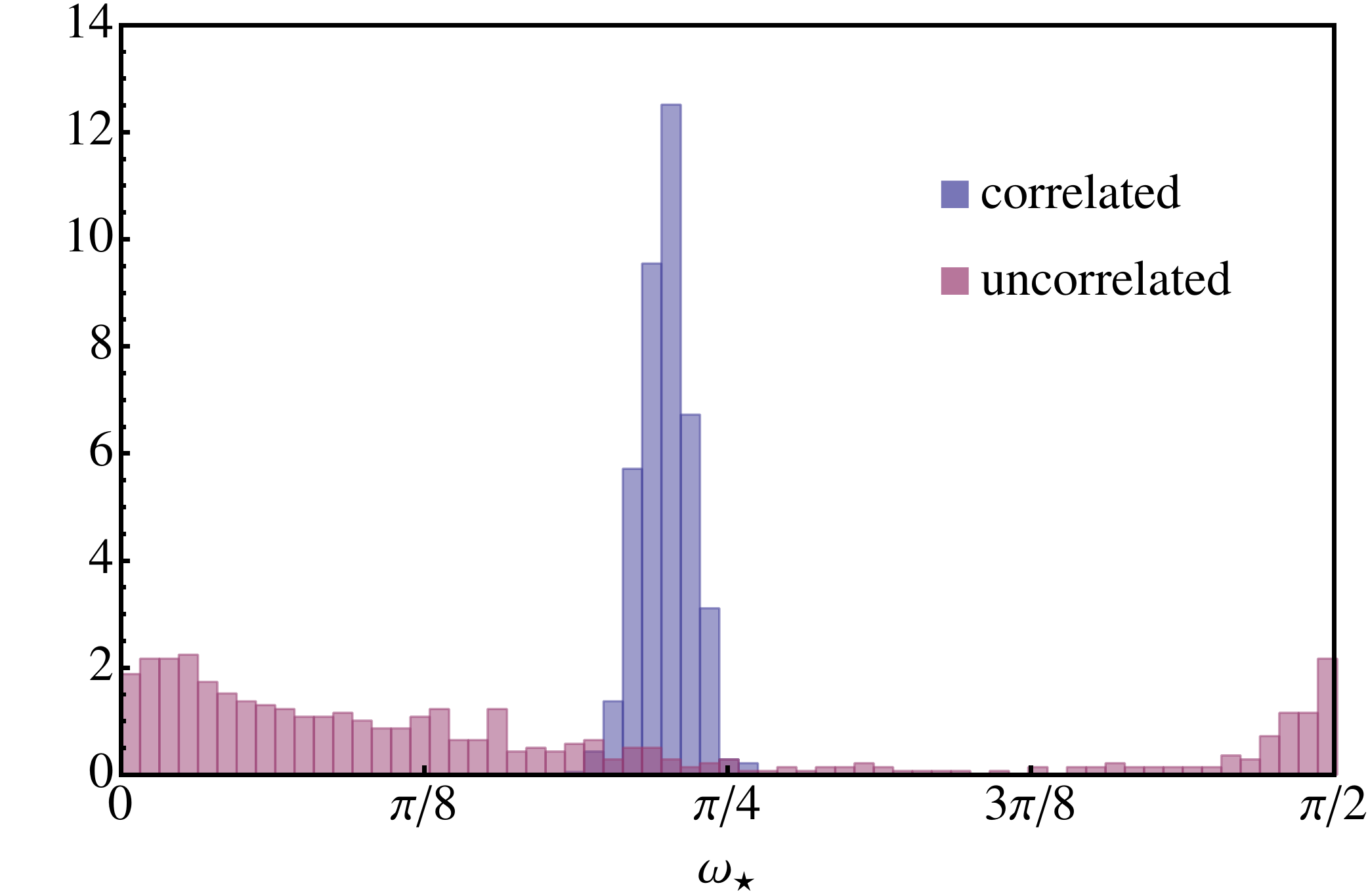}
\includegraphics[width=\figwidth]{\figdir 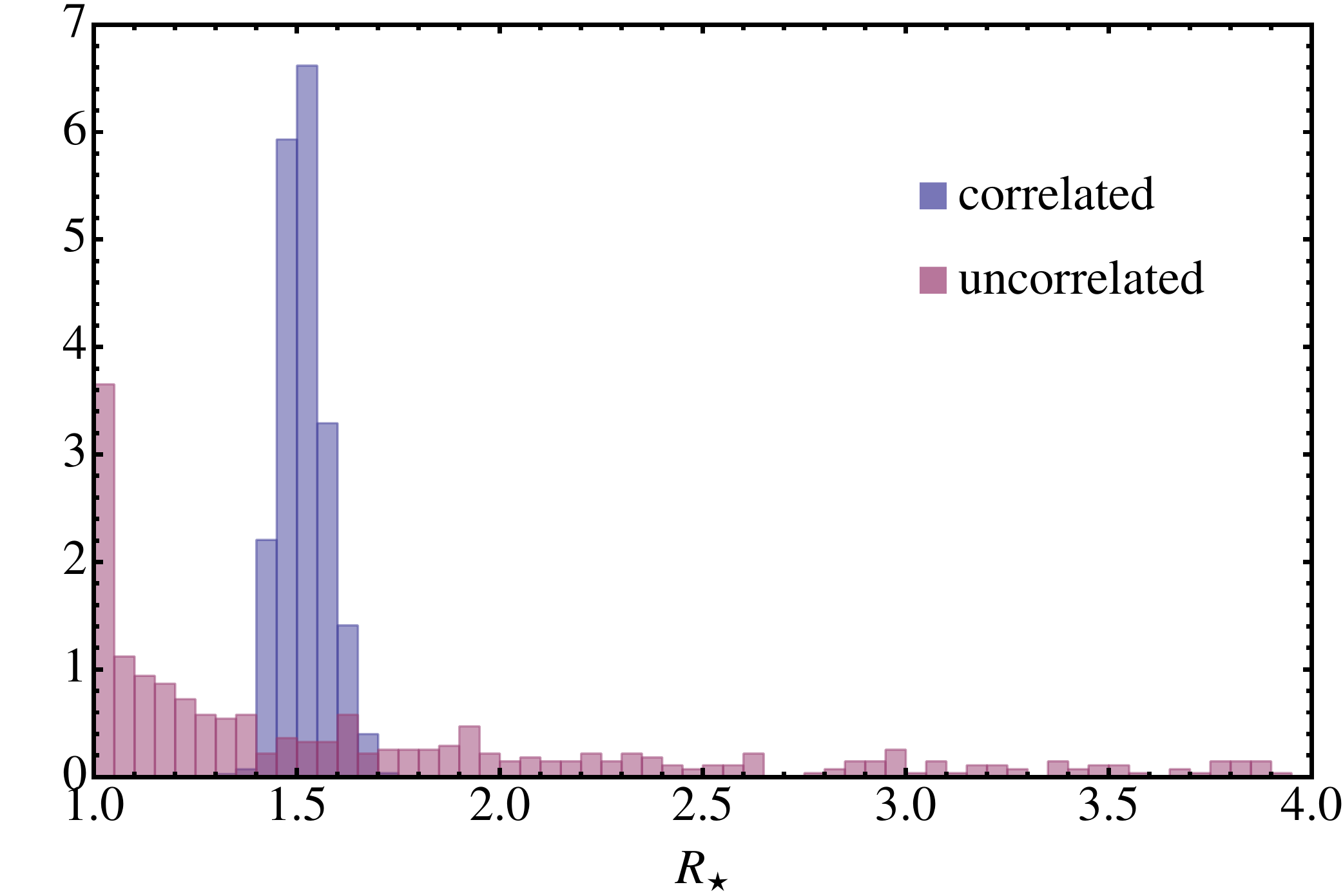}
\caption{\label{fig:rho_InterpNA_histograms_Star}%
Histogram of $\omega_\star$ (left) and $R_\star$ (right) obtained by extremizing the signal/noise $\hat\theta_\star(\omega,0)$ for the two-state rho correlator on each bootstrap ensemble.
The analysis is performed by retaining the correlations between the matrix elements of $C$ (correlated) and disposing such correlations (uncorrelated).
}
\end{figure}

In \Sec{toy_model}, we argued the importance of retaining correlations between matrix elements in order to achieve an enhancement in signal/noise using the optimization techniques introduced in \Sec{correlator_optimization_strategies}.
To see explicitly the role of such correlations, we may compare the signal/noise optimization in the case of the two-state rho system by retaining them and also by discarding them in our analysis.
We illustrate this effect by considering the distributions for $\omega_\star$ and $R_\star$ obtained by extremizing estimates of $\theta_\star(\omega,0)$ on bootstrap resampled correlators.
Bootstrap ensembles were generated such that the matrix elements retained their correlations (correlated) and such that the correlations were removed by an independent resampling of each matrix element (uncorrelated).
On each bootstrap ensemble, an estimator for $\theta_\star(\omega,\delta)$ was constructed and maximized to obtain estimates of $\omega_\star$, $R_\star$ and $\delta_\star$.
The distributions for $\omega_\star$ and $R_\star$ generated from the bootstrap analysis are plotted in \Fig{rho_InterpNA_histograms_Star} for both the correlated and uncorrelated cases (by construction of the analysis, $\delta_\star=0$ in all cases).
We find that in the former case, the distributions are well-localized about a central value which is consistent with the results of \Fig{rho_InterpNA_landscape_Star}, whereas in the later case they are not.
Interestingly, one finds that an enhancement in signal/noise always possible (i.e., $R_\star >1$) for uncorrelated matrix elements, however, such an enhancement on average is less that that of the correlated case and may no longer be attributed to the tuning of overlap factors, but rather the tuning of statistical fluctuations which are ensemble and time-slice specific.

\subsection{Comparison of strategies for single hadron correlators}
\label{sec:comparison_of_strategies}

Here, we provide a comparison of strategies for determining the energies of hadron states using some of the methods outlined in \Sec{correlator_optimization_strategies}.
In particular, we consider correlators constructed from source-optimized source and sink vectors, a source-optimized source and signal/noise-optimized sink vector, and signal/noise-optimized source and sink vectors.
In all cases, we focus on the ground state energy, and in addition, we consider the first excited state energy for the rho.
In the case of the $26\times 26$ rho correlator matrix, we determine the source optimized vectors using the variational techniques described in \Appendix{source_optimization} with $\tau_0=3$ and $\tau_1=14$.
In the case of the pion, proton and delta correlator matrices, the variational technique produces ground state correlators that are statistically indistinguishable from those produced by using the least contaminated diagonal element of the correlator matrix, indicating that the different Gaussian smearings that were used in the sources and sinks were not sufficiently orthogonal for the variational approach to work well.
The correlators constructed from the variationally determined source and sink vectors, however, possessed considerably larger uncertainties by comparison.
For these systems, we therefore opt to use the least contaminated matrix element as a proxy for a correlator constructed from source-optimized source and sink vectors in our analysis.

\begin{figure}
\includegraphics[width=\figwidth]{\figdir 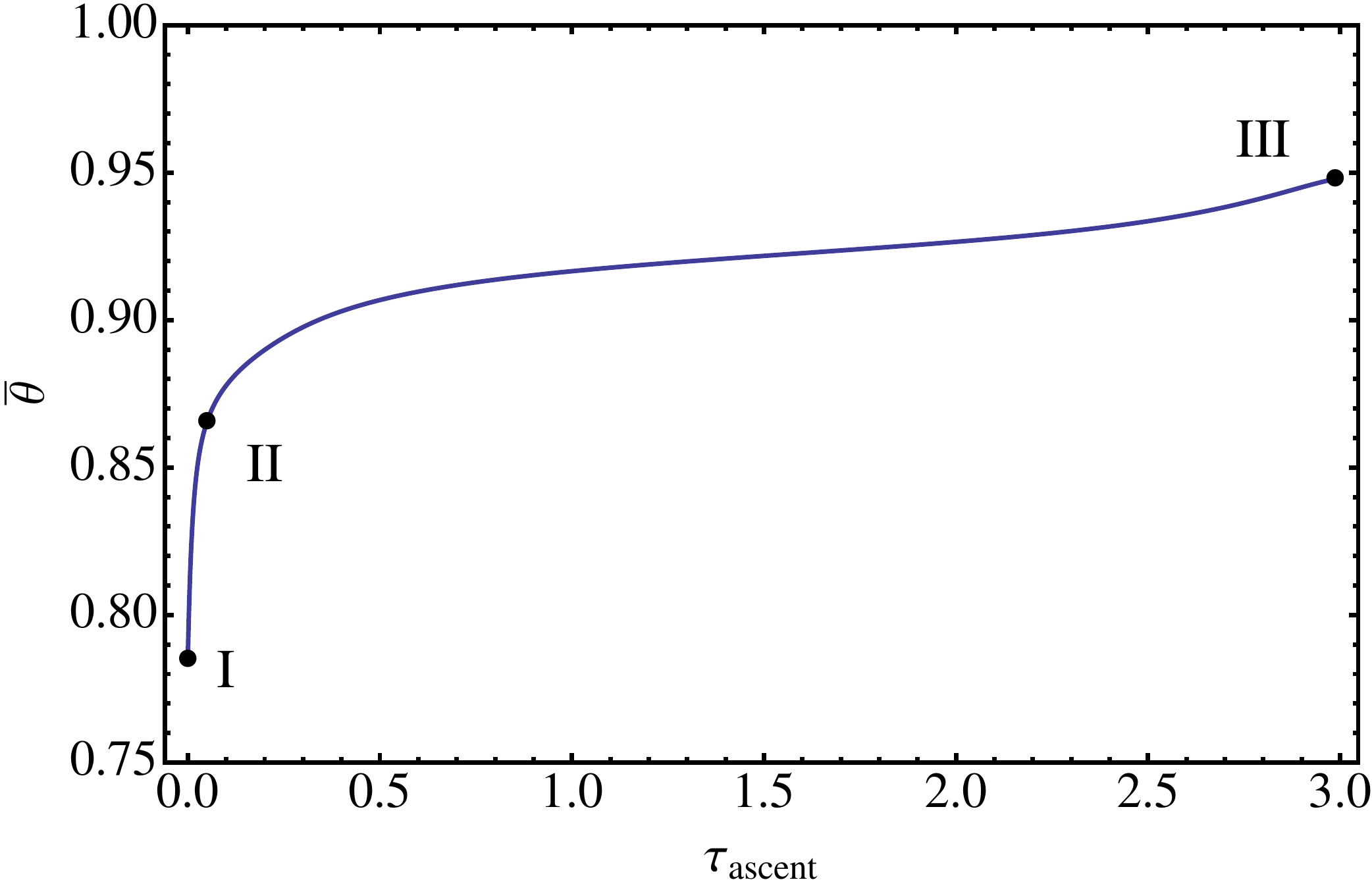}
\includegraphics[width=\figwidth]{\figdir 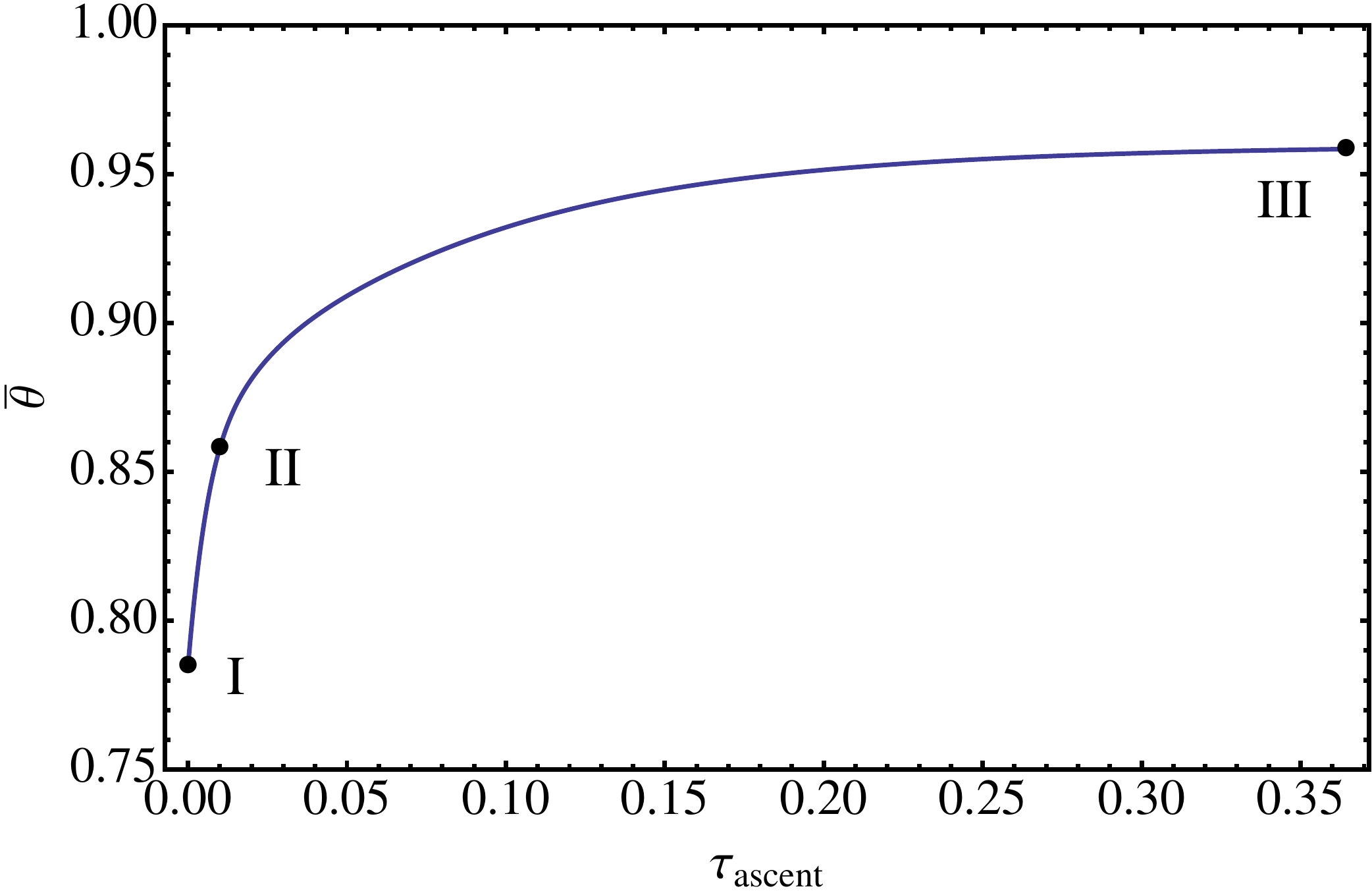}
\includegraphics[width=\figwidth]{\figdir 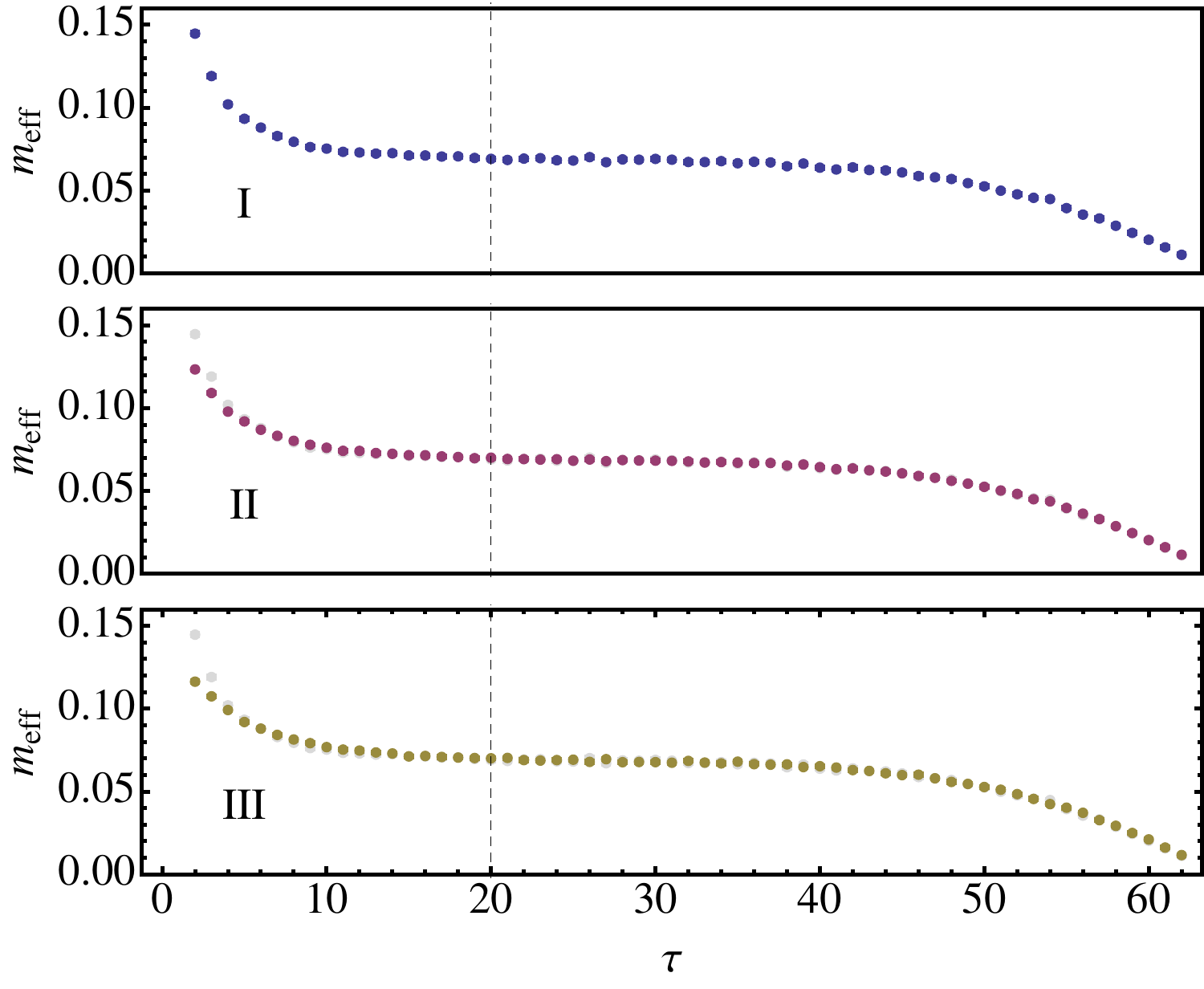}
\includegraphics[width=\figwidth]{\figdir 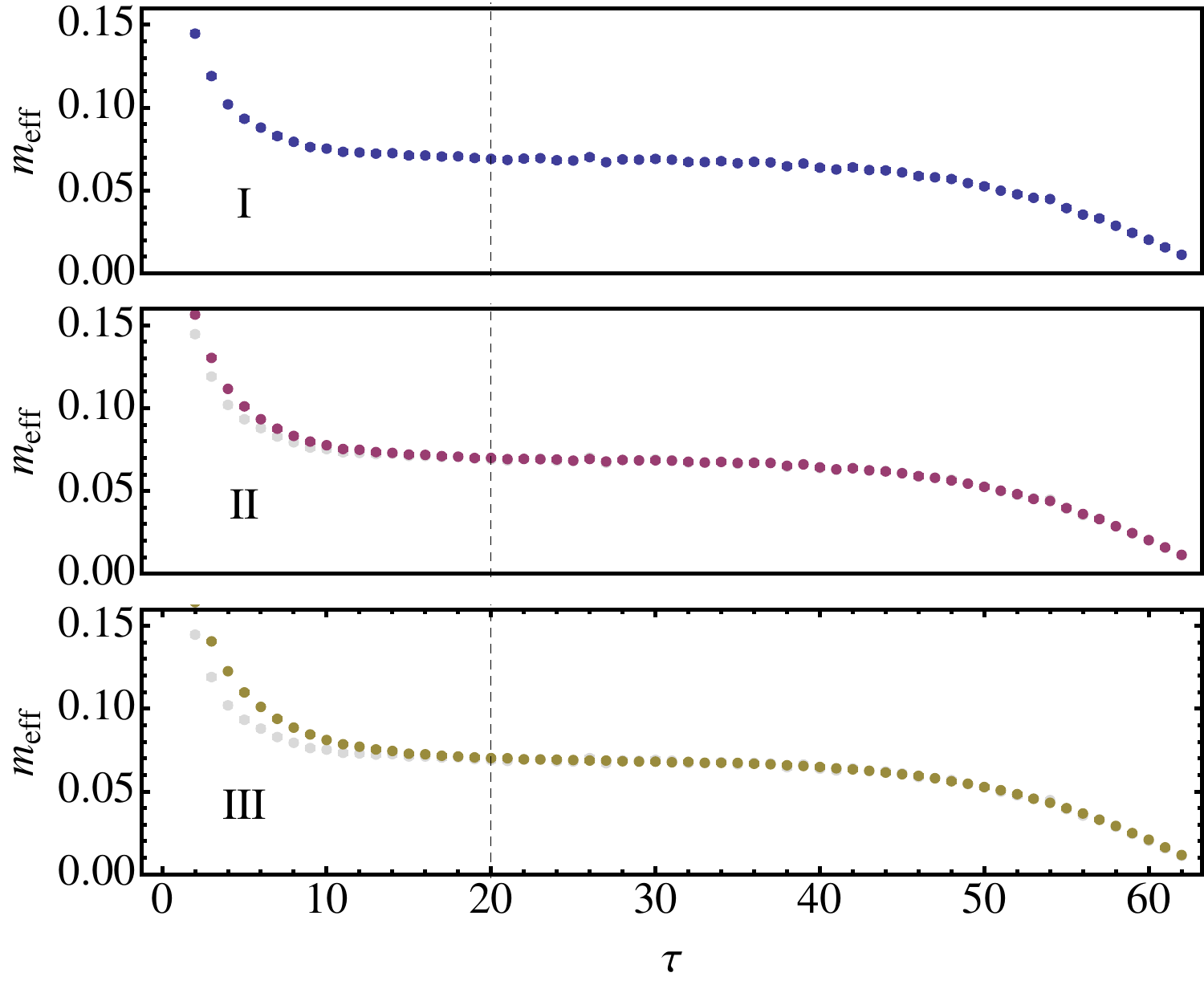}
\includegraphics[width=\figwidth]{\figdir 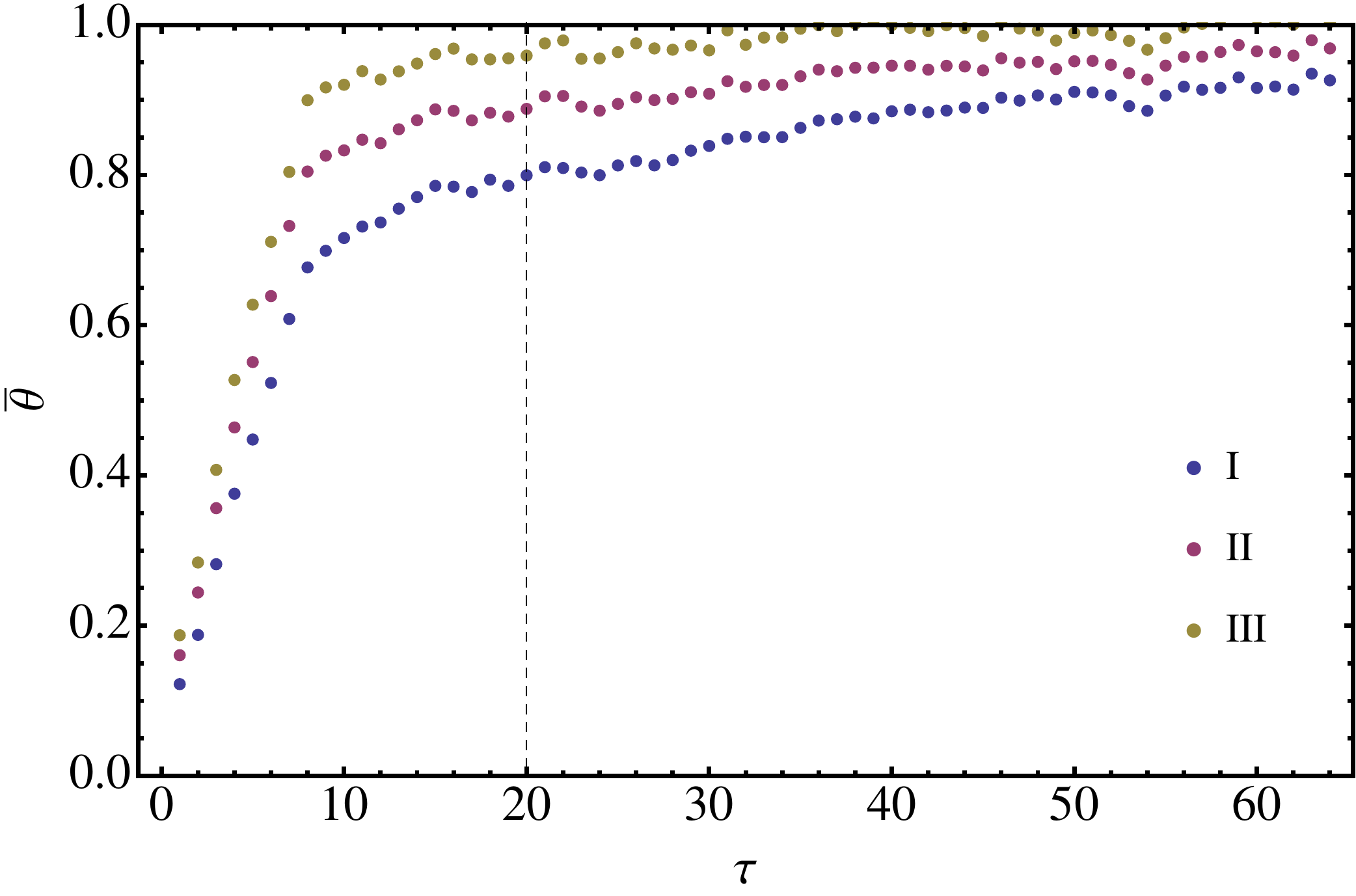}
\includegraphics[width=\figwidth]{\figdir 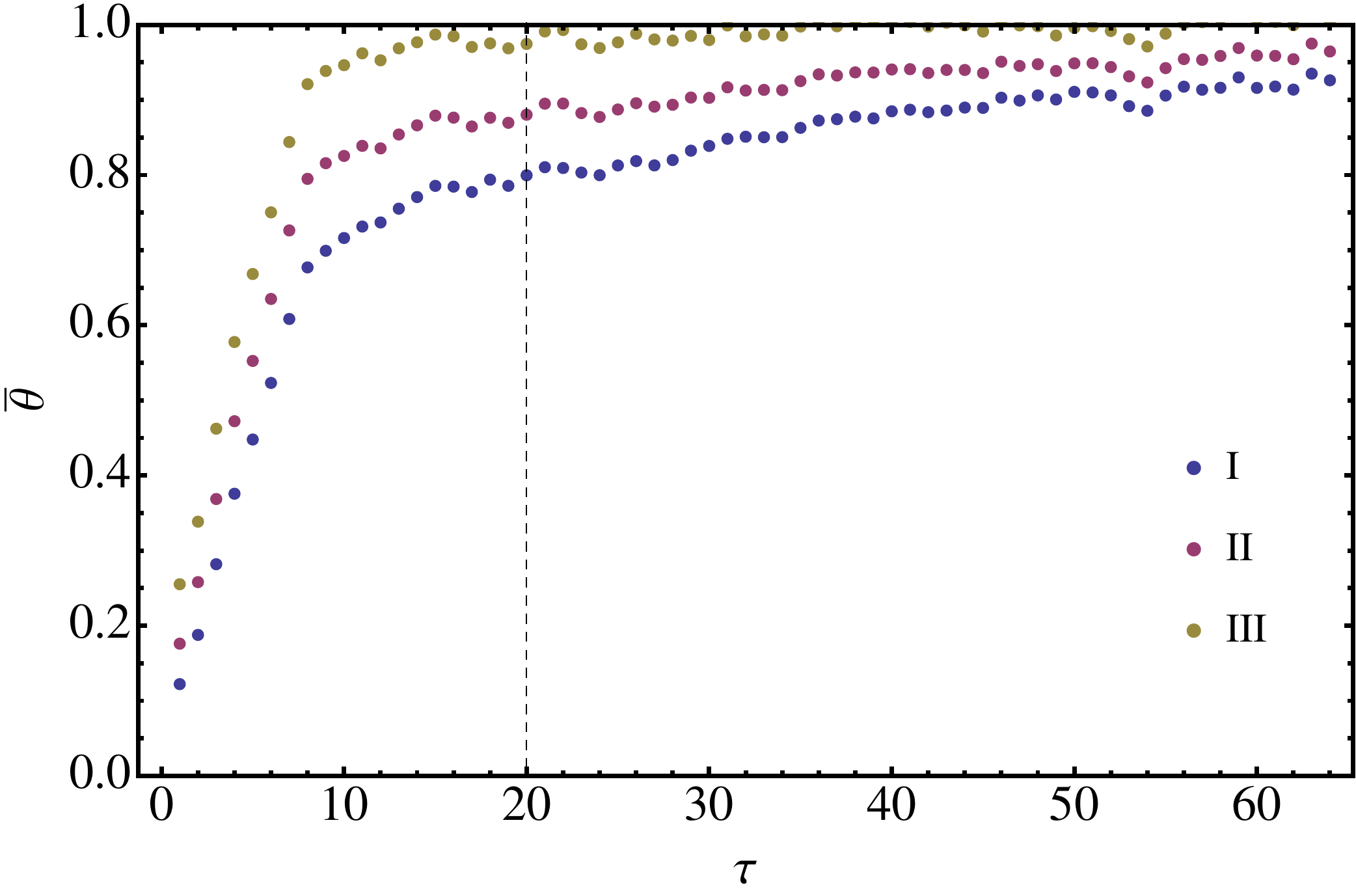}
\caption{\label{fig:pion_Interp1_optPlots}%
Top: Normalized signal/noise for the pion correlator at a fixed $\tau_\textrm{s}=\tau_\textrm{n}$ along a path of steepest ascent as a function of the ascent time. 
Trajectories are shown for a fixed source and unconstrained sink vector (left) and for unconstrained, but equal, source and sink vectors (right).
In both plots, point (I) corresponds to a correlator constructed from equal source and sink vectors chosen to produce an optimal ground state, point (II) corresponds to some intermediate points along the trajectory, and point (III) corresponds to signal/noise optimized correlators.
Center: Effective mass plots for the correlators described above as a function of time separation, $\tau$ (shown in color; the effective mass associated with point (I) is shown in light gray in each plot for comparison).
The dashed line indicates the value of $\tau_\textrm{s}=\tau_\textrm{n}$ used for the signal/noise optimization.
Bottom: Corresponding normalized signal/noise as a function of the time separation.
}
\end{figure}

\begin{figure}
\includegraphics[width=\figwidth]{\figdir 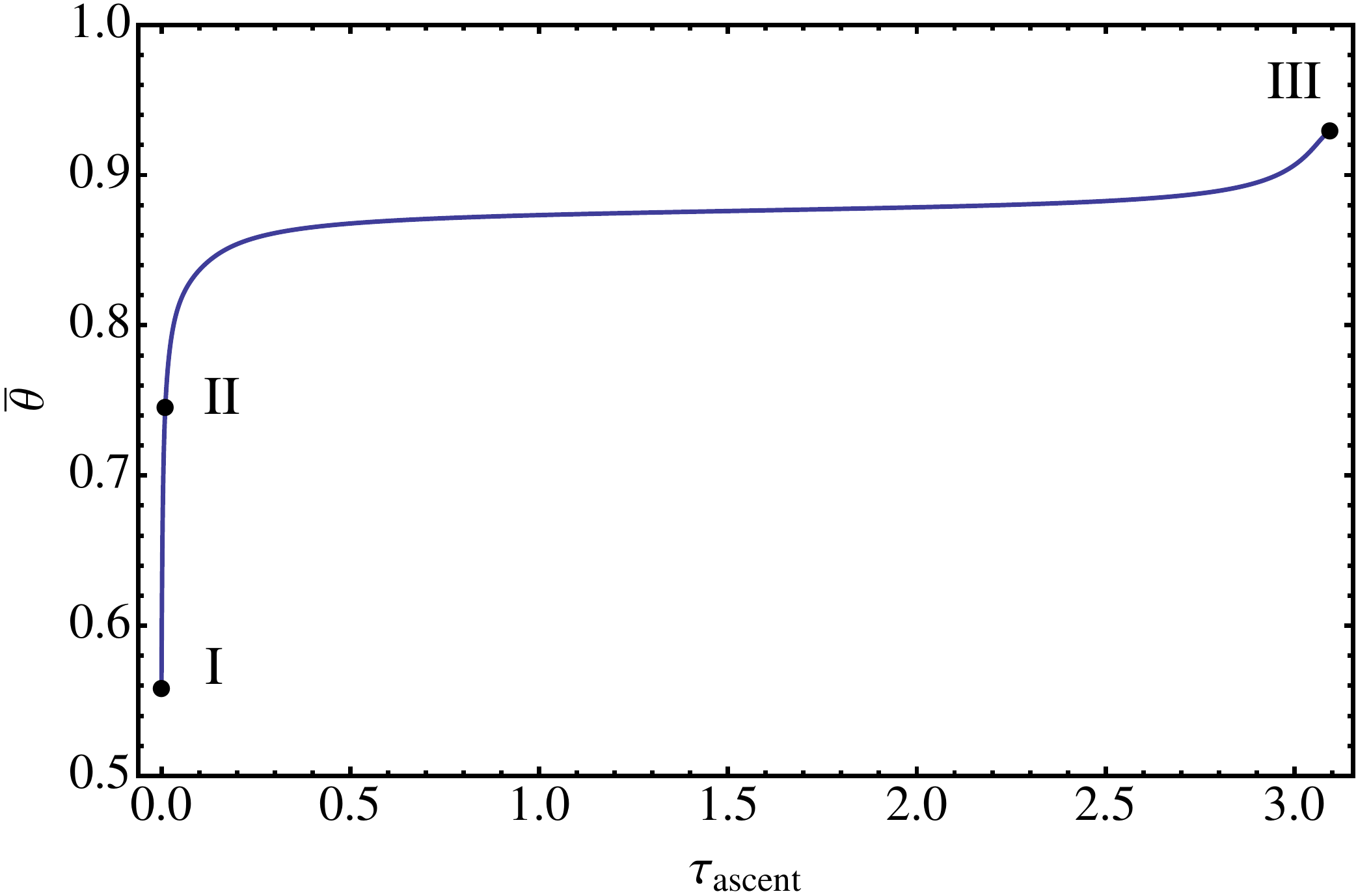}
\includegraphics[width=\figwidth]{\figdir 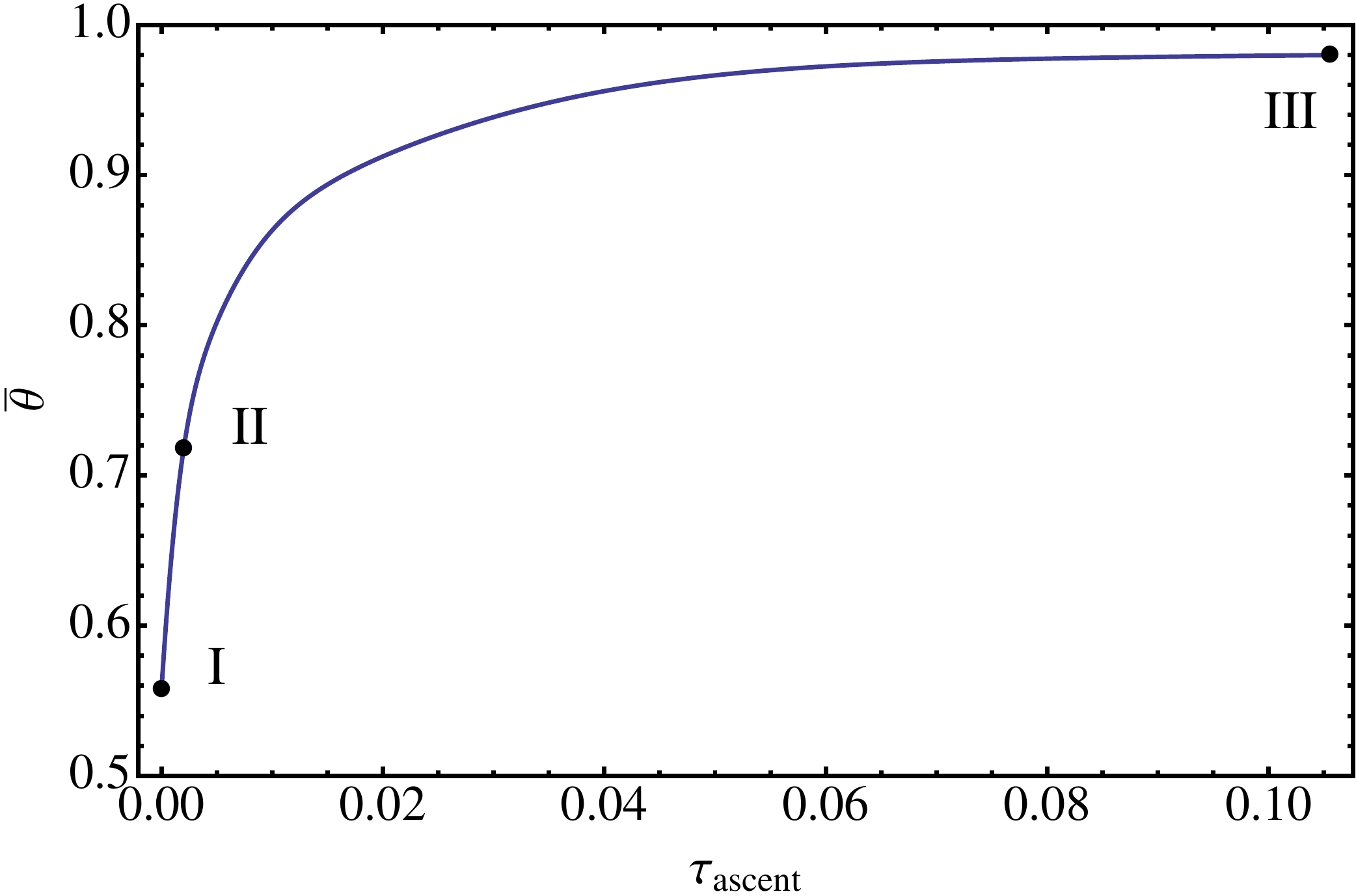}
\includegraphics[width=\figwidth]{\figdir 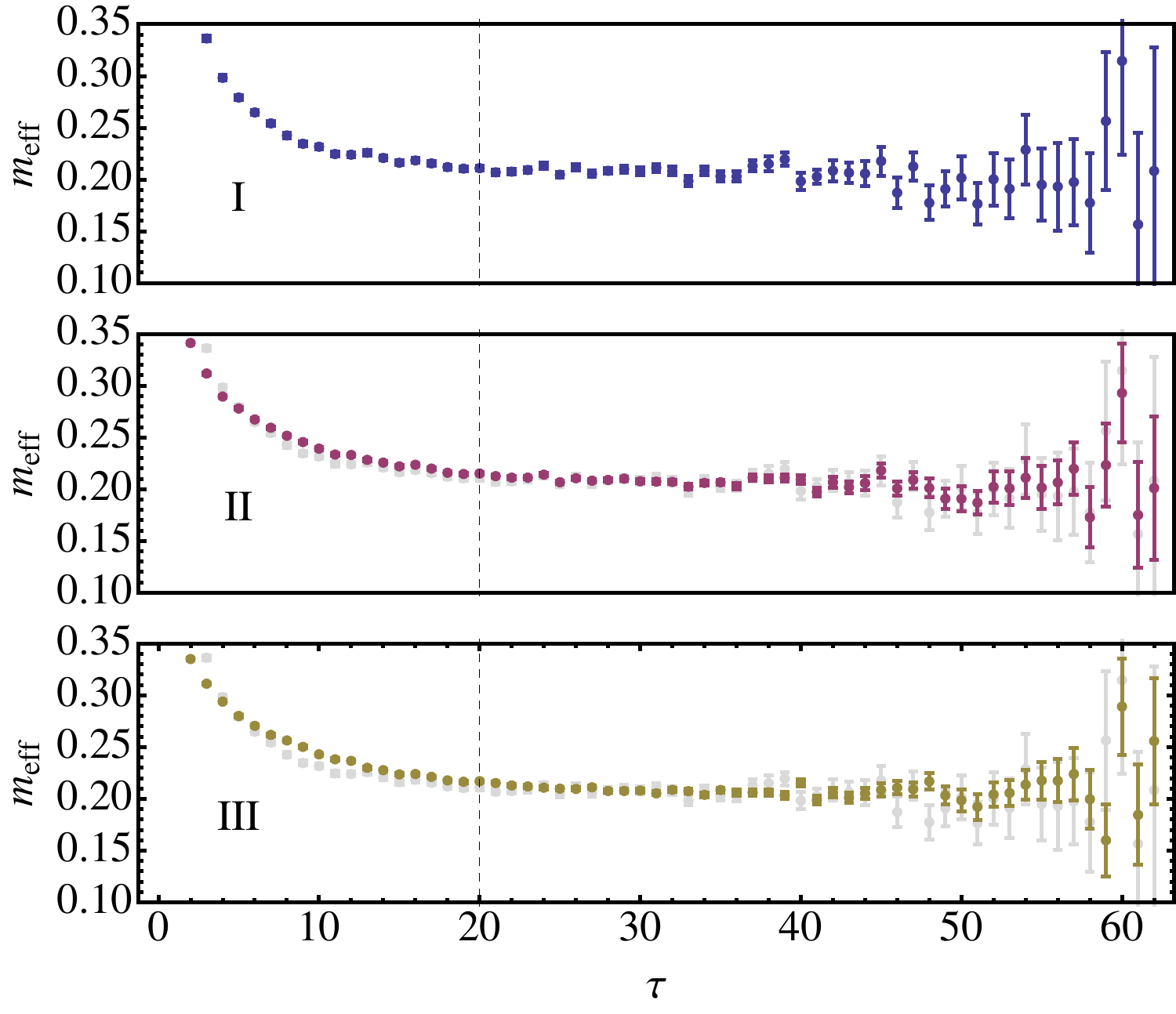}
\includegraphics[width=\figwidth]{\figdir 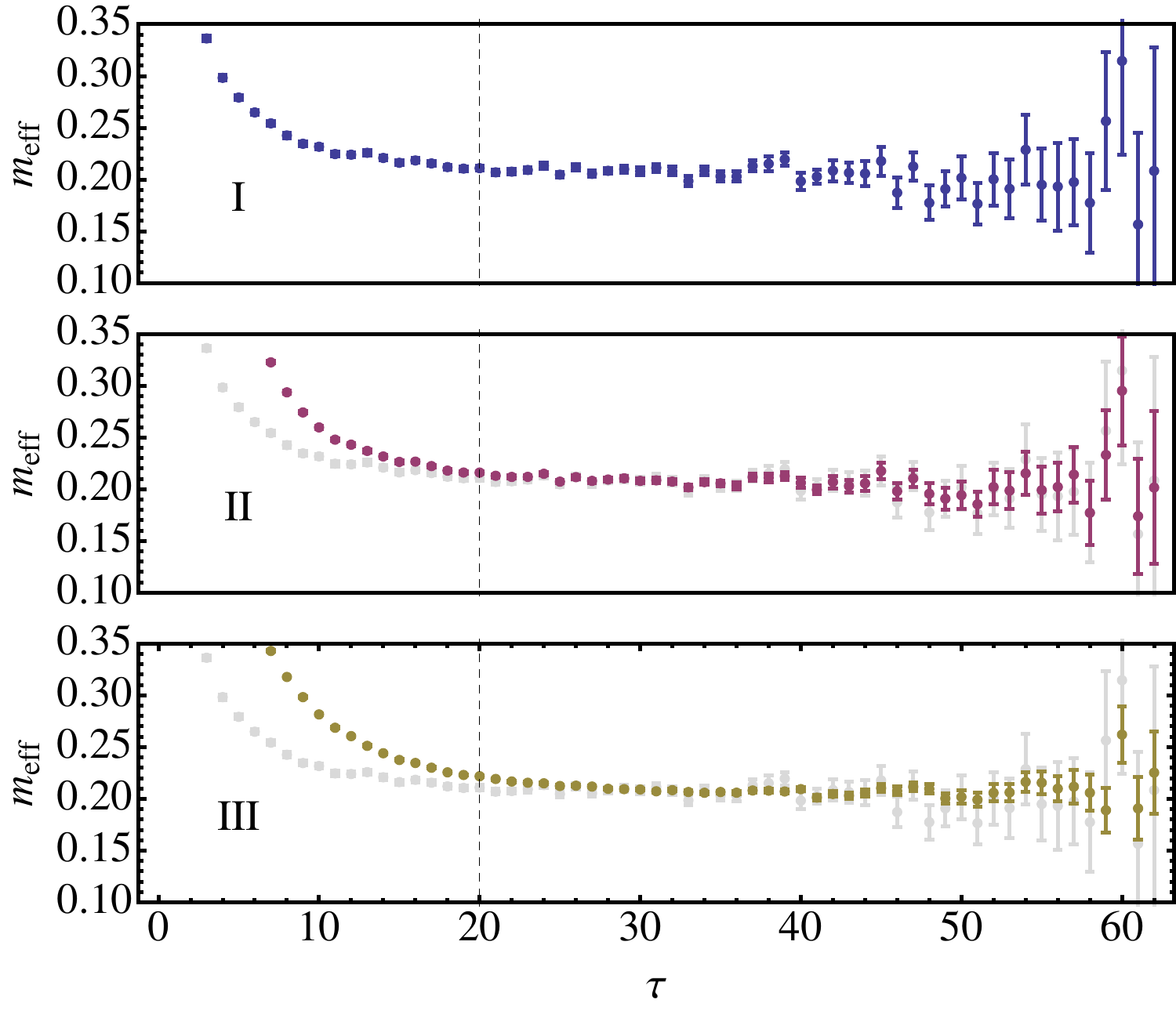}
\includegraphics[width=\figwidth]{\figdir 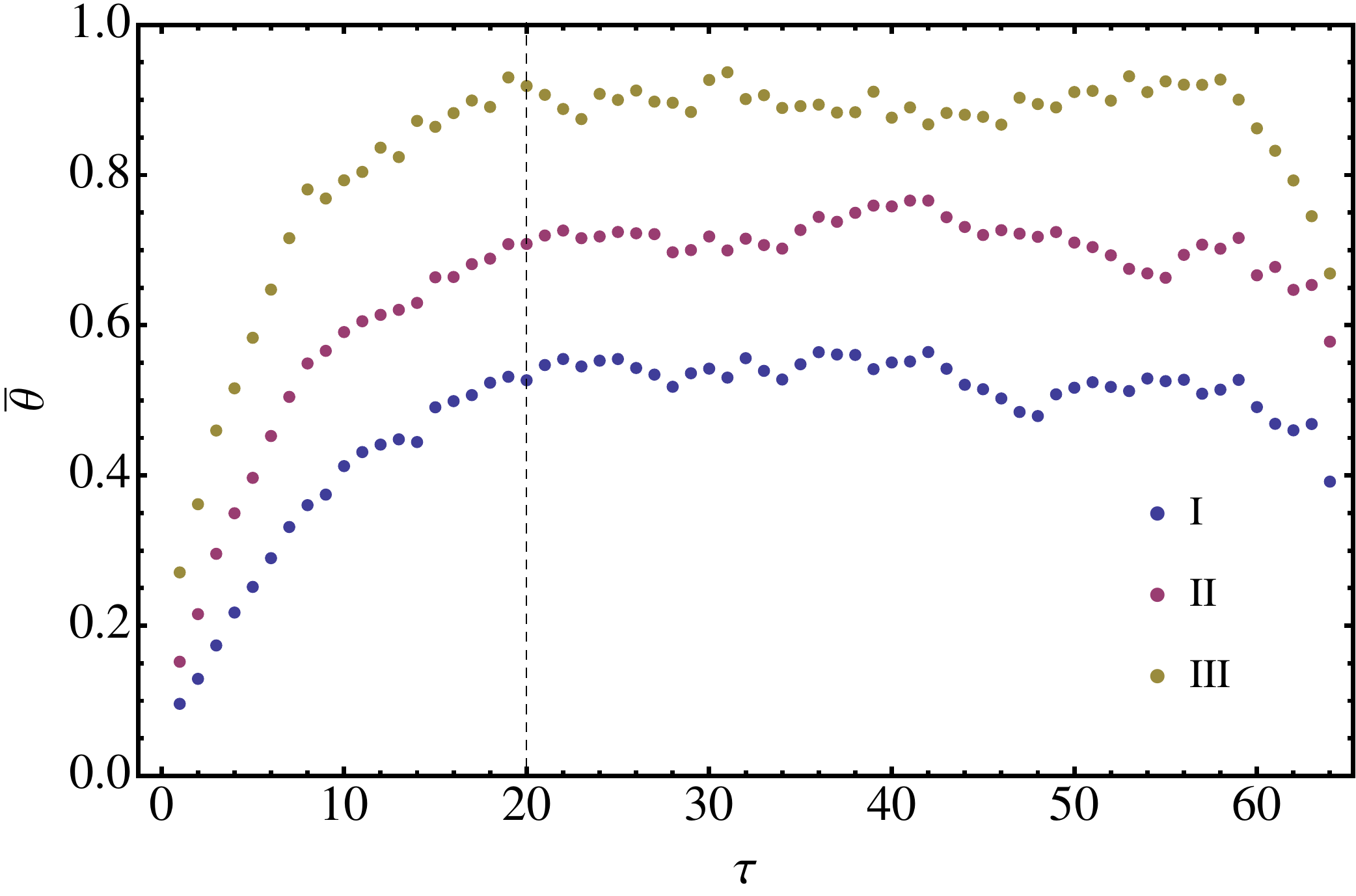}
\includegraphics[width=\figwidth]{\figdir 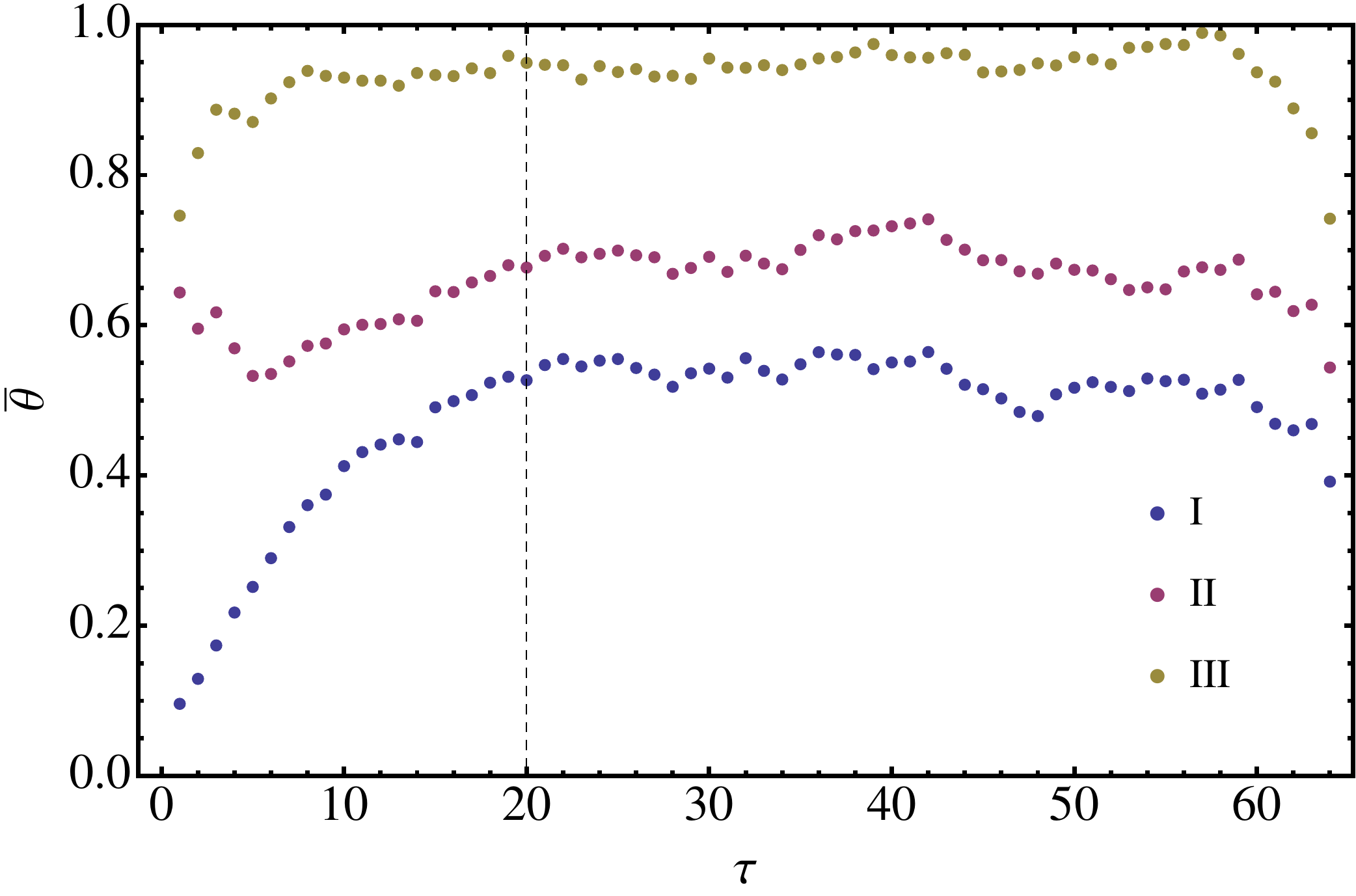}
\caption{\label{fig:proton_Interp4_optPlots}%
Same as \Fig{pion_Interp1_optPlots}, for the proton ground state.
}
\end{figure}

\begin{figure}
\includegraphics[width=\figwidth]{\figdir 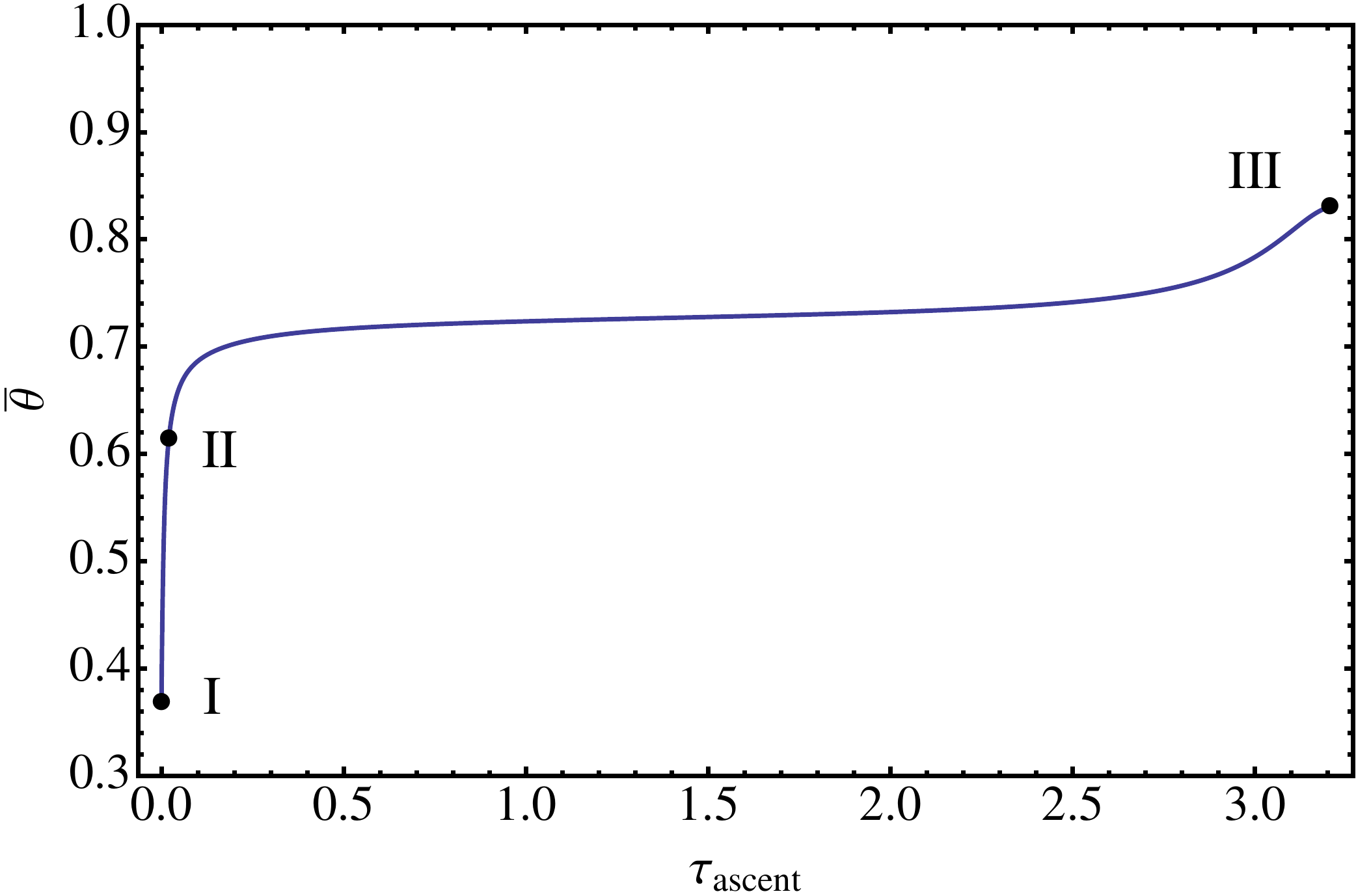}
\includegraphics[width=\figwidth]{\figdir 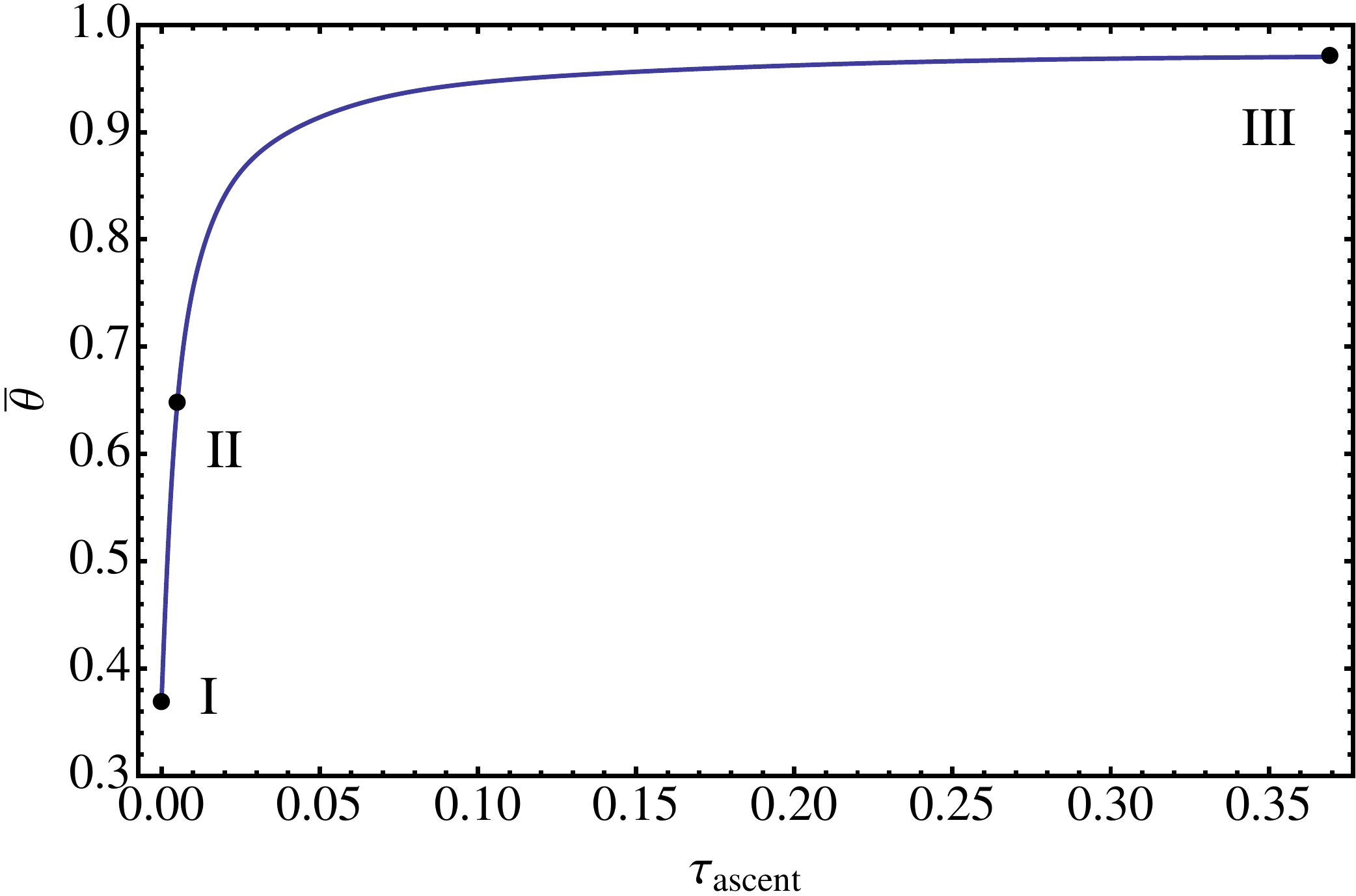}
\includegraphics[width=\figwidth]{\figdir 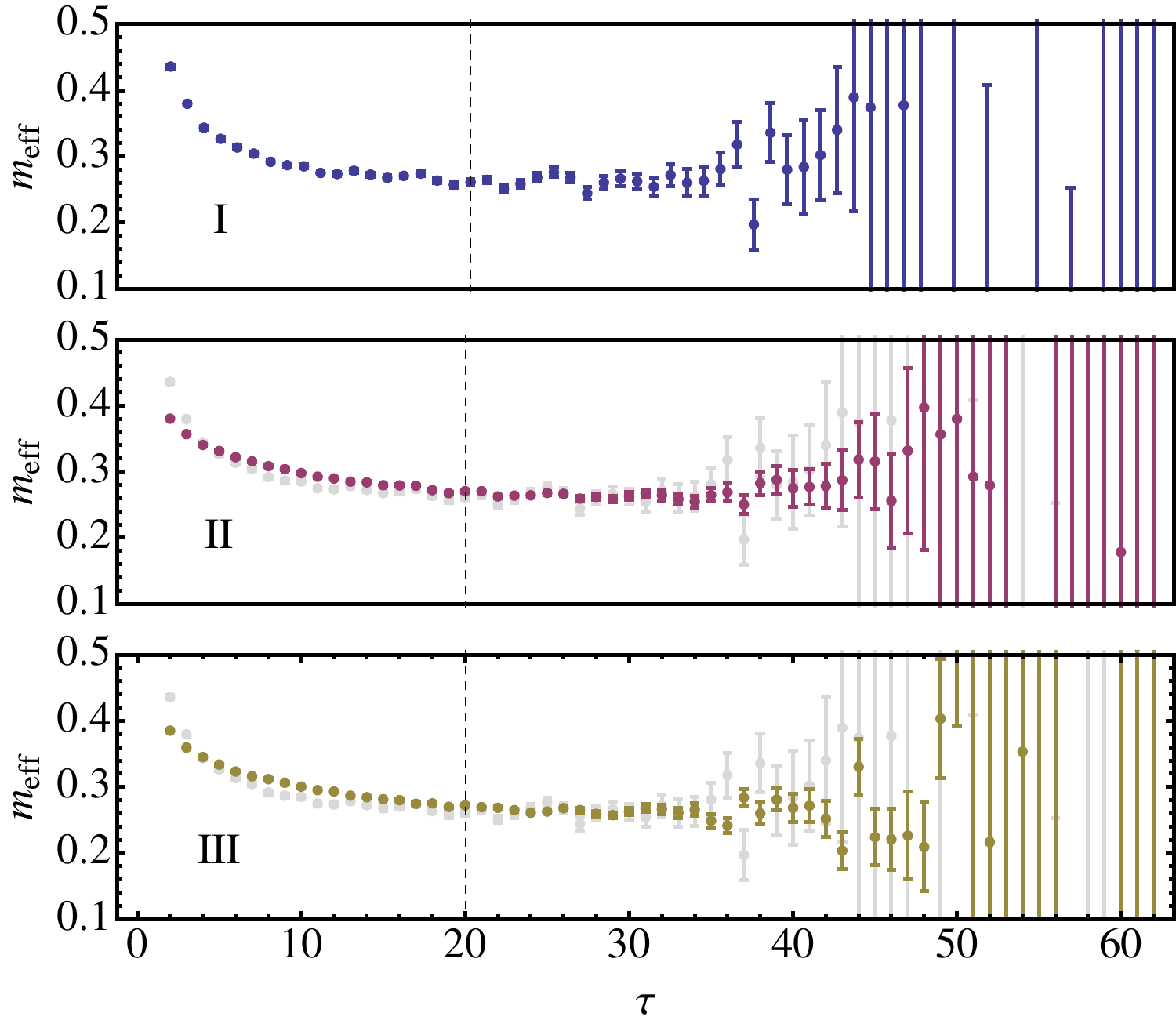}
\includegraphics[width=\figwidth]{\figdir 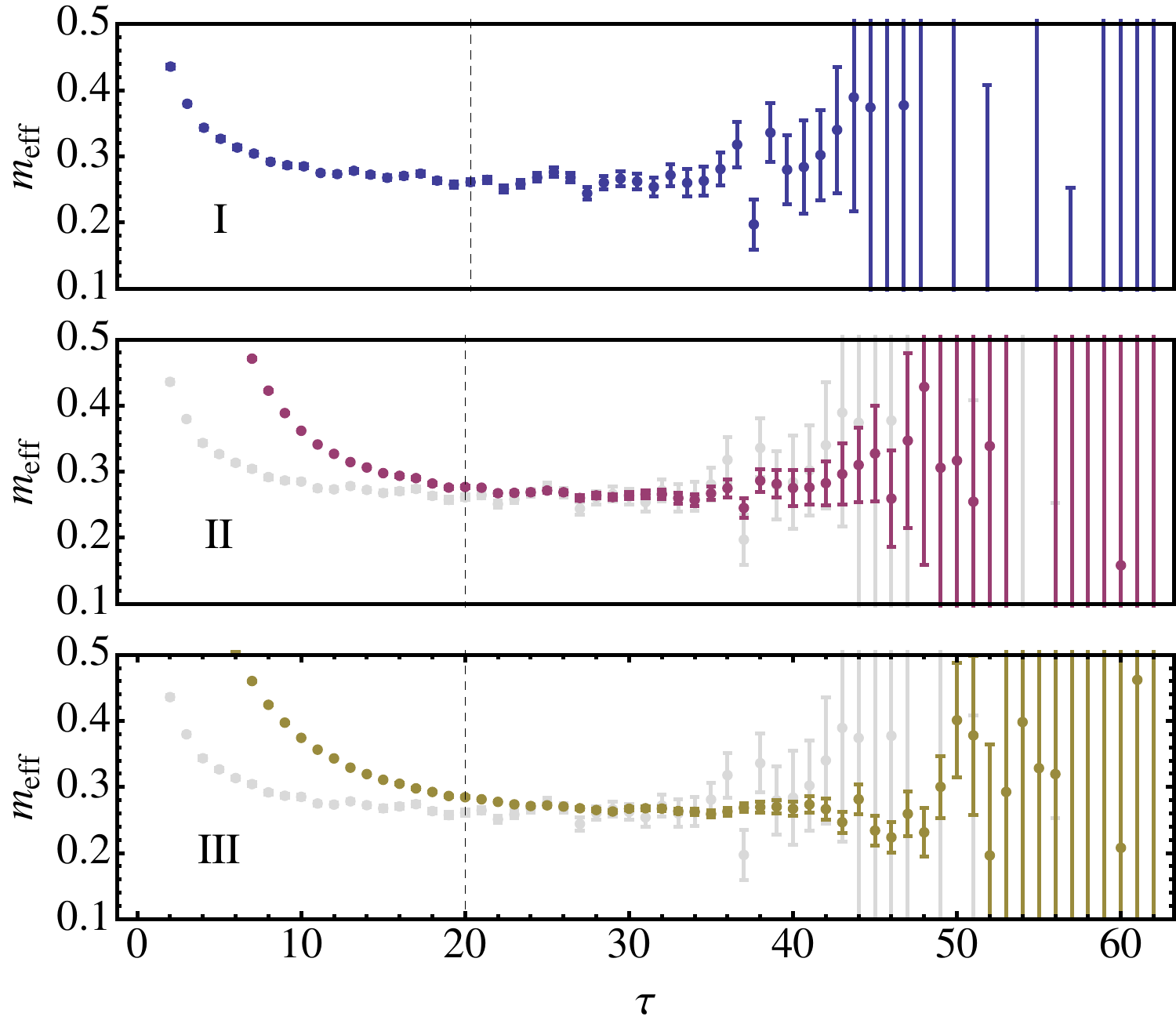}
\includegraphics[width=\figwidth]{\figdir 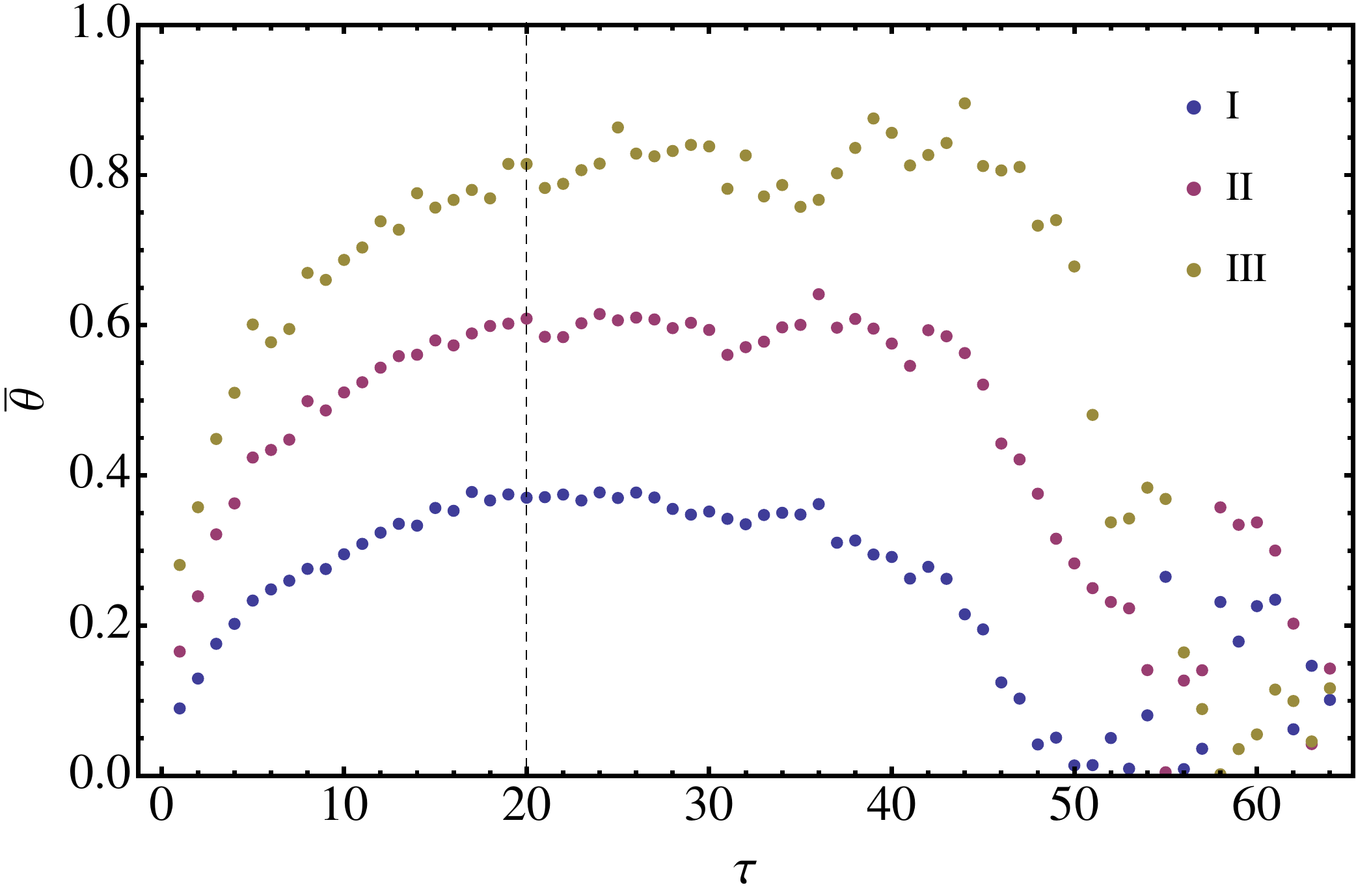}
\includegraphics[width=\figwidth]{\figdir 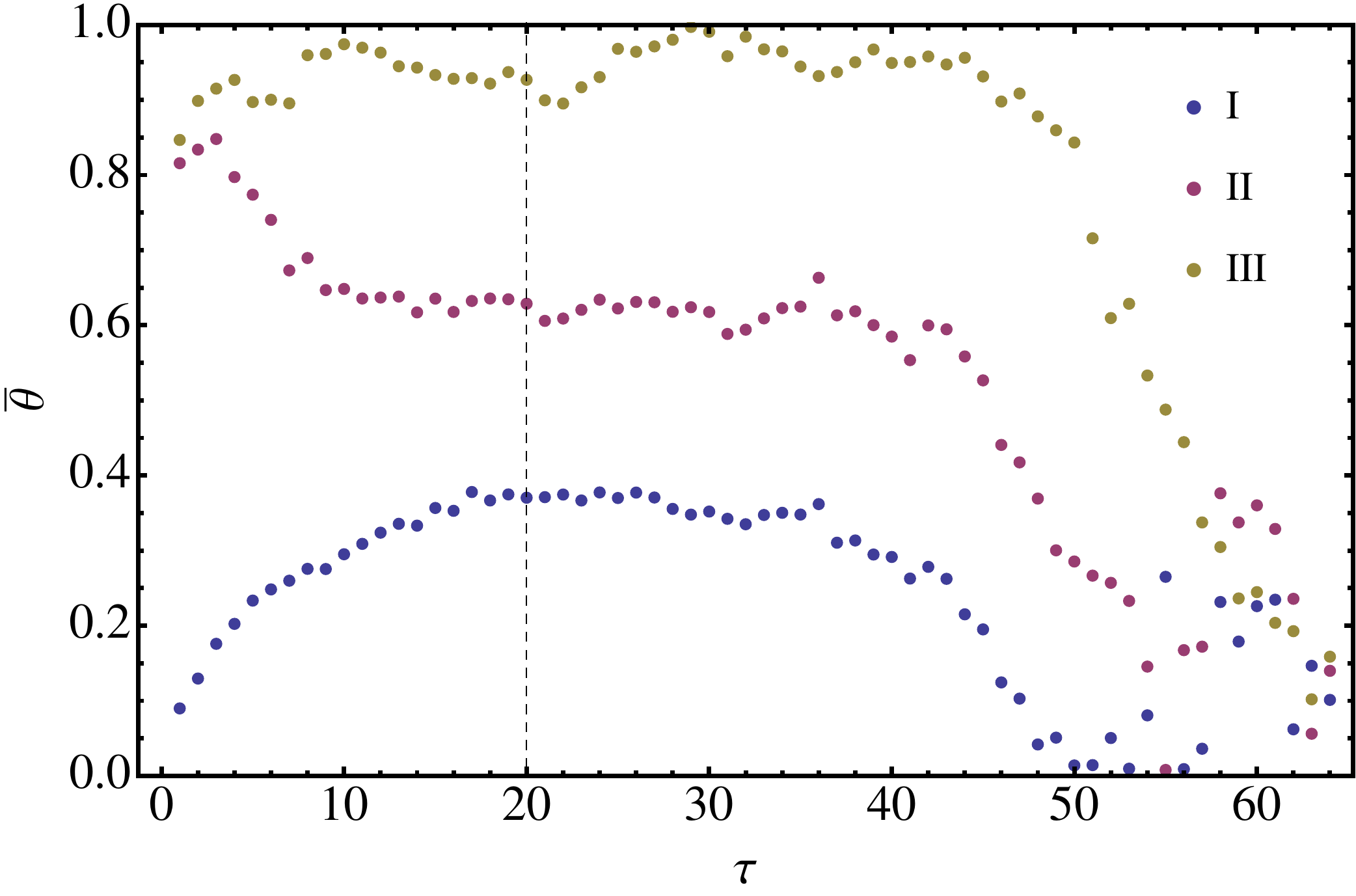}
\caption{\label{fig:delta_Interp1_optPlots}%
Same as \Fig{pion_Interp1_optPlots}, for the delta ground state.
}
\end{figure}

\begin{figure}
\includegraphics[width=\figwidth]{\figdir 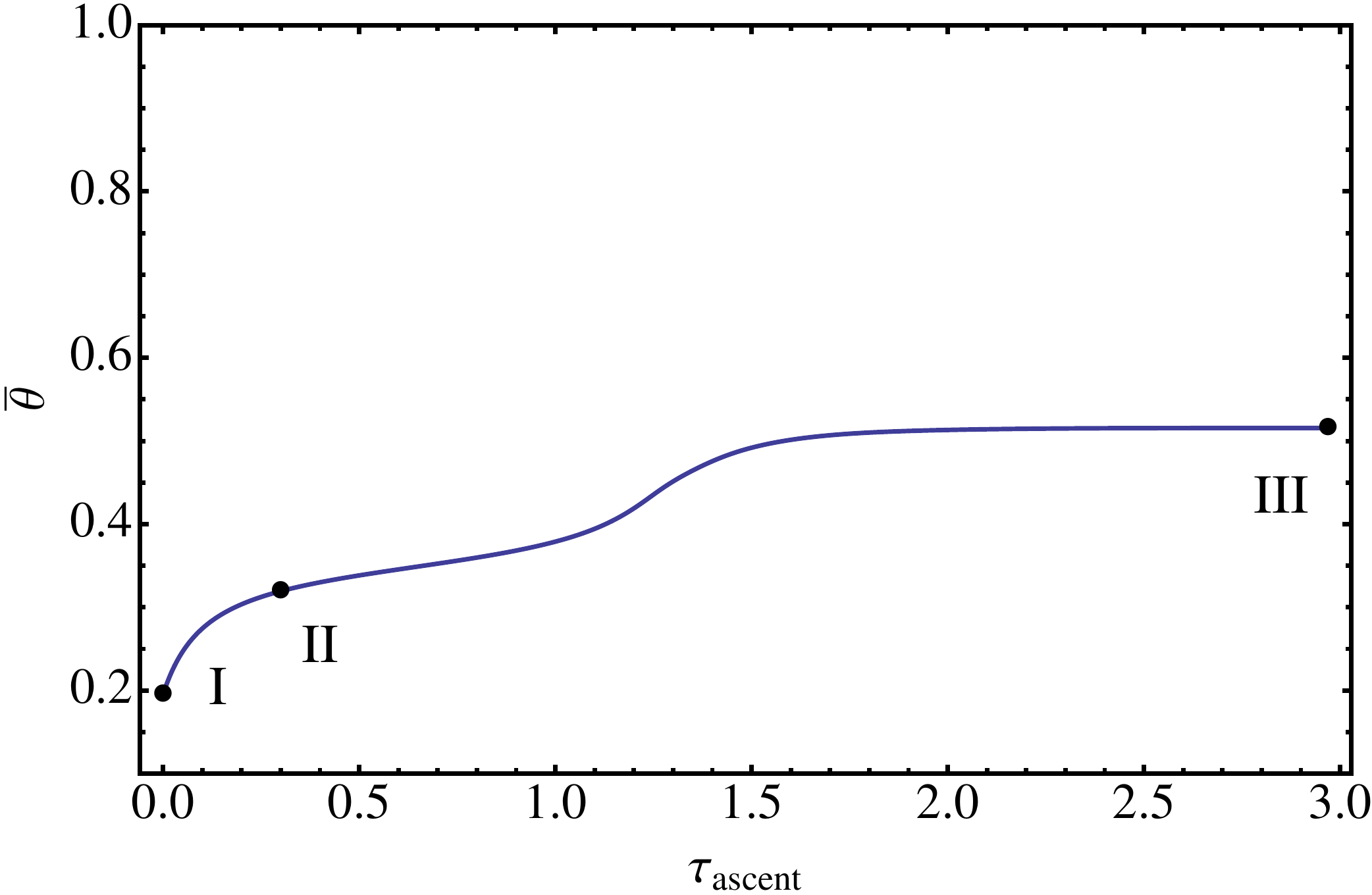}
\includegraphics[width=\figwidth]{\figdir 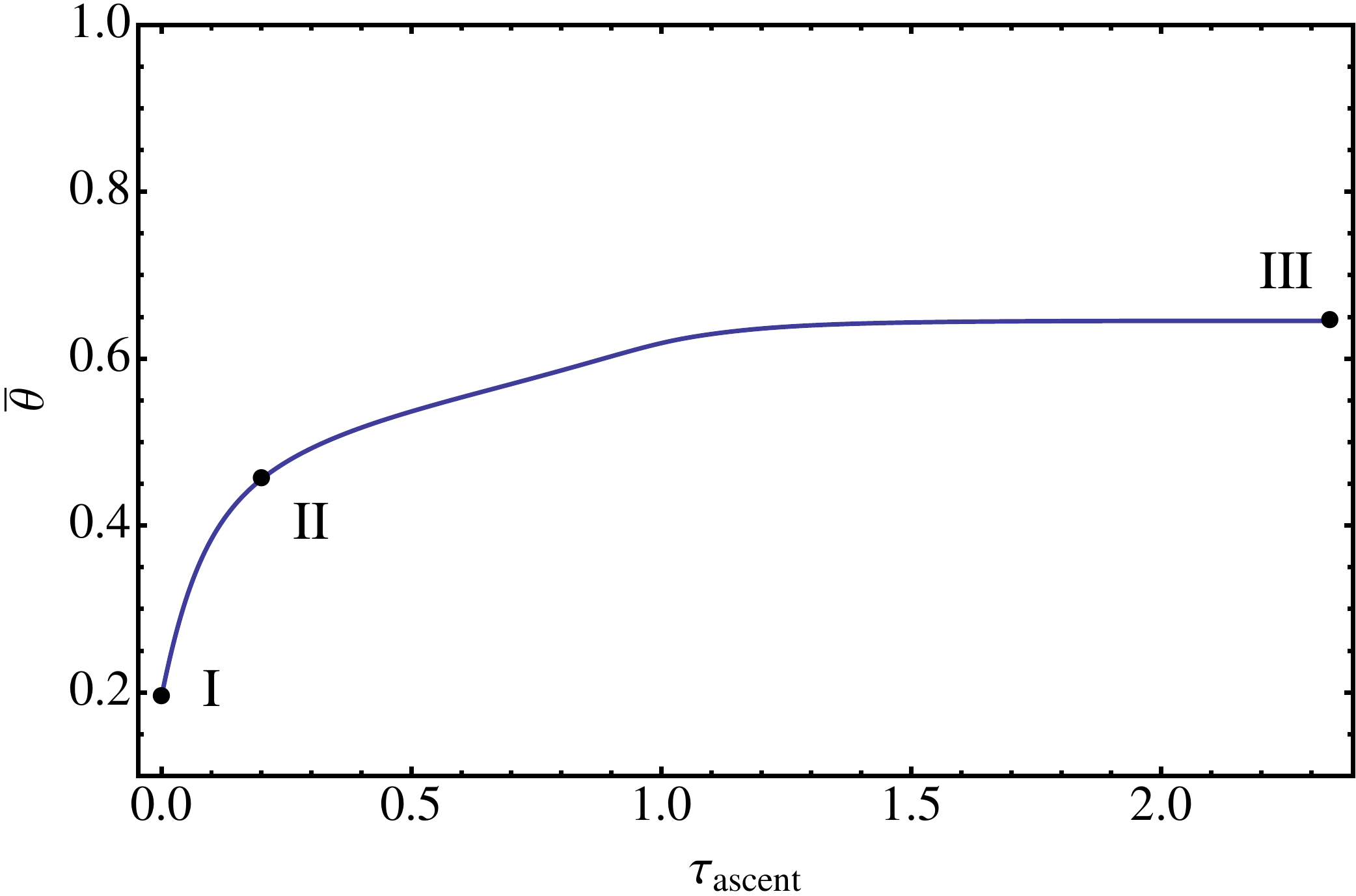}
\includegraphics[width=\figwidth]{\figdir 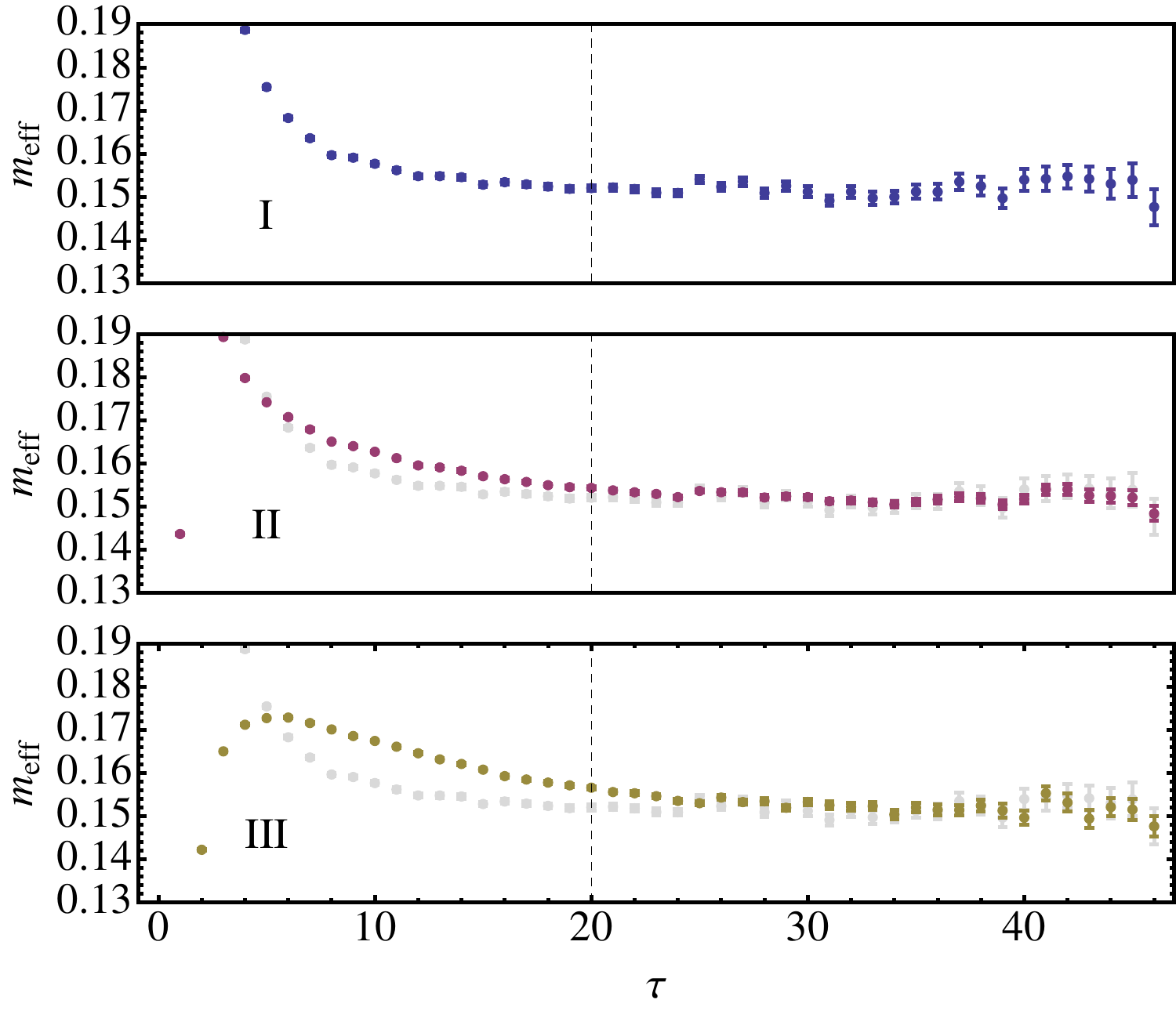}
\includegraphics[width=\figwidth]{\figdir 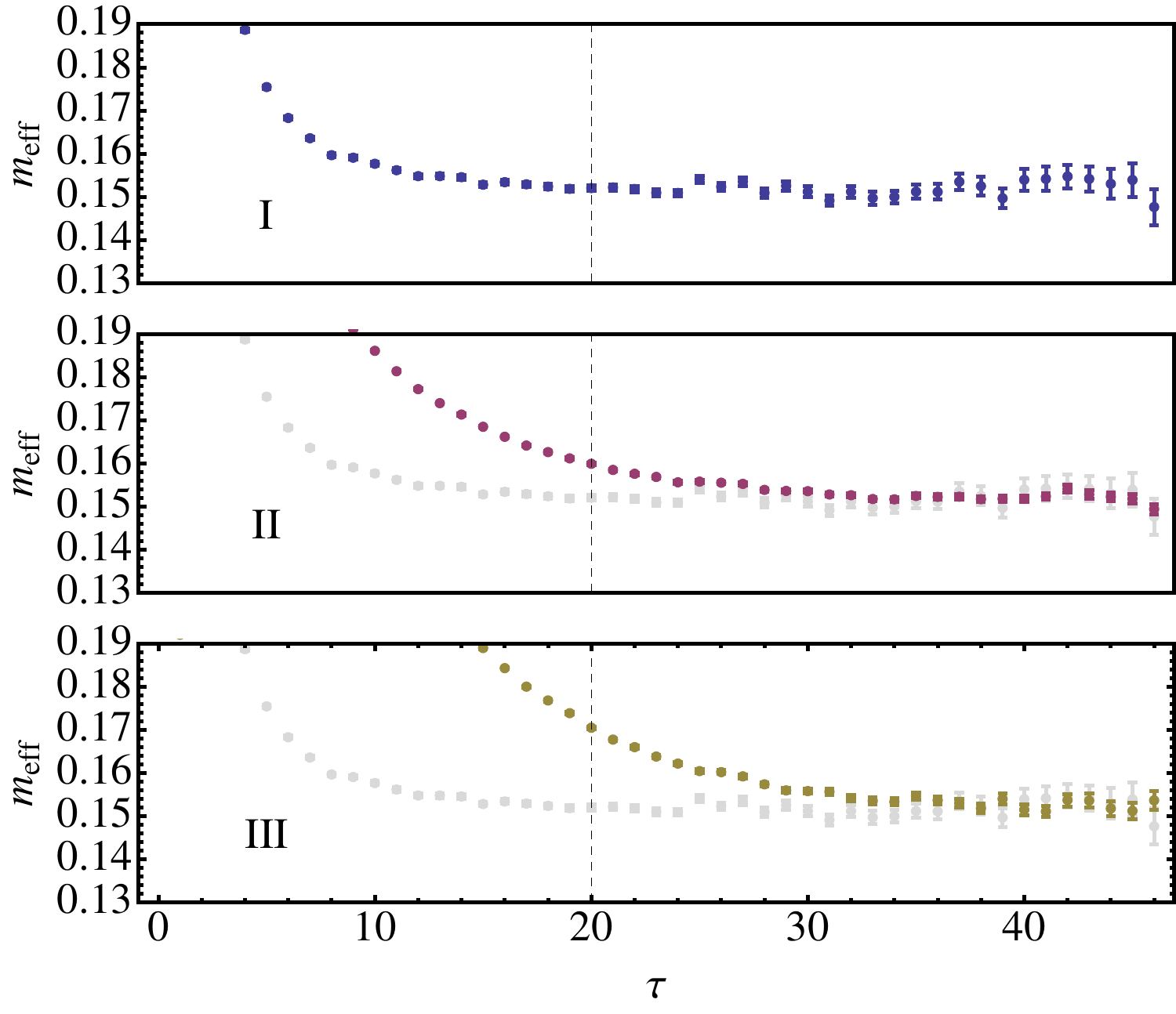}
\includegraphics[width=\figwidth]{\figdir 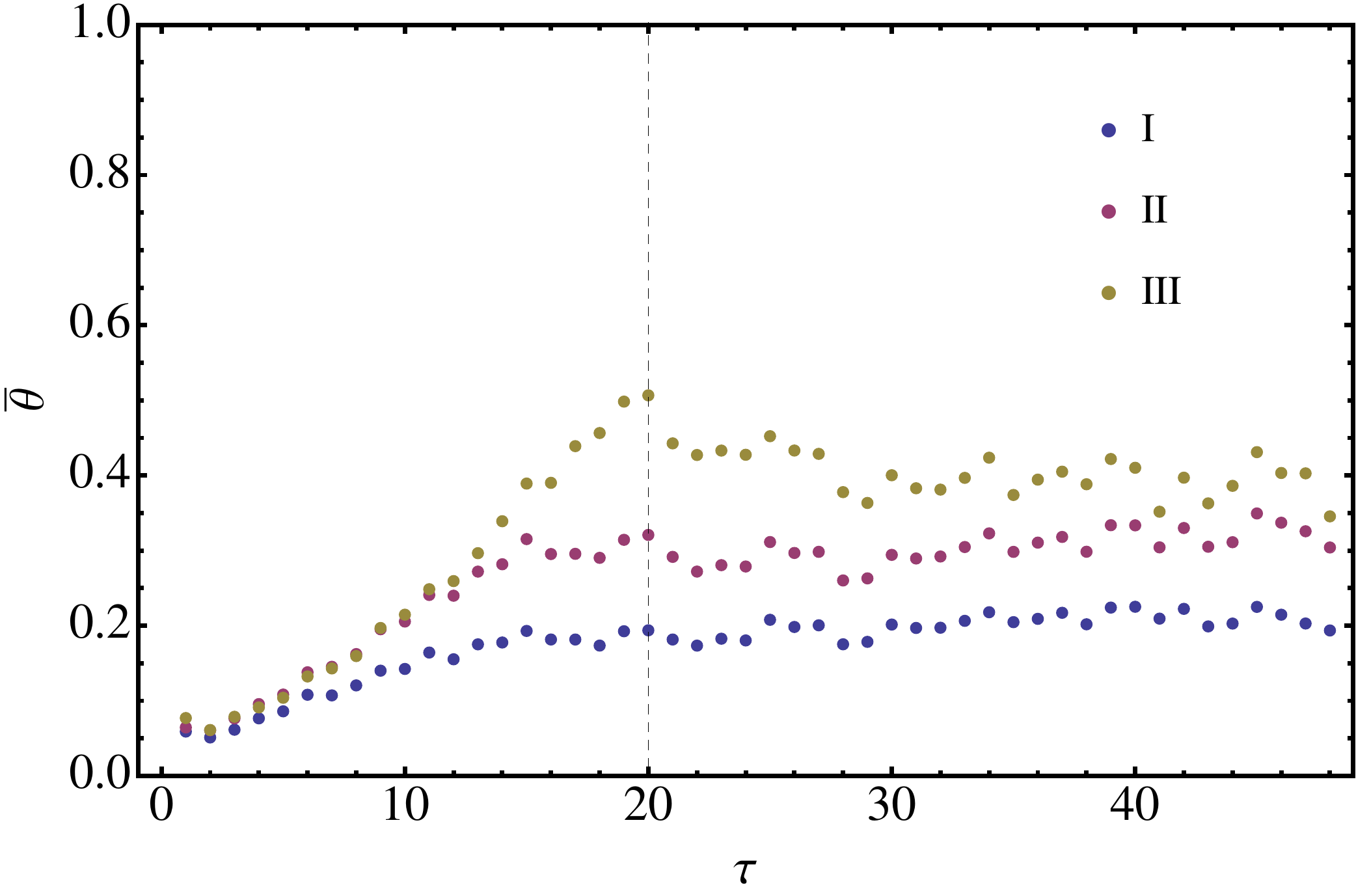}
\includegraphics[width=\figwidth]{\figdir 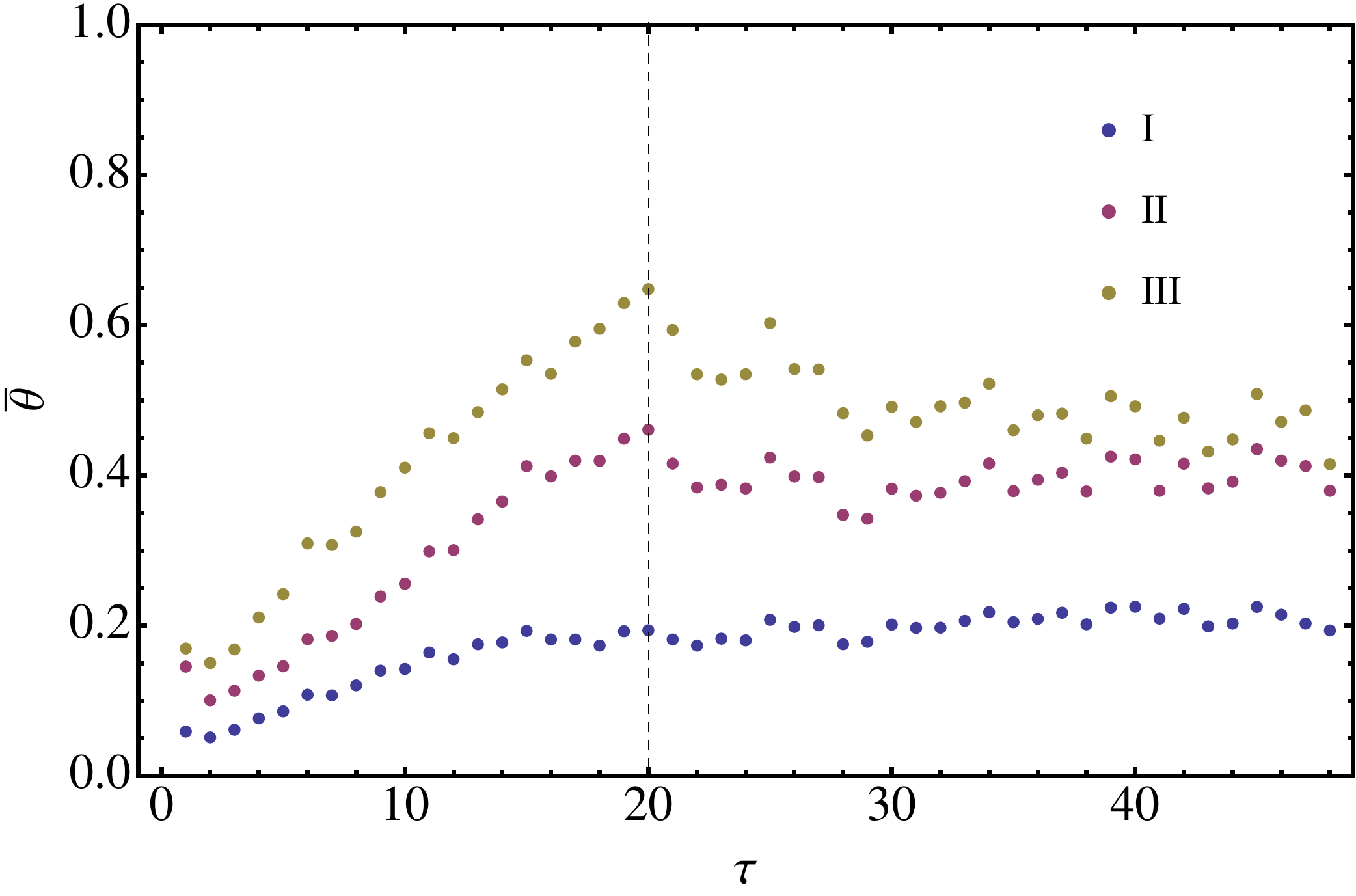}
\caption{\label{fig:rho_InterpNA_optPlots}%
Same as \Fig{pion_Interp1_optPlots}, for the rho ground state.
}
\end{figure}

Let us begin with the ground state correlators.
In \Fig{pion_Interp1_optPlots} - \Fig{rho_InterpNA_optPlots}, we consider correlators generated from source and sink vectors obtained at points along a path of steepest ascent on the signal/noise landscape, following the methods described in \Sec{steepest_ascent}.
The trajectory of ascent begins with source and sink vectors which have been source optimized.
We consider ascents on two different signal/noise landscapes.
In one case (left), the landscape is defined by the space of all sink vectors, while holding the source vector fixed, whereas in the other case (right) we consider the landscape obtained by constraining the source and sink vectors to be equal.
By considering the trajectory of ascent in these ways we are able to continuously interpolate between source-optimized correlators and signal/noise-optimized correlators.
We consider three points along the trajectory: the source optimized starting point (I), some arbitrarily chosen intermediate point (II) and the signal/noise optimized end point (III). 
The signal/noise ratios as a function of the trajectory time, $\tau_\textrm{ascent} = \epsilon n$ (where $\epsilon = 0.0001$ is the step size and $n$ corresponds to the $n$th iteration of \Eq{steepest_ascent}), are illustrated in the figures (top), with the points (I), (II), and (III) indicated along the trajectories.
The signal/noise is normalized in the figures by the maximum signal/noise, $\theta_\textrm{max}$, obtained by using the optimization procedure discussed in \Sec{max_sn}.
The optimization along a path of steepest ascent is performed in all cases using $\tau_\textrm{s}=\tau_\textrm{n}=20$.

\begin{figure}
\includegraphics[width=\figwidth]{\figdir 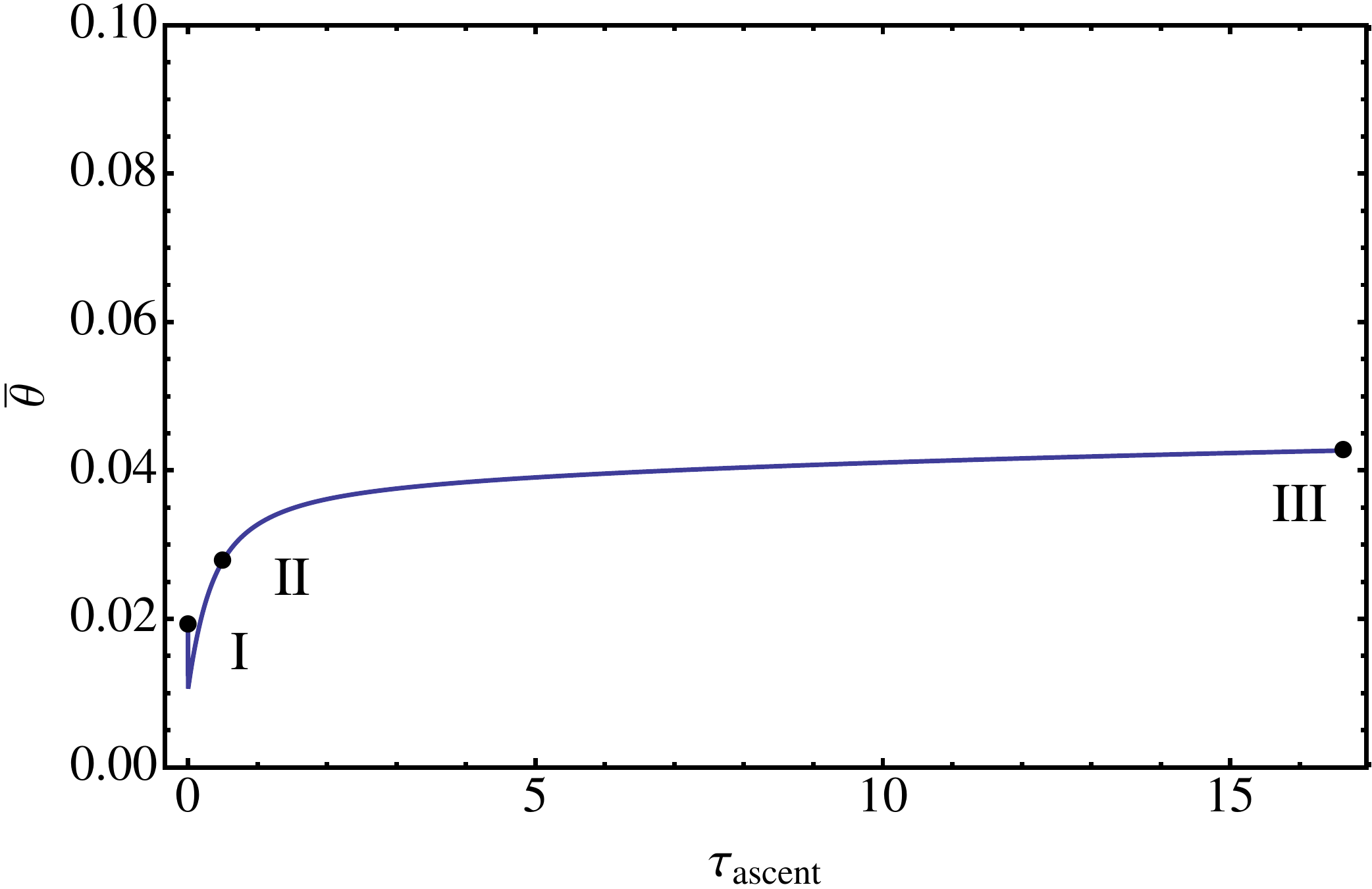}
\includegraphics[width=\figwidth]{\figdir 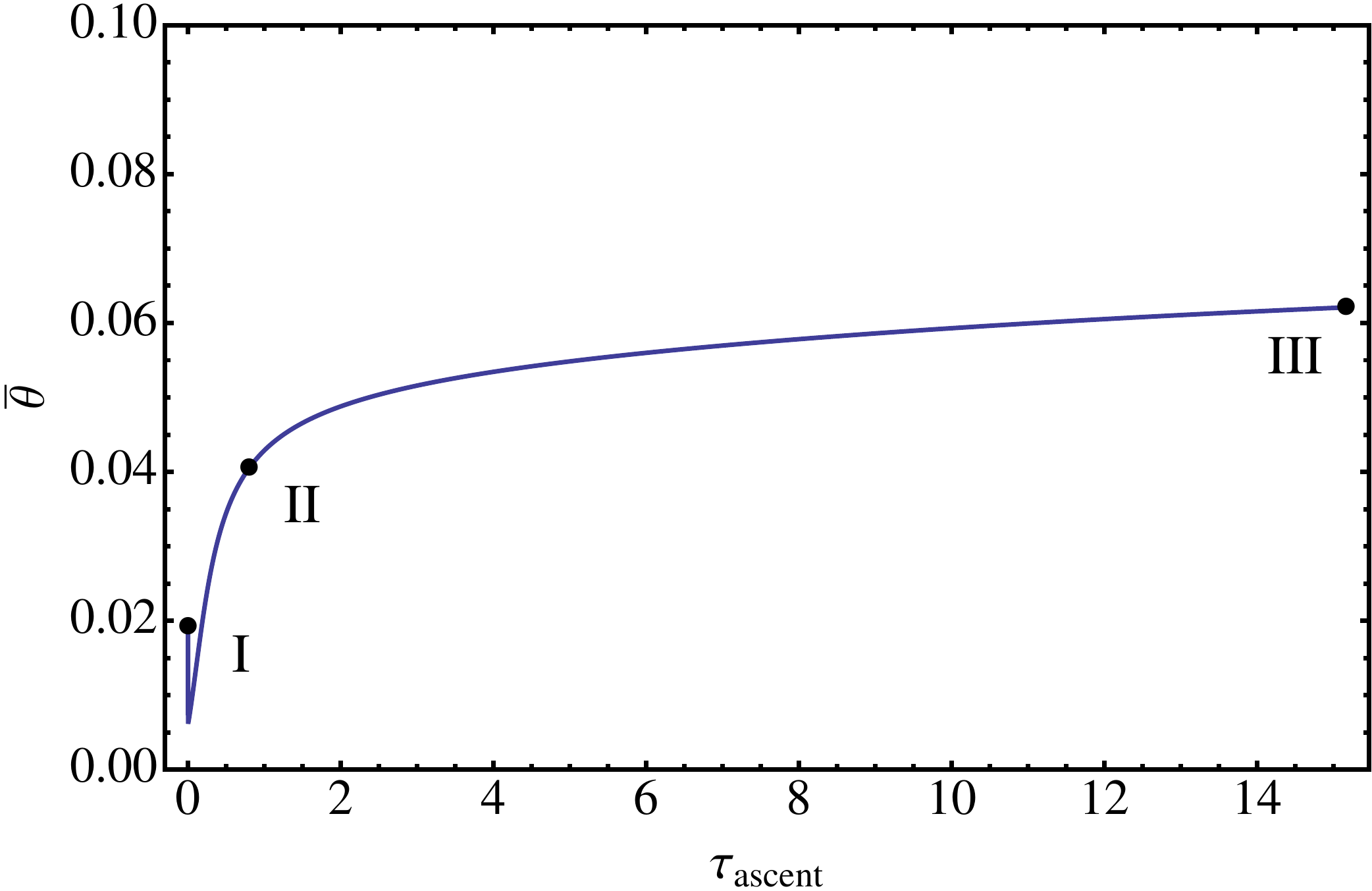}
\includegraphics[width=\figwidth]{\figdir 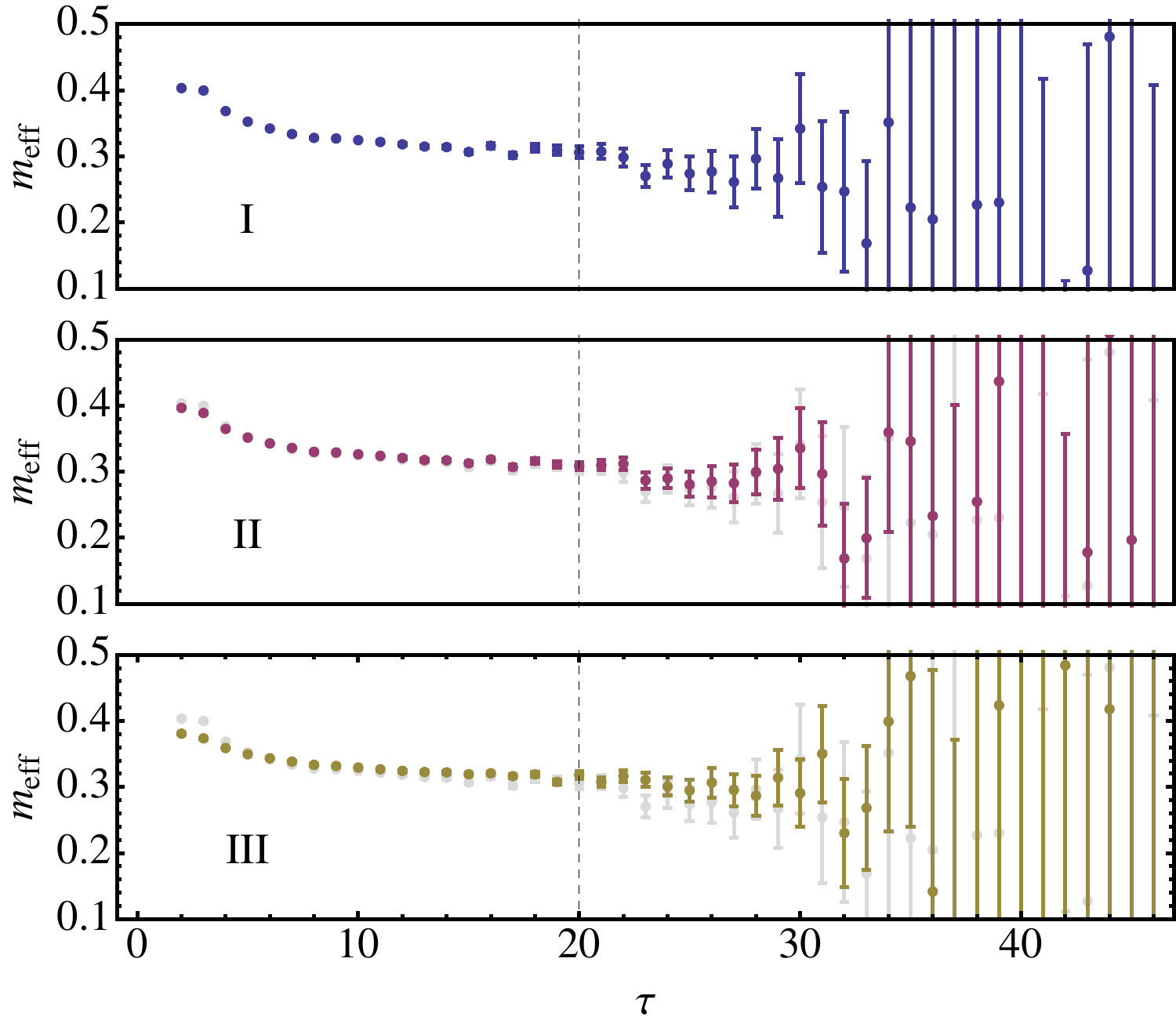}
\includegraphics[width=\figwidth]{\figdir 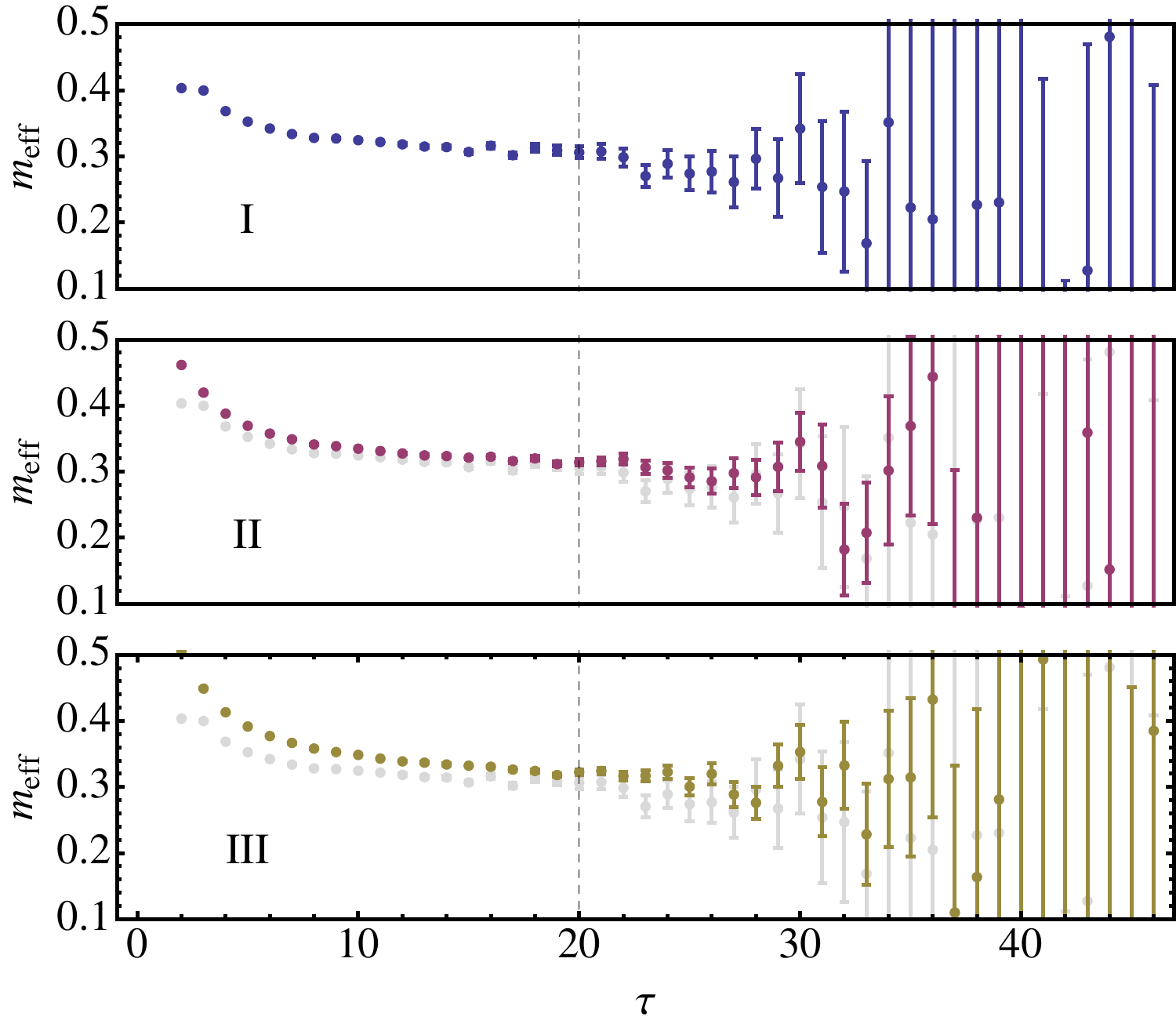}
\includegraphics[width=\figwidth]{\figdir 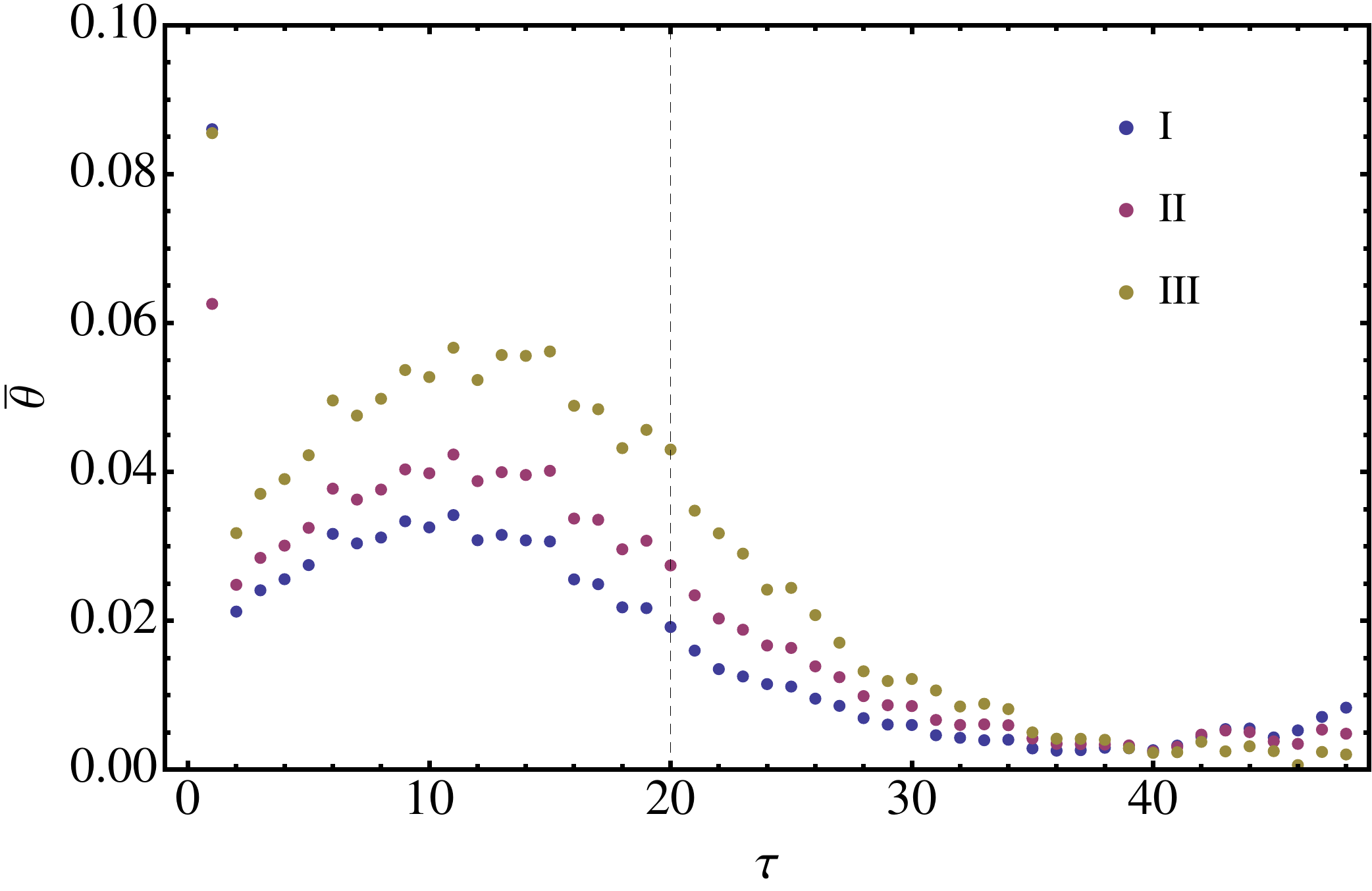}
\includegraphics[width=\figwidth]{\figdir 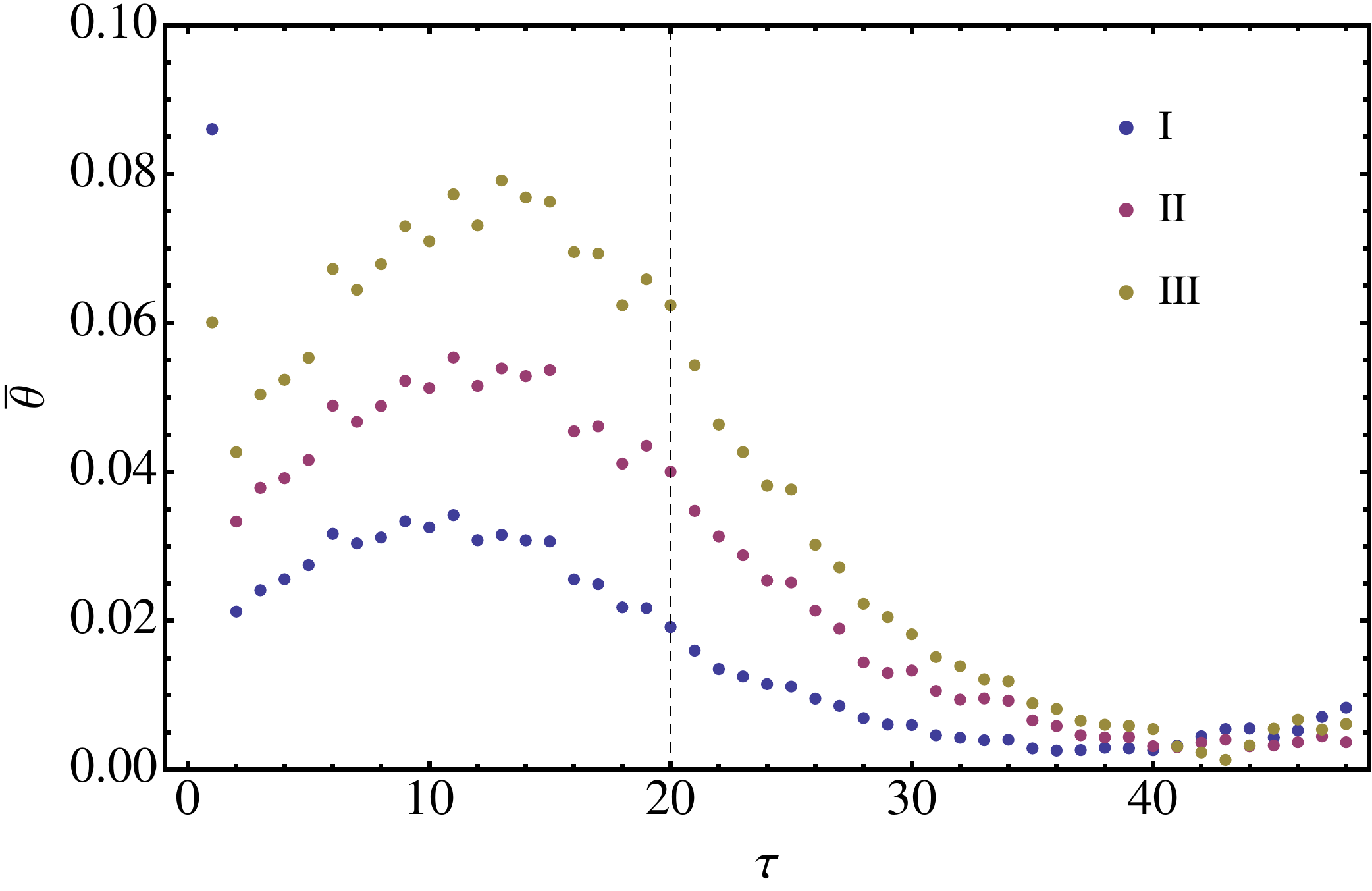}
\caption{\label{fig:rho2_InterpNA_optPlots}%
Same as \Fig{pion_Interp1_optPlots}, for the first rho excited state.
}
\end{figure}

For the first rho excited state, shown in \Fig{rho2_InterpNA_optPlots}, we follow the same procedure as the ground states.
In addition, we impose constraints on the sink and/or source, requiring that they remain orthogonal to all ground state vectors obtained variationally at time slices $\tau=13-20$.
Such constraints prevent the correlator from potentially becoming ground-state dominated within that time interval as a result of the presence of statistical fluctuations.
Note that point (I) along the steepest ascent trajectory corresponds to the variationally obtained source and sink vectors for the first excited state.
As such, the vector only approximately respects the orthogonality constraints imposed.
All subsequent points along the trajectory have the constraint fully imposed, however, giving rise to the discontinuity in the signal/noise trajectory at the initial ascent time.

The effective mass associated with correlators constructed from  source and sink vectors at positions (I), (II), and (III) are displayed in \Fig{pion_Interp1_optPlots} - \Fig{rho2_InterpNA_optPlots} (center), along with the corresponding signal/noise as a function of time separation (bottom).
Note that the normalized signal/noise associated with the correlators may slightly exceed unity because the uncertainties were determined via a bootstrap analysis of the correlation functions, which are subject to statistical variation, whereas the normalization factor was determined by extremizing the signal/noise ratio.
In the case of the pion ground state, a sustained signal/noise enhancement of approximately 1.2 is achieved for the correlator, whereas for the proton, delta, and rho ground states, a sustained enhancement ranges from approximately 2 to 3.
In each case, significant excited state contamination is introduced when source and sink vectors are left unfixed in the optimization.
Such contamination is suppressed for the pion, proton and delta, and to a lesser extent the rho, when the source is held fixed and equal to a source optimized source vector.

\begin{figure}
\includegraphics[width=\figwidth]{\figdir 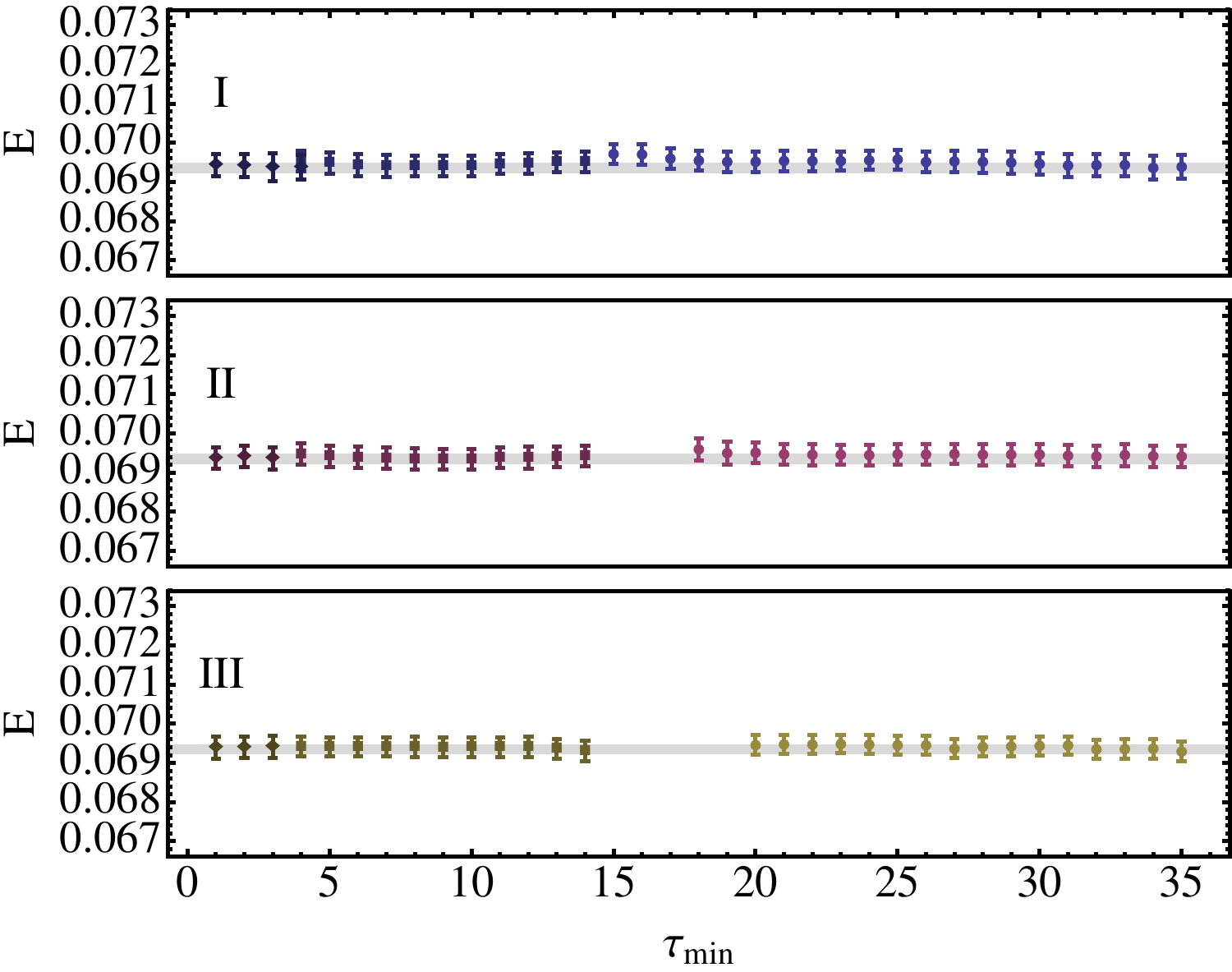}
\includegraphics[width=\figwidth]{\figdir 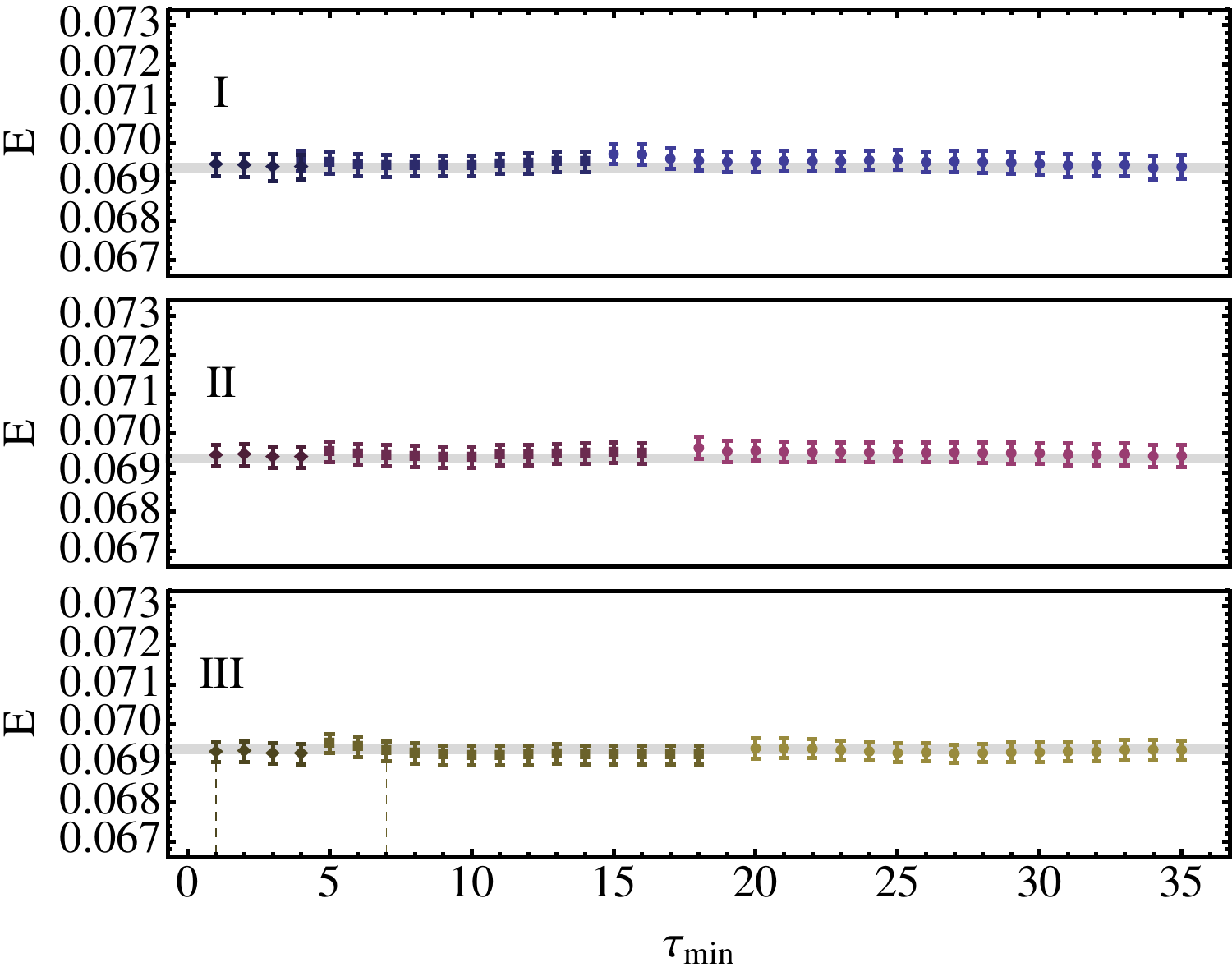}
\includegraphics[width=\figwidth]{\figdir 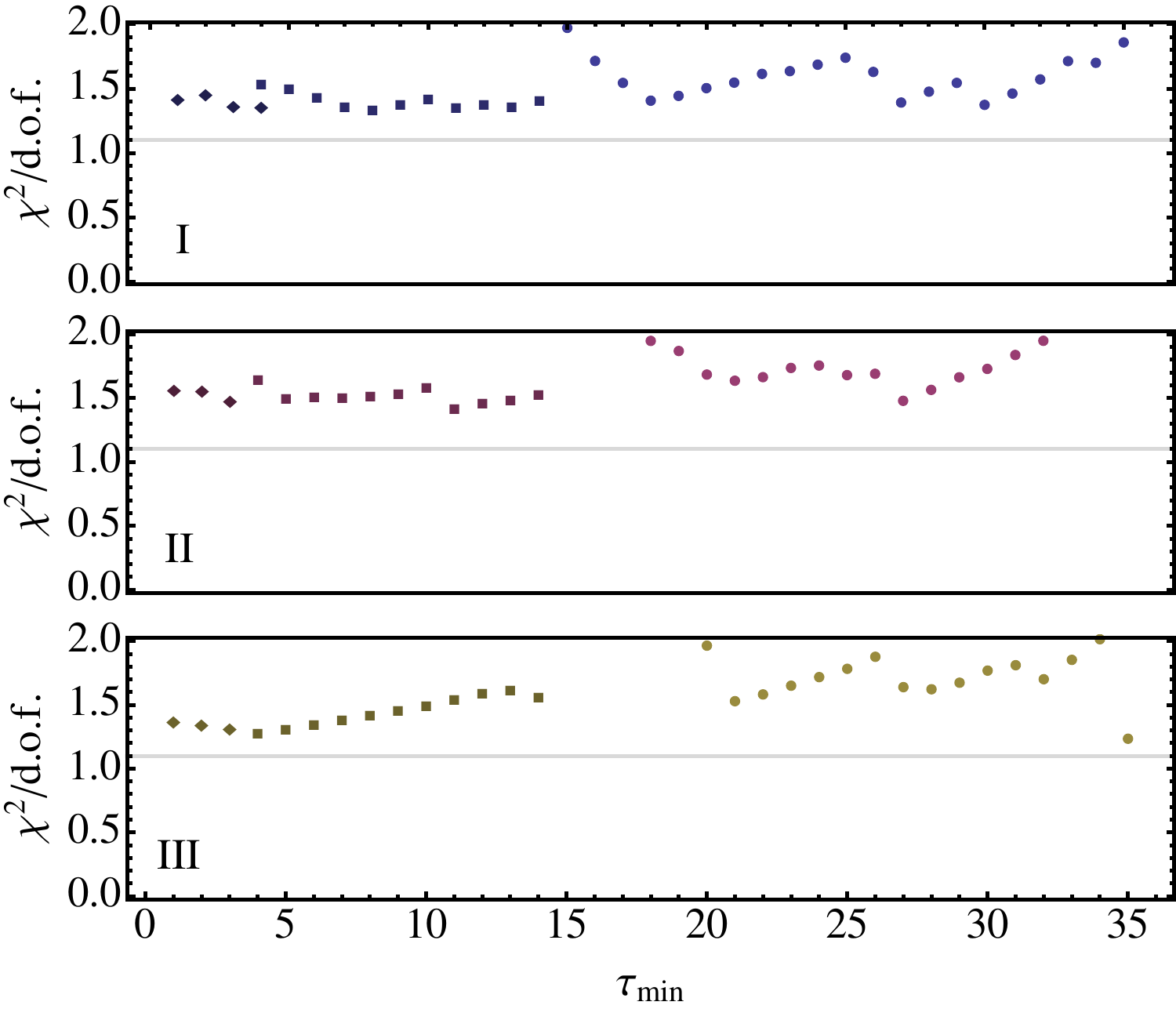}
\includegraphics[width=\figwidth]{\figdir 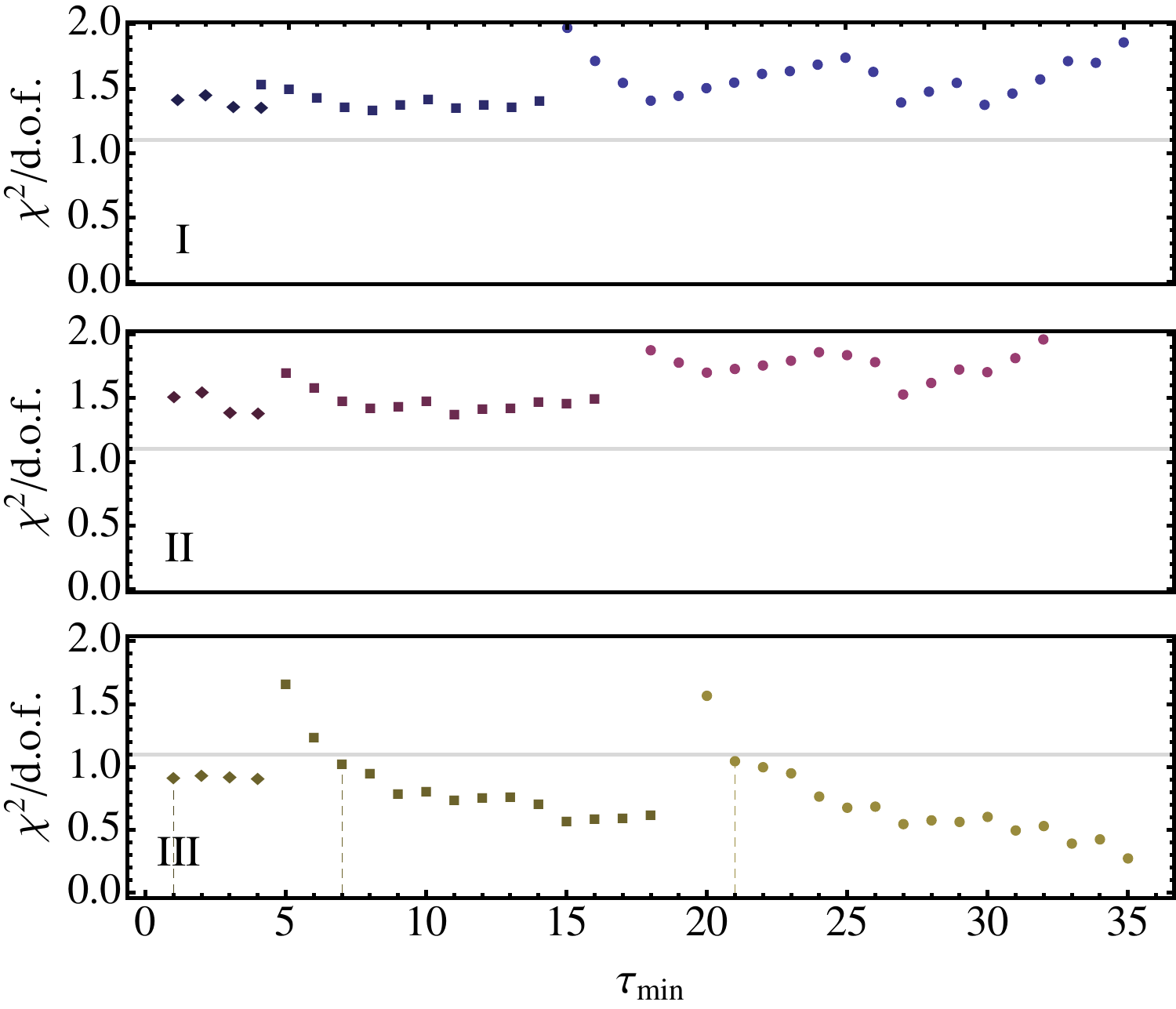}
\includegraphics[width=\figwidth]{\figdir 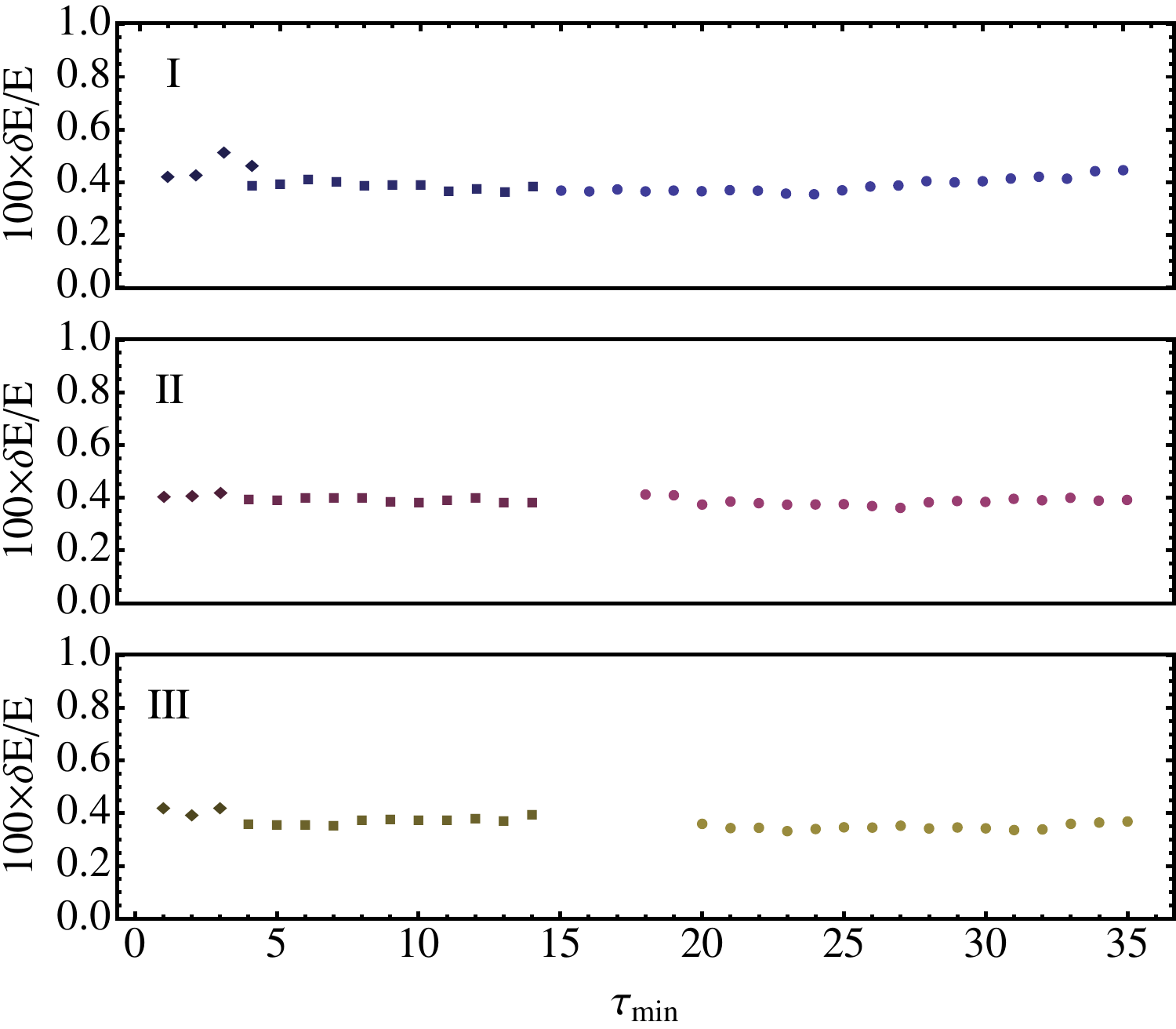}
\includegraphics[width=\figwidth]{\figdir 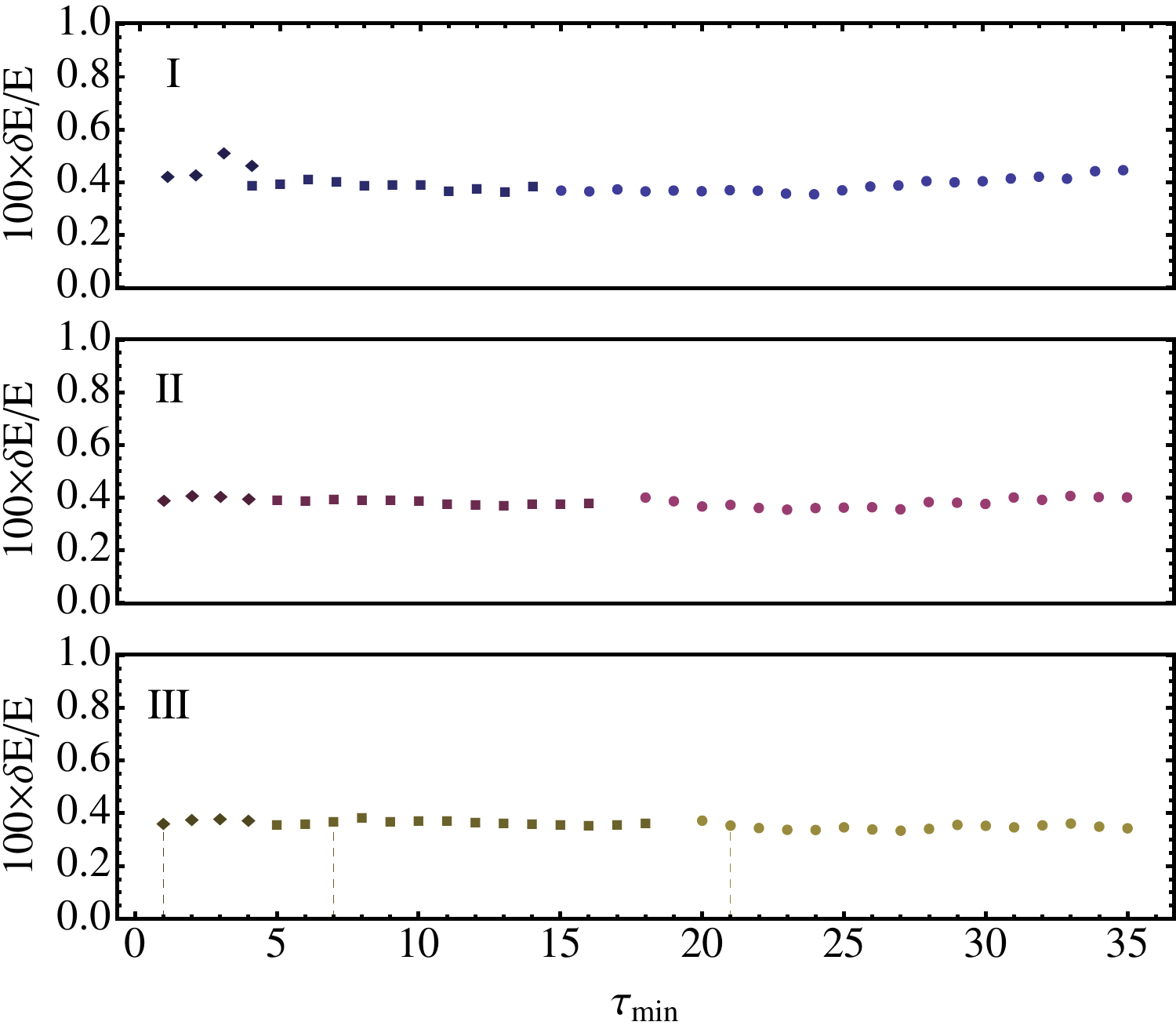}
\caption{\label{fig:pion_Interp1_results}%
Top: Extracted energy of the pion, from single- (circle), double- (square), and triple-exponential (diamond) fits to the correlators shown in \Fig{pion_Interp1_optPlots} (center) as a function of $\tau_\textrm{min}$.
Results are shown for signal/noise-optimized correlators with a fixed (left) and unconstrained (right) source, and for the respective cases labeled (I), (II) and (III).
A horizontal band indicates the $1\sigma$ error about the central value estimate of the energy quoted in \cite{Beane:2009kya}.
Center: Corresponding $\chi^2/\textrm{d.o.f.}$ obtained for each fit.
The horizontal line indicates the $\chi^2/\textrm{d.o.f.} = 1.1$ threshold used to define an acceptable fit.
Fit results associated with the earliest $\tau_\textrm{min}$ satisfying $\chi^2/\textrm{d.o.f.} < 1.1$ are indicated with vertical dashed lines, and provided in \Tab{pion_fit_results}.
Bottom: Relative uncertainties on the extracted energy as a function of $\tau_\textrm{min}$.
}
\end{figure}

\begin{figure}
\includegraphics[width=\figwidth]{\figdir 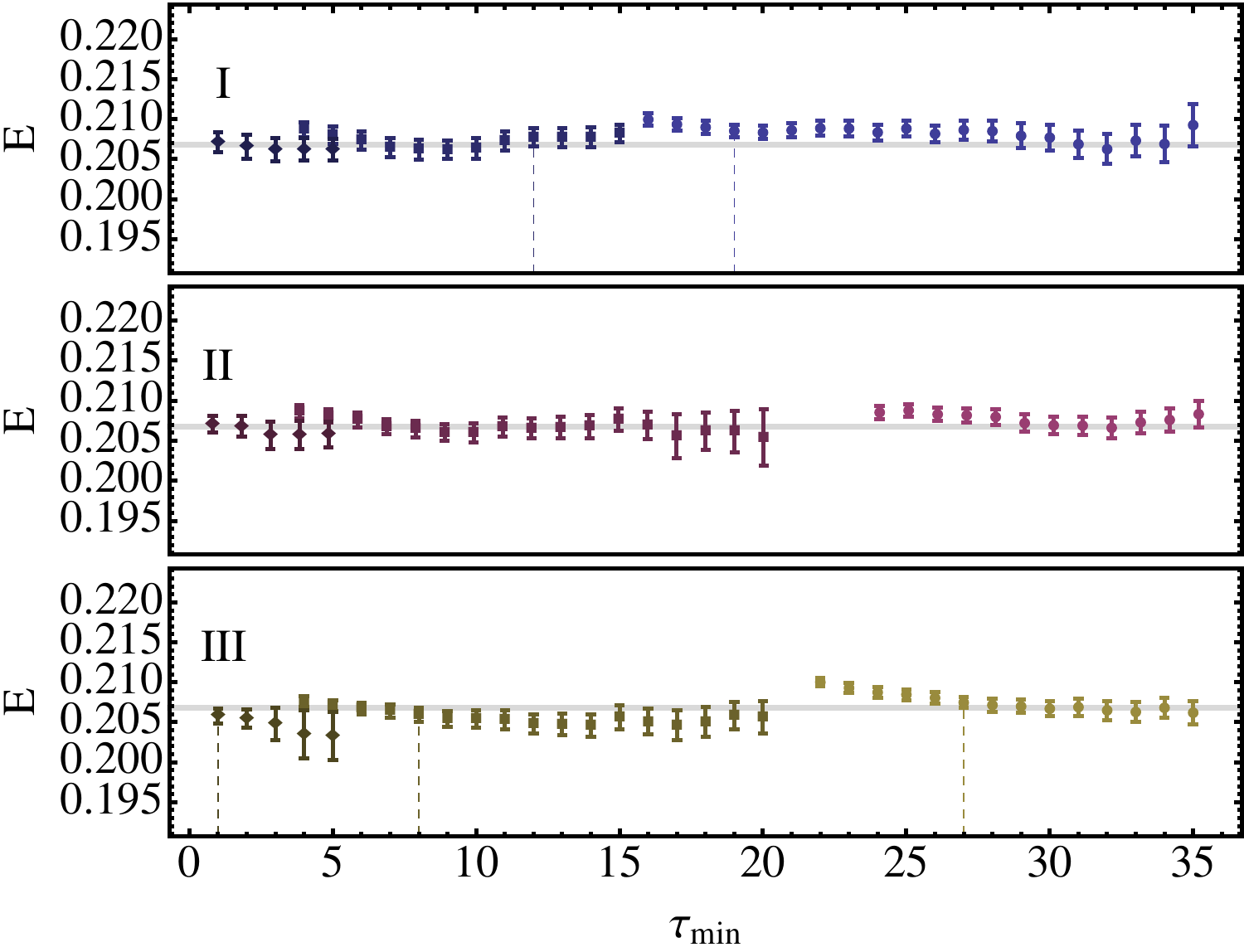}
\includegraphics[width=\figwidth]{\figdir 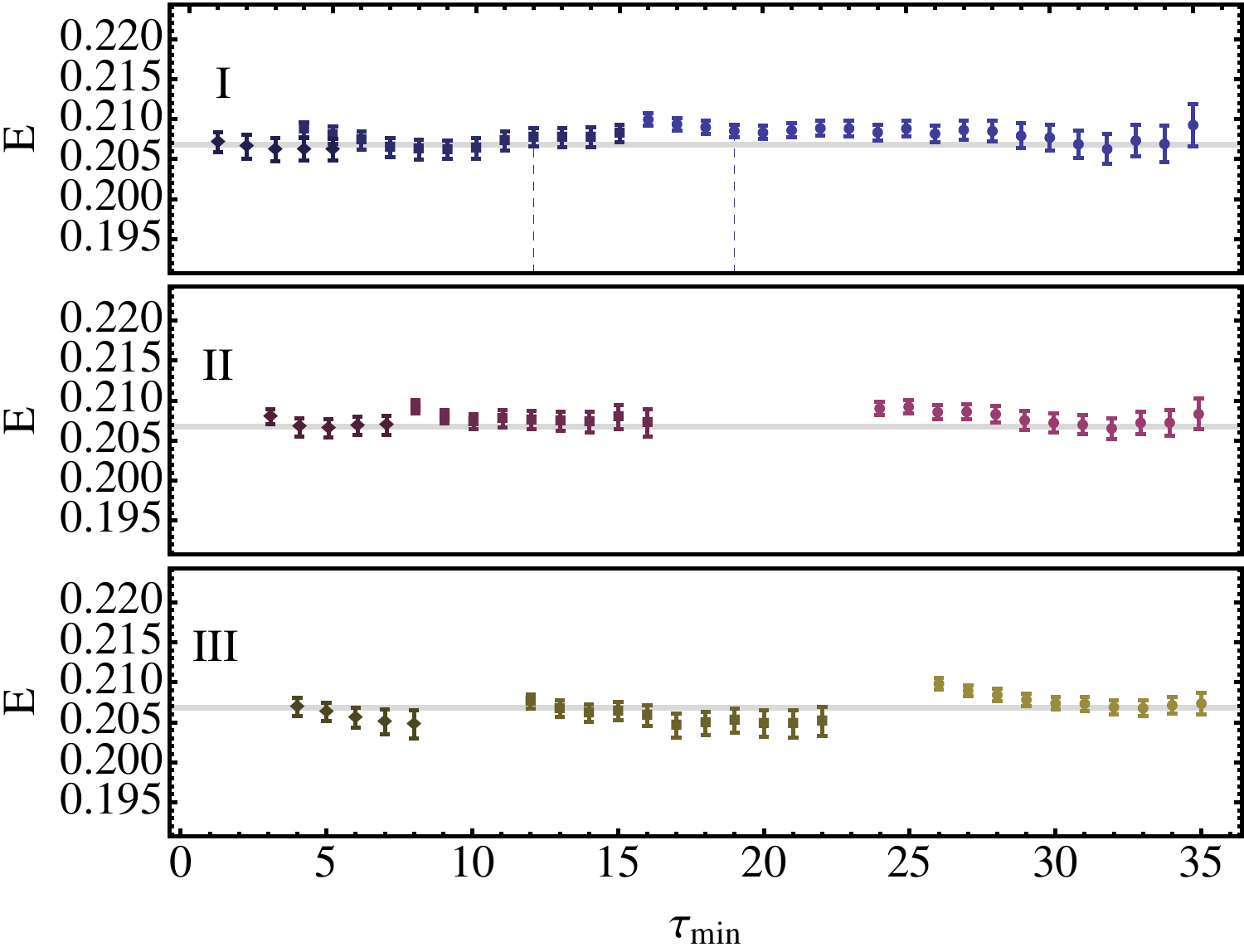}
\includegraphics[width=\figwidth]{\figdir 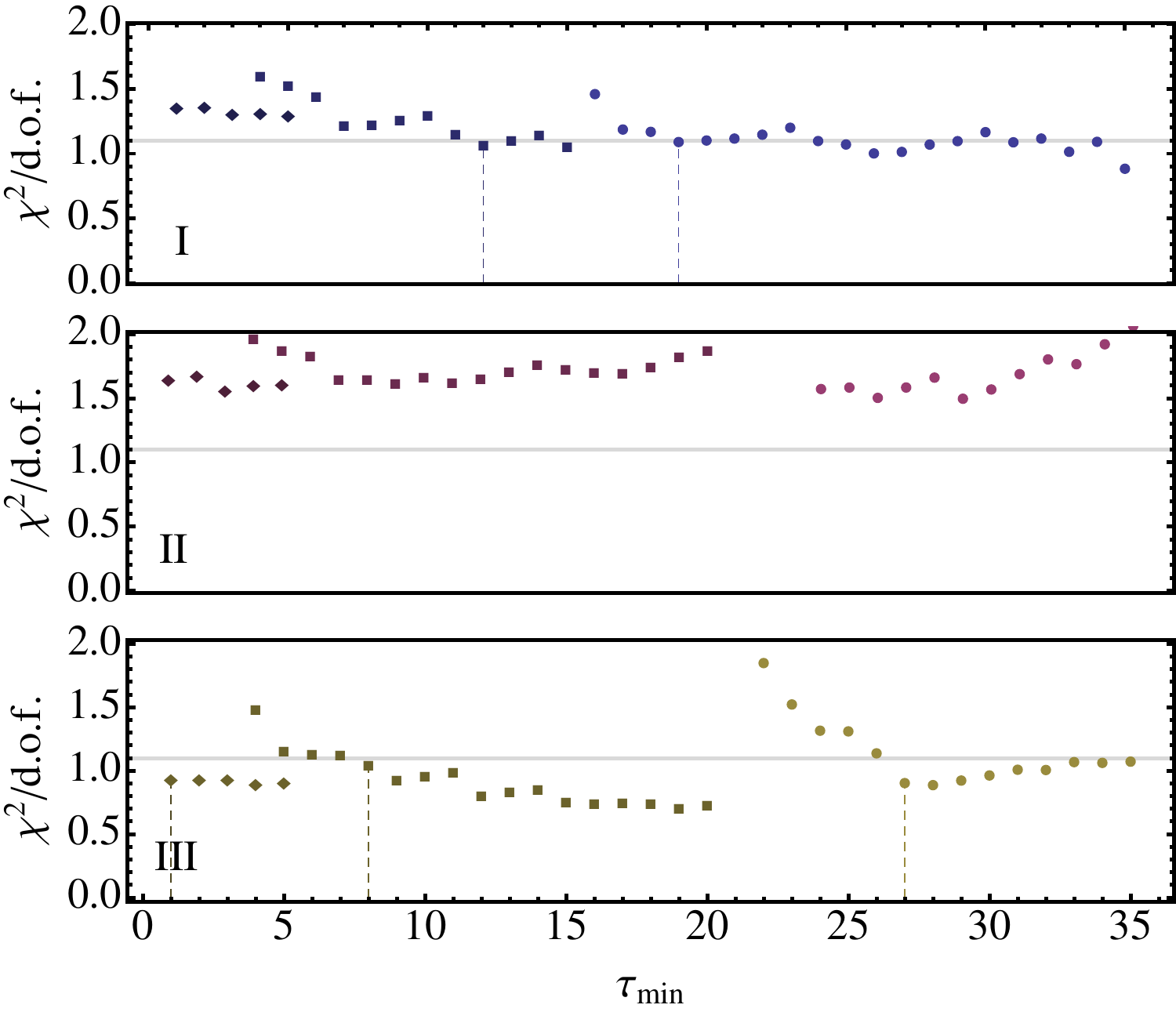}
\includegraphics[width=\figwidth]{\figdir 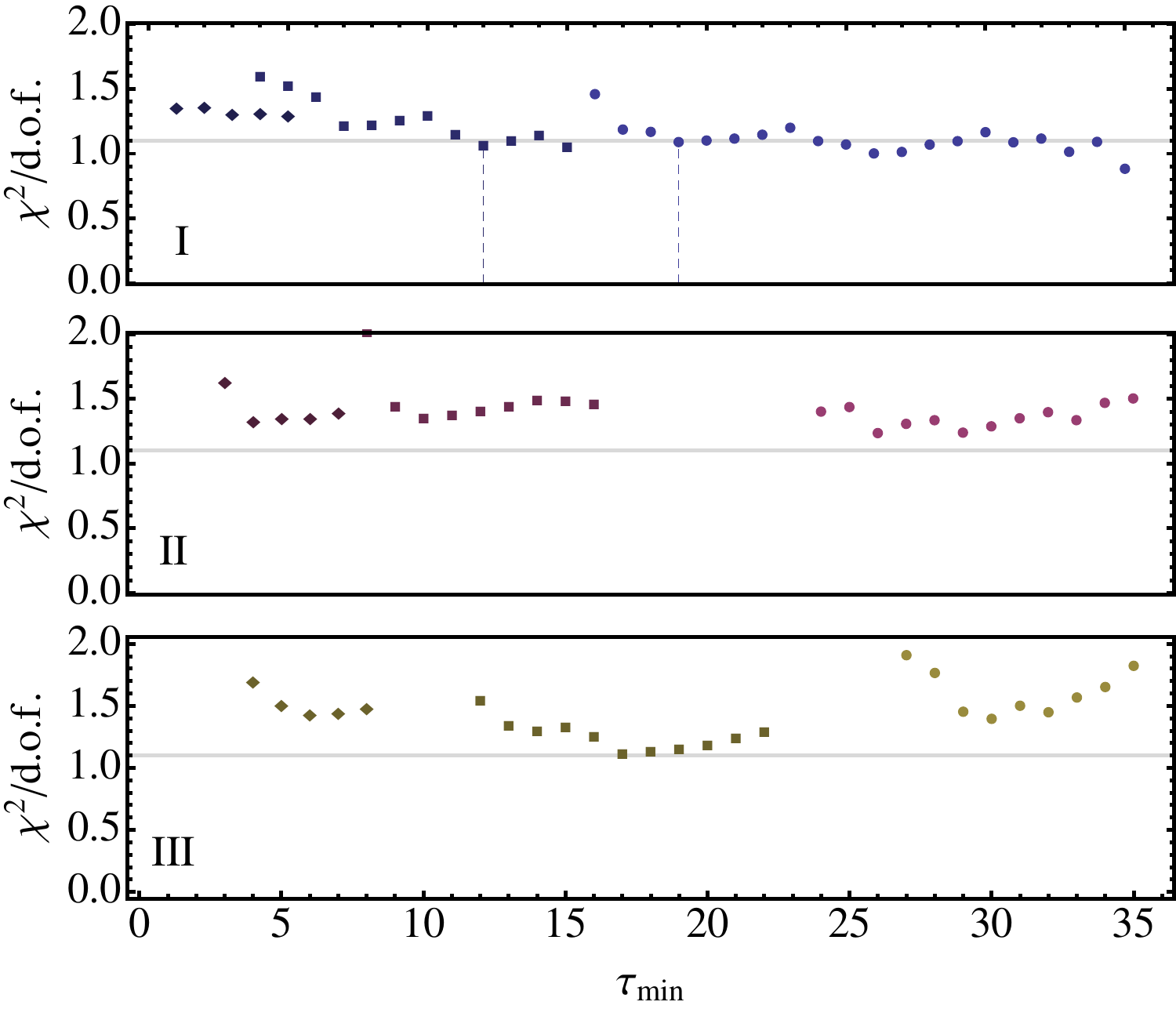}
\includegraphics[width=\figwidth]{\figdir 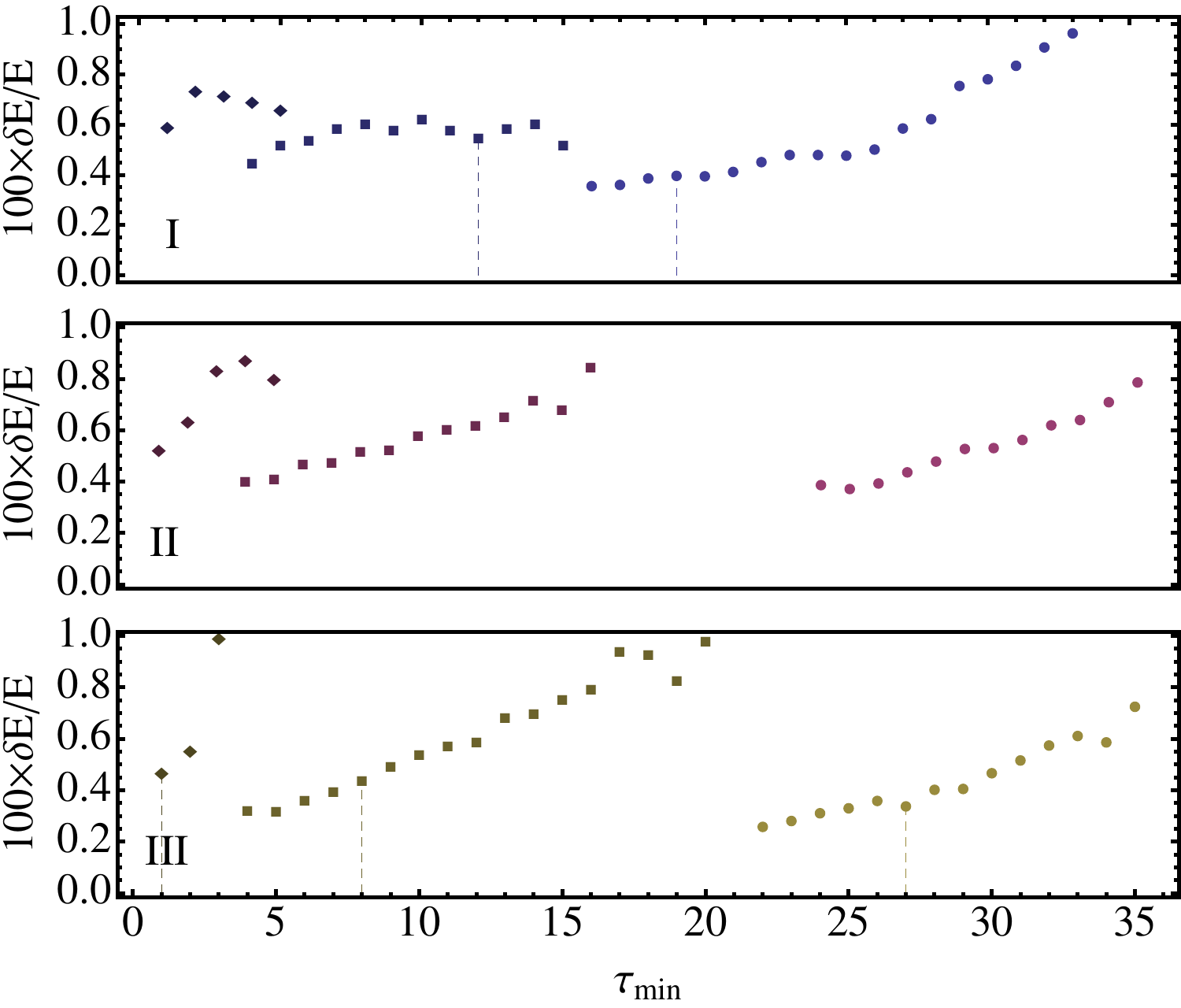}
\includegraphics[width=\figwidth]{\figdir 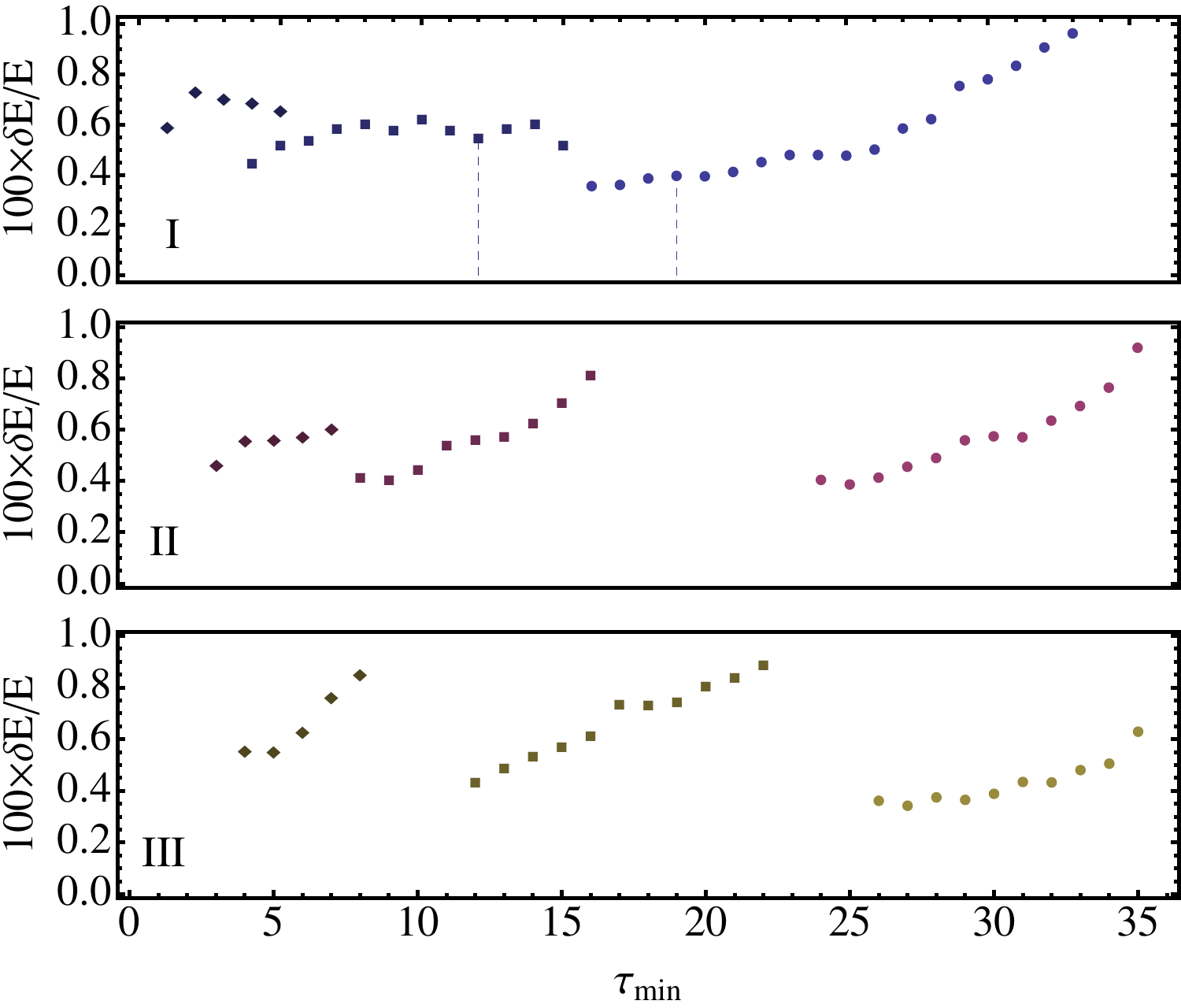}
\caption{\label{fig:proton_Interp4_results}%
Same as \Fig{pion_Interp1_optPlots}, for the proton ground state.
Fit results associated with the earliest $\tau_\textrm{min}$ satisfying $\chi^2/\textrm{d.o.f.} < 1.1$ are indicated with vertical dashed lines, and provided in \Tab{proton_fit_results}.
}
\end{figure}

\begin{figure}
\includegraphics[width=\figwidth]{\figdir 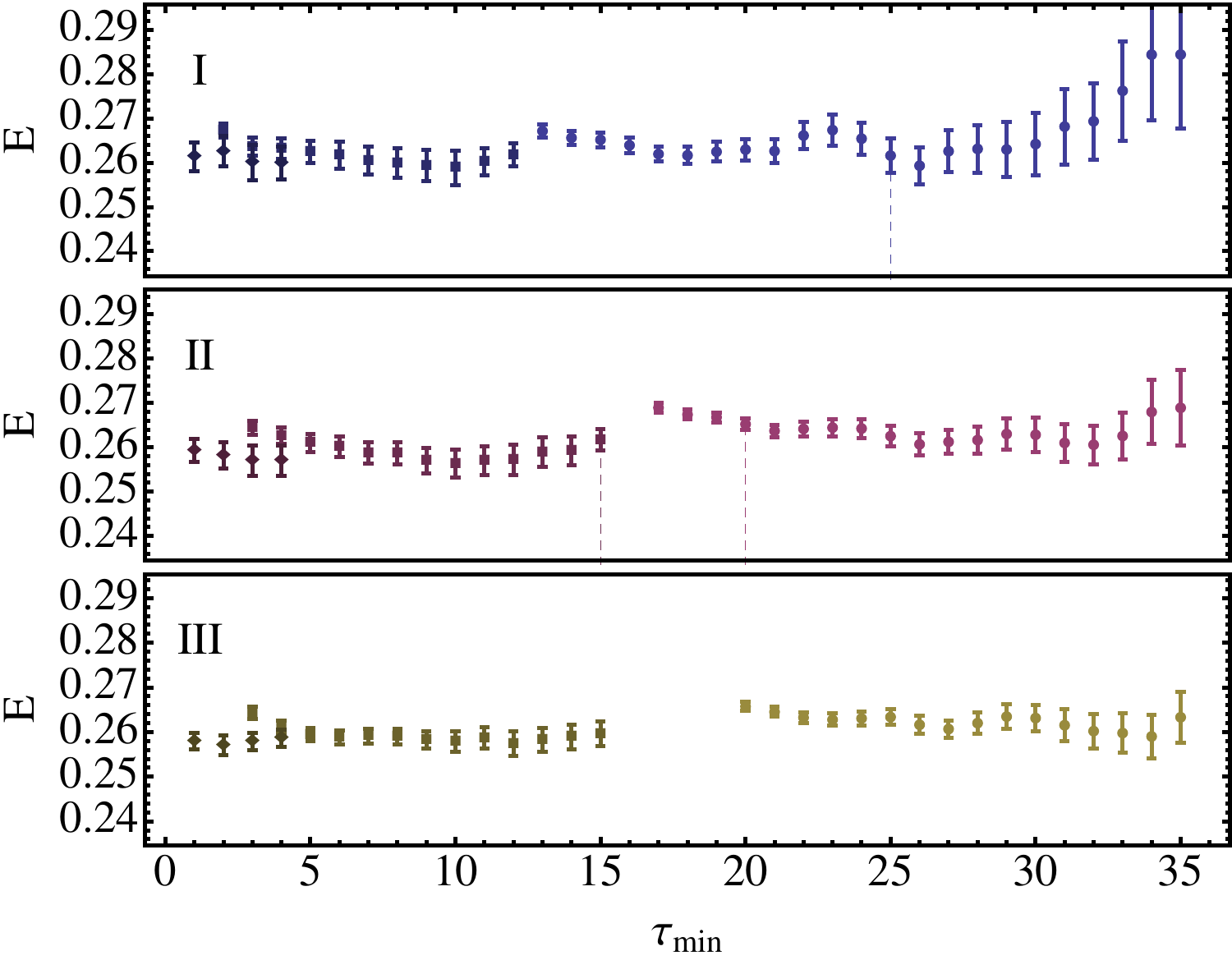}
\includegraphics[width=\figwidth]{\figdir 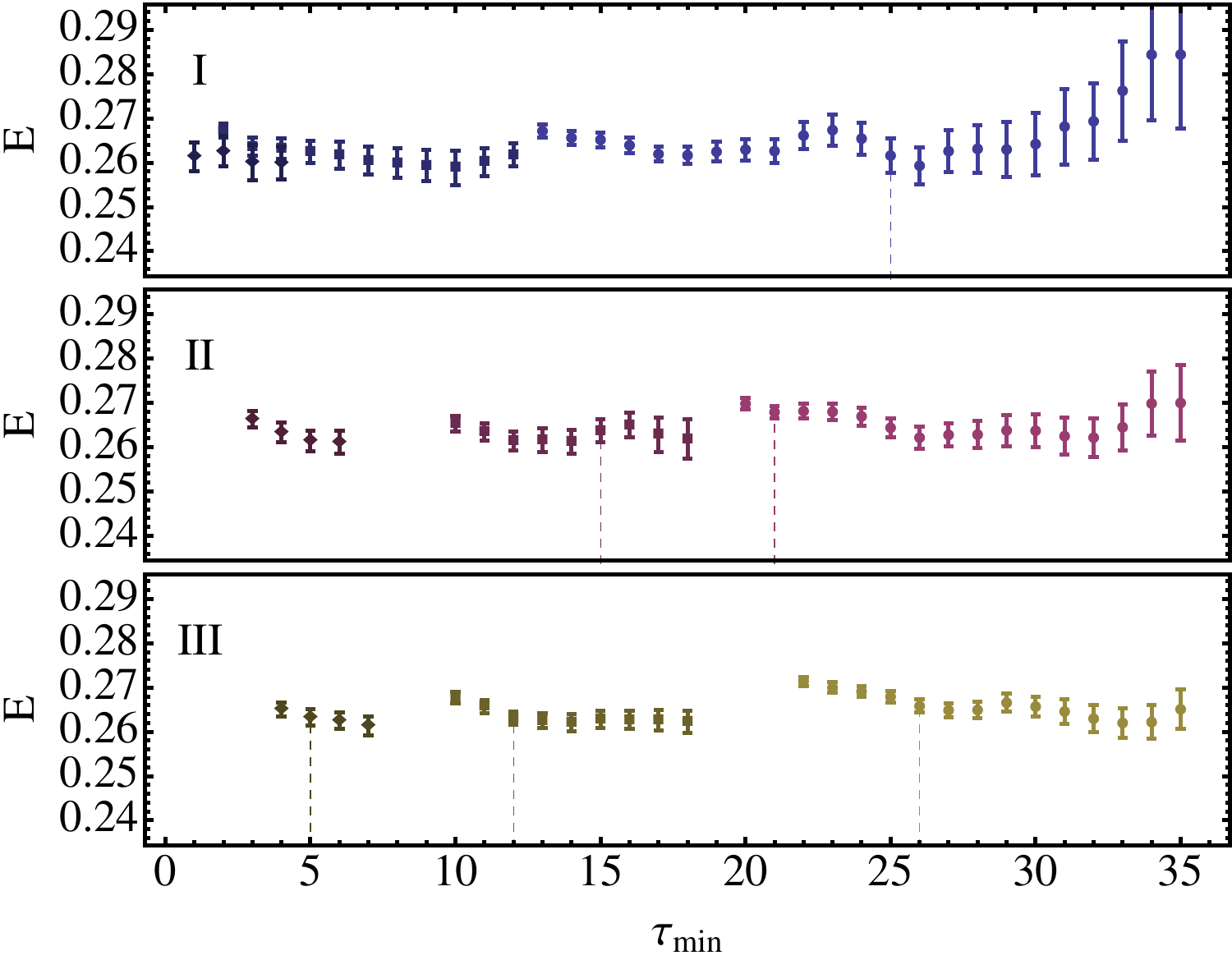}
\includegraphics[width=\figwidth]{\figdir 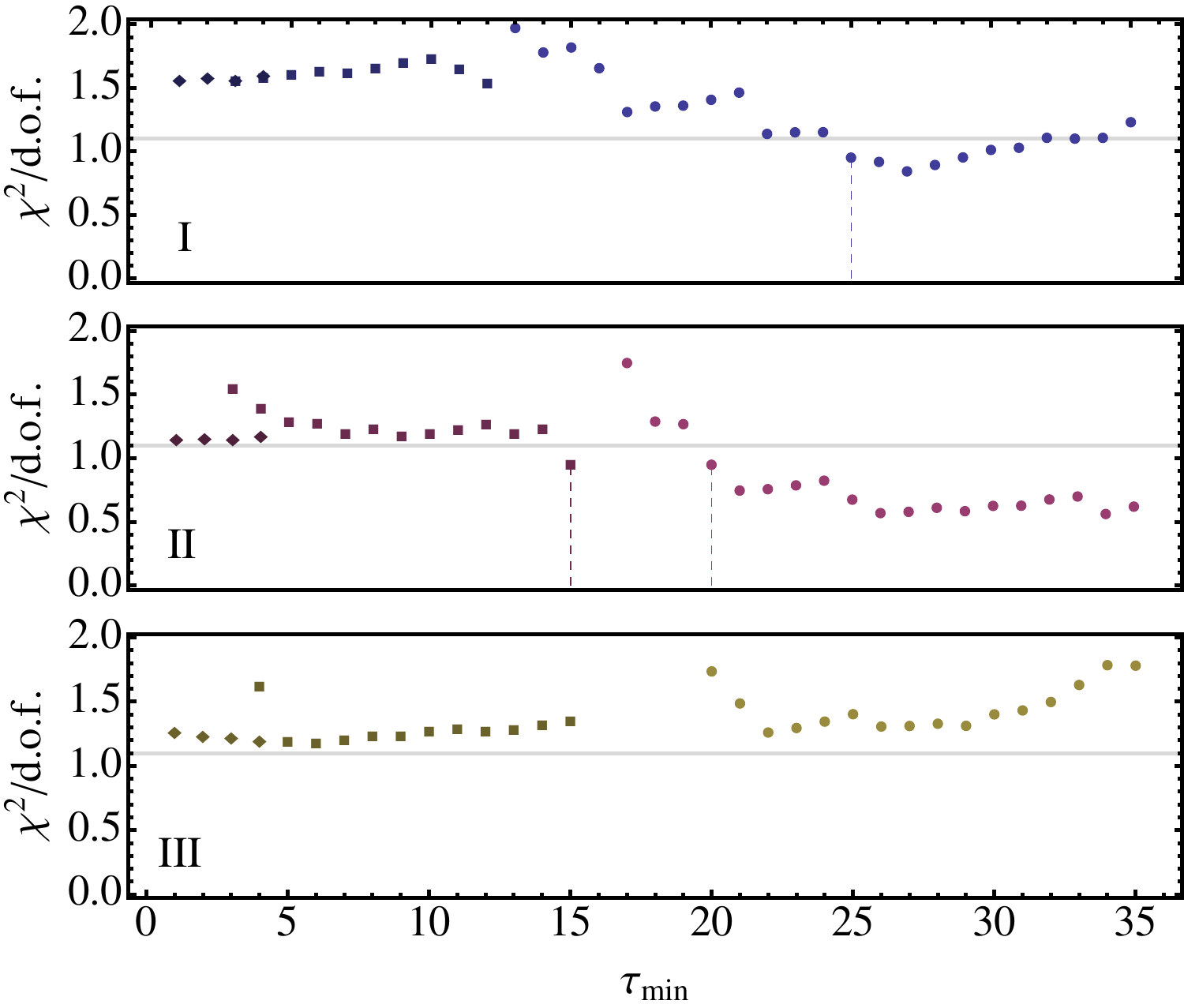}
\includegraphics[width=\figwidth]{\figdir 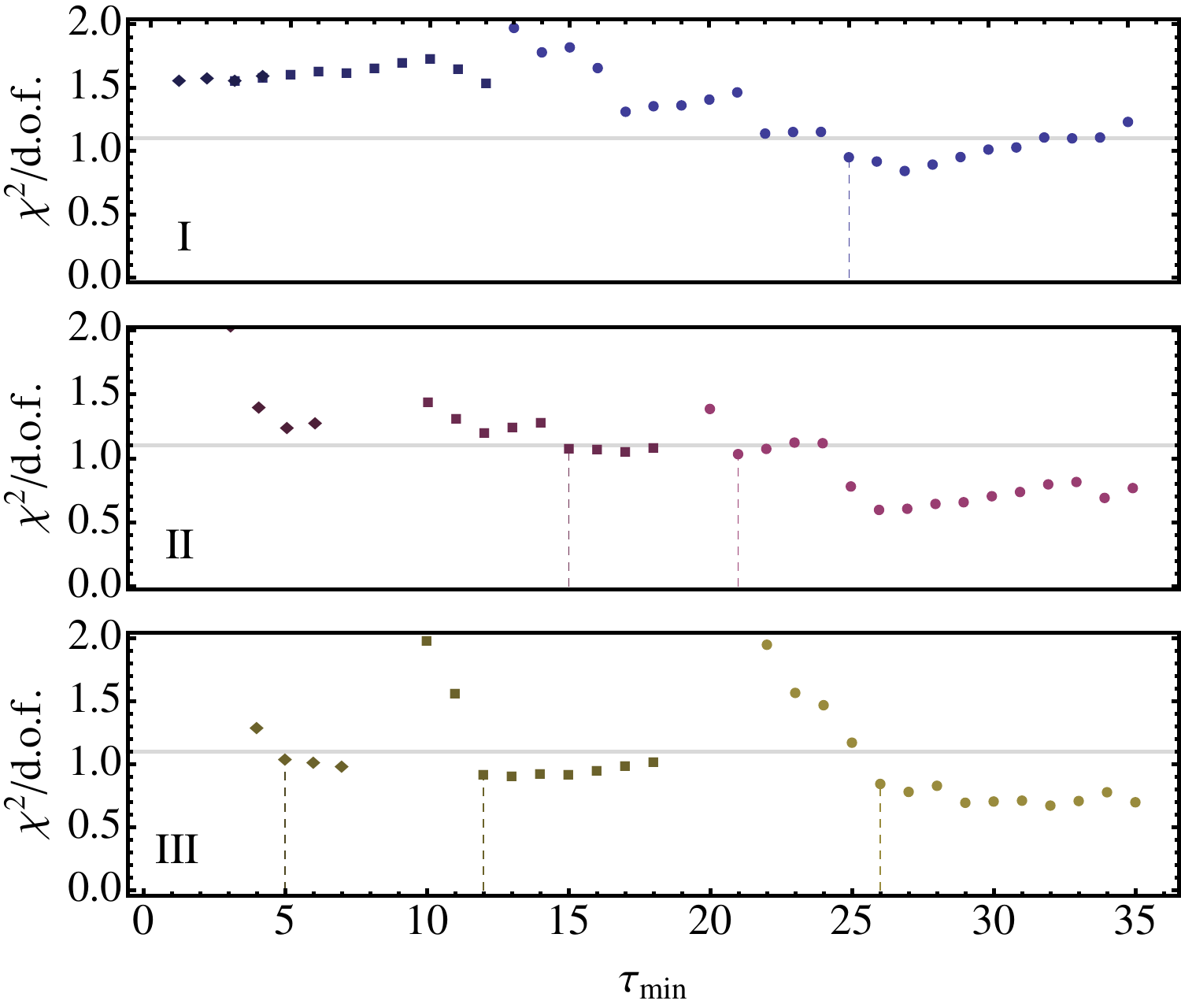}
\includegraphics[width=\figwidth]{\figdir 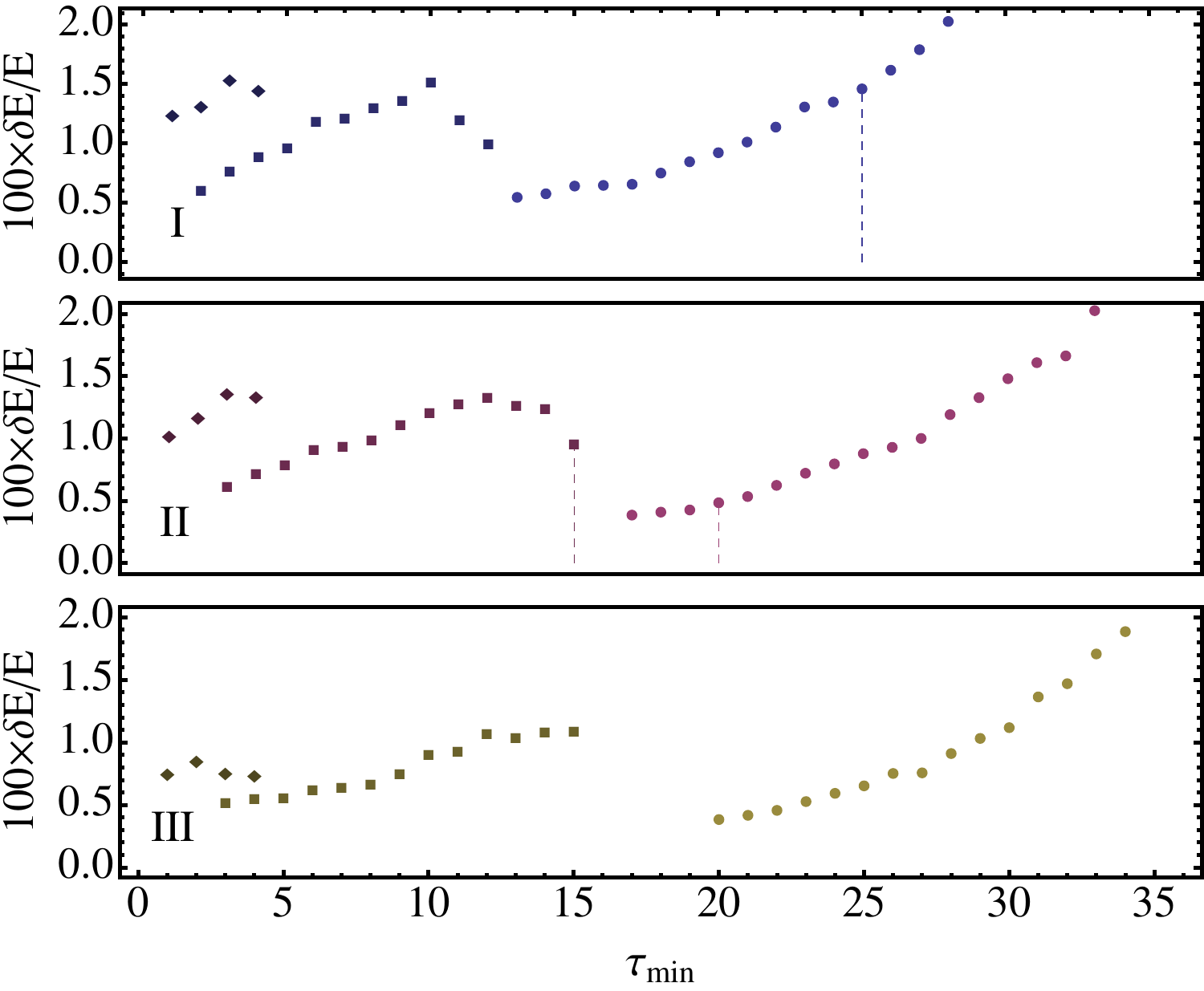}
\includegraphics[width=\figwidth]{\figdir 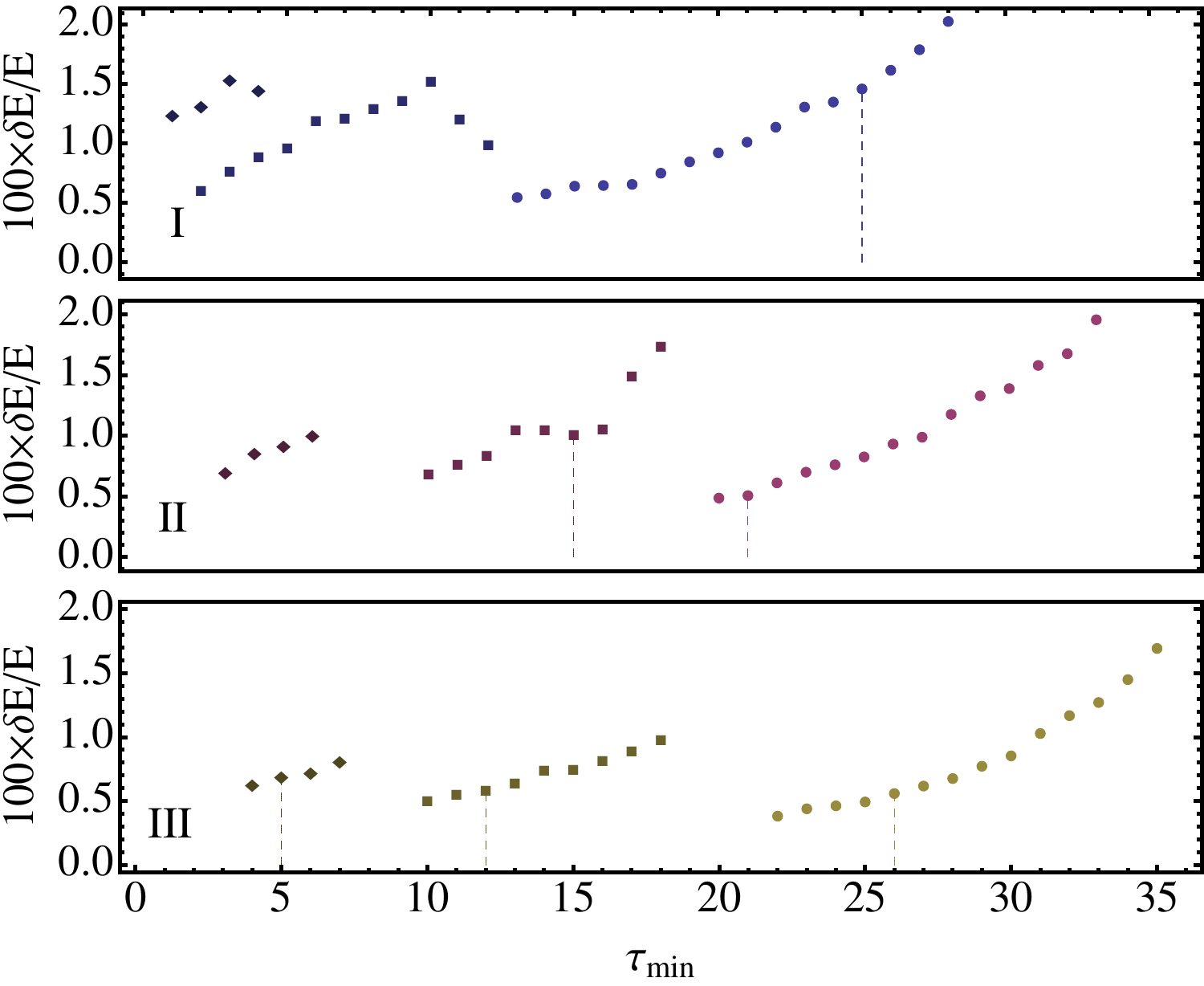}
\caption{\label{fig:delta_Interp1_results}%
Same as \Fig{pion_Interp1_optPlots}, for the delta ground state.
Fit results associated with the earliest $\tau_\textrm{min}$ satisfying $\chi^2/\textrm{d.o.f.} < 1.1$ are indicated with vertical dashed lines, and provided in \Tab{delta_fit_results}.
}
\end{figure}

\begin{figure}
\includegraphics[width=\figwidth]{\figdir 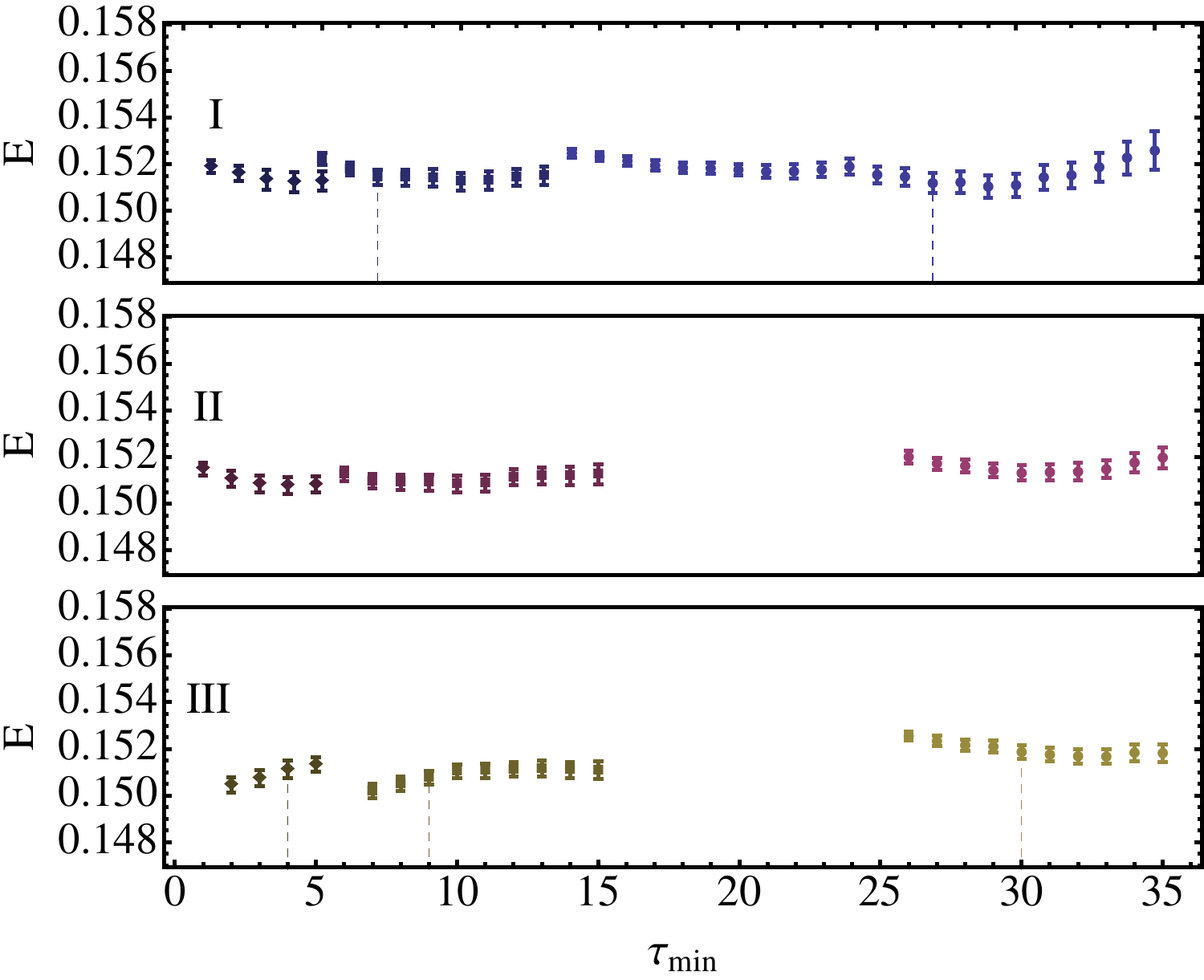}
\includegraphics[width=\figwidth]{\figdir 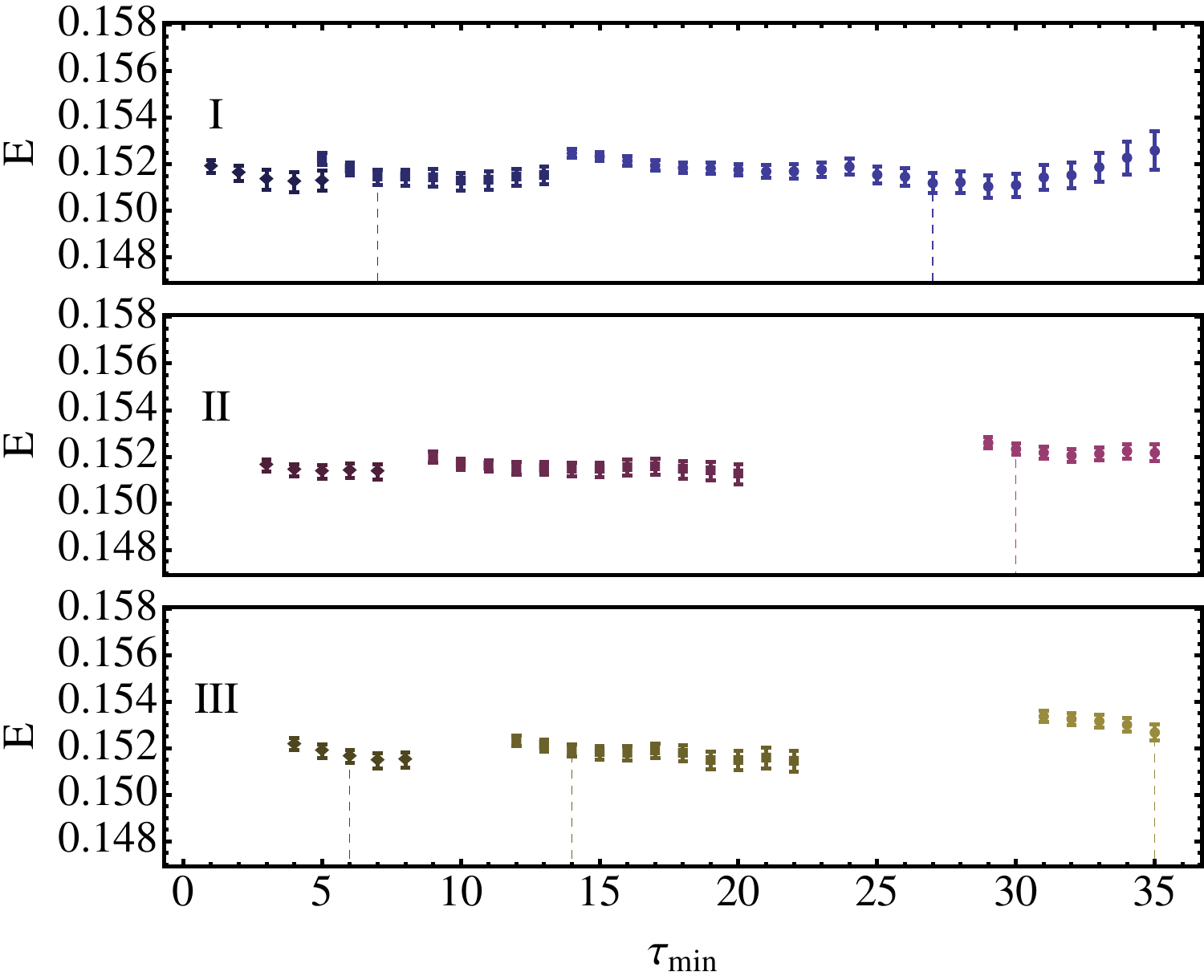}
\includegraphics[width=\figwidth]{\figdir 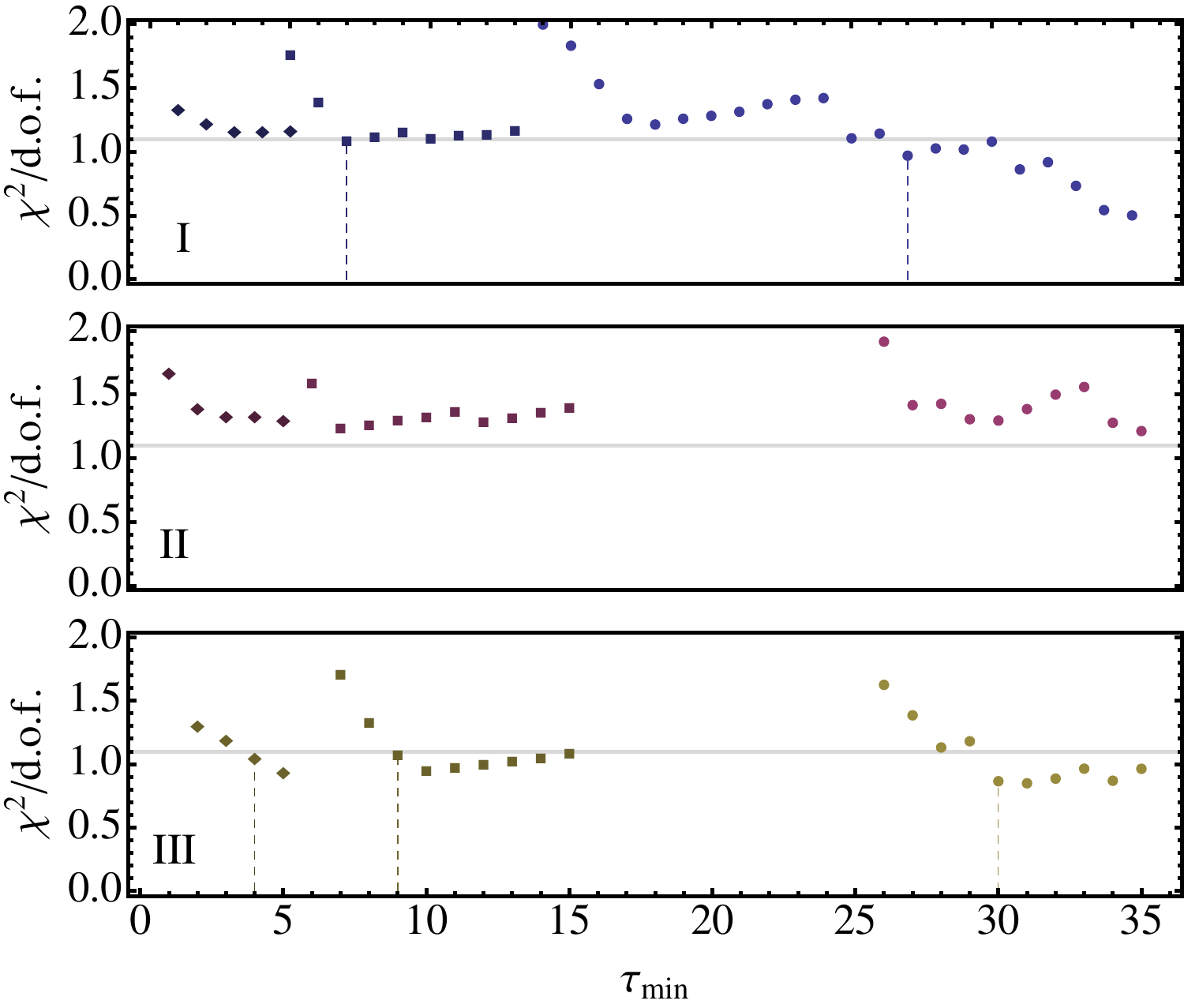}
\includegraphics[width=\figwidth]{\figdir 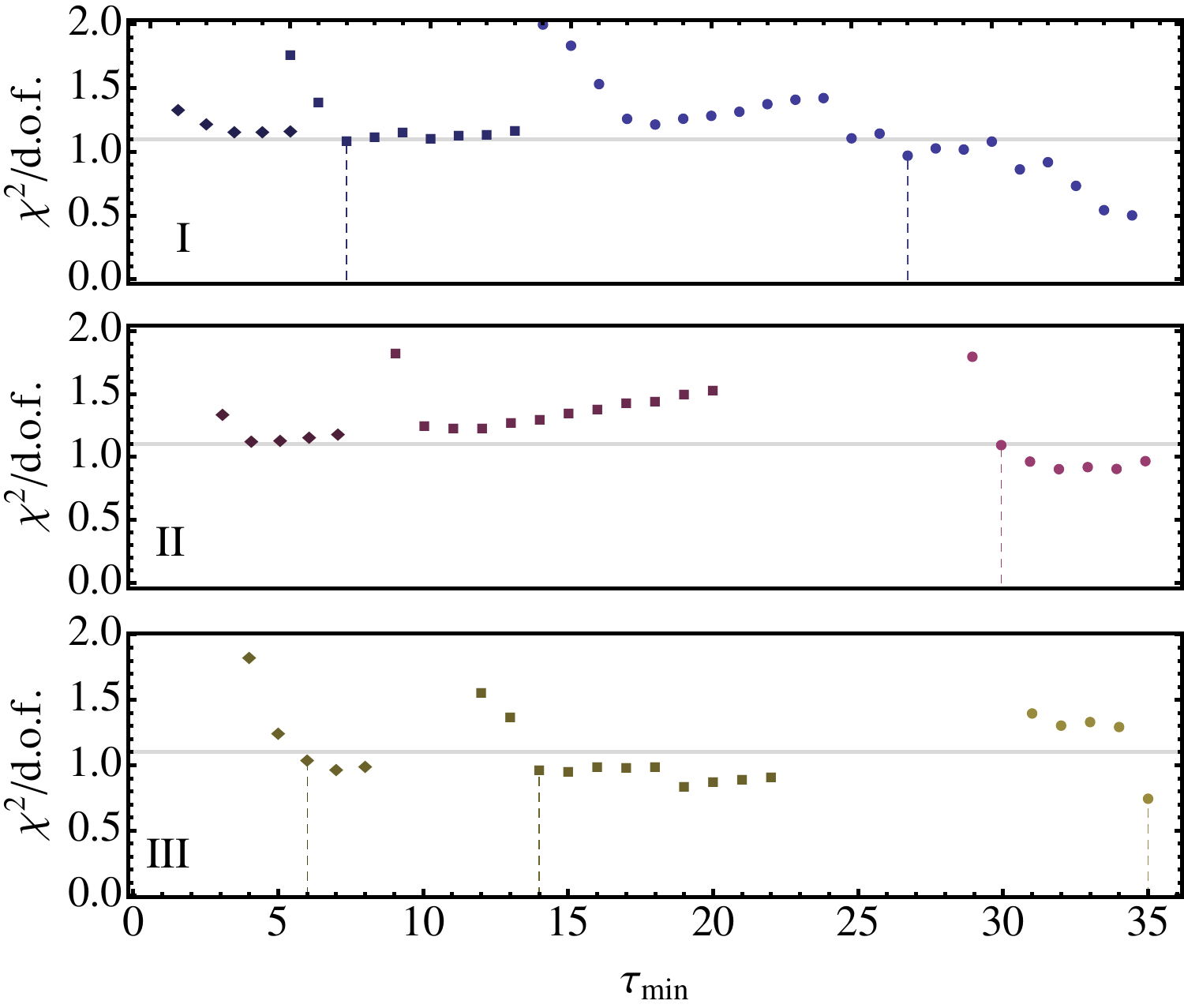}
\includegraphics[width=\figwidth]{\figdir 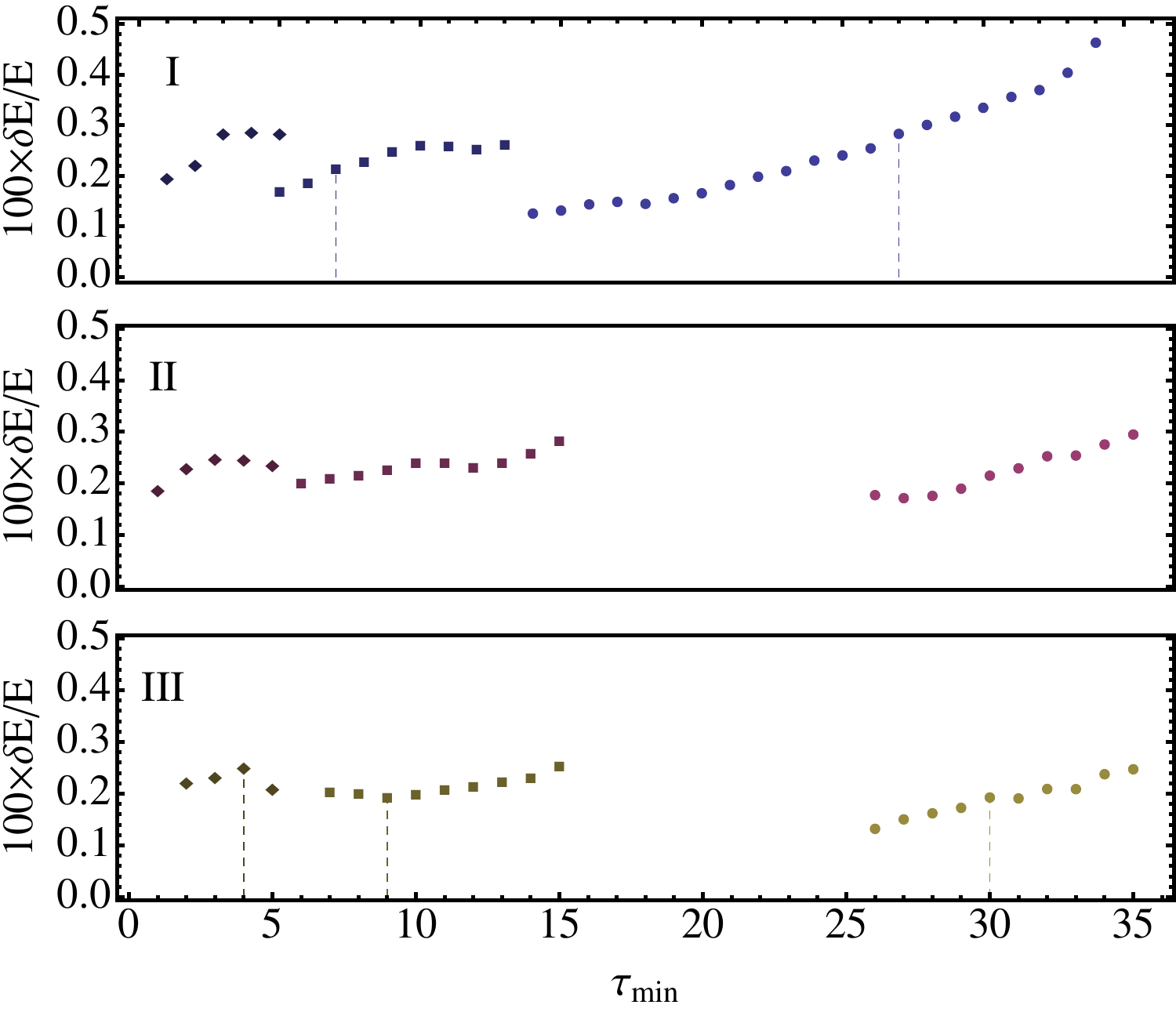}
\includegraphics[width=\figwidth]{\figdir 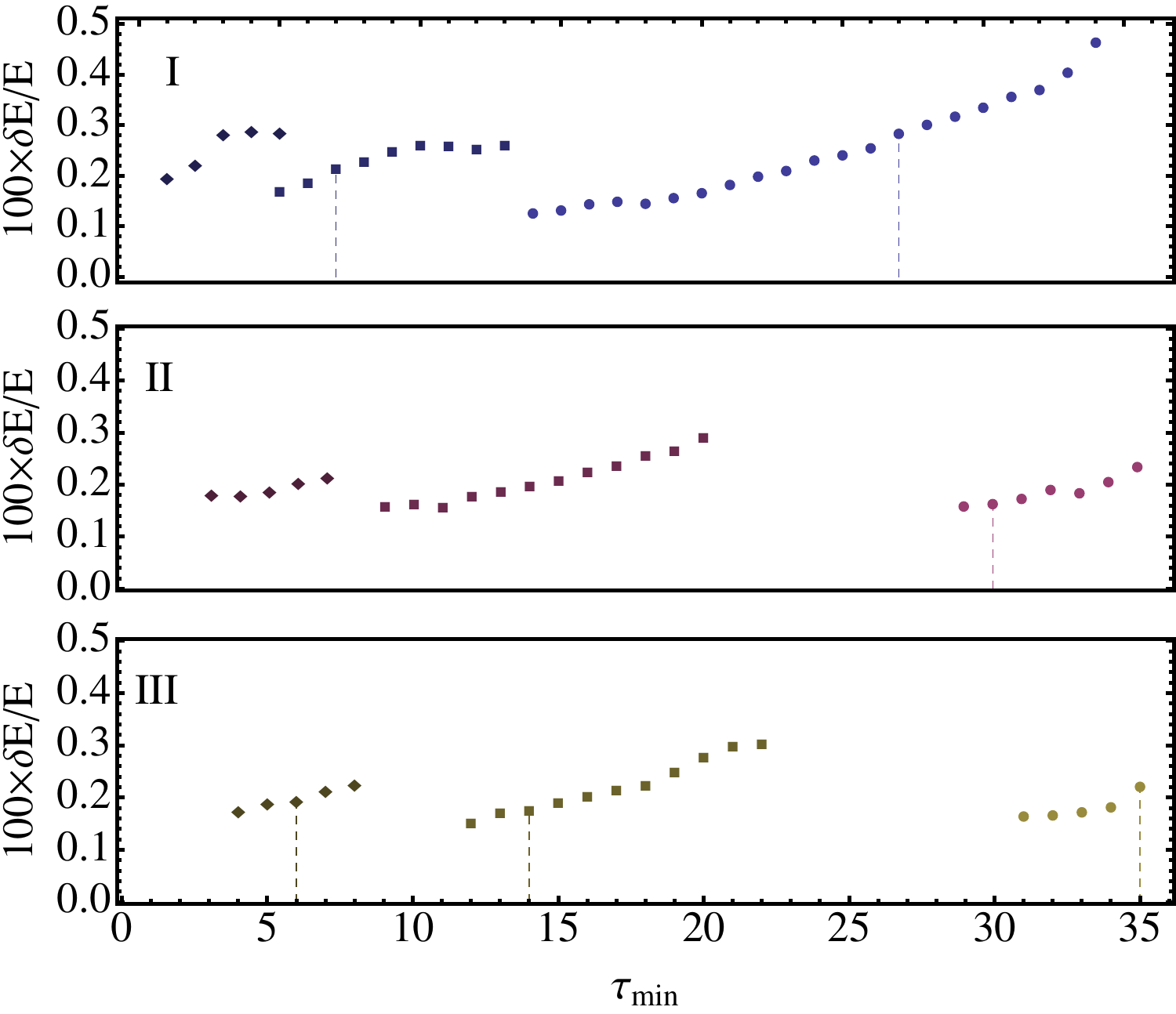}
\caption{\label{fig:rho_InterpNA_results}%
Same as \Fig{pion_Interp1_optPlots}, for the rho ground state.
Fit results associated with the earliest $\tau_\textrm{min}$ satisfying $\chi^2/\textrm{d.o.f.} < 1.1$ are indicated with vertical dashed lines, and provided in \Tab{rho_fit_results}.
}
\end{figure}

\begin{figure}
\includegraphics[width=\figwidth]{\figdir 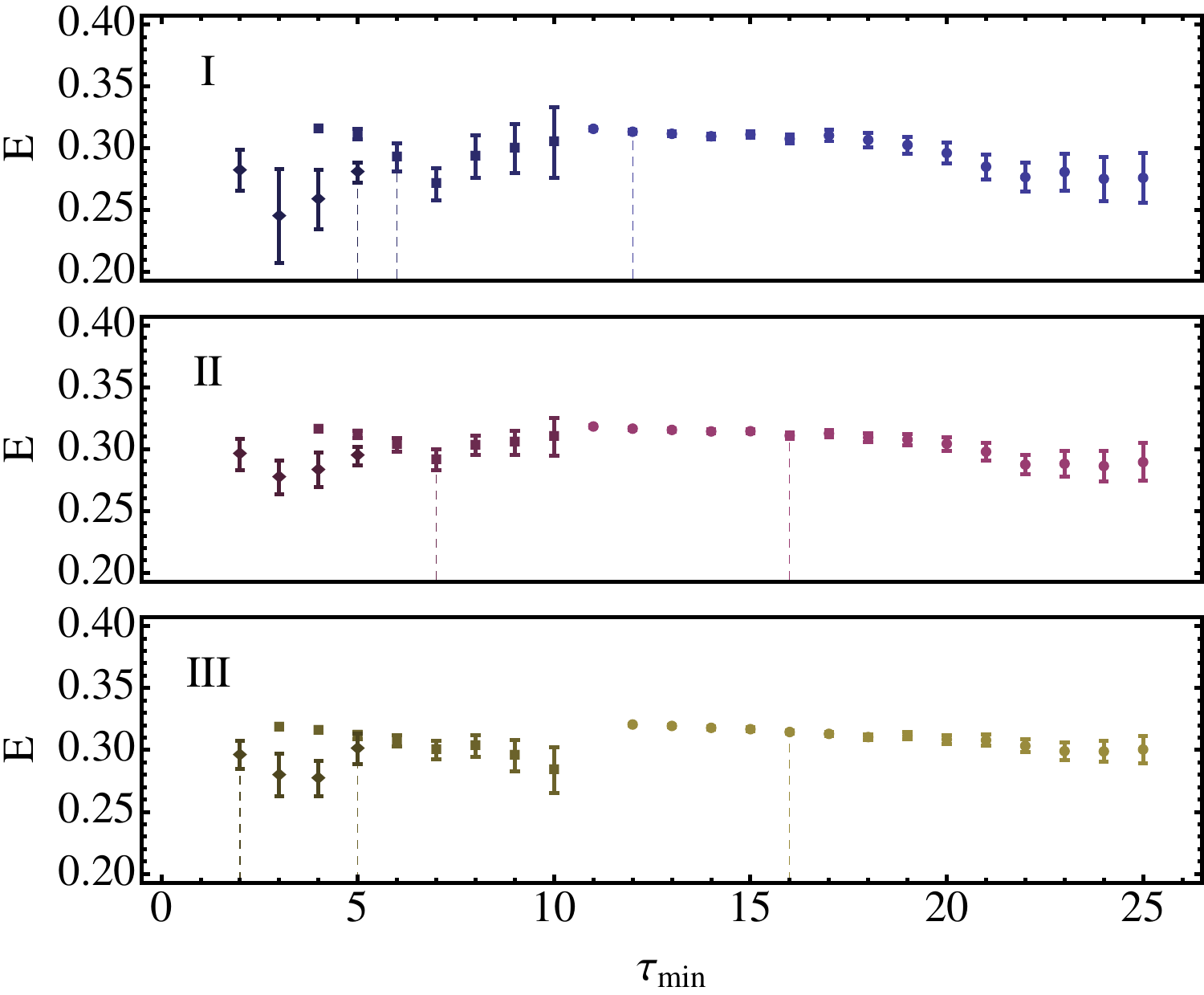}
\includegraphics[width=\figwidth]{\figdir 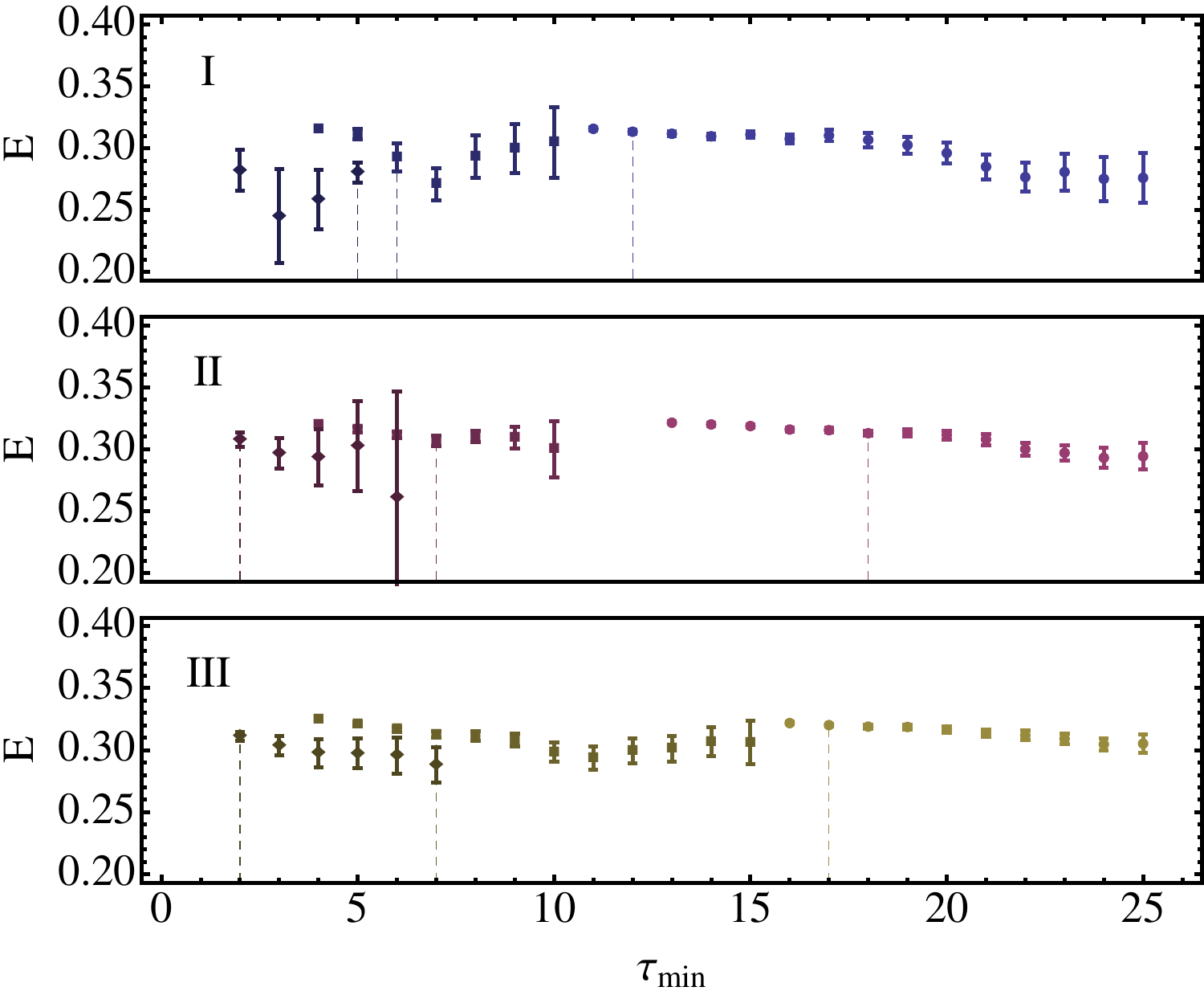}
\includegraphics[width=\figwidth]{\figdir 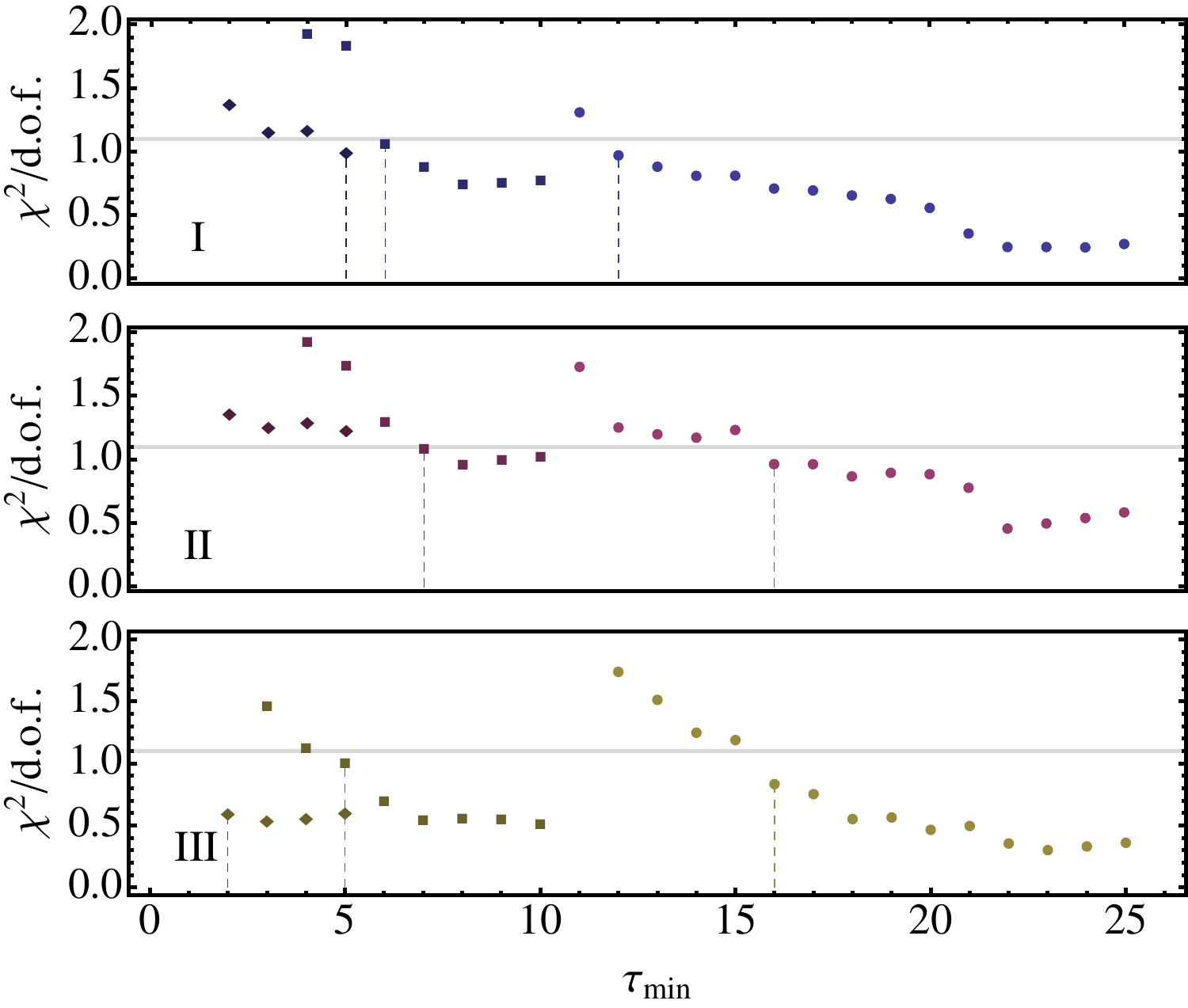}
\includegraphics[width=\figwidth]{\figdir 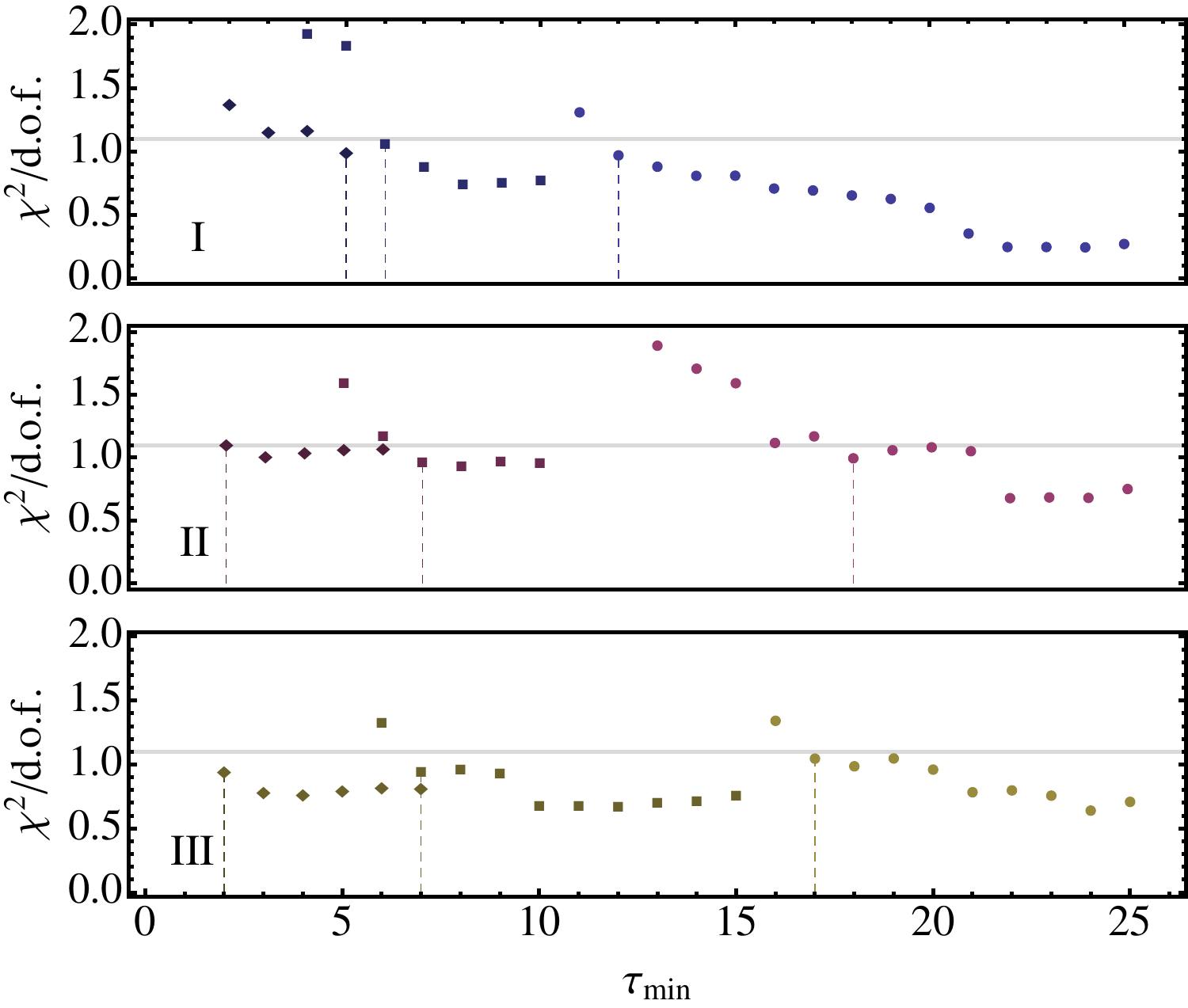}
\includegraphics[width=\figwidth]{\figdir 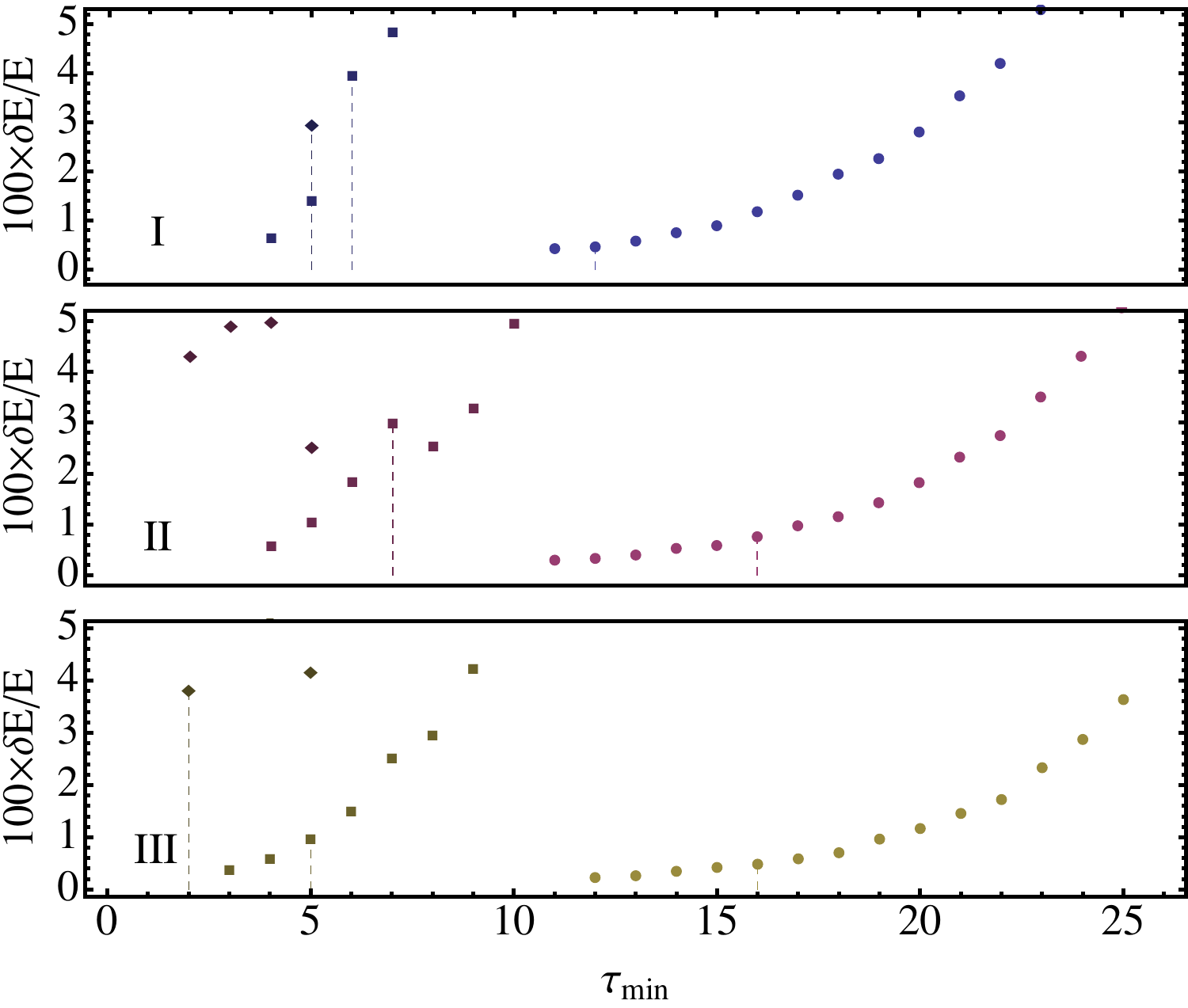}
\includegraphics[width=\figwidth]{\figdir 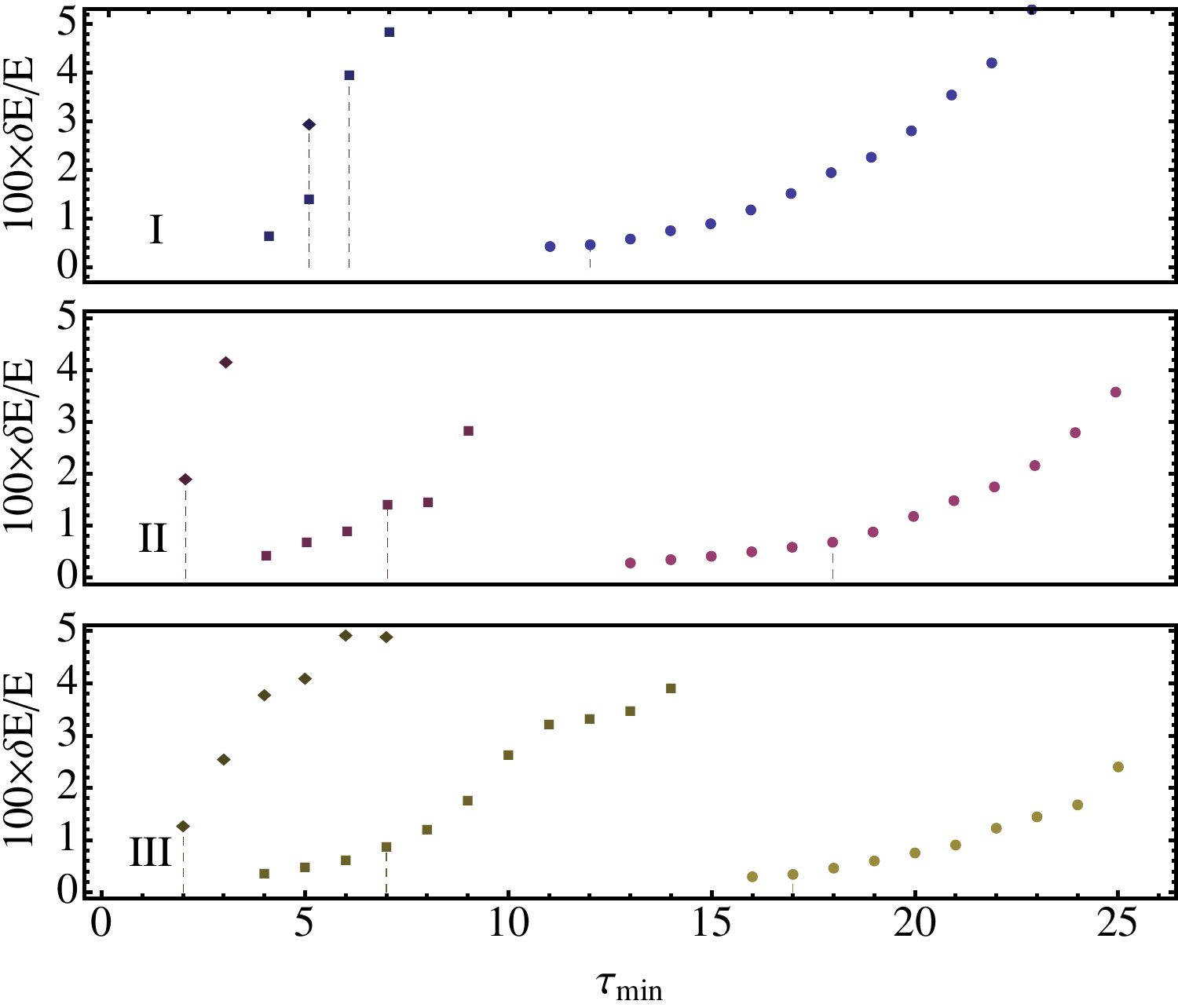}
\caption{\label{fig:rho2_InterpNA_results}%
Same as \Fig{pion_Interp1_optPlots}, for the first rho excited state.
Fit results associated with the earliest $\tau_\textrm{min}$ satisfying $\chi^2/\textrm{d.o.f.} < 1.1$ are indicated with vertical dashed lines, and provided in \Tab{rho2_fit_results}.
}
\end{figure}

For each ground state correlation function, we perform fully correlated, multiexponential fits to the data, scanning in $\tau_{\textrm{min}}$ for fixed $\tau_{\textrm{max}} = 40$ and $45$; for the first rho excited state, we consider $\tau_{\textrm{max}} = 30$ and $35$.
Fits were performed using single-, double- and triple-exponential model fit functions.
Results obtained for the extracted energies in each fit are shown in \Fig{pion_Interp1_results} - \Fig{rho2_InterpNA_results} (top), along with previously determined estimates of the extracted energies obtained from the significantly higher statistics calculations of \cite{Beane:2009kya} ($1\sigma$ uncertainties are indicated by a horizontal band), where available.
As in \Fig{pion_Interp1_optPlots} - \Fig{rho2_InterpNA_optPlots}, results are shown for correlators obtained at points (I), (II) and (III) along the paths of steepest ascent for fixed (left) and unconstrained (right) sources.
Also shown in \Fig{pion_Interp1_results} - \Fig{rho2_InterpNA_results} (center) is the corresponding goodness of fit, as quantified by the $\chi^2$ per degree of freedom (d.o.f).
Variation in the fit results with $\tau_{\textrm{max}}$ is found to be negligible, and therefore all results are displayed for $\tau_{\textrm{max}} = 45$ (ground states) and $\tau_{\textrm{max}} = 35$ (excited rho state).
The specific values of $\tau_{\textrm{min}}$ used for the fits may be inferred from these plots.
Only fit results yielding a $\chi^2/\textrm{d.o.f.} < 2$ and a statistically-meaningful, stable extraction of the energy are displayed.

In \Fig{pion_Interp1_results} - \Fig{rho2_InterpNA_results} (bottom), we show the percent deviation associated with the $1\sigma$ statistical uncertainties, $\delta E$, on the extracted energies, $E$, obtained from a bootstrap analysis of each fit.
We may use this quantity to directly compare the noise reduction associated with the fit results of each correlation function, given a choice of $\tau_{\textrm{min}}$. 
For a fixed $\tau_{\textrm{min}}$ chosen to yield good fit results (i.e., $\tau_{\textrm{min}}$ values for which $\chi^2/\textrm{d.o.f.} \lesssim 1$ when possible, and for which the fit result is robust against small variations about $\tau_{\textrm{min}}$) for all three correlators, (I), (II), and (III), along the path of steepest ascent, we find that statistical errors on the extracted energies decrease by an amount which is commensurate with the enhancement of the correlator signal/noise.

It is often possible to achieve good fits at earlier values of $\tau_{\textrm{min}}$ for the source-optimized correlators (I) compared to the intermediate correlator (II) and signal/noise-optimized correlator (III).
To make a fair comparison of the uncertainties in energies extracted from correlators at different $\tau_\textrm{min}$, we start by defining a threshold $\chi^2/\textrm{d.o.f.}$, below which, the fit is deemed acceptable.
Note that the precise value is not relevant, so long as it is consistent among all the correlators, and is small enough to reasonably ignore possible systematic effects (particularly those attributed to excited state contamination at early $\tau_\textrm{min}$).
For this study, we choose the threshold value to be $1.1$.
For each correlator, we define the best acceptable fit result to be one which has a $\chi^2/\textrm{d.o.f.}$ less than the threshold value, and the earliest possible $\tau_{\textrm{min}}$.
This typically corresponds to the acceptable fit with the smallest uncertainty.
The selected fits are indicated in \Fig{pion_Interp1_results} - \Fig{rho2_InterpNA_results} (vertical dashed lines), and the numerical results are also provided in \Tab{pion_fit_results} - \Tab{rho2_fit_results}.
Note that for some correlators, the $\chi^2/\textrm{d.o.f.}$ obtained from the fits exceed the chosen threshold value for all $\tau_\textrm{min}$, and therefore fit results for those correlators are excluded from the analysis.
In the tables, we use bold typeface to identify the best-fit energy result yielding the smallest statistical uncertainties for each hadron type.
In all cases, we find that the energies have smaller statistical uncertainties when extracted from correlators obtained using some form of signal/noise optimization (either fixed or unconstrained).
Although we find only a modest improvement in uncertainties for energies extracted for the pion\footnote{The majority of pion correlators produce an unacceptable fit, although the ground state energies are in good agreement with each other, and with the high precision determination of \cite{Beane:2009kya}. The improvement in the uncertainties of the signal/noise-optimized correlators is marginal compared to the source-optimized correlator in all cases.}, proton, rho ground state, and first rho excited state using signal/noise optimization, the delta exhibits a threefold enhancement in signal/noise compared to that obtained from source-optimized correlators.

\begin{figure} 
\includegraphics[width=\figwidth]{\figdir 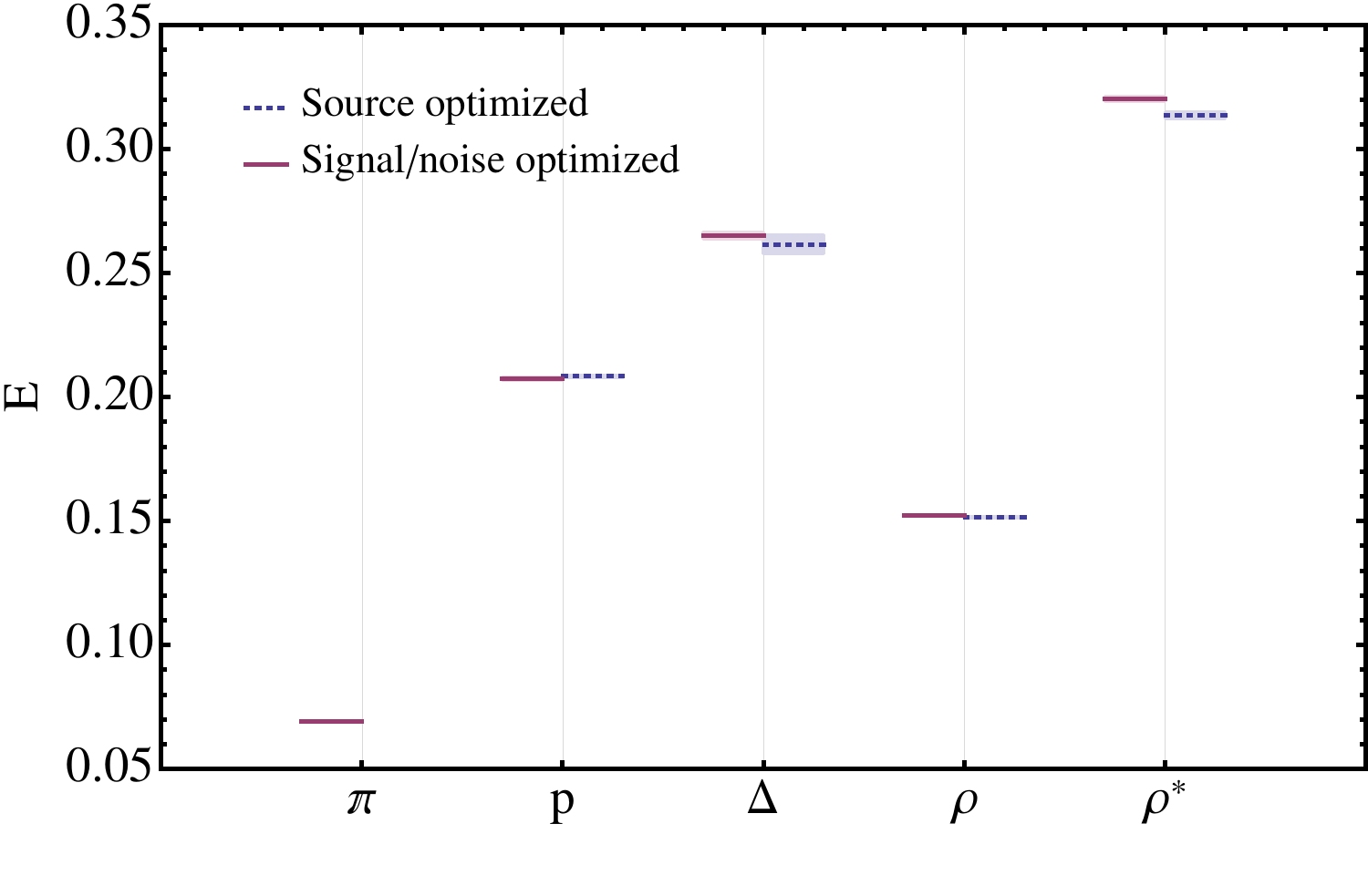}
\includegraphics[width=\figwidth]{\figdir 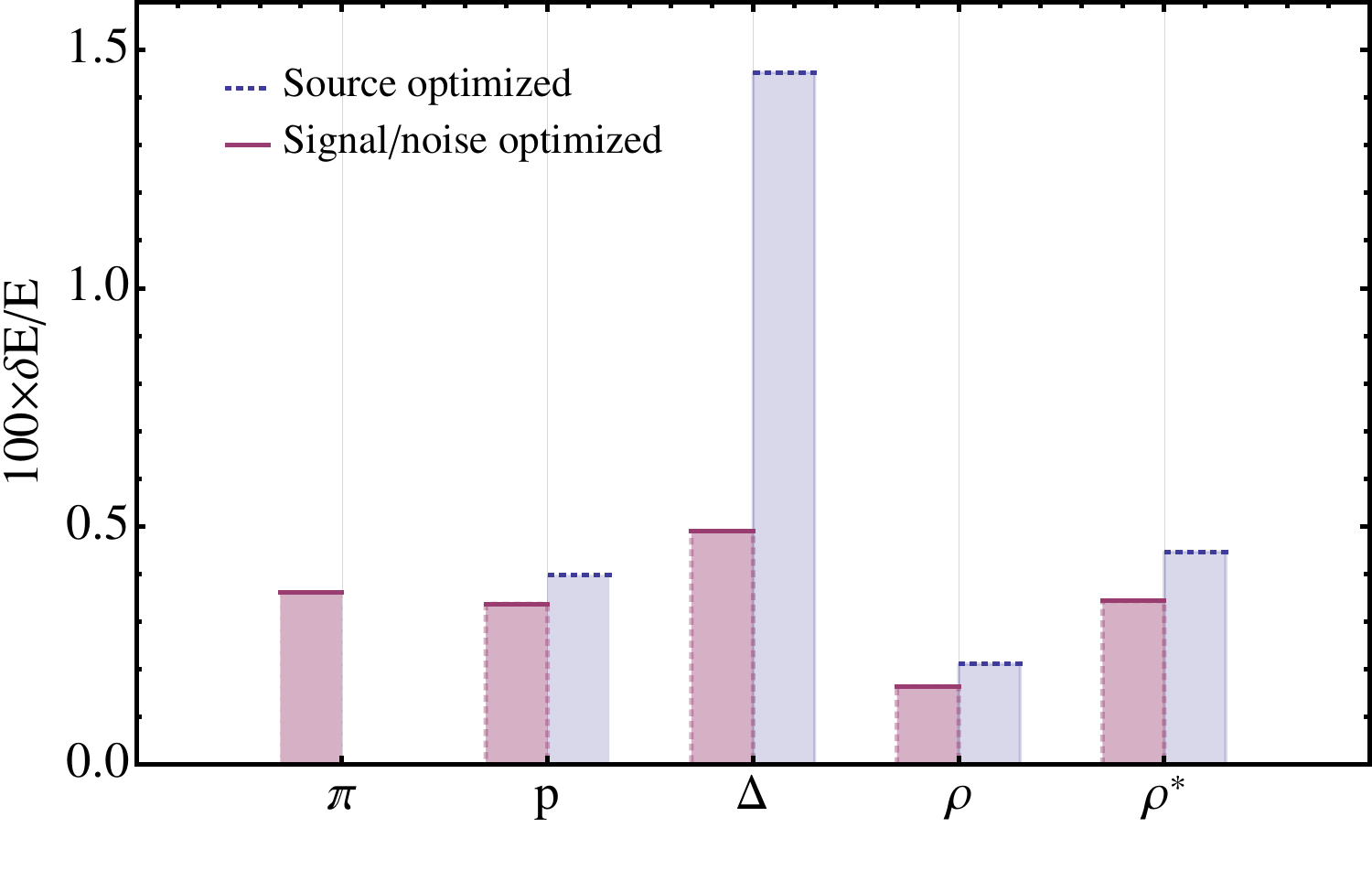}
\caption{\label{fig:SummaryPlots}%
Left: Extracted energies for the pion ($\pi$), proton ($p$), delta ($\Delta$), rho ground state ($\rho$), and first rho excited state ($\rho^*$).
Energies are determined using the best fit result among correlators of type I (source-optimized) and best fit result among correlators of type II and III (signal/noise-optimized) obtained from \Tab{pion_fit_results}-\Tab{rho2_fit_results}; the latter are indicated in bold typeface in the tables.
Right: Corresponding relative errors for each of the selected fit results.
}
\end{figure}

We summarize the best fit results for energies extracted from source-optimized correlators and signal/noise optimized correlators in \Fig{SummaryPlots} (left).
The former is defined as the fit result obtained for the energy with smallest uncertainties, among correlators of type (I) listed in \Tab{pion_fit_results} - \Tab{rho2_fit_results}.
The latter correspond to energies in \Tab{pion_fit_results} - \Tab{rho2_fit_results} listed in bold typeface, and are either of type (II) or (III).
With exception to the first rho excited state energy, all fit results in \Fig{SummaryPlots} (left) are consistent within statistical uncertainties, and systematic uncertainties from reasonable variations of the fit range are negligible.
The first rho excited state energy has significant systematic uncertainties attributed to a drift in the energy as a function of $\tau_\textrm{min}$ (see \Fig{rho2_InterpNA_results}), but these uncertainties are comparable in magnitude for both source- and signal/noise-optimized correlators, and therefore we omit them from the comparison.
In \Fig{SummaryPlots} (right) we show a summary plot of the corresponding relative uncertainties obtained for each of the extracted energies.

Let us conclude this section by noting that a more significant reduction in noise for the extracted energies may be possible by performing a combined fit of the signal/noise-enhanced correlator(s) in the late time regime, and source-optimized correlators in the early time regime.
It is presently unclear whether the marginal improvements in signal/noise for some of these correlators is a result of the nature of the system itself, or whether it is a reflection of the specifics of the correlators currently at our disposal.

\begin{table}
\caption{%
\label{tab:pion_fit_results}%
Multiexponential (i.e., $n_\textrm{exp}=1,2$ and $3$ exponential) least-squares fit results for the pion ground state energy, $E$, and associated statistical uncertainties, $\delta E$, $\chi^2$/d.o.f, and fit quality, $Q$.
Results are provided for both fixed (left) and unconstrained (right) sources, corresponding to correlators of type (I), (II) and (III) displayed in \Fig{pion_Interp1_optPlots}.
The fit result yielding the smallest statistical uncertainty is indicated in bold typeface (in cases of a tie, the fit with the lowest $\chi^2$/d.o.f was selected).
Fit results were omitted in cases where the $\chi^2$/d.o.f. exceeded a threshold value of $1.1$ for all $\tau_\textrm{min}$.
Tabulated fit results are indicated in \Fig{pion_Interp1_results}.
}
\begin{ruledtabular}
\begin{tabular}{cc|cccc|cccc}
label & $n_\textrm{exp}$ & $\tau$ range  & $E$($\delta E$) & $\chi^2$/d.o.f  & Q & $\tau$ range  & $E$($\delta E$) & $\chi^2$/d.o.f  & Q \\
\hline
 I   & 1 & -     & -            & -    & -    & -     & -            & -    & -    \\
 I   & 2 & -     & -            & -    & -    & -     & -            & -    & -    \\
 I   & 3 & -     & -            & -    & -    & -     & -            & -    & -    \\
\hline                                                                           
 II  & 1 & -     & -            & -    & -    & -     & -            & -    & -    \\
 II  & 2 & -     & -            & -    & -    & -     & -            & -    & -    \\
 II  & 3 & -     & -            & -    & -    & -     & -            & -    & -    \\
\hline                                                               
 III & 1 & -     & -            & -    & -    & 21-45 & 0.06938(25)  & 1.05 & 0.40 \\
 III & 2 & -     & -            & -    & -    &  7-45 & 0.06930(25)  & 1.02 & 0.44 \\
 III & 3 & -     & -            & -    & -    &\bf  1-45 &\bf 0.06928(25)  &\bf 0.90 &\bf 0.64 \\
\end{tabular}
\end{ruledtabular}
\end{table}

\begin{table}
\caption{%
\label{tab:proton_fit_results}%
Same as \Tab{pion_fit_results}, for the proton ground state.
Tabulated fit results are indicated in \Fig{proton_Interp4_results}.
}
\begin{ruledtabular}
\begin{tabular}{cc|cccc|cccc}
label & $n_\textrm{exp}$ & $\tau$ range  & $E$($\delta E$) & $\chi^2$/d.o.f  & Q & $\tau$ range  & $E$($\delta E$) & $\chi^2$/d.o.f  & Q \\
\hline
 I   & 1 & 19-45 & 0.20850(83)  & 1.09 & 0.34 & 19-45 & 0.20850(83)  & 1.09 & 0.34 \\
 I   & 2 & 12-45 & 0.20770(112) & 1.06 & 0.38 & 12-45 & 0.20770(112) & 1.06 & 0.38 \\
 I   & 3 & -     & -            & -    & -    & -     & -            & -    & -    \\
\hline                                                                              
 II  & 1 & -     & -            & -    & -    & -     & -            & -    & -    \\
 II  & 2 & -     & -            & -    & -    & -     & -            & -    & -    \\
 II  & 3 & -     & -            & -    & -    & -     & -            & -    & -    \\
\hline                                                                              
 III & 1 &\bf 27-45 &\bf 0.20747(70)  &\bf 0.90 &\bf 0.57 & -     & -            & -    & -    \\
 III & 2 &  8-45 & 0.20591(89)  & 1.04 & 0.41 & -     & -            & -    & -    \\
 III & 3 &  1-45 & 0.20581(95)  & 0.92 & 0.62 & -     & -            & -    & -    \\
\end{tabular}
\end{ruledtabular}
\end{table}

\begin{table}
\caption{%
\label{tab:delta_fit_results}%
Same as \Tab{pion_fit_results}, for the delta ground state.
Tabulated fit results are indicated in \Fig{delta_Interp1_results}.
}
\begin{ruledtabular}
\begin{tabular}{cc|cccc|cccc}
label & $n_\textrm{exp}$ & $\tau$ range  & $E$($\delta E$) & $\chi^2$/d.o.f  & Q & $\tau$ range  & $E$($\delta E$) & $\chi^2$/d.o.f  & Q \\
\hline
 I   & 1 & 25-45 & 0.2616(38) & 0.95 & 0.52 & 25-45 & 0.2616(38) & 0.95 & 0.52 \\
 I   & 2 & -     & -            & -    & -    & -     & -            & -    & -    \\
 I   & 3 & -     & -            & -    & -    & -     & -            & -    & -    \\
\hline
 II  & 1 &\bf 20-45 &\bf 0.2652(13) &\bf 0.95 &\bf 0.53 & 21-45 & 0.2679(14) & 1.03 & 0.42 \\
 II  & 2 & 15-45 & 0.2617(25) & 0.94 & 0.55 & 15-45 & 0.2637(26) & 1.07 & 0.37 \\
 II  & 3 & -     & -            & -    & -    & -     & -            & -    & -    \\
\hline                                        
 III & 1 & -     & -            & -    & -    & 26-45 & 0.2658(15) & 0.85 & 0.65 \\
 III & 2 & -     & -            & -    & -    & 12-45 & 0.2630(15) & 0.91 & 0.60 \\
 III & 3 & -     & -            & -    & -    &  5-45 & 0.2633(18) & 1.03 & 0.41 \\
\end{tabular}
\end{ruledtabular}
\end{table}

\begin{table}
\caption{%
\label{tab:rho_fit_results}%
Same as \Tab{pion_fit_results}, for the rho ground state.
Tabulated fit results are indicated in \Fig{rho_InterpNA_results}.
}
\begin{ruledtabular}
\begin{tabular}{cc|cccc|cccc}
label & $n_\textrm{exp}$ & $\tau$ range  & $E$($\delta E$) & $\chi^2$/d.o.f  & Q & $\tau$ range  & $E$($\delta E$) & $\chi^2$/d.o.f  & Q \\
\hline
 I   & 1 & 27-45 & 0.15120(43) & 0.97 & 0.49 & 27-45 & 0.15120(43) & 0.97 & 0.49 \\
 I   & 2 &  7-45 & 0.15144(32) & 1.08 & 0.34 &  7-45 & 0.15144(32) & 1.08 & 0.34 \\
 I   & 3 & -     & -           & -    & -    & -     & -           & -    & -    \\
\hline                                       
 II  & 1 & -     & -           & -    & -    &\bf 30-45 &\bf 0.15233(25) &\bf 1.09 &\bf 0.36 \\
 II  & 2 & -     & -           & -    & -    & -     & -           & -    & -    \\
 II  & 3 & -     & -           & -    & -    & -     & -           & -    & -    \\
\hline                                                                           
 III & 1 & 30-45 & 0.15187(29) & 0.87 & 0.60 & 35-45 & 0.15268(34) & 0.74 & 0.67 \\
 III & 2 &  9-45 & 0.15078(29) & 1.07 & 0.37 & 14-45 & 0.15190(26) & 0.96 & 0.53 \\
 III & 3 &  4-45 & 0.15113(37) & 1.03 & 0.41 &  6-45 & 0.15165(29) & 1.03 & 0.42 \\
\end{tabular}
\end{ruledtabular}
\end{table}

\begin{table}
\caption{%
\label{tab:rho2_fit_results}%
Same as \Tab{pion_fit_results}, for the first rho excited state.
Tabulated fit results are indicated in \Fig{rho2_InterpNA_results}.
}
\begin{ruledtabular}
\begin{tabular}{cc|cccc|cccc}
label & $n_\textrm{exp}$ & $\tau$ range  & $E$($\delta E$) & $\chi^2$/d.o.f  & Q & $\tau$ range  & $E$($\delta E$) & $\chi^2$/d.o.f  & Q \\
\hline
 I   & 1 & 12-35 & 0.3135(14)  & 0.97 & 0.50 & 12-35 & 0.3135(14)  & 0.97 & 0.50 \\
 I   & 2 &  6-35 & 0.2928(115) & 1.05 & 0.39 &  6-35 & 0.2928(115) & 1.05 & 0.39 \\
 I   & 3 &  5-35 & 0.2802(82)  & 0.98 & 0.49 &  5-35 & 0.2802(82)  & 0.98 & 0.49 \\
\hline                         
 II  & 1 & 16-35 & 0.3110(24)  & 0.96 & 0.50 & 18-35 & 0.3129(21)  & 0.99 & 0.46 \\
 II  & 2 &  7-35 & 0.2914(87)  & 1.07 & 0.36 &  7-35 & 0.3066(43)  & 0.96 & 0.53 \\
 II  & 3 & -     & -           & -    & -    &  2-35 & 0.3077(58)  & 1.09 & 0.34 \\
\hline                         
 III & 1 & 16-35 & 0.3145(15)  & 0.83 & 0.66 &\bf 17-35 &\bf 0.3202(11)  &\bf 1.05 &\bf 0.40 \\
 III & 2 &  5-35 & 0.3116(29)  & 1.00 & 0.47 &  7-35 & 0.3128(27)  & 0.94 & 0.55 \\
 III & 3 &  2-35 & 0.2959(112) & 0.58 & 0.96 &  2-35 & 0.3111(39)  & 0.93 & 0.57 \\
\end{tabular}
\end{ruledtabular}
\end{table}

\section{Discussion and conclusion}
\label{sec:conclusion}

In this paper, we have developed a set of strategies for increasing the signal/noise in stochastic estimates of correlation functions that, counterintuitively, involve a tuning of source and sink interpolating operators away from those that have maximal overlap with the eigenstates of the system.
These strategies are implemented primarily via linear algebra and are computationally inexpensive to apply.
We have demonstrated in a two-state toy model that, with such tuning, it is possible to achieve dramatic enhancements in signal/noise compared to that achieved using source-optimized sources and sinks.
Furthermore, such enhancements can occur without introducing significant excited state contamination in correlators at early times, either as a result of fortune, or by imposing appropriate constraints on the source and/or sink vectors.

The methods we advocate are applicable to both ground and excited states.
We provide explicit formulas for the signal/noise-optimized source and sink vectors expressed in terms of a given correlator matrix and its associated noise correlator.
The results allow for arbitrarily imposed constraints on either the source vector, sink vector, or both.
We furthermore describe possible ways of combining source-optimization and signal/noise-optimization strategies, which may be of use for extracting better estimates of energies from current correlator data. 
We apply some of the proposed methods to examples of QCD data, specifically focusing on single hadron correlation functions.
Although the signal/noise enhancement in the correlators themselves show promise, ranging from enhancement factors of approximately $1.2$ for pions and 2 to 3 for the proton, delta and rho, the associated enhancements found in the energies extracted from multiexponential fits of the correlators are, in most cases, less impressive (one notable exception is the delta ground state energy, which exhibited a threefold enhancement over that extracted from a source-optimized correlator).
The loss in enhancement for energies is traced to the introduction of excited state contamination as early times as a result of a finite operator basis and/or contamination from all states due to statistical noise.
At present, it remains unclear whether the situation might be improved upon by considering a larger, or improved, basis of interpolating operators, or whether those QCD correlators fundamentally lack the qualities required for achieving the large enhancements that appear possible in the toy model.
Had the enhancements been more substantial in the correlator data, it is likely that more significant gains in signal/noise for extracted energies would follow.
Despite the varied results, we remain optimistic that the methods may be profitably applied to other systems, possibly including multihadron correlators, where a greater freedom in the kinds of interpolating operators used might be exploited.
An interesting avenue to investigate further is whether forming correlators with interpolating operators carrying unequal quantum numbers (for example, differing in momentum, spin, parity or any other quantum number that is not conserved for a fixed, stochastically generated background field configuration) might yield an additional signal/noise enhancement through the techniques we have discussed.
Although the correlations between operators of differing quantum numbers vanish in the limit of infinite statistics, nontrivial contributions to the noise correlator arise, which may, in turn, be profitably exploited. 

We have argued that the maximum achievable signal/noise enhancement in correlators is determined, in part, by the number and choice of interpolating operators used in their construction.
However, even in the limit that the basis of interpolating operators is complete, there remains a fundamental limit on the amount of enhancement that is possible, determined entirely by the lowest energy signal and noise states.
In the context of lattice simulations, an interesting direction to consider is whether it is possible to favorably manipulate the properties of the noise states by introducing finite volume artifacts via appropriately chosen boundary conditions, or discretization artifacts through the introduction of higher dimension operators (which would have no effect on the continuum limit of the theory).
In the former case, for example, the boundary conditions may be chosen so as to eliminate the lowest energy noise state, resulting in modest gains in signal/noise (see, e.g., the references \cite{Bedaque:2007pe,Bedaque:2008hn}).
In the latter case, the presence of discretization artifacts can lead to a reduction in the energy splittings between signal and noise ground states, thereby decreasing the decay rate for signal/noise degradation.
Such an approach has been used to significantly reduce signal/noise in the case of heavy quarks \cite{DellaMorte:2003mn,Detmold:2008ww}.
In these approaches, the finite volume and lattice spacing effects which are present in measured quantities may subsequently be removed by conventional infinite volume and continuum limit extrapolations.
It would be interesting to explore whether finite volume or lattice artifacts can be exploited, in the same spirit, in order to favorably alter the character of the overlap factors appearing in the signal/noise ratio, ultimately allowing for a greater enhancements than what is inherently allowed by the physical (i.e., continuum limit and infinite volume) signal and noise states.

It is interesting to consider the origin of noise in correlation functions, here specifically focusing on QCD.
Given the ultimate goal of finding a correlation function in which signal/noise remains constant in time, so that a significant increase in statistical precision is feasible with a polynomial increase in computational effort, it is important to ascertain whether this is practical, or indeed possible.
To achieve such a correlator, we would require that the associated noise correlator have vanishing, or exponentially small, overlap onto the eigenstates of the noise system with energies below twice the energy of the state that we are attempting to extract.
That is, we would need to project against a subspace in the noise correlator spanned by such states.
Note that such a projection, if it were possible, would coincide with signal/noise optimization in the limit of exponential enhancement at late times.
This connection was explicitly made for the toy model in \Sec{toy_model}, were we demonstrated how a large signal/noise enhancement for the ground state is, in part, limited by the subleading contributions to the noise correlator; orthogonalizing the sources against these subleading noise states results in the removal of such a cutoff on the enhancement.

Let us explore the feasibility of this goal further, by focusing on the case of $A$-nucleon correlators with a ground state of energy, $E_{AN}$.
In this case, the most relevant degrees of freedom contributing to the noise are the pions, as they are the lightest states.
At a fixed volume, $V=L^3$, and quark mass, $m_q$, there are a finite number of noise-creating pion states, with gaps between the states controlled (up to the effects of hadronic interactions) by $m_\pi$ and $L$, and energies $\tilde{E}_n \approx \sqrt{E_{3A\pi}^2 +(2\pi/L)^2 n}$ up to some $\tilde{E}_{n_\textrm{max}} \alt 2 E_{AN}$, where $E_{3A\pi}$ is the ground state energy for $3A$ pions at rest.
The multiplicity of these states is given by the number of ways of partitioning the integer $n$ into $3A$ sums of three squares.
One may explicitly evaluate the total number of such states with vanishing total momentum between the energies $E_{3A\pi}$ and $2E_{AN}$, and find that it scales asymptotically as $\sim( L \sqrt{\delta E^2})^{3(3A-1)}$ at large volume, where $\delta E^2 = 4 E_{AN}^2-E_{3A\pi}^2$.
Note that this estimate neglects isospin-dependent prefactors which can only further increase the estimate by a volume-independent multiplicative amount.
As the volume becomes large, or as $m_q/\Lambda_{QCD}$ decreases (thereby making pions relatively lighter than typical hadrons), the number of such states dramatically increases with the aforementioned scaling (for a purely fermionic system, this scaling is cut off by the total number of states that can be represented on the discrete set of points), and consequently, the task of constructing a source and sink interpolating operator that is orthogonal to these noise states becomes increasingly difficult.
Indeed, it is clear that the problem scales exponentially in the atomic number $A$, explicitly showing the close connection to the sign problem that hampers calculations at nonzero baryon chemical potential \cite{PhysRevLett.94.170201}. 

In \cite{Beane:2009gs}, it was argued that by using sinks that individually project baryons in a multiple baryon correlation function to zero momentum, the overlap of the noise correlator onto the state with three pions at rest is suppressed by the volume (the quarks in the individual correlator, $\calC$, and the antiquarks in $\calC^*$ are near each other only $1/V$ of the time relative to forming baryon and antibaryon states).
This seems sensible, but as the volume becomes large, the exponential proliferation of states that contribute to noise will overwhelm the polynomial suppression of the overlap onto individual states, making extraction of signals increasingly problematic.
Even if the space of states that we want to avoid was of manageable dimension, a critical problem is that one does not directly control the source and sink interpolators for the noise correlator, and as discussed at length in the current work, they arise as the outer products of the interpolators in the original correlator.
Unless the eigenstates of the noise correlator are also outer product states, and there is no reason that they should be, then it is in principle not possible to fully orthogonalize against them.
The best one can do along these lines is reduce the overlap, and explore the interplay between the resulting signal/noise enhancement and contamination from undesired states, as advocated in this study.\footnote{It is perhaps interesting to note that in many low dimensional quantum systems, the ground states are reasonably represented as matrix product states of low bond dimension, so it is possible that the noisiest eigenstates are not far from outer products.}

\begin{acknowledgments}

We would like to gratefully acknowledge S. Meinel for the use of his QMBF and XMBF fitting software.
We would also like to acknowledge J. Dudek for generously sharing rho correlator data from the Hadron Spectrum Collaboration, and K. Orginos and A. Walker-Loud for generously sharing pion, proton and delta correlator data.
The latter data were computed with Teragrid resources and local resources at the College of William and Mary.
This study was supported by the U. S. Department of Energy under cooperative research agreement Contract No. DE-FG02-94ER40818, the U. S. Department of Energy Early Career Research Award No. DE-SC0010495, and the Solomon Buchsbaum Fund at MIT.

\end{acknowledgments}

\bibliography{noise}

\appendix

\section{Source optimization}
\label{appendix:source_optimization}

A standard approach for analysis of correlators is the variational method \cite{Michael:1982gb,Michael:1985ne,Luscher:1990ck,Blossier:2009kd}, a linear algebra-based approach for finding linear combinations of interpolating fields having maximal overlap with the low-lying states of the system.
The method can be framed as an optimization problem similar to that for signal/noise, and we refer to this as ``source optimization''.
Let us specialize to the case where $N^\prime=N$, and $C = C^\dagger$.
Given equal source and sink vectors, $\psi^\prime=\psi$, the correlation function is positive-definite and a monotonically decreasing function of time separation.
We may therefore find source and sink vectors having maximal overlap with the eigenstates of the system by extremizing the function
\begin{eqnarray}
\rho(\psi,\xi) = - \log\left( \frac{\psi^\dagger C \psi}{\psi^\dagger C_0 \psi} \right) - \xi \left( \psi^\dagger \psi -1 \right)
\end{eqnarray}
with respect to $\psi$ and $\xi$ at every time slice, given $C_0 = C(\tau_0)$ evaluated at a reference time slice, $\tau_0$.
There are $N$ such critical points corresponding to the $N$ lowest energy states of the system.
Explicitly, the critical points at each time slice, $\tau$, are given by solutions to the generalized eigenvalue problem:
\begin{eqnarray}
C(\tau) \psi_n(\tau) = \lambda_n(\tau) C(\tau_0) \psi_n(\tau) \ ,\qquad n = 0,\cdots,N-1\ .
\label{eq:generalized_ev}
\end{eqnarray}
One can prove that the eigenvalues obtained in this approach satisfy
\begin{eqnarray}
\lambda_n(\tau) = e^{-E_n (\tau - \tau_0)}\ ,
\end{eqnarray}
up to relative corrections of order $e^{- (E_N - E_n) \tau}$ within the time interval $\tau_0 < \tau < 2 \tau_0$, and for $\tau_0$ sufficiently large \cite{Blossier:2009kd}.
Furthermore, the correlator overlap factors and source vectors satisfy
\begin{eqnarray}
Z^\dagger_m \psi_n = \delta_{m,n}\ ,
\end{eqnarray}
up to exponentially suppressed corrections for $\tau$ within this regime.
In addition to the excited state contamination which arises from the use of a limited basis of interpolating fields, a secondary source of contamination coming from all states may be attributed to fluctuations in the estimate of the correlator matrix.
In \Sec{noisy_excisions}, we have outlined how one might reduce the overall uncertainties in variational calculations in cases when the statistical uncertainties dominate the systematic uncertainties.

\section{Derivation of \Eq{sn_basis_solns}}
\label{appendix:app1}

In this section, we derive \Eq{sn_basis_solns} from \Eq{sn_basis_eqns}.
We focus only on the solution for $\psi^\prime_\alpha$, since the solution for $\psi_\alpha$ follows from an identical analysis.
To begin, note that by left-multiplying both sides of the first equation in \Eq{sn_basis_eqns} by ${\psi^\prime_\alpha}^\dagger$ one finds that $\xi_{\alpha\alpha} = 0$.
Combining this observation with the fact that $\psi^\prime_0 \propto C\psi_0$, we may reexpress the equation as
\begin{eqnarray}
\frac{ \psi^\prime_0 }{  {\psi^\prime_\alpha}^\dagger \psi^\prime_0 } = \frac{\psi^\prime_\alpha }{ {\psi^\prime_\alpha}^\dagger \sigma^2_{\psi_0} \psi^\prime_\alpha } + \sum_{\beta=0}^{\alpha-1} \xi^\prime_{\alpha\beta} \sigma^{-2}_{\psi_0}  \psi^\prime_\beta \ .
\end{eqnarray}
From the structure of this formula, one can verify by induction that $\sigma^{-2}_{\psi_0} \psi^\prime_\alpha$ is a linear combination of vectors $\psi^\prime_\beta$ for $\beta < \alpha + 1$.
The most general expression for $\psi^\prime_\alpha$ must therefore be of the form
\begin{eqnarray}
\psi^\prime_\alpha = \sum_{\beta=0}^{\alpha-1} y_{\alpha\beta} \psi^\prime_\beta + y_\alpha \sigma^{-2}_{\psi_0}  \psi^\prime_{\alpha-1} 
\end{eqnarray}
for some undetermined coefficients $y_{\alpha\beta}$ and $y_\alpha$.
By exploiting the orthonormality of $\psi^\prime_\alpha$, one finds
\begin{eqnarray}
y_{\alpha\beta} + y_\alpha   {\psi^\prime_\beta}^\dagger  \sigma^{-2}_{\psi_0}\psi^\prime_{\alpha-1} = 0\ ,\qquad \beta<\alpha\ .
\end{eqnarray}
Plugging the result for $y_{\alpha\beta}$ back into the above expression, one arrives at
\begin{eqnarray}
\psi^\prime_\alpha = y_\alpha \left[ 1 - \sum_{\beta=0}^{\alpha-1} \psi^\prime_\beta {\psi^\prime_\beta}^\dagger  \right] \sigma^{-2}_{\psi_0}  \psi^\prime_{\alpha-1} \ .
\end{eqnarray}
The term in brackets is identified with $\calQ^\prime_{\alpha-1}$ and the overall coefficient $y_\alpha$ is identified with the normalization factor $A_\alpha(\psi_0)$.


\end{document}